\documentclass[reprint,amsmath,amssymb,
prb,floatfix
]{revtex4-2}
\usepackage{graphicx}
\usepackage[usenames,dvipsnames]{xcolor}
\usepackage{multirow}
\usepackage{comment}
\usepackage{dcolumn}
\usepackage{bm}
\usepackage{physics}
\usepackage{xcolor,colortbl}
\usepackage{tabularx,booktabs}
\usepackage{pifont}
\newcommand{\cmark}{\ding{51}}
\newcommand{\xmark}{\ding{55}}
\usepackage{makecell}
\usepackage[normalem]{ulem}
\newcommand{\orcid}[1]{\href{https://orcid.org/#1}{\includegraphics[width=8pt]{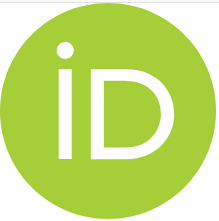}}}
\usepackage{nicematrix}
\usepackage[colorlinks,urlcolor=goodgreen,citecolor=blue,linkcolor=goodred]{hyperref}
\definecolor{orange}{rgb}{1,0.5,0}
\definecolor{goodgreen}{rgb}{0.1,0.5,0}
\definecolor{goodred}{rgb}{0.7,0,0}
\newcolumntype{Y}{>{\centering\arraybackslash}X}
\newcolumntype{a}{>{\columncolor{goodgreen!20}}Y}
\newcolumntype{f}{>{\columncolor{orange!20}}Y}
\newcolumntype{d}{>{\columncolor{purple!20}}Y}

\let\oldepsilon\epsilon \let\epsilon\varepsilon \let\varepsilon\oldepsilon

\renewcommand\vec{\boldsymbol}

\begin{document}
\title{Full counting statistics for unconventional superconductor junctions
}

\author{Tim Kokkeler\orcid{0000-0001-8681-3376}}
\email{tim.kokkeler@dipc.org}
\affiliation{Donostia International Physics Center (DIPC), 20018 Donostia--San Sebasti\'an, Spain}
\affiliation{University of Twente, 7522 NB Enschede, The Netherlands}

\author{Alexander Golubov\orcid{0000-0001-5085-5195}}
\affiliation{University of Twente, 7522 NB Enschede, The Netherlands}
\author{F. Sebastian Bergeret\orcid{0000-0001-6007-4878}}
\affiliation{Centro de F\'isica de Materiales (CFM-MPC) Centro Mixto CSIC-UPV/EHU, E-20018 Donostia-San Sebasti\'an,  Spain}
\affiliation{Donostia International Physics Center (DIPC), 20018 Donostia--San Sebasti\'an, Spain}
\author{Yukio Tanaka\orcid{0000-0003-1537-4788}}
\affiliation{Department of Applied Physics, Nagoya University, 464-8603 Nagoya, Japan\\
}
\affiliation{Research Center for Crystalline Materials Engineering, Nagoya University, 464-8603 Nagoya Japan}
\begin{abstract}
        Noise and current measurements are key tools for studying mesoscopic systems, revealing insights beyond conductance alone.  For instance, noise measurements show that transport carriers in conventional superconductors have charge 2e. The noise power also depends on   junction type, distinguishing different transport processes.  Existing theories focus primarily   on zero temperature shot noise in tunnel junctions with conventional superconductors, where transport is carried by quasiparticles and  Cooper pairs.  Here we develop a full counting statistics theory for  unconventional superconductor / normal metal junctions of different types, incorporating  the effect of thermal noise on the differential Fano factor, the ratio of differential noise power and conductance. In these junctions there is a third type of transport carrier, surface Andreev bound states.
    Our study reveals 
    that junctions with dispersionless surface Andreev bound states exhibit negative differential Fano factor at finite temperatures. 
    In contrast,  in the presence of dispersive surface Andreev bound states, the noise power always increases with voltage, but there are local minima in the differential noise at those voltages corresponding to the extrema of the surface Andreev bound state spectrum. For normal metals and conventional superconductors the voltage dependence of the differential Fano factor is similar in all types of junctions, including tunnel junctions and diffusive barriers. However, significant differences arise  with  unconventional superconductors, making distinct junction types valuable tools for identifying pairing symmetries.
    Our results also highlight the importance of finite temperature effects in noise power measurements for potential unconventional superconductors, offering new means to determine pairing symmetries in  topological superconductors.
\end{abstract}
\maketitle
\section{Introduction}
Measurements of noise power and higher order current correlation functions provide insight into the mechanisms of transport and the elementary quasiparticles involved. Therefore, noise \cite{blanter2000shot,nazarov2012quantum,lesovik1993negative,oberholzer2002crossover,beenakker1997random,lesovik1993negative,khlus1987current,lesovik1989excess,buttiker1990scattering,blanter2000shot,reznikov1998quantum,kobayashi2021shot,chen2012excess,jehl2000detection,tikhonov2016andreev,steinbach1996observation,liefrink1994experimental,kozhevnikov2000observation,henny1999shot,choi2005shot,banerjee2022anomalous,sahu2019enhanced,golub2004fano,mei2024identifying,dieleman1997observation,braggio2011superconducting,dejong1997shot,constantinian2003shot,morten2008full,belzig2003full2,bastiaans2019imaging,massee2018atomic,niu2024why,massee2019noisy,bastiaans2018charge,chen2012excess,thupakula2022coherent,ge2023single,burset2017current,tanaka2000interface,strubi2011interferometric,marquardt2004influence,landauer1993solid,tworzydlo2006sub,nagaev2001semiclassical,camino2005shot} and full counting statistics (FCS) \cite{levitov1996electron,belzig2001fullSN,belzig2003full,belzig2003full2,morten2008full} have been studied since long in mesoscopic hybrid systems.  Noise arises in any electronic system, due to thermal fluctuations and the discrete nature of electrical charge. The charge carriers approach the barriers from both sides and can either be transmitted or reflected, and may even undergo virtual processes in which they tunnel back and forth via a forbidden state. On a microscopic scale, this leads to rapid fluctuations of the current as a function of time \cite{blanter2000shot}.  

However, while the time dependence of the current is random, its statistical properties are well defined,  and they are determined by the nature of the charge, the type of junction and the occupation of the states.  All statistical properties of the current together form the FCS \cite{levitov1996electron,blanter2000shot,belzig2001fullSN}. 

 There are  several theoretical works discussing noise and FCS in mesoscopic junctions with conventional superconductors \cite{blanter2000shot,belzig2001fullSN,belzig2000spin,beenakker1997random}.  Noise power calculations have also been extended to tunnel junctions with unconventional superconductors, such as topological superconductors hosting Majorana bound states \cite{li2023shot,burset2017current}.  However, these works focus on the zero temperature limit, and therefore neglect the contributions of thermal noise. Moreover, the extension of present theories to the FCS of unconventional superconductor (USC) tunnel junctions, and the extension of USC transport theories to junctions with different geometries, is highly nontrivial, and not yet  studied. 

The aim of the present work is twofold. On the one hand,  we develop the theory of FCS for unconventional superconductors at finite temperatures in tunnel junctions and  various types of  interfaces.  Our approach is based on  the quasiclassical Keldysh technique \cite{nazarov1999novel,nazarov2012quantum,belzigfull2001,belzig2003full,tanaka2003circuit}, since this formalism allows for the treatment of nonequilibrium and finite temperature
effects.

On the other hand, we use our formalism to study the transport properties of unconventional superconducting junctions at finite temperatures and voltages.
We show that the finite temperature dependence of the noise power provides new ways to detect symmetry properties of unconventional superconductors. Indeed, we find that dispersionless zero energy surface Andreev bound states (ZESABS) lead to negative differential noise power. On the other hand, the differential Fano factors in junctions with superconductors that host dispersive surface Andreev bound states (SABSs) shows local minima without sign change at voltages that correspond  to the extrema of the SABS spectrum. In short, we show how conductance and noise measurements can be used to infer the spectral properties of the SABS and, consequently, the pairing symmetry of the host superconductor. 

Our main results are summarized in Tables \ref{tab:Dispersions} and \ref{tab:FanoDiffusive}. As shown in the  Table \ref{tab:Dispersions}, the nonzero temperature differential Fano factor strongly depends on the type of SABSs. This demonstrates that finite temperature noise reveals distinct signatures of SABSs, making it a valuable tool for identifying the symmetry of the pair potential in unconventional superconductors, addressing a long-standing issue. In fact, our theory is applicable to several types of superconductors, such as spin triplet superconductors \cite{chiu2021observation,ran2019nearly,yang2021spin,Aoki-2019, saxena2000superconductivity,aoki2001coexistence,Hardy-2005,huy2007super,Mineev_2017}, noncentrosymmetric superconductors \cite{bauer2012non,bauer2004heavy,amano2004superconductivity,akazawa2004pressure,togano2004Superconductivity,tateiwa2005novel,kimura2005pressure,sugitani2006pressure,honda2010pressure,SETTAI2007844,Bauer2010Unconventional, xie2020captas,iniotakis2007andreev,eschrig2010theoretical,annunziata2012proximity,rahnavard2014magnetic, mishra2021effects,ikegaya2021proposal,daido2017majorana,chiu2023tuning,chiu2024electronic,asano2011josephson,klam2014josephson,wu2009tunneling,fujimoto2009unamiguous,borkje2006tunneling,goryo2012possible}, time-reversal symmetry broken superconductors \cite{wenger1993dwave,rokhsar1993pairing,covington1997observation,sigrist1991phenomenological,sigrist1991phenomenology,laughlin1994tunneling,laughlin1998magnetic,hillier2009evidence,wakatsuki2017nonreciprocal,kivelson2020proposal,suh2020stabilizing,xia2006high,luke1998time,sigrist1998time,ghosh2020recent,ghosh2022time,lee2009pairing,farhang2023revealing,ajeesh2023fate,shang2020simultaneous,grinenko2021split,willa2021inhomogeneous,maisuradze2010evidence,roising2022heat,movshovich1998low,balatsky1998spontaneous,belyavsky2012chiral,black2014chiral,fisher2014chiral}, and anapole superconductors \cite{kanasugi2022anapole,kitamura2022quantum,chazono2022piezoelectric,goswami2014axionic,mockli2019magnetic,kokkeler2023proximity}. Moreover, we extend the effective action describing transport through the unconventional superconductor-normal junction to account for multiple conduction modes with an arbitrary distribution of transmissions. This enables us to study the noise and FCS in diffusive and chaotic junctions. A summary of the  results for the different types of junctions is shown  in Table \ref{tab:FanoDiffusive}.

The article is organized  as follows.  In the next section, Sec. \ref{sec:BasicConceptsMainResults}, we present an overview of the key concepts needed to understand our findings and summarize the main results of the paper. In  Sec. \ref{sec:FCSaction} we develop the formalism used for the calculation of FCS and noise power in unconventional superconductor junctions. We continue in Sec. \ref{sec:Tunnel} with a detailed discussion of our results. First, we focus on tunnel junctions, which can be experimentally realized using techniques such as scanning tunneling microscopy (STM) \cite{bastiaans2019imaging,massee2018atomic,niu2024why,massee2019noisy,bastiaans2018charge,chen2012excess,thupakula2022coherent,ge2023single}. We calculate  the noise power in several distinct types of unconventional superconductors at nonzero temperatures and determine the dependence of the differential Fano factor on voltage.  
In Sec. \ref{sec:OtherJunctions} we present the extension of our FCS model to systems  described by a distribution of transmission eigenvalues. This extension allows us to study diffusive barrier, chaotic cavity and double barrier junctions. Our results show that each of these cases offers distinct advantages for determining pairing symmetry. In Sec. \ref{sec:Discussion} a summary and discussion of our results as well as possibilities for extension of our work are discussed.

\section{Basic concepts and main results}\label{sec:BasicConceptsMainResults}

We focus on two types of noise: thermal and shot noise. Thermal noise arises because of  thermal excitations at finite temperature. At zero temperature, however, these excitations freeze out and shot noise  dominates \cite{schottky1918spontane}. Shot noise is a consequence of the fact that  transport is probabilistic, and carried by discrete charges \cite{beenakker1997random,lesovik1993negative,khlus1987current,lesovik1989excess,buttiker1990scattering,blanter2000shot,beenakker2003quantum}. Each electron that approaches an interface can either be transmitted, with a given probability given by the transmission coefficient $\Tilde{\mathcal{T}}$, or reflected, with probability $1-\Tilde{\mathcal{T}}$. Thus, while the average contribution of an electron to the current is $e\Tilde{\mathcal{T}}$, each electron can only contribute $0$ or $e$ to the current. Consequently, not all electrons have the same contribution, which implies the variance, i.e. the noise power, is finite \cite{ross2014introduction}. Indeed, the electrons either contribute $e^{2}$ or $0$ to $I^{2}$, and hence  $\langle I^{2}(t)\rangle\propto e^{2}\Tilde{\mathcal{T}}+0(1-\Tilde{\mathcal{T}}) = e^{2}\Tilde{\mathcal{T}}$, where the angle brackets denote a time average. 
Meanwhile $\langle I(t)\rangle^{2} \propto e^{2}\Tilde{\mathcal{T}}^{2}$, and consequently $\langle I^{2}(t)\rangle\neq \langle I(t)\rangle^{2}$, that is, the  noise power is finite.   

For normal metals, the current $I$ and noise power $P_{N}$ through a contact with a single channel for spin up and a single channel for spin down, with spin-independent transparency $\mathcal{\Tilde{T}}$,  can be expressed compactly as follows \cite{beenakker1997random,lesovik1993negative,khlus1987current,lesovik1989excess,buttiker1990scattering,blanter2000shot,reznikov1998quantum,kobayashi2021shot}:
\begin{align}\label{eq:ConductanceNormal}
   I& = \frac{e}{\pi \hbar}\int_{-\infty}^{\infty} dE \Tilde{\mathcal{T}}(E)(n_{l}-n_{r})\;, \\
   P_{N}&= \label{eq:NoisePowerNormal}\frac{2e^{2}}{\pi\hbar}\int_{-\infty}^{\infty} dE \Tilde{\mathcal{T}}^{2}(E)(n_{l}(1-n_{l})+n_{r}(1-n_{r}))\nonumber\\&+\Tilde{\mathcal{T}}(E)(1-\Tilde{\mathcal{T}}(E))(n_{l}(1-n_{r})+n_{r}(1-n_{l}))\;,
\end{align}
where $n_{l,r}$ are the occupation numbers in the left and right electrodes, given by $(1+\text{exp}(E-\mu_{l,r})/k_{B}T)$, where $\mu_{l,r}$ are the chemical potentials of the left and right electrodes respectively, and $\Tilde{\mathcal{T}}(E)$ is the transparency of the junction, which may depend on the energy of the electrons.  Here, we consider materials in which the Fermi level lies well within a band and we may assume $\Tilde{\mathcal{T}}(E)$ does not depend on energy but attains a constant value $\Tilde{T}$.

 The first term in Eq. (\ref{eq:NoisePowerNormal}) represents thermal fluctuations. It is second order in transparency since it involves processes in which electrons tunnel back and forth between the two electrodes. It is only nonvanishing if there exist energies for which $n_{l}$ or $n_{r}$ is neither 0 nor 1, that is if, at the same energy, there are both occupied states from which the electron may start and empty states to which it may return after tunneling.
 This is only the case at finite temperatures.
 At zero temperature only the second contribution to $P_{N}$ survives. It  describes the shot noise \cite{beenakker1997random,lesovik1993negative,khlus1987current,lesovik1989excess,buttiker1990scattering,blanter2000shot,beenakker2003quantum}. It is simple to check from Eq. (\ref{eq:NoisePowerNormal})  that 
 in two limits  transport is not probabilistic and the shot noise power vanishes: $\Tilde{T} = 0$ and $\Tilde{T} = 1$. If $\Tilde{T} = 0$ none of the electrons is transmitted, and hence both current and noise power vanish. On the other hand, for perfect transmission, the current is nonzero, but the noise power vanishes, because any particle entering the junction is transmitted. 

The noise power is not always independent of the conductance. Indeed,  
transport through  tunneling junctions between normal metals is described by 
Poissonian statistics, and the noise power scales linearly with the current \cite{ross2014introduction}.  To distinguish different types of transport it is useful to consider the ratio of the differential noise power and conductance, the so-called differential Fano factor \cite{blanter2000shot}: \begin{align}
F = \frac{1}{2e}\frac{\frac{\partial P_{N}}{\partial V}}{\frac{\partial I}{\partial V}}\; . \label{eq:FanoDef}
\end{align}
where $V = \frac{1}{e}(\mu_{r}-\mu_{l})$ is the applied voltage. The factor $(2e)^{-1}$ exactly cancels the difference in the prefactors in Eqs. (\ref{eq:ConductanceNormal}) and (\ref{eq:NoisePowerNormal}). There are two limits in which the differential Fano factor takes a particularly simple form. In the fully transparent limit, $\Tilde{T} =1$, the noise power vanishes while the current does not. Consequently, $F = 0$ in this limit. On the other hand, in the limit of low transparency, we may approximate $1-\Tilde{T}\approx 1$. At zero temperature, all states below the chemical potential are occupied, while all states above the chemical potential are empty. Thus, the occupation numbers satisfy
 $n_{l,r} = 1$ for $|eV|<\mu_{l,r}$ and $n_{l,r} = 0$ for $|eV|>\mu_{l,r}$. Hence we deduce from Eq. (\ref{eq:ConductanceNormal}) that $I = \frac{e^{2}}{\pi\hbar}\Tilde{T}V$ and from Eq. (\ref{eq:NoisePowerNormal}) that $P_{N} = \frac{2e^{3}}{\pi\hbar}\Tilde{T}|V|$.  Using (\ref{eq:FanoDef}) we conclude that  $F\approx 1$ in low transparency normal metal / normal metal junctions. Values of $F$ different from unity indicate a  deviation from the tunneling regime.

Now consider the case in which one of the materials is a conventional \textit{s} - wave superconductor. If the voltage difference is smaller than the superconducting gap, electrons   can only be transferred  from the normal metal to the superconductor via Andreev reflection \cite{Andreev1964thermal}: the incoming electron is reflected at the hybrid interface as a hole, and a Cooper pair is transmitted to the superconductor. Since this is equivalent to a two-electron process, the probability of such an event for low transparencies is of the order of $\Tilde{T}^{2}$.  
Moreover,  the transmitted charge is either $0$, or $2e$. This implies that $\langle I(t)\rangle \propto 2e \Tilde{T}^2$. Similarly, the contribution to $\langle I^{2}(t)\rangle$ of a single event is now either $0$ or $4e^{2}$, which means $\langle I^{2}(t)\rangle \propto 4e^{2}\Tilde{T}^2$. If we compare with single electron transport, we observe that while $\langle I(t)\rangle$ has an extra factor 2$\Tilde{T}$, the shot noise contribution to $\langle I^{2}(t)\rangle$ has a factor 4$\Tilde{T}$. Thus, transport by Cooper pairs is twice as noisy as transport by single electrons, and consequently, $F\approx 2$ for low transparencies at zero temperature. Just like for normal metals, if the transparency is increased, the differential Fano factor decreases. However, this decrease is slower, and consequently the differential Fano factor in the superconducting state is more than twice as large as in the normal state.

For unconventional superconductors, the dependence of the differential Fano factor on voltage is significantly different. First of all, in nodal gap superconductors, next to Cooper pair tunneling also quasiparticle tunneling is finite for all voltages. Since quasiparticle tunneling is first order in transmission, while Cooper pair tunneling is second order, shot noise of the quasiparticle current is the dominant source of noise and consequently in the tunneling limit $F\approx 1$ for such superconductors.

In addition, unconventional superconductors, such as $\text{\textit{p}}_{\text{x}}$ - wave, $\text{\textit{d}}_{\text{xy}}$ - wave, helical \textit{p} - wave and chiral \textit{p} - wave superconductors (see  Table \ref{tab:Deltadefs}), may host SABSs because Cooper pairs experience different pair potentials before and after reflection \cite{buchholtz1981identification,hara1987quasiclassical,hu1994midgap,tanaka1995theory,bruder1990andreev,kashiwaya1996theory,kashiwaya2000tunneling}. 
These SABSs provide a resonance, and the effective transparency is $1$ when the voltage equals the energy of a SABS. As we have seen before, if the transparency equals $1$, the noise vanishes. Thus, these resonances contribute to the current, but not to the shot noise power, and the differential Fano factor is strongly suppressed when the applied voltage equals the SABS energy \cite{zhu1999shot,tanakainterface2000,burset2017current,pal2023transport,bolech2007observing,cuevas1999shot,cuevas2002shot,lofwander2003shot,li2023shot,perrin2021identifying,mei2024identifying}.
Such suppressed differential Fano factors have been observed experimentally \cite{tikhonov2016andreev}.

 In summary, in tunneling junctions, if the current is carried by Cooper pairs, $F = 2$, if quasiparticle transport  dominates the noise resembles  that of a normal junction with $F = 1$.  Quasiparticle transport through a SABSs  is  noiseless, and hence $F = 0$ for such contributions. In intermediate cases, the differential Fano factor reflects the balance between various contributions and can assume intermediate values. Thus, noise power measurements can be used to identify SABSs and hence to distinguish different types of superconductivity. However, preexisting theories only have limited applicability.

 First of all, so far, efforts focus on the zero temperature limit and do not consider the influence of thermal noise. For example, even in normal junctions, the differential Fano factor vanishes at zero voltage for nonzero temperatures \cite{blanter2000shot}. Indeed, within quasiclassics the noise power is an even function of voltage because the density of states is constant \cite{eilenberger1968transformation}, and it is smooth for nonzero temperatures, because the Fermi-Dirac distribution function is smooth for $T>0$. Consequently, its voltage derivative vanishes at $eV = 0$. Thus, to verify any prediction made based on zero temperature calculations, one must always verify that $k_{B}T\ll |eV|$. Near zero voltage, which is a particularly important regime for searches for Majorana fermions, this imposes a strong restriction and a careful analysis of the interplay between shot noise and thermal noise is required. Such  analysis has not yet been presented for junctions with unconventional superconductors.

\begin{figure}
    \centering
    \includegraphics[width=8.6cm]{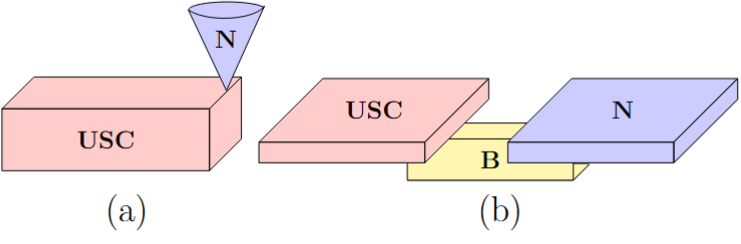}
    \caption{(a): Schematic of an STM measurement on an unconventional superconductor, which may be described as a tunnel junction, and two electrodes, one of which is an unconventional superconductor (USC), one of which a normal metal (N). The tunnel junction has low transparency $\Tilde{T}$. (b): Two electrodes, one USC, one N connected by a boundary region (B) in a lateral or planar junction, which may either be a tunnel barrier, a double barrier, a diffusive barrier  or a chaotic cavity.}
    \label{fig:Setups}
\end{figure}

Moreover, in addition to STM measurements \cite{braggio2011superconducting,chen2012excess,massee2018atomic,massee2019noisy,thupakula2022coherent,bastiaans2018charge,bastiaans2019imaging,niu2024why,ge2023single}, as schematically shown in Fig. \ref{fig:Setups}(a), transport through lateral junctions, like the one shown in Fig. \ref{fig:Setups}(b), is also experimentally explored. In this case, the normal (N) and unconventional superconductor (USC) electrodes are connected through a finite link that, depending on the degree of disorder, can be considered a tunnel junction, a chaotic cavity, a diffusive barrier, or a double barrier junction. Noise in diffusive barrier junctions has  been studied experimentally  in Refs. \cite{jehl2000detection,tikhonov2016andreev,steinbach1996observation,sahu2019enhanced,sahu2021quantized}. 
 Description of transport through such junctions requires accounting for transmission eigenvalues that are rapidly varying functions of the incoming angle, and the randomization of in-plane momenta due to scattering. Existing theories of transport in junctions with unconventional superconductors \cite{tanaka2003circuit,tanaka2004theory,tanaka2005theory} do not consider this randomization, and therefore an extension of these theories is required.

Throughout the rest of this paper, we address these issues and present the theory of noise at finite temperature for various types of junctions.  
Specifically, we derive the effective actions describing the FCS of hybrid junctions between various types of unconventional superconductors and a normal metal.   
FCS describes not only the current and the noise, but also all order current correlation functions \cite{belzig2001fullSN,belzig2003full,belzig2003full2,belzigfull2001, levitov1996electron} $\langle I^{n}(t)\rangle$ with $n\geq 1$. These  are obtained from  the  generating function $Z(\chi) = e^{iS(\chi)} = \langle e^{-\frac{i\chi}{e} \int_{0}^{t_{0}}I(t) dt}\rangle
$ \cite{belzig2001fullSN}, where   $S$ and $\chi$ are  the action of FCS and the counting field respectively. From this generating function the current correlation functions, can be calculated from $(\partial_{\chi})^{n} Z (\chi = 0)$ at $\chi = 0$ \cite{belzig2001fullSN,belzig2003full}.

\begin{table*}[ht]
    \centering
    \begin{tabular}{|c|c|c|c|c|c|}
    \hline
    \multirow{2}{*}{Dimension}&\multirow{2}{*}{\hspace{0.5cm}SABS\hspace{0.5cm}}&\multirow{2}{*}{\hspace{0.2cm}$F_{00}$\hspace{0.2cm}}&Negative $F$&Minima of $F$ &\multirow{2}{*}{Examples}\\
    &&&for $T>0$&for $|eV|,T>0$&\\
    \hline
    \multirow{2}{*}{1D}&None&2&\xmark&\xmark&1D \textit{s} - wave\\
&\hspace{0.2cm}\includegraphics[width = 0.1cm]{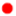}&0&\cmark&$|eV|\approx k_{B}T$& 1D \textit{p} - wave, Kitaev chain\\
\hline
       \multirow{3}{*}{2D}&None&2&\xmark&\xmark&2D \textit{s} - wave\\
       & \includegraphics[width = 0.3cm]{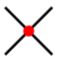} &  $1$&\xmark&$|eV|\approx\Delta_{-}$&2D helical \textit{p} - wave, 2D chiral \textit{p} - wave\\
        & \includegraphics[width = 0.5cm]{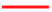} & 0&\cmark &$|eV|\approx k_{B}T$&2D $\text{\textit{p}}_{\text{x}}$ - wave, 2D $\text{\textit{d}}_{\text{xy}}$ - wave\\
        \hline
        \multirow{4}{*}{3D}&None&2&\xmark&\xmark&3D \textit{s} - wave\\
        & \includegraphics[width = 0.6cm]{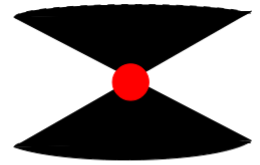}& $2$&\xmark&$|eV|\approx\Delta_{-}$& B-W\\
        &  \includegraphics[width = 0.5cm]{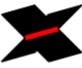}& $1$&\xmark&$|eV|\approx\Delta_{-}$& 3D chiral \textit{d} - wave ($k_{z}(k_{x}+ik_{y})$ or $k_{y}(k_{z}+ik_{x})$)\\
        &  \includegraphics[width = 0.5cm]{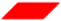}&0&\cmark&$|eV| \approx k_{B}T$&3D $\text{\textit{p}}_{\text{x}}$ - wave, 3D chiral \textit{d} - wave ($k_{x}(k_{y}+ik_{z})$)\\
        \hline
    \end{tabular}
    \caption{Characteristics of the differential Fano factor in tunneling junctions between a normal metal and a superconductor. We consider superconductors with different types of dispersion of  surface Andreev bound states (SABSs) in the three different dimensions, with zero energy surface Andreev bound states (ZESABSs) indicated in red. Schematics of the SABS energy as a function of the in-plane momentum are shown, with the ZESABSs shown in red. In one dimension, there is no angle dependence and there are only two options, no ZESABS or ZESABS. In two dimensions there are three options, if the ZESABSs exist they may form a line or a point in momentum space. If the zero temperature zero voltage differential Fano factor, $F_{00} = \text{lim}_{eV\xrightarrow{}0}F(T = 0)$, equals $ 0$, the differential Fano factor becomes negative at finite temperatures. For linear dispersion  $F_{00} = 1$. Moreover, in this case, at finite temperatures additional local minima appear at finite voltages, determined by the highest and lowest energies of the SABS spectrum. In three dimensions there are four possibilities, there are no ZESABSs or the ZESABSs form a point, a line or a plane in in-plane momentum space. The type of SABSs determines the differential Fano factor.}
    \label{tab:Dispersions}
\end{table*}

In what follows we derive the FCS for all junctions in Fig. \ref{fig:Setups} with unconventional superconductors. 

\begin{table*}[ht]
    \centering
    \begin{tabular}{|c|c|c|c|c|c|c|}
     \hline
         Type&$\Delta_{1}(\phi)$&$\Delta_{2}(\phi)$&$\vec{d}(\phi)$&TRS&Dimension&SABS  \\
         \hline
         \textit{s} - wave& $\Delta_{0}$& $\Delta_{0}$&-&\cmark&1D,2D,3D&None\\
         $\text{\textit{p}}_{\text{x}}$ - wave&$\Delta_{0}\cos\phi$&$-\Delta_{0}\cos\phi$&(0,0,1)&\cmark&2D,3D&\includegraphics[width = 0.5cm]{figures/DispersionlessSchematicRed.png}\includegraphics[width = 0.5cm]{figures/Dispersionless3DSchematicRed2.png}\\
         $\text{\textit{d}}_{\text{x}^{2}-\text{y}^{2}}$ - wave&$\Delta_{0}\cos2\phi$&$\Delta_{0}\cos2\phi$&-&\cmark&2D,3D&None\\
         $\text{\textit{d}}_{\text{xy}}$ - wave&$\Delta_{0}\sin2\phi$&$\Delta_{0}\sin2\phi$&-&\cmark&2D,3D&\includegraphics[width = 0.5cm]{figures/DispersionlessSchematicRed.png}\includegraphics[width = 0.5cm]{figures/Dispersionless3DSchematicRed2.png}\\
         chiral \textit{p} - wave&$\Delta_{0}e^{i\phi}$&$-\Delta_{0}e^{i\phi}$&$(0,0,1)$&\xmark&2D,3D&\includegraphics[width = 0.3cm]{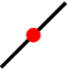}\includegraphics[width = 0.4cm]{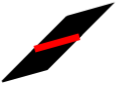}\\
         helical \textit{p} - wave&$\Delta_{0}$&$-\Delta_{0}$&$(\cos\phi,\sin\phi,0)$&\cmark&2D,3D&\includegraphics[width = 0.3cm]{figures/DispersiveSchematicRed.png}\includegraphics[width = 0.4cm]{figures/1DispersionSchematicRed.png}\\
         B-W (\textit{p} - wave)&$\Delta_{0}$&$-\Delta_{0}$&$(\cos\phi, \sin\phi\cos\theta,\sin\phi\sin\theta)$&\cmark&3D&\includegraphics[width = 0.5cm]{figures/ConeBroad.png}\\
         2D chiral \textit{d} - wave &$\Delta_{0}e^{2i\phi}$&$\Delta_{0}e^{2i\phi}$&-&\xmark&2D&\includegraphics[width = 0.6cm]{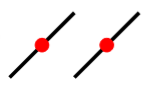}\\
         3D chiral \textit{d} - wave &$\Delta_{0}\cos\theta e^{i\phi}$&$\Delta_{0}\cos\theta e^{i\phi}$&-&\xmark&3D&\includegraphics[width = 0.5cm]{figures/1DispersionSchematicRed.png}\\
         \hline
    \end{tabular}
    \caption{Different types of pair potentials of conventional superconductors, their behavior under time reversal, the dimensions in which they may appear and their SABS spectrum.}
    \label{tab:types}
\end{table*}

\begin{figure*}[hb]
    \centering
    \includegraphics[width = 8.6cm]{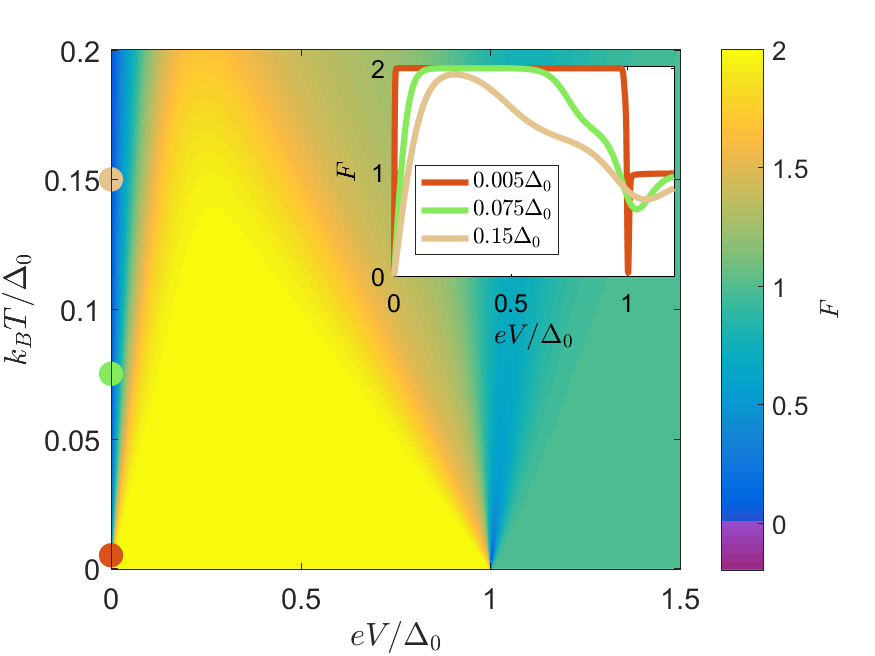}
    \includegraphics[width = 8.6cm]{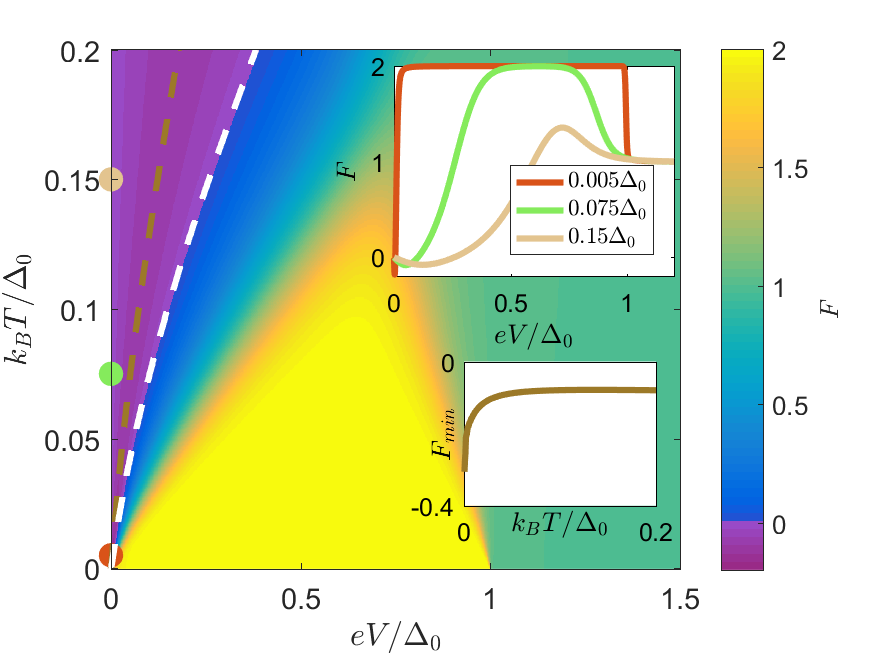}
    \includegraphics[width = 8.6cm]{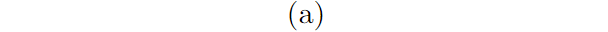}
    \includegraphics[width = 8.6cm]{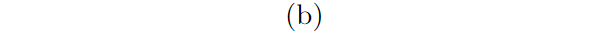}
    \caption{The voltage and temperature dependence of the differential Fano factor in one dimensional S/N junctions with \textit{s} - wave (a) and \textit{p} - wave (b) superconductors. In both cases, $F(eV = 0,T>0) = 0$,  while $F(|\Delta_{0}|>|eV|\gg k_{B}T) = 2$ and $F(|eV|-|\Delta_{0}|\gg k_{B}T) = 1$. The dependence for $0<|eV|<k_{B}T$ however is different. For the \textit{s} - wave superconductor (a), $F>0$, while for \textit{p} - wave  superconductors (b) $F<0$ . The voltage for which the minimum differential Fano factor is attained is indicated with a brown dashed line, the voltage for which the differential Fano factor crosses from negative to positive with a white dashed line . Inset in (a), Upper inset in (b): Dependence of the differential Fano factor on voltage at selected temperatures, which are indicated on the axis of the main figure using large dots of the same color. Lower inset in (b): The minimal obtainable of the differential Fano factor $F_{min}$ for different temperatures.}
    \label{fig:FinNoiseSdom}
\end{figure*}
\begin{figure*}[hb]
    \centering
   \includegraphics[width = 8.6cm]{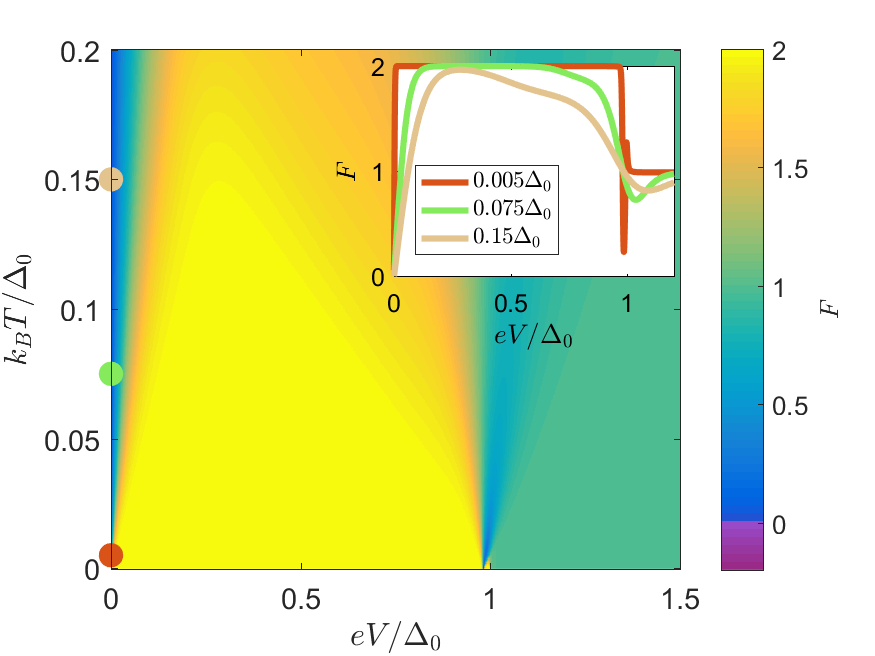}
    \includegraphics[width = 8.6cm]{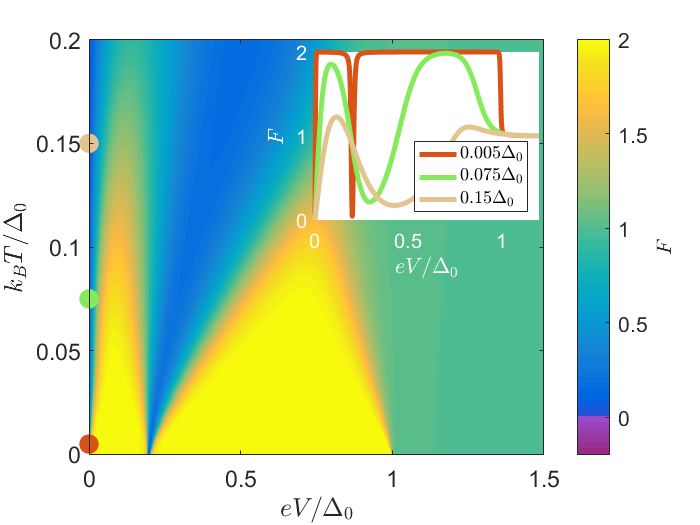}
    \includegraphics[width = 8.6cm]{figures/A.png}
    \includegraphics[width = 8.6cm]{figures/B.png}
    \caption{The voltage and temperature dependence of the differential Fano factor in S/N junctions with a one-dimensional i\textit{s} + \textit{p} - wave superconductor. The ratio $r$ between the \textit{p} - wave and \textit{s} - wave components of the pair potential is given by $r = 0.2$ (a) and $r = 5$ (b). Due to the time-reversal symmetry breaking the ZESABS disappears even for $r>1$. At low temperatures a sharp local minimum appears due to a SABS at $|eV| = \Delta_{s}$ (Table \ref{tab:Deltadefs}), which corresponds to $0.98\Delta_{0}$ for $r = 0.2$ and $0.20\Delta_{0}$ for $r = 5$. At finite temperatures this local minimum becomes broader but remains deep, providing a clear signature of this symmetry of the superconductor pair potential. At very low temperatures the differential Fano factor for $|eV|\approx\Delta_{s}$ is even negative, however, this temperature window is so small it is impossible to see from this graph. Insets: Dependence of the differential Fano factor on voltage at selected temperatures, which are indicated on the axis of the main figure using large dots of the same color. }
    \label{fig:FinNoiseSdomISP}
\end{figure*}

\section{Full counting statistics action}\label{sec:FCSaction}
In this section, we present the FCS for tunnel junctions with unconventional superconductors. The FCS is most conveniently derived using the method pioneered by Nazarov \cite{nazarov1999novel,nazarov2012quantum}, in which one writes a matrix current expressed in terms of the quasiclassical Green's functions describing the bulk electrodes. To this end it is assumed that the Fermi energy is the largest energy scale in the electrodes. For a junction between a USC and a normal electrode, the generalized expression for the charge current is given by the so-called Tanaka-Nazarov boundary condition \cite{tanaka2003circuit,tanaka2004theory,tanaka2022theory}:
\begin{align}\label{eq:TanakaNazarov}
    I &= -\frac{e N_{\text{ch}}}{16\pi\hbar}\int_{-\infty}^{\infty} dE \int_{-\frac{\pi}{2}}^{\frac{\pi}{2}}\cos\phi d\phi\text{Tr}\Bigg(\tau_{K}\Big((4-2\Tilde{T}(\phi))\mathbf{1}\nonumber\\&+\Tilde{T}(\phi)\{G_{N},C(\phi)\}\Big)^{-1}\Tilde{T}(\phi)[G_{N},C(\phi)]\Bigg)\; , 
\end{align}
where $\tau_{K} = \rho_{1}\tau_{3}$, $\rho_{1}$ is the first Pauli matrix in Keldysh space and $\tau_{3}$ is the third Pauli matrix in Nambu space, $\mathbf{1}$ is the identity matrix in Keldysh-Nambu-spin space and $N_{\text{ch}}$ is the number of conduction  channels of the junction.  The transparency $\Tilde{T}(\phi)$ is given by 
\begin{align}\Tilde{T}(\phi) = \frac{\cos^{2}\phi}{\cos^{2}\phi+z^{2}}\;,\label{eq:BTKtransparency}\end{align} 
where $z$ is the Blonder-Tinkham-Klapwijk (BTK) parameter \cite{blonder1982transition}. For $z = 0$ the transparency is $1$ for all angles, that is, the contact is fully transparent, while for $z>0$ the transparency depends on the angle of incidence and it is smaller than unity for all angles. $z\gg1$ corresponds to the  tunnel limit. $G_{N}$ is the bulk Green's function in the normal electrode. 
 Its retarded (R), advanced (A) and Keldysh (K) component are given by $G_{N}^{R} = -G_{N}^{A} = \tau_{3}$ and $G_{N}^{K} = 2f_{T}+2f_{L}\tau_{3}$, where $f_{L} = \frac{1}{2}\tanh{\frac{E+eV}{2k_{B}T}}+\frac{1}{2}\tanh{\frac{E-eV}{2k_{B}T}}$ and $f_{T} = \frac{1}{2}\tanh{\frac{E+eV}{2k_{B}T}}-\frac{1}{2}\tanh{\frac{E-eV}{2k_{B}T}}$ are the longitudinal (L) and transversal (T) distribution functions related to the occupation number of the right electrode via $2n_{r} = 1-(f_{L}+f_{T})$.
 
$C(\phi)$ denotes  the surface Green's function of the superconductor which can be calculated from the bulk superconductor Green's function via \cite{tanaka2022theory}
\begin{align}
    C(\phi) &= H_{+}(\phi)^{-1}(\mathbf{1}-H_{-}(\phi))\;,\\
    H_{\pm}(\phi) &= \frac{1}{2}\Big(G_{S}(\phi)\pm G_{S}(\pi-\phi)\Big)\;.
\end{align}
where $G_{S}$ is the bulk Green's function of an unconventional superconductor with angle of incidence $\phi$ and pair potential $\Delta(\phi)$ \cite{tanaka2003circuit,tanaka2004theory,tanaka2022theory}. It consists of retarded, advanced and Keldysh components, which are related by $G_{S}^{A} = -\tau_{3}(G_{S}^{R})^{\dagger}\tau_{3}$ and $G^{K} = f_{S}(G_{S}^{R}-G_{S}^{A})$, where $f_{S} = \tanh\frac{E}{2k_{B}T}$ is the equilibrium distribution function of the superconductor. 

The exact form of the retarded part depends  on the type of superconductor, but using the Nambu space basis of the bispinor $(\psi_{\uparrow},\psi_{\downarrow},\psi^{\dagger}_{\downarrow},-\psi^{\dagger}_{\uparrow})$, it can always be written in the form
\begin{widetext}
\begin{align}
    G_{S}^{R}(\phi) = \frac{1}{2}(\mathbf{1}+\vec{d}(\phi)\cdot\vec{\sigma})\frac{1}{\sqrt{E^{2}-|\Delta_{1}(\phi)|^{2}}}\otimes\begin{bmatrix}
        E&\Delta_{1}(\phi)\\-\Delta_{1}^{*}(\phi)&-E
    \end{bmatrix}+\frac{1}{2}(\mathbf{1}-\vec{d}(\phi)\cdot\vec{\sigma})\frac{1}{\sqrt{E^{2}-|\Delta_{2}(\phi)|^{2}}}\otimes\begin{bmatrix}
        E&\Delta_{2}(\phi)\\-\Delta_{2}^{*}(\phi)&-E
    \end{bmatrix}\;,
\end{align}
\end{widetext}
where the d-vector $\vec{d}(\phi)$ is a unit vector and $\Delta_{1,2}(\phi)$ are two energy scales associated with the pair potential. For any spin singlet superconductor, $\Delta_{1}(\phi) = \Delta_{2}(\phi)$ and therefore the choice of $\vec{d}(\phi)$ is irrelevant. 
For example, for  conventional \textit{s} - wave superconductors we have  $\Delta_{1}(\phi) = \Delta_{0}$. 
For spin triplet superconductors, such as $\text{\textit{p}}_{\text{x}}$ - wave superconductors,  $\Delta_{1}(\phi) = -\Delta_{2}(\phi)$ and $\vec{d}(\phi)$ corresponds to the d-vector of the spin triplet correlations. 

In this article,  we consider several superconducting pair potentials.  They are summarized in Table \ref{tab:types} together with the structure of the pair potential and a schematic of their SABS spectrum. We also use mixed parity superconductors in which the pair potential is a mixtures of the spin singlet and spin triplet pair potentials above. In this case, denoting the spin singlet component of the pair potential by $\Delta_{s1,2}$ and the spin triplet component by $\Delta_{t1,2}$ we have $\Delta_{1,2}(\phi) = \frac{e^{i\frac{\pi}{2}\phi_{t}}\Delta_{s1,2}(\phi)+r\Delta_{t1,2}(\phi)}{\sqrt{1+r^{2}}}$, where $r$ is called the mixing parameter and $\phi_{t}$ is the relative phase difference between the spin singlet and spin triplet components. We focus on time-reversal symmetric non-centrosymmetric superconductors ($\phi_{t} = 0$) \cite{bauer2012non,bauer2004heavy,amano2004superconductivity,akazawa2004pressure,togano2004Superconductivity,tateiwa2005novel,kimura2005pressure,sugitani2006pressure,honda2010pressure,SETTAI2007844,Bauer2010Unconventional, xie2020captas,iniotakis2007andreev,eschrig2010theoretical,annunziata2012proximity,rahnavard2014magnetic, mishra2021effects,ikegaya2021proposal,daido2017majorana,chiu2023tuning,asano2011josephson,klam2014josephson,wu2009tunneling,fujimoto2009unamiguous,borkje2006tunneling} and  anapole superconductors ($\phi_{t} = 1$), in which both time-reversal and inversion symmetries are broken but their product is preserved \cite{kanasugi2022anapole,kitamura2022quantum,chazono2022piezoelectric,goswami2014axionic,mockli2019magnetic,kokkeler2023proximity}. For mixed parity superconductors $\vec{d}$ is equal to the d-vector of the spin triplet component.

In the case of a conventional \textit{s} - wave superconductor the Tanaka-Nazarov boundary condition reduces to the Nazarov boundary condition \cite{nazarov1999novel,kuprianov1988influence}. 
For such conventional superconductors, in the tunneling limit ($\Tilde{T}\ll 1$) the denominator of Eq. (\ref{eq:TanakaNazarov}) becomes trivial and the expressions reduce to the Kuprianov-Luckichev boundary condition \cite{kuprianov1988influence}. On the other hand, for unconventional superconductors the denominator of Eq. (\ref{eq:TanakaNazarov}) remains nontrivial even in the limit of low transparency.
The reason for this is that 
unconventional superconductors may host SABSs \cite{buchholtz1981identification,hara1987quasiclassical,hu1994midgap,tanaka1995theory}. These SABSs lead to resonant tunneling and appear as poles in the surface Green's function $C$. Thus, for unconventional superconductors the Tanaka-Nazarov boundary condition does not reduce to the Kuprianov-Luckichev boundary condition at low transparencies but additionally contains a resonant tunneling mode at the energy of the SABS \cite{tanaka2022theory}. Thus, it is required to take into account the full expression of the Tanaka-Nazarov boundary conditions, even in the tunneling limit. 

\begin{table*}
    \centering
    \begin{tabular}{|l|l|}
    \hline
       Quantity  & Importance\\
       \hline
        \multirow{2}{*}{$\Delta_{0}$} & Gap magnitude in \textit{s} - wave, i\textit{s}+$\text{\textit{p}}_{\text{x}}$ - wave i\textit{s} + helical \textit{p} - wave and i\textit{s} + B-w superconductors\\
        
        &Maximum of SABS energy spectrum in i\textit{s} + helical \textit{p} - wave and i\textit{s} + B-W superconductors\\
        \hline
        $|\Delta_{+}| = \frac{|e^{i\frac{\pi}{2}\psi_{t}}+ r|}{\sqrt{1+r^{2}}}\Delta_{0}$& Maximum gap magnitude in \textit{s} + $\text{\textit{p}}_{\text{x}}$ - wave, \textit{s} + helical \textit{p} - wave and \textit{s} + B-W \textit{p} - wave superconductors\\
        \hline
        \multirow{3}{*}{$|\Delta_{-}| = \frac{|e^{i\frac{\pi}{2}\psi_{t}}-r|}{\sqrt{1+r^{2}}}\Delta_{0}$}
        & Minimum gap magnitude\\&  Maximum of SABS energy spectrum in \textit{s} + helical  \textit{p} - wave and \textit{s} + B-W superconductor\\
        &Energy of SABS in \textit{s} + $\text{\textit{p}}_{\text{x}}$ - wave superconductors\\
        \hline
        \multirow{3}{*}{$\Delta_{s} = \frac{\Delta_{0}}{\sqrt{1+r^{2}}}$}&\textit{s} - wave component of pair potential\\& Minimum of SABS energy spectrum in i\textit{s} + helical \textit{p} - wave and i\textit{s} + B-W superconductors\\&Energy of SABS in i\textit{s} + $\text{\textit{p}}_{\text{x}}$ - wave superconductors\\\hline
        $\Delta_{p} = \frac{r\Delta_{0}}{\sqrt{1+r^{2}}}$&\textit{p} - wave component of pair potential\\
        \hline
    \end{tabular}
    \caption{Definition of relevant energy scales in the junctions considered in this article and their relevance. $r$ denotes the ratio between \textit{s} - wave and \textit{p} - wave components, $\chi_{t}$ their relative phase difference.}
    \label{tab:Deltadefs}
\end{table*}
Noise power and current can be calculated as the first and second moments of a generating function. To construct it, it is convenient to introduce the counting field $\chi$, 
which after using a proper transformation can be absorbed in the Green functions of the normal electrode \cite{belzig2003full}:
\begin{align}
\label{eq:Gchi}
    G_{\chi} = e^{-i\chi\frac{1}{2}\tau_{K}}G_{N}e^{i\chi\frac{1}{2}\tau_{K}}\;.
\end{align}
 Following Ref. \cite{nazarov2012quantum}, we construct the action  such that its saddle point is the current, so that
\begin{align}\label{eq:IfromS}
    I = i\frac{e}{t_{0}}\partial_{\chi}S(\chi)\;,
\end{align}
where $I$ is the current in Eq. (\ref{eq:TanakaNazarov}), $e$ is the elementary charge and $t_{0}$ is the measuring time. For time-invariant systems, such as the one here, the choice of $t_{0}$ is irrelevant for the observables such as current and noise power.

We used an analogous derivation to Ref. \cite{belzig2001fullSN}, for details see Appendix  \ref{sec:ProofAction}. In 2D the action that leads to the current, Eq. (\ref{eq:TanakaNazarov}),  via the relation in Eq. (\ref{eq:IfromS}) is  given by:
\begin{widetext}
\begin{align}
    S(\chi) = -\frac{t_{0}N_{\text{ch}}}{16\pi\hbar}\int_{-\infty}^{\infty} dE\int_{-\frac{\pi}{2}}^{\frac{\pi}{2}}\cos\phi d\phi \text{Tr ln}\Big(4-2\Tilde{T}(\phi)+\Tilde{T}(\phi)\{G_{\chi},C(\phi)\}\Big)\;.\label{eq:Action}
\end{align}

The noise power can be calculated from the second derivative of the action with respect to the counting field \cite{levitov1996electron,belzigfull2001,belzig2003full,belzig2001fullSN} via
\begin{align}\label{eq:PfromS}
    P_{N} = \frac{2e^{2}}{t_{0}}\partial^2_{\chi\chi}S\;.
\end{align}  Thus, from Eqs. (\ref{eq:Gchi},\ref{eq:Action},\ref{eq:PfromS})  we obtain:
\begin{align}\label{eq:PNdef}
    P_{N} &= -\int_{-\infty}^{\infty} dE \int_{-\frac{\pi}{2}}^{\frac{\pi}{2}} \cos(\phi)d\phi\frac{e^{2}N_{\text{ch}}}{8\pi\hbar}\text{Tr}\Bigg(\frac{\Tilde{T}(\phi)}{2}(4-2\Tilde{T}(\phi)+\Tilde{T}(\phi)\{C,G_{N}\})^{-1}(\{C,\tau_{K}G_{N}\tau_{K}\}-\{C,G_{N}\})\nonumber\\&+\frac{\Tilde{T}(\phi)^{2}}{4}\Big((4-2\Tilde{T}(\phi)+\Tilde{T}(\phi)\{C,G_{N}\})^{-1}\{C,[\tau_{K},G_{N}]\}\Big)^{2}\Bigg)\; ,
\end{align}
\end{widetext}
where the transparency $\Tilde{T}(\phi)$ is given by Eq. (\ref{eq:BTKtransparency}).
This expression describes the noise power in a junction with arbitrary transparency  between any USC from Table \ref{tab:Deltadefs} and a normal electrode in 2D.
The corresponding expressions for 1D and 3D can be found by replacing $\frac{1}{2}\int_{-\frac{\pi}{2}}^{\frac{\pi}{2}}\cos\phi d\phi$ by  $1$ and $\frac{1}{\pi}\int_{0}^{\frac{\pi}{2}}d\phi\int_{0}^{2\pi}d\theta\cos\phi\sin\phi$ respectively.
In the next section, we use Eq. (\ref{eq:PNdef}) to compute the noise power in  different junctions.

\section{Noise in tunnel junctions}\label{sec:Tunnel}

Several recent experiments to measure noise power are done via STM measurements \cite{bastiaans2019imaging,massee2018atomic,niu2024why,massee2019noisy,bastiaans2018charge,chen2012excess,thupakula2022coherent,ge2023single}. In such measurements the barrier resistance is very large and  the BTK parameter $z\gg 1$. Hence the proximity effect can be ignored, that is, $G_{N}$ is the Green's function of a bulk normal metal. 
It turns out that this is the best regime to distinguish between different symmetries of pair potentials, see Appendix \ref{sec:1Dnumerical} , and therefore we ignore the proximity effect in what follows, and hence set the BTK parameter $z$ in Eq. (\ref{eq:BTKtransparency}) to $z=10$.

With the help of Eqs. (\ref{eq:FanoDef}), (\ref{eq:TanakaNazarov}) and (\ref{eq:PNdef}) we compute the voltage and temperature dependence of the differential Fano factor in normal metal / unconventional superconductor junction and discuss how it depends on the presence of a SABS. For illustrative purposes, we focus on \textit{s}, \textit{p}, \textit{d}, \textit{s} + \textit{p} - wave and i\textit{s} + \textit{p} - wave superconductors, and elaborate on how these considerations can be generalized to different types of superconductors.  In the discussion of our results, we refer to several relevant energy scales, such as the minimum gap of the pair potential. A summary of those quantities, including a description of their relevance, can be found in Table \ref{tab:Deltadefs}.

We focus on  the tunneling limit $\Tilde{T}\ll 1$  and superconducting electrodes of three different types, 
 1D nanowires, 2D layers and 3D superconductors. We discuss the results for the 1D \textit{s}, \textit{p} and i\textit{s} + \textit{p} - wave superconductors, the 2D \textit{d} - wave and (i)\textit{s} + helical \textit{p} - wave superconductors and the 3D (i)\textit{s} + B-W superconductors, because with these  superconductors we may illustrate the differential Fano factors for the four different types of SABSs that may appear in superconductors. 

\subsection{1D}
We first consider the one-dimensional case, which can be experimentally realized via tunneling 
measurements in nanowires \cite{mikkelsen2004direct,kolmer2017two}. In 1D,  there are only  two possible cases, namely (i) No SABS or (ii) a SABS appears at one specific energy. 
In the zero temperature limit, 
one can obtain analytical expressions for current and noise power, see Appendix \ref{sec:1Danalytical},  Eqs. (\ref{eq:IBelowGap}) - (\ref{eq:FAbovegap}), (\ref{eq:Iresonance}) - (\ref{eq:Fresonsance}) for the \textit{s} - wave case and (\ref{eq:IbelowgapP}) - (\ref{eq:FabovegapP}), (\ref{eq:IresonanceP}) - (\ref{eq:FresonanceP}) for the \textit{p} - wave case. Here we discuss the main outcomes summarized in Table \ref{tab:Dispersions}. If the \textit{s} - wave component is dominant, for voltages above the minimum gap ($|eV|>\Delta_{-} = \frac{1-r}{\sqrt{1+r^{2}}}\Delta_{0}$) transport, in leading order in transparency,  is mainly due to quasiparticles, i.e.  single-electron transport, and hence the differential Fano factor equals $1$.
Below the gap the current is second order in transparency and dominated by Cooper pairs and therefore the differential Fano factor is 2.

In 1D the \textit{p} - wave superconductor is nodeless, hence if the \textit{p} - wave component is dominant, there is also a transition between $F = 2$ and $F = 1$ at the minimum gap due to the transition between Cooper pair and quasiparticle transport. However, additionally, there exists a ZESABS and hence the elements of $C$ diverge at $E = 0$ \cite{tanaka2022theory}. In this limit the current in Eq. (\ref{eq:TanakaNazarov}) can be written as the energy integral over $-\frac{e}{8\pi\hbar}\text{Tr}\Big(\{C,G_{N}\}^{-1}[C,G_{N}]\Big)$, that is, the transparency drops out. Since there is only one channel per spin, all transport at zero voltage goes via this ZESABS. Therefore, at zero voltage the junction is  effectively fully transparent, and hence the noise power and consequently the differential Fano factor vanishes. Because for small energies the elements of $C$ are of order $\frac{\Tilde{T}\Delta_{0}}{E}$, the local minimum has a width of order $\Tilde{T}\Delta_{0}$, and the differential Fano factor approaches $2$ for $|eV|\gg \Tilde{T}\Delta_{0}$.
\begin{figure*}[!htb]
    \centering
    \includegraphics[width = 8.6cm]{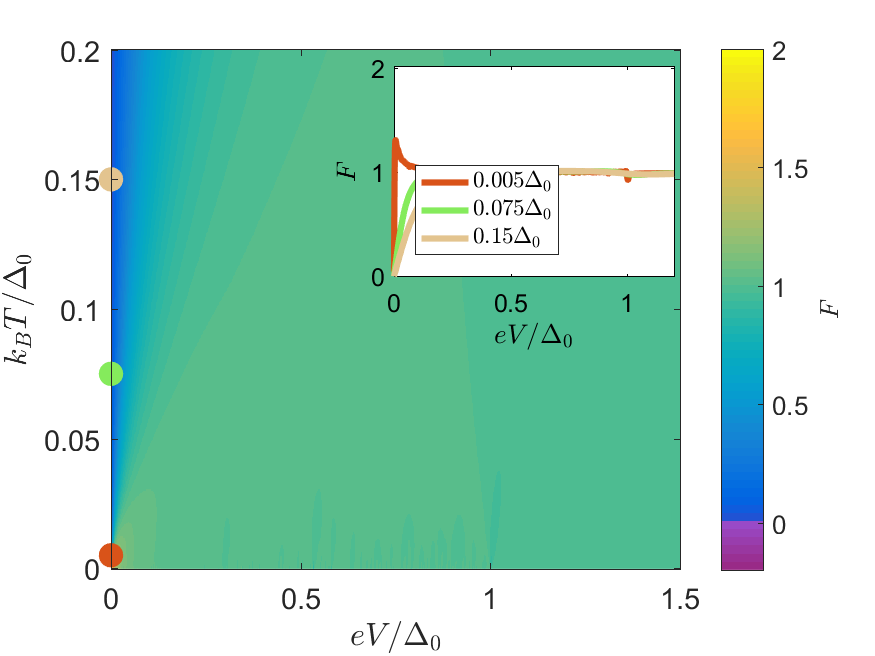}
    \includegraphics[width = 8.6cm]{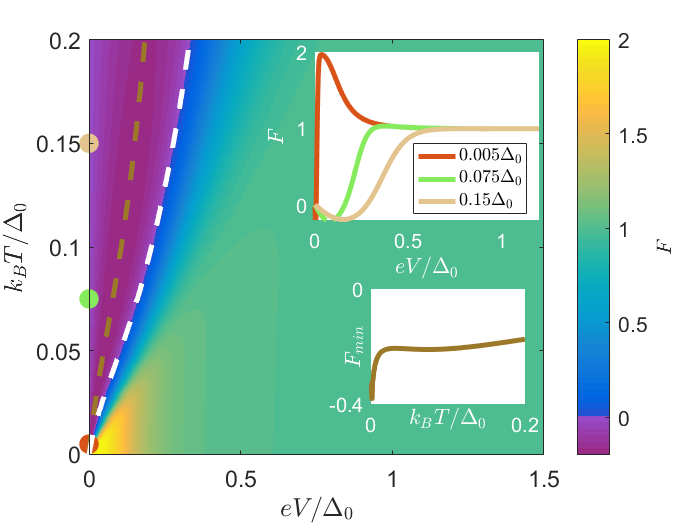}
    \includegraphics[width = 8.6cm]{figures/A.png}
    \includegraphics[width = 8.6cm]{figures/B.png}
    \caption{(a): The voltage and temperature dependence of the differential Fano factor in an S / N junction in which the superconductor is a \textit{d} - wave superconductor with its lobe oriented along the normal of the interface, i.e. it is a $\text{\textit{d}}_{\text{x}^{2}-\text{y}^{2}}$ - wave superconductor. There are no SABSs, however, there exist quasiparticles for any energies, it is a nodal gap superconductor. Hence the differential Fano factor is determined by the relative amount of channels for which there is Cooper pair transport and it approaches two only for low voltages. At finite temperatures $F = 2$ is not reached. There is a smooth transition to $F\approx 1$ over a small voltage window.
    (b): The voltage and temperature dependence of the differential Fano factor in an S / N junction in which the normal to the interface corresponds to a node direction of the pair potential, i.e. a $\text{\textit{d}}_{\text{xy}}$ - wave superconductor. Because of the dispersionless ZESABSs the differential Fano factor is negative at finite temperatures. The BTK parameter was set to $z = 10$.}
    \label{fig:FanoD1}
\end{figure*}
\begin{figure*}
    \centering
    \includegraphics[width = 8.6cm]{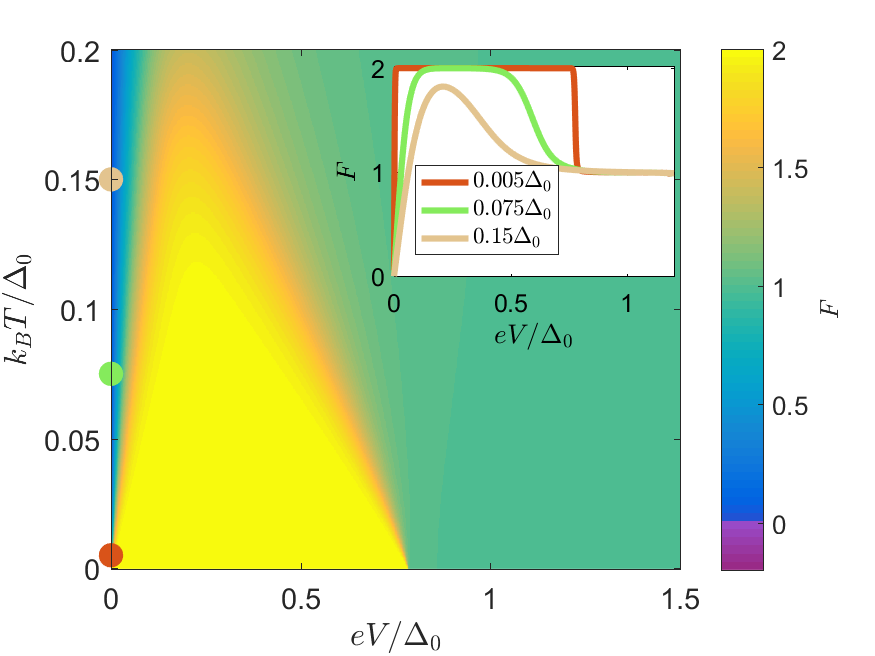}
    \includegraphics[width = 8.6cm]{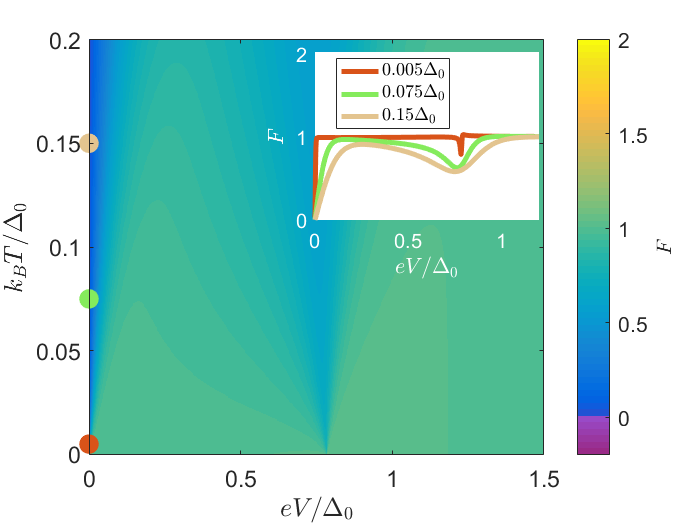}
    \includegraphics[width = 8.6cm]{figures/A.png}
    \includegraphics[width = 8.6cm]{figures/B.png}
    \caption{The voltage and temperature dependence of the differential Fano factor in two dimensional \textit{s} + helical \textit{p} - wave  superconductors. The ratio $r$ between the \textit{p} - wave and \textit{s} - wave components of the pair potential is given by $r = 0.2$ (a) and $r = 5$ (b). The differential Fano factor is nonnegative. If the \textit{s} - wave component is dominant (a), the noise power spectrum is very similar to that of a conventional \textit{s} - wave superconductor with gap of size $\Delta_{s}$ (Table \ref{tab:Deltadefs}). If the \textit{p} - wave component is dominant, the differential Fano factor does not reach two, but instead is close to 1, except for $|eV|\approx\Delta_{-}$ (Table \ref{tab:Deltadefs}), for which there is a local minimum because it is the maximum energy for which there is a SABS. Its width is of order $k_{B}T$. Insets: The dependence of the differential Fano factor on voltage for three different temperatures, which are indicated on the axis of the main figure using large dots of the same color.}
    \label{fig:FinNoiseSH}
\end{figure*}
\begin{figure*}
    \centering
    \includegraphics[width = 8.6cm]{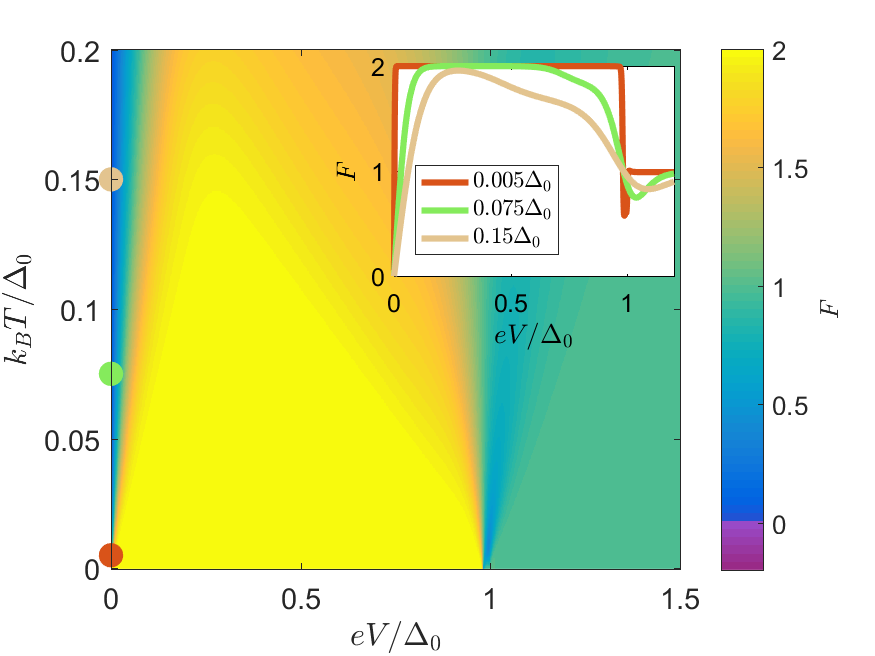}
    \includegraphics[width = 8.6cm]{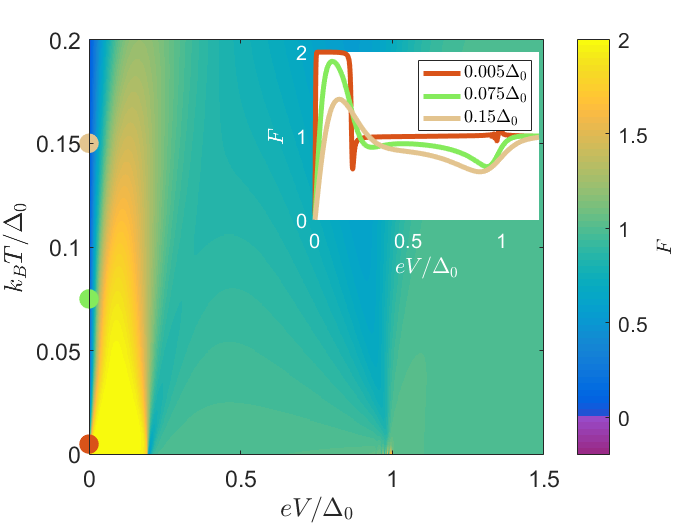}
    \includegraphics[width = 8.6cm]{figures/A.png}
    \includegraphics[width = 8.6cm]{figures/B.png}
    \caption{The voltage and temperature dependence of the differential Fano factor in two dimensional i\textit{s} + helical \textit{p} - wave  superconductors. The ratio $r$ between the \textit{p} - wave and \textit{s} - wave components of the pair potential is given by $r = 0.2$ (a) and $r = 5$ (b). If the \textit{s} - wave component is dominant (a), the differential Fano factor features a local minimum at $|eV| = \Delta_{s}$ (Table \ref{tab:Deltadefs}). With an increase of temperature this local minimum becomes broader and extends for $|eV|>\Delta_{0}$. Even if the \textit{p} - wave component is dominant (b), the differential Fano factor approaches 2 for $k_{B}T\ll |eV|\ll \frac{\Delta_{0}}{\sqrt{r^{2}+1}}$. It features a pronounced local minimum just above this value and for $|eV|\approx\Delta_{0}$, which are the lowest and highest positive energies for which there is a SABS. Insets: The dependence of the differential Fano factor on voltage for three different temperatures, which are indicated on the axis of the main figure using large dots of the same color.}
    \label{fig:FinNoiseISH}
\end{figure*}

Beyond the ideal tunneling limit  of low transparency and no inelastic scattering, the distinction between the different types of superconductors and the distinction between superconductors and normal metals is more difficult. In the presence of inelastic scattering, which may be characterized by a finite Dynes parameter $\eta$, the density of states below the gap in the superconductor is nonzero and  of order $\eta/\Delta_{0}$. Quasiparticle tunneling into these states gives a current contribution of $O( \Tilde{T}\eta/\Delta_{0})$. Therefore the differential Fano factor approaches 1 if this transport dominates over the transport by Cooper pairs, that is, if $\eta/\Delta_{0}\gg \Tilde{T}$ \cite{lofwander2002interplay}. 
Moreover, in the other limiting case of transparency, {\it i.e.} for a fully transparent junction, the differential Fano factor vanishes regardless of the materials that constitute the junction. Numerical analyses of these effects can be found in Appendix \ref{sec:1Dnumerical}.

At finite temperatures, the distribution functions of the electrons in the material are not given by step functions. This increases the computational costs of the evaluation of Eqs. (\ref{eq:TanakaNazarov}) and (\ref{eq:PfromS}). However, as shown in Appendix \ref{sec:FiniteTemperature}, we may exploit the Keldysh structure to reduce the computational costs of numerical evaluation.
We focus here on \textit{s}, \textit{p} and i\textit{s} + \textit{p} - wave superconductors. The results for 1D superconductors at finite temperature  are shown in Figs. \ref{fig:FinNoiseSdom} and \ref{fig:FinNoiseSdomISP}.  Within our assumption that the Fermi energy is much larger than any other energy scale,  conductance and noise power are even functions of voltage. Hence, we show only the results for $eV>0$. As shown in Fig. \ref{fig:FinNoiseSdom}(a), our calculations predict that in \textit{s} - wave superconductors at finite temperatures the differential Fano factor vanishes at $eV = 0$ even in the \textit{s} - wave dominant case, just as in normal metal / normal metal junctions \cite{blanter2000shot,beenakker1997random} due to the interplay between thermal noise and shot noise. The differential Fano factor approaches 2 for $|eV|\gg k_{B}T$ and hence the width of this dip is controllable via temperature. For \textit{p} - wave  superconductors, Fig. \ref{fig:FinNoiseSdom}(b), we predict a distinctly different dependence of the differential Fano factor on voltage and temperature. While the differential Fano factor still vanishes at $eV = 0$, it is negative for $0<|eV|\lesssim k_{B}T$. 
This feature can be attributed to the increase of effective transparency due to the resonance of the ZESABS. Due to this ZESABS there is a sharp peak in the effective transparency as a function of energy at $E = 0$. Therefore, we may use an analogy with a model in which the transparency $\Tilde{\mathcal{T}}(E)$ of a junction between two normal metal has a sharp peak at $E = 0$. As shown in \cite{lesovik1993negative} such sharp peaks lead to a decrease in the noise power at finite voltages. This can be understood using Eq. (\ref{eq:NoisePowerNormal}). At zero voltage the peak in the transparency $\Tilde{\mathcal{T}}(E)$ and the peak in $n_{l}(1-n_{l})$ are both centered at $E = 0$. However, at finite voltage, the peak in $n_{l}(1-n_{l})$ splits into two peaks at $E = \pm eV$. Thus, the overlap between the peaks decreases upon application of a voltage and consequently, the noise power decreases.
If the decay of the transmission peak is sharp one may obtain a negative differential noise power \cite{lesovik1993negative}. In our calculations, even though $\Tilde{T}$ itself does not depend on energy,  for $|E|\ll \Delta_{0}$ we have $C\propto \frac{\Tilde{T}\Delta_{0}}{E}$, see Eq. (\ref{eq:CdefappGP}),  and hence there is a sharp peak in the effective transparency at $E = 0$, which leads to the same effect. If we include an \textit{s} - wave component in the pair potential, these results remain the same until the \textit{s} - wave component becomes dominant. The transition is sharp because of the topological phase transition when their magnitudes are equal \cite{tanaka2009theory,sato2017topological}.

If we consider a one-dimensional junction with an i\textit{s} + \textit{p} - wave superconductor, this topological protection is absent due to the time reversal symmetry breaking \cite{kokkeler2023proximity}. Therefore, the inclusion of an \textit{s} - wave component shifts the SABS to nonzero energies. For this reason, the zero temperature differential Fano factor is finite at $eV = 0$ but vanishes for a finite voltage, specifically $|eV| = \Delta_{s} =  \frac{\Delta_{0}}{\sqrt{1+r^{2}}}$, as shown in Fig \ref{fig:FinNoiseSdomISP}(a) for the \textit{s} - wave dominant case and \ref{fig:FinNoiseSdomISP}(b) for the \textit{p} - wave dominant case. If there is a nonzero \textit{s} - wave component in the pair potential the temperature window in which the differential Fano factor is negative becomes very small and almost impossible to detect, for larger temperatures a nonnegative local minimum, centered at $|eV| = \Delta_{s}$, that becomes broader with increasing temperature can be observed.  
Because of the absence of a topological quantum phase transition \cite{sato2017topological} for i\textit{s} + \textit{p} - wave superconductors, there is no qualitative difference between \textit{s} - wave dominant and \textit{p} - wave dominant i\textit{s} + \textit{p} - wave superconductors, as shown in Fig. \ref{fig:FinNoiseSdomISP}.

A summary of the 1D results for the differential Fano factor can be found in the corresponding rows of Table \ref{tab:Dispersions}.
\subsubsection{Higher order current correlation functions}\label{HigherOrder}
Using the mean and variance, one cannot determine the full probability distribution of the current fluctuations. However, it is possible to determine higher order moments and eventually the FCS \cite{tobiska2004josephson}  by using the higher derivatives of the action, Eq. (\ref{eq:Action}), with respect to the counting field. 
Since we are interested in the voltage dependence of observables we consider the voltage derivative of the action. 
If we substitute the Green's function for the \textit{p} - wave superconductor in Eq. (\ref{eq:Action}) and compute the voltage derivative, see Appendix \ref{sec:FullCountingStatistics}, we obtain
\begin{align}\label{eq:dSdVP}
    \frac{\partial S}{\partial V}(T = 0,V\xrightarrow{}0)&=-i\frac{t_{0}N_{\text{ch}}}{\pi\hbar}\chi\;,
\end{align}
where the notation $(T = 0,V\xrightarrow{}0)$ was used to indicate we first take the zero temperature limit and subsequently the zero voltage limit.
Eq. (\ref{eq:dSdVP}) contains a few interesting features. First of all, Eq. (\ref{eq:dSdVP}) is completely independent of $\Tilde{T}(\phi)$, the transmission of the barrier does not at all influence transport via the ZESABS. Secondly, Eq. (\ref{eq:dSdVP}) is linear in the counting field $\chi$. This means all higher order derivatives vanish at $\chi = 0$. Consequently, not only the differential noise, but also the voltage derivative of any higher order current characteristic vanishes at $eV = 0$.

This can be compared to the equivalent expressions for \textit{s} - wave superconductors, for which the FCS has been discussed in detail in Ref. \cite{belzig2003full}. In this case, the voltage derivative of the action depends on transparency and generally all higher order correlation functions exist.

However, as shown in detail in Appendix \ref{sec:FullCountingStatistics}, for fully transparent junctions ($\Tilde{T} = 1$), the voltage derivative of the action reduces to Eq. (\ref{eq:dSdVP}). Thus, in fully transparent junctions, \textit{s} - wave and \textit{p} - wave superconductors can not be distinguished by any current correlation function, the FCS is the same in both cases. This highlights the importance of the low transparency limit.
\subsection{2D}

 Next to the one dimensional junction, we also consider junctions with two or three dimensional superconductors. 
 In two-dimensional junctions we may distinguish two types of SABSs.
 First of all, there exist superconductors with dispersionless ZESABSs, for example $\text{\textit{p}}_{\text{x}}$ - wave superconductors or $\text{\textit{d}}_{\text{xy}}$ - wave superconductors, which are protected by a topological quantum number for any fixed in-plane momentum \cite{sato2011topology}. Next to this, there exist superconductors with dispersive SABSs, for example  chiral \cite{furusaki2001spontaneous,matsumoto1999quasiparticle,read2000paired} or helical \cite{volovik2003universe,iniotakis2007andreev,tanaka2009theory} \textit{p} - wave superconductors, which are usually associated with topological Chern numbers and whose ZESABS forms Majorana bound states \cite{qi2009time,tanaka2024theory}. 

 First we discuss the differential Fano factor with either no ZESABS or dispersionless ZESABSs. These two have analogs in the 1D case. Indeed, in superconductors with no ZESABS, there is only Cooper pair tunneling and quasiparticle tunneling in the continuum, while for superconductors with dispersionless ZESABSs, all transport at $eV = 0$ is via a ZESABS and hence the results should be qualitatively similar to that of the 1D \textit{p} - wave superconductor. 
 
 The results for these two cases are illustrated using \textit{d} - wave superconductors in  Fig. \ref{fig:FanoD1}. For $\text{\textit{d}}_{\text{x}^{2}-\text{y}^{2}}$ - wave superconductors, shown in Fig. \ref{fig:FanoD1}(a), there are no SABSs \cite{tanaka1995theory,kashiwaya2000tunneling}, while the magnitude of the gap depends on the angle of incidence. Therefore, for any voltage there is a mixture of Cooper pair tunneling and quasiparticle tunneling via the continuum. For small voltages, for almost all angles there is Cooper pair transport, and consequently $F\approx 2$. On the other hand, with increase of voltage quasiparticle tunneling quickly becomes dominant because it is of lower order in $\Tilde{T}$. If the \textit{d} - wave superconductor is rotated by $\pi/4$, a  $\text{\textit{d}}_{\text{xy}}$ - wave superconductor is obtained, which hosts dispersionless ZESABSs \cite{hu1994midgap,tanaka1995theory,kashiwaya2000tunneling,sato2011topology}. Consequently, as shown in Fig. \ref{fig:FanoD1}(b), at finite temperatures a negative differential Fano factor may be observed. Because of the angle dependence of the gap, the differential Fano factor does not reach 2 at finite temperatures, but instead has a strongly temperature dependent maximum. 
 
Using $\text{\textit{p}}_{\text{x}}$ - wave superconductors we verified that the same features as for the $\text{\textit{d}}_{\text{xy}}$ - wave superconductor are observed for spin triplet superconductors with dispersionless SABS \cite{hara1987quasiclassical,yamashiro1998theory,sato2011topology,kwon2004fractional}, and that the addition of an \textit{s} - wave component to the pair potential has a similar effect as in 1D, see Figs. \ref{fig:FanoPXS} and \ref{fig:FanoPxiS} in Appendix \ref{sec:FiniteTemperature} for more details.  

Next we consider superconductors with dispersive SABSs, which do not have an analog in 1D. Specifically, we consider those superconductors whose SABS spectrum is linear around normal incidence, i.e. $E_{BS}\propto\phi$ for small $\phi$. Examples of such superconductors are chiral and helical \textit{p} - wave superconductors \cite{yamashiro1998theory,matsumoto1999quasiparticle,furusaki2001spontaneous,volovik2003universe}. In this case transport at $eV = 0$ is not noiseless, because there is only a single ZESABS, hence there are many momenta that do not feel the ZESABS. For these momenta, there is only Cooper pair transport, which enhances the shot noise compared to the normal state. According to our calculations, at zero temperature these two effects effectively cancel each other and the differential Fano factor is almost equal to $1$ for small voltages, in agreement with Refs. \cite{perrin2021identifying,ge2023single,mei2024identifying}. 

In fact, it can be shown analytically that $F_{00}  = \text{lim}_{eV\xrightarrow{}0} F(T =0) = 1,$ see Eqs. (\ref{eq:ConductanceHelical}), (\ref{eq:NoiseHelical}) and (\ref{eq:FHelical}) in Appendix  \ref{sec:ZeroFano}. For voltages close to the superconducting gap, the resonance shifts to oblique angles. 
Since the channels are determined by constant spacing in in-plane momentum, the density of channels is smaller for oblique angles, and therefore the contribution for channels with oblique incidence is suppressed by a factor $\cos\phi$ \cite{tanaka2004theory}. Consequently, the differential Fano factor increases, see also Fig. \ref{fig:CHPdom} in Appendix \ref{sec:2Dresults}. This is a clear signature of dispersive SABSs, however, due to its very small width, it is hard to observe and easily destroyed by perturbations.

Because there is no energy for which all transport is via a SABS, the finite temperature differential Fano factor in chiral and helical \textit{p} - wave superconductors remains nonnegative. However, the temperature dependence of $F$ can still be used to distinguish this type of superconductor from superconductors with SABSs. As shown in Fig. \ref{fig:FinNoiseSH}(a), the differential Fano factor is qualitatively similar for \textit{s} - wave dominant \textit{s} + helical \textit{p} - wave superconductors and conventional \textit{s} - wave superconductors. However, for \textit{p} - wave dominant \textit{s} + helical \textit{p} - wave superconductors, shown in Fig. \ref{fig:FinNoiseSH}(b), at finite temperatures, the differential Fano factor does not approach 2, because there is a combination of Cooper pair transport and quasiparticle transport via a SABS at all voltages below the gap. Moreover, a local minimum in the differential Fano factor appears around $|eV| = |\Delta_{-}| = \frac{|1-r|}{\sqrt{1+r^{2}}}\Delta_{0}$. 

The appearance of this local minimum can be understood via the suppression of the differential Fano factor by a peak in the effective transparency \cite{levitov1996electron}, in similar fashion to the negative differential Fano factor for superconductors with dispersionless ZESABSs. SABSs exist for $|E|<|\Delta_{-}|$. Therefore, for $|E|<|\Delta_{-}|$ there are channels with a resonance, while for $|E|>|\Delta_{-}|$ there are not. Thus, there is a sharp decay of the effective transparency at $|E| = |\Delta_{-}|$. Following \cite{lesovik1993negative} this leads to a suppression of the differential noise power, though in this case the change in effective transparency is not strong enough to yield negative differential Fano factors. The width of the local minimum is of the order of $k_{B}T$.  Just as for the 1D \textit{p} - wave and $\text{\textit{p}}_{\text{x}}$ - wave, the transition between \textit{s} - wave dominant and \textit{p} - wave dominant behavior is sharp because of a topological phase transition \cite{sato2017topological}.

For the time-reversal symmetry broken i\textit{s} +  helical \textit{p} - wave superconductor additional features appear. Indeed, due to the time-reversal symmetry breaking a gap in the SABS spectrum around zero energy appears, the energy of the SABSs is given by $E(\phi) = \Delta_{0}\sqrt{1-\frac{r^{2}}{1+r^{2}}\cos^{2}\phi}$. Therefore, for $|eV| < \Delta_{s}$  transport is only carried by Cooper pairs and thus $F\approx 2$. This holds for both \textit{s} - wave and \textit{p} - wave dominant pair potentials, as shown in Fig. \ref{fig:FinNoiseISH}. Moreover, since the SABS spectrum is now bounded both from above and below, two local minima appear, one centered at the lower bound ($\Delta_{s}$) and one centered at the upper bound $(\Delta_{0})$. 

If the \textit{s} - wave component is significantly larger than the \textit{p} - wave component, as for the results shown in  Fig. \ref{fig:FinNoiseISH}(a), $\Delta_{s}$ is of similar magnitude as $\Delta_{0}$ and these local minima merge already for low temperatures, leaving only this feature and the global minimum at $eV = 0$. Similarly, for \textit{p} - wave dominant superconductors in which the \textit{p} - wave component is much larger than the \textit{s} - wave components the local minima at $|eV| = \Delta_{s}$ and $eV = 0$ become indistinguishable. This however, does not happen unless $r$ is very large. Indeed, as shown for \ref{fig:FinNoiseISH}(b) the local minima at $eV = 0$, $|eV| = \Delta_{s}$ and $|eV| = \Delta_{0}$ can be clearly distinguished for $r = 5$.

So far we have considered the cases of no dispersion ($E_{BS} = 0$) and dispersion that is linear near to normal incidence ($E_{BS}\propto\phi$). 
We may also consider more exotic types of dispersion, which  are likely to appear in the phase diagram at the boundary between different topological states \cite{yamakage2012theory,kobayashi2015fragile,schnyder2012types,schnyder2015topological}. In such cases it may happen that close to normal incidence $\phi^{n}$, where $n>1$. Examples of superconductors with such type of SABSs are combinations of \textit{p}, \textit{f} and \textit{h} - wave pair potentials, see Appendix \ref{sec:2Dresults} for details of the construction. In this case the energy of modes at almost normal incidence is highly suppressed compared to the case of linear dispersion. Thus, there are more channels close to a resonance at $E = 0$. Therefore, noiseless transport via SABSs dominates and the differential Fano factor at $eV = 0$ is suppressed. We have verified numerically and analytically in Appendices \ref{sec:2Dresults} and  \ref{sec:AnaFrac} that at zero temperature $F_{00} =  \frac{1}{n}+O(\Tilde{T})$. For $|eV|\gg \Tilde{T}\Delta_{0}$ we numerically obtain $F\approx 1$ due to the finite dispersion for nonzero angles.  Following our previous discussions, to observe the fractional differential Fano factor one needs to access the regime $k_{B}T \ll |eV|\ll \Tilde{T}\Delta_{0} \ll \Delta_{0}$ to avoid finite temperature and finite transparency effects. 

 A summary of the three different cases that may appear in pair potentials with a single component are summarized in the 2D rows of Table \ref{tab:Dispersions}.

\subsection{3D}
 In 3D, there exists a fourth possibility for the dispersion of the SABS, namely $E\propto \sqrt{k_{y}^{2}+k_{z}^{2}}$, for which there is a ZESABS only at $k_{y} = k_{z} = 0$ \cite{buchholtz1981identification,asano2003a,nagato2009strong}.
Thus, in three dimensions the ZESABSs may form a plane, a line or a point in momentum space. If the ZESABSs form a plane, like for the 3D $\text{\textit{p}}_{\text{x}}$ -wave superconductor, or a line, like for the 3D chiral \textit{d} - wave pair potential, the results are similar to that of a 1D \textit{p} - wave or a 2D chiral \textit{p} - wave superconductor respectively. Therefore, we focus on superconductors in which a ZESABS only appears for normal incidence. 

\begin{figure}[b]
    \centering
    \includegraphics[width = 8.6cm]{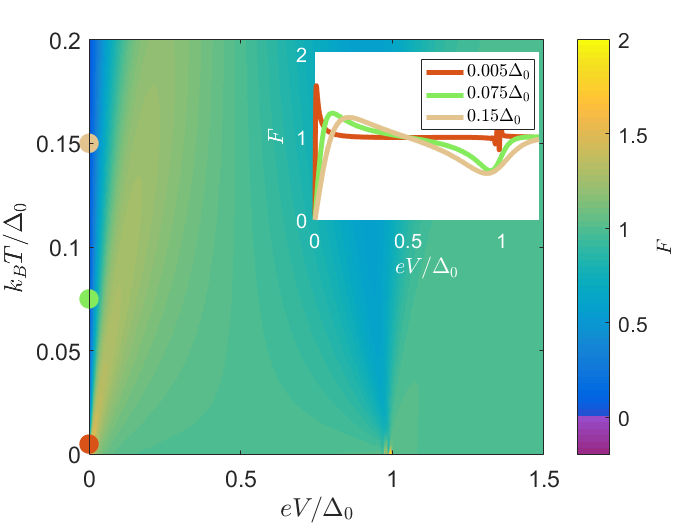}
    \caption{The differential Fano factor in the junction between a normal metal and a B-W superconductor as a function of voltage and temperature. For voltages of the order $k_{B}T$, Cooper pair transport dominates over SABS transport and therefore the differential Fano factor reaches almost two. For larger in-gap voltages there is an almost perfect balance between quasiparticle transport and Cooper pair transport and hence $F = 1$. At finite temperatures a minimum with a width of order $k_{B}T$ develops. Just as for the 2D superconductors this feature can be attributed to the difference in effective transparency below and above the gap.}
    \label{fig:BWMain}
\end{figure}

If there is only one ZESABS per spin, the transport by Cooper pairs dominates over the noiseless transport via SABSs and hence the differential Fano factor at zero temperature and zero voltage equals $F_{00} = 2$, see Eqs. (\ref{eq:ConductanceBW}), (\ref{eq:NoiseBW}) and (\ref{eq:FBW}) in Appendix \ref{sec:ZeroFano}. If the dispersion is nonlinear, with a dispersion $\phi^{n}$ around the point, the noiseless transport via SABSs is more important and in this case $F_{00} = \frac{2}{n}$, as discussed in Appendix \ref{sec:ZeroFano}. A summary of those findings can be found in Table \ref{tab:Dispersions}. For nonzero energies, the SABSs form a circle in momentum space instead of a point. Therefore, for nonzero voltages the B-W superconductor is similar to a helical \text{p} - wave superconductor and $F\approx 1$ for $\Tilde{T}\Delta_{0}\ll |eV|\ll\Delta_{0}$, as shown in Fig. \ref{fig:BWMain}.
Also in this case local minima in the differential Fano factor can be found when the voltage equals the highest energy of the SABS spectrum, the results are qualitatively similar to the two-dimensional case. The inclusion of an \textit{s} - wave component of the pair potential has similar effects as for the helical \textit{p} - wave case, as detailed in Figs. \ref{fig:SBW1}-\ref{fig:BWI} in Appendix \ref{sec:FiniteTemperature}.

The results for the four different types of dispersion in 3D are summarized in the last rows of Table \ref{tab:Dispersions}.

\section{Diffusive, chaotic and double barrier junctions}\label{sec:OtherJunctions}
In previous sections, we considered transport through a single interface described by a set of ballistic conduction channels with transmissions $\Tilde{T}$ as defined in Eq. (\ref{eq:BTKtransparency}). In such junctions momentum parallel to the interface is conserved, and moreover, the transmission eigenvalues vary slowly as a function of the incident angle.

In this section, we focus on junctions in which instead the transmission is a rapidly varying function of incident angle. Such junctions can be realized  in  lateral structures, such as the one shown in Fig. \ref{fig:Setups}(b), where the unconventional superconductor and the normal metal electrode are connected via a conducting link. This link may either conserve in-plane momentum, for example if it is a double barrier junction, or it may alter the in-plane momentum of the electrons, \textit{e.g. }if it is a diffusive barrier junction or a chaotic cavity junction.

If the transmission eigenvalues are rapidly varying, junctions are well characterized by assuming a probability distribution of transmission eigenvalues instead of a ballistic model like Eq. (\ref{eq:BTKtransparency}). 
For the double barrier, diffusive barrier and chaotic cavity, the distributions can be written as 
\begin{align}\label{eq:Formofdistributions}
    \rho(\Tilde{T})  = A(p) \Tilde{T}^{-\frac{p}{2}}(1-\Tilde{T})^{-\frac{1}{2}}\;,
\end{align}
where $A(p)$ is a normalization constant which guarantees that $\int_{0}^{1}\rho(\Tilde{T})d\Tilde{T} = 1$ and $p = 1$ corresponds to the symmetric chaotic cavity \cite{silvestrov2003noiseless}, $p = 2$ to the diffusive conductor \cite{dorokhov1982transmission} and $p = 3$ to the double barrier junction \cite{dejong1996distribution}. We have $A(1) = \frac{1}{\pi}$. However, for $p = 2$ and $p = 3$ the distribution in Eq. (\ref{eq:Formofdistributions}) is not normalizable on the interval $(0,1)$. Therefore, a low-transmission cut-off $\Tilde{T}_{l}$ is required. In terms of this low-transmission cut-off, $A(2) = 2\text{arctanh}(\sqrt{1-\Tilde{T}_{l}})$ and $A(3) = 2\sqrt{\frac{1}{\Tilde{T}_{l}}-1}$. 

We note that, while the normal state conductance and noise depend on the cut-off $\Tilde{T}_{l}$ through the normalization constant and in the case $p = 2$ and $p = 3$ vanish as $\Tilde{T}_{l}\xrightarrow{}0$ were found to vanish due to the divergence of the normalization constant, in the calculation of the differential Fano factor, or the ratio of the conductance to the normal state conductance, both numerator and denominator contain this same normalization constant, and hence they cancel out. Moreover, all other integrals involved converge in the limit $\Tilde{T}_{l}\xrightarrow{} 0$, as shown in Appendix \ref{sec:AnalyticalDiffFano}, except in the presence of dispersionless ZESABS. In this latter case the dependence on $\Tilde{T}_{l}$ can be found analytically, as discussed below. Therefore, below we discuss the results to lowest order in $\Tilde{T}_{l}$. The conductance and differential Fano factor at zero temperature and voltage are summarized in Table \ref{tab:FanoDiffusive}.

Below, to compare the different types of junctions we focus on the zero temperature limit. The generalization of these results to finite temperatures goes in a manner similar to the tunnel junction. A brief discussion of these features can be found at the end of this section. 
\subsection{Formalism}
For conventional superconductors, the current, noise and action in Eqs. (\ref{eq:TanakaNazarov},\ref{eq:Action},\ref{eq:PNdef}) can be generalized to chaotic cavity, diffusive barrier and double barrier junctions directly by assuming that the transmission eigenvalues $\Tilde{T}(\phi)$ that appear in Eqs. (\ref{eq:TanakaNazarov},\ref{eq:Action},\ref{eq:PNdef}) are distributed according to the eigenvalue distributions in Eq. (\ref{eq:Formofdistributions}) \cite{belzig2000spin}.

The generalization of this argument to junctions with unconventional superconductors is nontrivial, because not only the transparency, but also the pair potential is different for different channels. 

However, the pair potential is usually a slowly varying function of the incoming angle. On the other hand, as mentioned before, the transmission eigenvalues vary rapidly as a function of incoming angle. Therefore, under the condition that the number of channels is large, so that the variation of the surface Green's function \footnote{As in the derivation of the Tanaka-Nazarov boundary condition for tunnel junctions \cite{tanaka2003circuit,tanaka2004anomalous} we require a constriction to avoid suppression of the pair potential in the superconductor. If we do not have such constriction, the Green's function at the boundary may or may not be significantly altered \cite{ovchinnikov1969critical,buchholtz1986fermi,zhang1987order,zhang1988tunneling,nagato1998rough,barash1997quasiparticle,tanuma1998local,golubov1999rough,fogelstrom1997tunneling,golubov1999rough,bakurskiy2014anomalous}, and hence the Eilenberger equation should be solved for each specific case.} between neighboring channels is small, we may use the same distribution for each incident angle, as discussed in Appendix \ref{sec:Extension}. 

There is one important distinction between the different types of junctions we need to make. While for the tunnel junction and double barrier junctions in-plane momentum is conserved, in diffusive barrier junctions or chaotic cavities this is not the case. Instead,  for those two types of junctions we may assume that incoming and outgoing momentum are independent, which implies that we have to integrate over both incoming and outgoing angles. Therefore, as elaborated in Appendix \ref{sec:DiffusiveAction}, for diffusive barrier and chaotic cavity junctions we need to introduce the two-angle generalization of the surface Green's function $C$, given by
\begin{align}
    \mathcal{C}(\phi,\Tilde{\phi}) &= \mathcal{H}_{+}(\phi,\Tilde{\phi})^{-1}(\mathbf{1}-\mathcal{H}_{-}(\phi,\Tilde{\phi}))\;,\\
    \mathcal{H}_{\pm}(\phi,\Tilde{\phi})&= \frac{1}{2}(G_{S}(\phi)\pm G_{S}(\pi-\Tilde{\phi}))\;.
\end{align}
This quantity is related to the quantity $C(\phi)$ that appears in Eq. (\ref{eq:TanakaNazarov}) for tunnel barriers \cite{tanaka2022theory} via $\mathcal{C}(\phi,\Tilde{\phi} = \phi) = C(\phi)$. 

The importance of taking into account this randomization of incoming and outgoing momentum can be illustrated by comparing junctions with $\text{\textit{p}}_{\text{x}}$ and $\text{\textit{d}}_{\text{xy}}$ - wave superconductors. In a  junction that conserves in-plane momentum, for both superconductors the electrons experience an opposite pair potential after reflection on the interface for any incoming momentum. Hence, for each channel there is a ZESABS and consequently transport at $eV = 0$ is noiseless. Thus, in such junctions these two types of superconductors can not be distinguished via the differential Fano factor near $eV = 0$.  

On the other hand, if in-plane momentum is randomized at the boundary, for the $\text{\textit{p}}_{\text{x}}$ - wave superconductor any electron still experiences an opposite sign pair potential after reflection, because the pair potential is positive for any incoming momentum and negative for any outgoing momentum \cite{tanaka2004anomalous}. Therefore, $\mathcal{C}(\phi,\Tilde{\phi})$ has a pole at $E = 0$ for any choice of $\phi,\Tilde{\phi}$ and consequently there is a perfect resonance at $eV = 0$. Thus, in junctions with $\text{\textit{p}}_{\text{x}}$ - wave superconductors the noise vanishes, while the conductance is twice the conductance quantum multiplied by the number of channels. 

Meanwhile, for the $\text{\textit{d}}_{\text{xy}}$ - wave superconductor, half of the incoming and half of the outgoing angles have positive pair potential. If electrons can be reflected in any direction, in a $\text{\textit{d}}_{\text{xy}}$ - wave superconductor they have a chance of $\frac{1}{2}$ to experience a gap with opposite sign after reflection and a chance of $\frac{1}{2}$ to experience a gap with opposite sign \cite{tanaka2004anomalous,tanaka2005anomalous,tanaka2011symmetry}. Thus, $\mathcal{C}(\phi,\Tilde{\phi})$ has a pole for only half of the possible combinations of $\phi$ and $\Tilde{\phi}$. Only for those combinations transport is noiseless, for the other half the noise power does not vanish. Indeed, for such momenta there is only Cooper pair transport, just like for \textit{s} - wave superconductors.
Therefore, in junctions that do not preserve in-plane momentum, $\text{\textit{p}}_{\text{x}}$ - wave superconductors can be distinguished from $\text{\textit{d}}_{\text{xy}}$ - wave superconductors via the differential Fano factor near $eV = 0$. This shows the importance of taking into account randomization of in-plane momentum and indicates that junctions with different geometries provide additional tools to determine the pair potential.

Using these considerations, we obtain for diffusive and chaotic junctions the following action, see Appendix Sec. \ref{sec:DiffusiveAction} for the derivation:
\begin{widetext}
\begin{align}\label{eq:WithTwoIntegrals}
    S_{D}(\chi) = -\frac{t_{0}N_{\text{ch}}}{32\pi\hbar}\int_{0}^{1} d\Tilde{T} \rho(\Tilde{T}) \int_{-\frac{\pi}{2}}^{\frac{\pi}{2}}d\phi\int_{-\frac{\pi}{2}}^{\frac{\pi}{2}}d\Tilde{\phi}\cos{\phi}\cos{\Tilde{\phi}}\text{Tr ln} \Big(4-2\Tilde{T}+\Tilde{T}\{G_{\chi},\mathcal{C}(\phi,\Tilde{\phi})\}\Big)\;.
\end{align}
For double barrier junctions on the other hand, motion of the electrons between the barriers is assumed to be ballistic and in-plane momentum is conserved. Hence we may require $\Tilde{\phi} = \phi$ in this case. We obtain, see Appendix Sec. \ref{sec:DBAction},
\begin{align}\label{eq:WithOneIntegral}
    S_{DB}(\chi) = -\frac{t_{0}N_{\text{ch}}}{16\pi\hbar}\int_{0}^{1} d\Tilde{T} \rho(\Tilde{T}) \int_{-\frac{\pi}{2}}^{\frac{\pi}{2}}d\phi\cos{\phi}\text{Tr ln} \Big(4-2\Tilde{T}+\Tilde{T}\{G_{\chi},C(\phi)\}\Big)\;,
\end{align}
\end{widetext}
where $C(\phi)$ is the surface Green's function that also appears in Eq. (\ref{eq:TanakaNazarov}) for the tunnel junction. 
The corresponding expressions in 3D can be found by replacing $\frac{1}{2}\int_{-\frac{\pi}{2}}^{\frac{\pi}{2}}\cos\phi d\phi$ by  $\frac{1}{\pi}\int_{0}^{\frac{\pi}{2}}d\phi\int_{0}^{2\pi}d\theta\cos\phi\sin\phi$ and using a similar modification for $\Tilde{\phi}$. Since the existence of many channels is required for the derivation of the action, there is no 1D analog of the action for chaotic cavity, diffusive barrier and double barrier junctions.
Noise power and current are calculated in the same manner as before, by taking the appropriate derivatives of the action.
\subsection{Diffusive barrier junction}
We first consider the diffusive barrier junctions, which are characterized by the distribution in Eq. (\ref{eq:Formofdistributions}) 
with  $p = 2$ \cite{blanter2000shot,dorokhov1982transmission,jehl2000detection,tikhonov2016andreev,beenakker2003quantum,steinbach1996observation,liefrink1994experimental,kozhevnikov2000observation,henny1999shot}, while in-plane momentum is not conserved because of the scattering of electrons within the barrier. The zero voltage results for this junction are summarized in the diffusive barrier row of Table \ref{tab:FanoDiffusive}. This specific form is called the Dorokhov distribution \cite{dorokhov1982transmission}. Transport in diffusive junctions is qualitatively different from transport in tunnel  junctions. For junctions described by this distribution the normal state differential Fano factor is $F_{N} = \frac{1}{3}$ instead of $1$. Moreover, in S / N junctions, the conductance is nonzero even for voltages below the gap, as evidenced by Table \ref{tab:FanoDiffusive}. The zero bias conductance in the presence of an \textit{s} - wave superconductors equals the normal state conductance in this limit.  Still though, the differential Fano factor doubles compared to the normal state \cite{blanter2000shot}, which is similar to the tunneling case described in Sec. \ref{sec:Tunnel}. 

 First, we consider the zero bias conductance in diffusive junctions with different types of superconductors, summarized in the green part of Table \ref{tab:FanoDiffusive}. For the \textit{s} - wave superconductor, we verified that the zero bias conductance equals the normal state conductance. For  $\text{\textit{p}}_{\text{x}}$ - wave superconductors, there is still a perfect resonance and hence the conductance is $2N_{\text{ch}}\frac{e^{2}}{\pi\hbar}$, as required by topology \cite{ikegaya2016quantization}. Thus, the zero bias conductance is very large compared to the normal state conductance, which is suppressed logarithmically by $\Tilde{T}_{l}$ due to the divergence of the normalization constant $A(2)$ as $\Tilde{T}_{l}\xrightarrow{}0$. Thus, $\sigma(eV = 0)/\sigma_{N}$ diverges logarithmically as $\Tilde{T}_{l}\xrightarrow{}0$. This is consistent with the very large zero bias conductance peak in the presence of an anomalous proximity effect in a highly resistive metal \cite{tanaka2004theory,tanaka2005theory}. For $\text{\textit{d}}_{\text{xy}}$ - wave superconductors half of the channels have a perfect resonance. Since the contribution of the other half of the channels is much smaller, namely half the normal state conductance, the zero bias conductance is approximately $N_{\text{ch}}\frac{e^{2}}{\pi\hbar}$. Consequently, for $\Tilde{T}_{l}\ll 1$, the zero bias conductance in $\text{\textit{d}}_{\text{xy}}$ - wave superconductors is half as large as for $\text{\textit{p}}_{\text{x}}$ - wave superconductors, but still much larger than the normal state conductance.
 For the helical \textit{p} - wave superconductor, the ZESABS are dispersive and hence the zero bias conductance is larger than in the normal state, but smaller than for $\text{\textit{p}}_{\text{x}}$ - wave superconductors, it equals approximately $1.84\sigma_{N}$.

\begin{figure}
    \centering
    \includegraphics[width = 8.6cm]{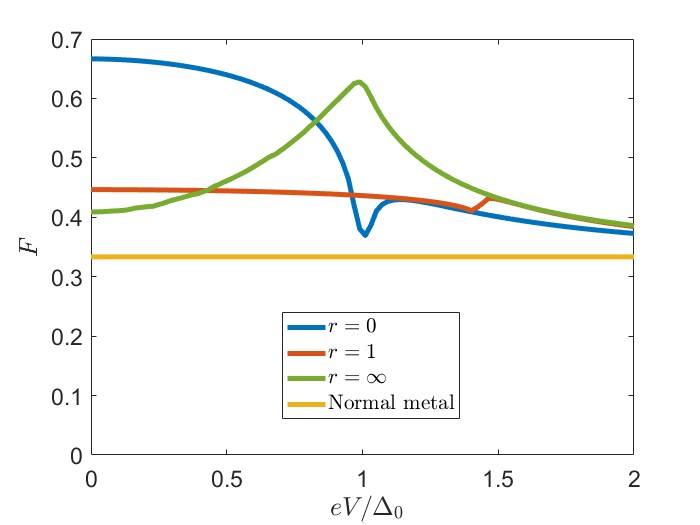}
    \caption{The zero temperature differential Fano factor for a diffusive barrier normal metal / 2D helical \textit{p} - wave superconductor junction, which is equivalent to those obtained by replacing the helical by a 2D chiral \textit{p} - wave superconductor. For \textit{s} - wave superconductors at $eV = 0$ the differential Fano factor is doubled compared to the normal state and attains $\frac{2}{3}$, for helical \textit{p} - wave superconductors $\text{lim}_{eV\xrightarrow{}0}F(T = 0) \approx 0.41$ and obtains a peak around $|eV| = \Delta_{0}$. For a perfect mixture $\text{lim}_{eV\xrightarrow{}0}F(T = 0) \approx 0.45$. Due to the presence of high transmission channels, the differential Fano factor in \textit{s} - wave superconductor decreases as a function of voltage already below the gap, while for the helical \textit{p} - wave superconductor the peak at $|eV| = \Delta_{0}$ is much broader than  in the tunnel junction.}
    \label{fig:DiffusiveHelical}
\end{figure}

Now that we have studied the current, we continue with the noise, whose zero voltage features are summarized in the yellow columns of Table \ref{tab:FanoDiffusive}. The zero temperature differential Fano factors for 2D \textit{s} + helical \textit{p} - wave superconductors are shown in Fig. \ref{fig:DiffusiveHelical}.  For an \textit{s} - wave superconductor our results confirm the usual doubling of the differential Fano factor due to the transport of current by Cooper pairs, i.e. $F(eV = 0) = \frac{2}{3}$. For the $\text{\textit{p}}_{\text{x}}$ - wave superconductor, the noise vanishes, while for the $\text{\textit{d}}_{\text{xy}}$ - wave superconductor the differential Fano factor is  suppressed by a factor $(\text{ln}\Tilde{T}_{l})^{-1}$ because of the divergence of $\rho(\Tilde{T})$ for small transparencies. In the presence of dispersive SABSs, Fig. \ref{fig:DiffusiveHelical}, we have $F(eV = 0)\approx 0.41$, which is good agreement with experimental measurements on a diffusive STI / TI system \cite{tikhonov2016andreev}.  Thus, the presence of SABSs with dispersion does suppress the differential Fano factor compared to the \textit{s} - wave case, but $F>F_{N} = \frac{1}{3}$ in this case, that is, the noise power is larger than in the normal state. This feature is qualitatively different compared to the tunnel junction described in Sec. \ref{sec:Tunnel}.

This difference can be understood from the competition between noiseless transport via SABSs and Cooper pair transport. As discussed before, noiseless transport via SABS is of order $\Tilde{T}^{0}$, while Cooper pair transport is of order $\Tilde{T}^{2}$. In tunnel junctions with helical \textit{p} - wave superconductors all momenta have low transparencies and there is exactly one ZESABS. As discussed in Sec. \ref{sec:Tunnel}, in this case noiseless transport via the SABS balances the Cooper pair transport by the many off-resonant modes, and hence the differential Fano factor is the same as in the normal state. 

Meanwhile, for diffusive junctions there are many momenta for which the transparency is small, but also momenta for which $\Tilde{T}$ is large. Now, since noiseless transport via the SABS is independent of $\Tilde{T}$, it is of the same magnitude in tunnel junctions and diffusive junctions. On the other hand, transport by Cooper pairs strongly depends on transparency. Therefore, Cooper pair transport is enhanced if the transparency is enlarged. This breaks the perfect balance between Cooper pair transport and noiseless transport via the SABS, in diffusive junctions with  helical \textit{p} - wave superconductors Cooper pair transport is more important than noiseless transport via the SABS. For this reason the differential Fano factor is larger than in the normal state.

The behavior for nonzero voltages depends on the ratio between the \textit{s} - wave and \textit{p} - wave components, as shown in Fig. \ref{fig:DiffusiveHelical}. For a helical \textit{p} - wave superconductor without \textit{s} - wave component the differential Fano factor has a peak around $|eV| = \Delta_{0}$ with $F(|eV| \approx\Delta_{0})\lesssim 0.65$, while for an s+\textit{p} - wave superconductor with equal magnitudes of the \textit{s} and \textit{p} - wave contributions this peak is absent.

\begin{table*}
    \centering\setlength{\tabcolsep}{-0.1pt}\def\arraystretch{1}
    \begin{tabularx}{\linewidth}
    {|d|a a a a a a |f f f f f f|}
    \hline
       \vspace{0.01pt}Type&\multicolumn{6}{c|}{$\sigma(eV = 0)/\sigma_{N}$\cellcolor{goodgreen!20}}&\multicolumn{6}{c|}{$F(eV = 0)$\cellcolor{orange!20}}\\
        & N  & \textit{s}&B-W & helical \textit{p} &  $\text{\textit{p}}_{\text{x}}$&$\text{\textit{d}}_{\text{xy}}$& N & \textit{s}&B-W & helical \textit{p} &  $\text{\textit{p}}_{\text{x}}$&$\text{\textit{d}}_{\text{xy}}$\\
       \hline
       Tunnel junction &\vspace{0.01pt}1&\vspace{0.01pt}0&\vspace{0.01pt}0&\vspace{0.01pt}$\frac{3\pi}{4}$&\vspace{0.01pt}$ 3z^{2}$&\vspace{0.01pt}$3z^{2}$&\vspace{0.01pt}1&\vspace{0.01pt}2&\vspace{0.01pt}2&\vspace{0.01pt}1&\vspace{0.01pt}0&\vspace{0.01pt}0\\
        Chaotic cavity &\vspace{0.01pt}1&\vspace{0.01pt}1.17 &\vspace{0.01pt}1.45&\vspace{0.01pt}1.88&\vspace{0.01pt} 4&\vspace{0.01pt}2.59&\vspace{0.01pt} $\frac{1}{4}$ &\vspace{0.01pt}$\frac{1+\sqrt{2}}{4}$&\vspace{0.01pt}0.51&\vspace{0.01pt}0.32&\vspace{0.01pt}0&\vspace{0.01pt}0.17\\
        Diffusive barrier &\vspace{0.01pt}1&\vspace{0.01pt}1&\vspace{0.01pt}1.30&\vspace{0.01pt}1.84&\vspace{0.01pt}$-2\text{ln}\frac{\Tilde{T}_{l}}{2}$&\vspace{0.01pt}$\frac{1}{2}-\text{ln}\frac{\Tilde{T}_{l}}{2}$ &\vspace{0.01pt}$\frac{1}{3}$&\vspace{0.01pt}$\frac{2}{3}$&\vspace{0.01pt}0.58&\vspace{0.01pt}0.41&\vspace{0.01pt}0&\vspace{0.01pt}\hspace{-0.1cm}$\frac{3}{\text{ln}\frac{2}{\Tilde{T}_{l}}}$\\
        Double barrier & \vspace{0.01pt}1&\vspace{0.01pt}0.71&\vspace{0.01pt}1.07&\vspace{0.01pt}1.76&\vspace{0.01pt}$\frac{4}{\pi\sqrt{\Tilde{T}_{l}}}$ &\vspace{0.01pt}$\frac{4}{\pi\sqrt{\Tilde{T}_{l}}}$ & \vspace{0.01pt}$\frac{1}{2}$ &\vspace{0.01pt}$\frac{3}{4}$&\vspace{0.01pt}0.67&\vspace{0.01pt}0.60&\vspace{0.01pt}0&\vspace{0.01pt}0\\
        \hline
    \end{tabularx}
    \caption{The zero temperature and zero voltage conductance $\sigma(T = 0,V = 0)/\sigma_{N}$ and differential Fano factor $F(T = 0, V\xrightarrow{}0)$ for S / N junctions whose transparency is described by a distribution of the form $\Tilde{T}^{-\frac{p}{2}}(1-\Tilde{T})^{-\frac{1}{2}}$ and for the tunnel junction, where $\Tilde{T}(\phi) = \frac{\cos^{2}\phi}{\cos^{2}\phi+z^{2}} $ and $z\gg 1$. 
    We consider the superconductor in the normal state, and in the superconducting state. We consider superconductors without SABSs, such as an \textit{s} - wave superconductor (s), as well as superconductors with dispersive SABSs, such as the 2D helical or 2D chiral \textit{p} - wave superconductors and superconductors with dispersionless ZESABSs such as the 1D \textit{p} - wave superconductor or the 2D $\text{\textit{p}}_{\text{x}}$ - wave superconductor. Next to this, in 3D we consider superconductors with only a single ZESABS, such as the B-W superconductor. $p = 1$ corresponds to the chaotic cavity, $p = 2$ to a diffusive barrier junction and $p = 3$ to a double barrier junction. For the tunnel junction all results were obtained analytically. For the other types of junctions the results in the normal state, \textit{s} - wave superconductor and $\text{\textit{p}}_{\text{x}}$ - wave superconductor were calculated analytically, for the other superconductors numerical calculations were used. For the $\text{\textit{p}}_{\text{x}}$ - wave superconductor all channels are fully transparent and therefore the conductance is twice the number of channels times the conductance quantum. The ratio with the normal state conductance thus depends on the cut-off transparency $\Tilde{T}_{l}$ of the distribution or the average transmission, which equals $\frac{2}{3z^{2}}$ for the tunnel junction. All other entries converge as $\Tilde{T}_{l}\xrightarrow{}0$. Numerically calculated values have been rounded to two decimals.}\label{tab:FanoDiffusive}
    \end{table*}
\subsection{Double barrier junction}    
 For the double barrier junction, which consists of two clean boundaries separated by a ballistic region,  the distribution is given by Eq. (\ref{eq:Formofdistributions}) with  $p = 3$ \cite{dejong1996distribution}, there are more low transparency channels than in the diffusive conductor, while in-plane momentum is conserved. The higher importance of low transparency channels has several consequences, as can be inferred from Table \ref{tab:FanoDiffusive}. First of all, since Cooper pair transport is of higher order in transparency than quasiparticle transport this implies that the zero bias conductance for a junction with an \textit{s} - wave superconductor is smaller than the normal state conductance. Specifically, their ratio is $\frac{1}{\sqrt{2}}$. 
 For the helical \textit{p} - wave superconductor, the zero bias conductance is larger than in the normal state due to the resonance, specifically, it equals $1.76\sigma_{N}$. 
 
 Another consequence of the smaller contribution of channels with almost unit transparency is that transport is more noisy, that is, the differential Fano factor is larger in double barrier junctions than in diffusive barrier junctions, as can be seen from the yellow columns. The normal state differential Fano factor is $\frac{1}{2}$, while for an \textit{s} - wave superconductor it is $\frac{3}{4}$. Thus, the ratio of the differential Fano factor in the superconducting state and in the normal state is less than 2, which is in agreement with previous literature \cite{fauchere1998finite,schep1997transport}. This can be understood as follows. While for low-transparent tunnel junctions and diffusive barrier junctions, transport by quasiparticles and Cooper pairs can be described in qualitatively the same way, respectively tunneling and diffusion, the Fabry-Perot like resonances that appear in a double barrier junction are qualitatively different for quasiparticles and Cooper pairs. Therefore, the charge of the carriers is not the only difference between transport in normal metal / normal metal on the one hand and normal metal / superconductor junctions on the other hand. Hence a factor 2 can not be expected.
 
 We may also consider double barrier junctions with USCs. Because the junction conserves in-plane momentum, for any superconductor with dispersionless SABSs, the conductance has a perfect resonance while the noise vanishes. Thus, $\text{\textit{p}}_{\text{x}}$ -wave superconductors can not be distinguished from $\text{\textit{d}}_{\text{xy}}$ - wave superconductors in double barrier junctions. In the presence of SABSs with dispersion, the differential Fano factor becomes $0.60$, which is approximately a factor $1.2$ larger in the normal state. Thus, compared to diffusive junctions, the difference between the normal state differential Fano factor and the differential Fano factor in junctions helical \textit{p} - wave superconductors is smaller. Indeed, in double barrier junctions low-transparency channels are more important, and hence there is more balance between Cooper pair transport and resonance transport via the SABS than in diffusive junctions. 
\subsection{Chaotic cavity junction}
Lastly, we consider a chaotic cavity. A chaotic cavity consists of fully reflective walls with two holes through which electrons can enter or leave. If these holes are small compared to the size of the cavity, electrons that entered the cavity bounce off the walls many times before they leave the cavity, that is, the dwell time is much larger than the scattering time. For such chaotic cavities, if the two holes have the same size and transparency, around half of the electrons that enter leave the cavity through the same hole, while the other half leave it through the other hole \cite{agam2000shot,oberholzer2002crossover}. It was shown that for such chaotic cavities the distribution is of the type in Eq. (\ref{eq:Formofdistributions}) with $p = 1$, \textit{i.e.} the divergence near $\Tilde{T} = 0$ is of the same order as the divergence near $\Tilde{T} = 1$, while due to the multiple reflection of the electrons on the walls the cavity in-plane momentum is not conserved.

Thus, compared to the diffusive and double barrier superconductors there is a larger fraction of channels with high transmission eigenvalues in chaotic cavities. Correspondingly, the differential Fano factor in such junctions is smaller, as can be seen from the yellow columns in the "chaotic cavity" row of Table \ref{tab:FanoDiffusive}. The differential Fano factor in the normal state is $\frac{1}{4}$, while in the superconducting state it is $\frac{1+\sqrt{2}}{4}$. As for the other cases, for the $\text{\textit{p}}_{\text{x}}$ - wave superconductor there is a perfect resonance. Because a chaotic cavity does not conserve in-plane momentum, there is no such perfect resonance for the $\text{\textit{d}}_{\text{xy}}$ - wave superconductor. Unlike for diffusive junctions, the  differential Fano factor is not suppressed by the cut-off $\Tilde{T}_{l}$, but instead converges to 0.17 at zero voltage. This happens because the contribution of channels with high transparency is larger than in diffusive junctions, which means the transport by Cooper pairs is more important, and consequently the noise power is larger.  This makes the chaotic cavity the most suitable junction to distinguish between $\text{\textit{p}}_{\text{x}}$ and $\text{\textit{d}}_{\text{xy}}$ - wave superconductors.  

In the presence of SABSs with dispersion, like for the helical \textit{p} - wave superconductor, the differential Fano factor decreases to $0.32$, which is approximately a factor $1.3$ larger than in the normal state. 

Interestingly, the zero bias conductance of junctions with an \textit{s} - wave superconductor is in chaotic cavities larger than the normal state conductance, by a factor $4-2\sqrt{2}\approx 1.17$. For the helical \textit{p} - wave superconductor, the zero bias conductance exceeds the normal state conductance by a factor $1.88$.

The results of this section are summarized in Table \ref{tab:FanoDiffusive}. For all junctions, SABSs decrease the differential Fano factor (yellow) and increase the conductance (green), but the extent to which they do so is qualitatively different in all of the cases, which shows that each type of junction can be used to probe different features of the pair potential.

At finite temperatures, the features are entirely similar to those observed for the tunnel junction, that is, for nonzero temperatures the differential Fano factor vanishes at $eV = 0$ and additional local minima appear that are due to SABSs. Moreover, if the differential Fano factor vanishes already at zero temperature, like for the $\text{\textit{p}}_{\text{x}}$ - wave superconductor, a negative differential Fano factor can be observed at finite temperatures, just as in tunnel junctions. 
\section{Discussion}\label{sec:Discussion}
We have presented an exhaustive method for the characterization of unconventional superconductors via noise power and higher order current correlations in unconventional superconductor / normal metal junctions. Using noise power, the structure of the SABS energy spectrum can be clarified, an important step towards the identification of topological superconductors.

To this end, we have developed a theory for the calculation of the FCS in tunnel junctions, diffusive barrier junctions, chaotic cavities and double barrier junctions at both zero and finite temperatures. In the presence of dispersionless ZESABSs the differential Fano factor becomes negative for nonzero temperatures and $|eV|\lesssim k_{B}T$. This feature may be used to distinguish peaks that are due to SABSs from those that appear regardless of SABSs, such as the Thouless energy peak. We have shown that the SABS spectrum and dimensionality determine the main features of the noise power spectrum, and found qualitatively different results in different dimensions. Our results provide another tool to identify SABSs and hence to determine  pairing symmetries in unconventional superconductors.

Next to this, our results show that for junctions that contain a superconductor that hosts dispersive SABSs, at finite temperatures local minima appear in the differential Fano factor as a function of voltage. These local minima appear for those voltages that correspond to extrema of the SABS spectra, for example, $|eV| = |\Delta_{-}|$ for \textit{s} + helical \textit{p} - wave superconductors and $|eV| = \Delta_{s}$ and $|eV| = \Delta_{0}$ for i\textit{s} + helical \textit{p} - wave superconductors. These local minima are inherently finite temperature effects, and can not be predicted from the zero temperature differential Fano factor, illustrating the importance of taking into finite temperature effects and thermal noise in the determination of pairing types using noise power measurements.

In junctions dominated by a ballistic interface, unconventional superconductivity can be most easily identified for low transparencies, and therefore we recommend to use  tunnel junctions. It is of vital importance that the residual density of states is very small, so that quasiparticle transport induced by a finite density of states in the superconductor is negligible. If this is not the case quasiparticle transport may wash out the distinguishing features in the differential Fano factor of unconventional superconductor junctions. 

Apart from tunnel junctions, we have also studied different types of junctions, such as the diffusive barrier junction, the double barrier junction and the chaotic cavity. To this end we have extended the Tanaka-Nazarov formalism beyond tunnel junctions. The developed formalism allows to study unconventional superconductors in more general types of junctions.

Our results show that the influence of SABSs on the differential Fano factor is qualitatively similar in all types of junctions, including tunnel junctions and diffusive barriers. For example, in $\text{\textit{p}}_{\text{x}}$ - wave superconductors the differential Fano factor vanishes at $eV = 0$ even in the zero temperature limit, for all of the junctions considered in this article. Therefore, the differential Fano factor always becomes negative at finite temperatures, irrespective of the type of junction under consideration. 

However, there are also important differences. First of all, while the $\text{\textit{p}}_{\text{x}}$ - wave and $\text{\textit{d}}_{\text{xy}}$ - wave superconductors are indistinguishable by noise measurements in any junction that conserves in-plane momentum, such as tunnel junctions and double barrier junctions, they show qualitatively different results in junctions that do not conserve in-plane momentum, such as diffusive barrier or chaotic cavity junctions. Next to this, we found that for superconductors with linearly dispersive SABSs the differential Fano factor is larger than in the normal state in diffusive barrier, double barrier and chaotic cavities. In these types of junctions this may serve as an additional tool to show the existence of SABSs.

There are several possible extensions of our work. First of all, our theory can be applied to other types of superconductivity, such as \textit{f} - wave or \textit{g} - wave superconductors. Next to this, our theory can be extended to  junctions in which the normal metal is replaced by an \textit{s} - wave superconductor, as is experimentally realized in STM measurements with superconducting tips.
In such junctions the time-dependence of the phase induced by the voltage should be taken into account. This is possible in the developed formalism, by replacing all matrix products by convolutions over time. Moreover, the formalism may be applied to spin-active interfaces as well. Next to this, extension of quasiclassical theory and the Tanaka-Nazarov boundary condition to multiband superconductors would allow for investigation of counting statistics in superconductors with even more exotic types of SABSs, such as Bogoliubov Fermi surfaces \cite{schnyder2012types,schnyder2015topological,brydon2018bogoliubov}.
\section{Acknowledgements}
We thank A. Brinkman for useful discussions.
T.K. and F.S.B. acknowledge financial support from Spanish MCIN/AEI/
10.13039/501100011033 through project PID2023-148225NB-C31 (SUNRISE)
and TED2021-130292B-C42,  the Basque Government through grant IT-1591-22, and
European Union’s Horizon Europe research and innovation programme under grant agreement No 101130224 (JOSEPHINE).
Y. T. acknowledges financial support from JSPS with Grants-in-Aid for
Scientific Research (KAKENHI Grants Nos. 23K17668,
24K00583, 24K00556, and 24K00578).

\bibliography{sources}
\clearpage
\appendix
\onecolumngrid\
\section{Calculating the derivatives of the action}\label{sec:ProofAction}
In this section we show how to obtain the current and noise power from the action and thereby prove the relations between Eqs. (\ref{eq:TanakaNazarov}), (\ref{eq:Action}) and (\ref{eq:PfromS}). Within this action we focus on the 1D case of a spinless superconductor. In the 2D case, $C$ depends on $\phi$ and an additional factor $\frac{N_{\text{ch}}}{2}\int_{-\frac{\pi}{2}}^{\frac{\pi}{2}} d\phi\cos\phi$, stemming from the density of channels at incident angle $\phi$, is required.
\subsection{Current}
We have
\begin{align}
    -\frac{8\pi\hbar}{t_{0}}S(\chi) = \int_{-\infty}^{\infty}dE\text{Tr ln}\Big(4-2\Tilde{T}+\Tilde{T}\{G_{\chi}C\}\Big)\;,
\end{align}
We may write this, up to a constant that only depends on $\Tilde{T}$, and not on $\chi$, and hence does not influence any of the physical observables, as
\begin{align}
    -\frac{8\pi\hbar}{t_{0}}S(\chi) = \int_{-\infty}^{\infty}dE\Big(\text{Tr ln}\Big(1+T_{1}^{2}+T_{1}\{C,G_{\chi}\}\Big) =\int_{-\infty}^{\infty}dE\Big( \text{Tr ln}(1+T_{1}CG_{\chi})+\text{Tr ln}(1+T_{1}G_{\chi}C)\Big)\;,
\end{align}
where $T_{1}$ satisfies
\begin{align}
    \frac{T_{1}}{1+T_{1}^{2}} = \frac{\Tilde{T}}{4-2\Tilde{T}}\;.\label{eq:T1TTildeT}
\end{align}
Since by cyclic permutation of the trace we have 
\begin{align}
    \text{Tr ln}\Big(1+T_{1}CG_{\chi}\Big) = -\text{Tr}\Big(\sum_{n = 0}^{\infty}\frac{(-1)^{n}}{n}(T_{1}CG_{\chi})^{n}\Big) =-\text{Tr}\Big(\sum_{n = 0}^{\infty}\frac{(-1)^{n}}{n}(T_{1}G_{\chi}C)^{n}\Big) = \text{Tr ln}\Big(1+T_{1}G_{\chi}C\Big)\;, 
\end{align}
we may write this as
\begin{align}
    -\frac{8\pi\hbar}{t_{0}}S(\chi) &= \int_{-\infty}^{\infty}dE2\text{Tr}\Big(1+T_{1}G_{\chi}C\Big) = -2\int_{-\infty}^{\infty} dE\sum_{n = 0}^{\infty}\frac{(-1)^{n}}{n}T_{1}^{n}\text{Tr}\Big((G_{\chi}C)^{n}\Big)\;.
\end{align}
Taking the derivative with respect to $\chi$ we obtain, taking into account the cyclic property of the trace again,
\begin{align}
    -\frac{8\pi\hbar}{t_{0}}\partial_{\chi}S(\chi) &= -\int_{-\infty}^{\infty}dEi\sum_{n = 1}^{\infty}(-1)^{n}T_{1}^{n}\text{Tr}\Big((G_{\chi}C)^{n-1}(G_{\chi}\tau_{K}C-\tau_{K}G_{\chi}C)\Big)\nonumber\\& = -i\int_{-\infty}^{\infty}dE\sum_{n = 1}^{\infty}(-1)^{n}T_{1}^{n}\text{Tr}\Big(\tau_{K}\big((CG_{\chi})^{n}-(G_{\chi}C)^{n}\big)\Big)\;.
\end{align}
The sums over $n$ can now be evaluated to give
\begin{align}
    -\frac{8\pi\hbar}{t_{0}}\partial_{\chi}S(\chi) = -\int_{-\infty}^{\infty}dEi\text{Tr}\Bigg(T_{1}\tau_{K}\Big((1+T_{1}CG_{\chi})^{-1}-(1+T_{1}G_{\chi}C)^{-1}\Big)\Bigg)\;. 
\end{align}
Multiplying the first term on the right by $(1+T_{1}G_{\chi}C)^{-1}(1+T_{1}G_{\chi}C)$ and the second term by $(1+T_{1}CG_{\chi})^{-1}(1+T_{1}CG_{\chi})$  we find
\begin{align}\label{eq:CurrentChi}
     -\frac{8\pi\hbar}{t_{0}}\partial_{\chi}S(\chi) &= -\int_{-\infty}^{\infty}dEi\text{Tr}\Big(\tau_{K}(1+T_{1}\{C,G_{\chi}\}+T_{1}^{2})^{-1}(1+T_{1}G_{\chi}C-1-T_{1}CG_{\chi})\Big)\nonumber\\& = -iT_{1}\int_{-\infty}^{\infty} dE\text{Tr}\Big(\tau_{K}(1+T_{1}\{C,G_{\chi}\}+T_{1}^{2})^{-1}[G_{\chi},C]\Big)\;.
\end{align}
Substituting $\Tilde{T}$ back instead of $T_{1}$ following Eq. (\ref{eq:T1TTildeT}) and setting $\chi = 0$ we obtain
\begin{align}
    I =  -\frac{e}{8\pi\hbar}\int_{-\infty}^{\infty}dE\text{Tr}\Big(\tau_{K}(4-2\Tilde{T}+\Tilde{T}\{C,G_{N}\})^{-1}\Tilde{T}[G_{N},C]\Big)\;.
\end{align}
In the two dimensional case one needs to insert an extra factor $\frac{N_{\text{ch}}}{2}\int_{-\frac{\pi}{2}}^{\frac{\pi}{2}} d\phi\cos\phi$, where $N_{\text{ch}}$ is the number of channels, which results in Eq. (\ref{eq:TanakaNazarov}) in the main text. 
 This confirms that the action in (\ref{eq:Action}) in the main text is the correct counting action for this boundary condition.

By noticing that $C,G_{N}$ both commute with the denominator
we note that Eq. (\ref{eq:CurrentChi}) may also be written as
\begin{align}\label{eq:CurrentChi2}
    -\int_{-\infty}^{\infty} dE \frac{i}{2}\text{Tr}\Big((1+T_{1}^{2}+T_{1}\{C,G_{N}\})^{-1}T_{1}\{C,[\tau_{K},G_{N}]\}\Big)\;.
\end{align}
This form is particularly convenient when deriving the noise power.

\subsection{Noise power}
To find the noise power, we are interested in the second derivative with respect to the counting field at $\chi = 0$. To find the second derivative, we substitute the Green's function modified up to first order in the counting field, $G_{\chi}\approx G_{N}-\frac{i}{2}[\tau_{K},G_{N}]$ into the first derivative. Using that $(1+T_{1}^{2}+T_{1}\{C,G_{N}+\delta G_{N}\})^{-1}\approx(1+T_{1}^{2}+T_{1}\{C,G_{N}\})^{-1}-(1+T_{1}^{2}+T_{1}\{C,G_{N}\})^{-1}(T_{1}\{C,\delta G_{N}\})(1+T_{1}^{2}+T_{1}\{C,G_{N}\})^{-1}$ to first order in $\delta G = -\frac{i\chi}{2}[\tau_{K},G_{N}]$ we obtain
\begin{align}
    \frac{8\pi\hbar}{t_{0}}\partial_{\chi}S(\chi) &= \int_{-\infty}^{\infty}dE\frac{iT_{1}}{2}\text{Tr}\Big((1+T_{1}^{2}+\{C,G_{N}\})^{-1}\{C,[\tau_{K},G_{N}]\}\Big)\nonumber\\&+
    \frac{\chi}{4}T_{1}\text{Tr}(1+T_{1}^{2}+\{C,G_{N}\})^{-1}\{C,[\tau_{K},[\tau_{K},G_{N}]]\}\nonumber\\&-\frac{\chi}{4}T_{1}^{2}\text{Tr}\Big((1+T_{1}^{2}+\{C,G_{N}\})^{-1}
    \{C,[\tau_{K},G_{N}]\}(1+T_{1}^{2}+\{C,G_{N}\})^{-1}\{C,[\tau_{K},G_{N}]\}\Big)\nonumber\\&+O(\chi^{2})\;.
\end{align}
With this we conclude that the second derivative is
\begin{align}
    &\frac{8\pi\hbar}{t_{0}}\partial_{\chi\chi}S(\chi) = \int_{-\infty}^{\infty}dE\text{Tr}\Bigg(-\frac{T_{1}}{2}(1+T_{1}^{2}+\{C,G_{N}\})^{-1}(\{C,\tau_{K}G_{N}\tau_{K}\}-\{C,G_{N}\})\nonumber\\&-\frac{T_{1}^{2}}{4}(1+T_{1}^{2}+T_{1}\{C,G_{N}\})^{-1}(C\tau_{K}G_{N}-CG_{N}\tau_{K}+\tau_{K}G_{N}C-G_{N}\tau_{K}C)(1+T_{1}^{2}+T_{1}\{C,G_{N}\})^{-1}\{C,[\tau_{K},G_{N}]\}\Bigg)\nonumber\\&+O(\chi)\;.
\end{align}
Thus, the noise power $P_{N} = \frac{2e^{2}}{t_{0}}\partial_{\chi\chi}S(\chi)$ for this action is given by
\begin{align}
    P_{N} &= -\frac{e^{2}}{4\pi\hbar}\int_{-\infty}^{\infty}dE\text{Tr}\Bigg(\frac{T_{1}}{2}(1+T_{1}^{2}+T_{1}\{C,G_{N}\})^{-1}(\{C,\tau_{K}G_{N}\tau_{K}\}-\{C,G_{N}\})\Bigg)\nonumber\\&+\text{Tr}\Bigg(\Big(\frac{T_{1}}{2}(1+T_{1}^{2}+T_{1}\{C,G_{N}\})^{-1}\{C,[\tau_{K},G_{N}]\}\Big)^{2}\Bigg)\;.
\end{align}
Reinserting $\Tilde{T}$ for $T_{1}$ using Eq. (\ref{eq:T1TTildeT}) we have
\begin{align}\label{eq:PNdefSupp}
    P_{N} &= -\frac{e^{2}}{4\pi\hbar}\int_{-\infty}^{\infty}\text{Tr}\Bigg(\frac{\Tilde{T}}{2}(4-2\Tilde{T}+\Tilde{T}\{C,G_{N}\})^{-1}(\{C,\tau_{K}G_{N}\tau_{K}\}-\{C,G_{N}\})\nonumber\\&+\frac{\Tilde{T}^{2}}{4}\Big((4-2\Tilde{T}+\Tilde{T}\{C,G_{N}\})^{-1}\{C,[\tau_{K},G_{N}]\}\Big)^{2}\Bigg)\;.
\end{align}
The generalization to 2D can be found by adding $\frac{N_{\text{ch}}}{2}\int_{-\frac{\pi}{2}}^{\frac{\pi}{2}} d\phi\cos\phi$ in front of the right hand side of Eq. (\ref{eq:PNdefSupp}). Using this, Eq. (\ref{eq:PNdef}) in the main text is obtained.
\clearpage
\section{One dimensional superconductors}\label{sec:1Danalytical}
In this section we show that for the one-dimensional junction analytical results can be obtained both in the low and and high transparency limits. The obtained values are in good correspondence with both our numerical results presented in Figs. \ref{fig:FinNoiseSdom} and \ref{fig:FinNoiseSdomISP} in the main text and those discussed in Appendix Sec. \ref{sec:1Dnumerical} and show the main characteristics of the current and noise power that we refer to in other parts in the supplemental material. In the 1D limit, we may omit the integration $\frac{1}{2}\int_{-\frac{\pi}{2}}^{\frac{\pi}{2}} d\phi\cos\phi$ over angles, and set $N_{\text{ch}} = 1, \phi = 0$. In each calculation, we use $G_{N}^{R} = -G_{N}^{A} = \tau_{3}$ and $G_{N}^{K} = 2f_{L}\tau_{3}+2f_{T}$ for the Green's function on the normal side. 

The noise power and current are given by
\begin{align}
    P_{N} &= \frac{e^{2}}{4\pi\hbar}\int_{-\infty}^{\infty} dE \mathcal{P}\;,\\
    I &= \frac{e}{8\pi\hbar}\int_{-\infty}^{\infty} dE \mathcal{I} \;,
\end{align}
and from Eqs. (\ref{eq:TanakaNazarov}) and (\ref{eq:PNdef}) in the main text we infer
\begin{align}
    \mathcal{I}\label{eq:MathcalI} &= -\text{Tr}\Bigg(\tau_{K}\Big((4-2\Tilde{T})\mathbf{1}+\Tilde{T}\{G_{N},C\}\Big)^{-1}\Tilde{T}[G_{N},C]\Bigg)
    \;.\\
    \mathcal{P}\label{eq:MathcalP}&= -\text{Tr}\Bigg(\frac{\Tilde{T}}{2}(4-2\Tilde{T}+\Tilde{T}\{C,G_{N}\})^{-1}(\{C,\tau_{K}G_{N}\tau_{K}\}-\{C,G_{N}\})+\frac{\Tilde{T}^{2}}{4}\Big((4-2\Tilde{T}+\Tilde{T}\{C,G_{N}\})^{-1}\{C,[\tau_{K},G_{N}]\}\Big)^{2}\Bigg)\;.
\end{align}
We are interested in the low transparency limit, for which, unless the elements of $C$ are very large, we have the following expansion up to second order in $\Tilde{T}$:
\begin{align}
    \mathcal{I}\label{eq:IlowT}&=-(\frac{\Tilde{T}(1+\frac{\Tilde{T}}{2})}{4})\text{Tr}\Big(\tau_{K}[G_{N},C]\Big)+\frac{\Tilde{T^{2}}}{16}\text{Tr}\Big(\tau_{K}\{G_{N},C\}[G_{N},C]\Big)+O(\Tilde{T}^{3})\;.\\
    \mathcal{P}\label{eq:PlowT} &= -\text{Tr}\Big(\frac{\Tilde{T}(1+\frac{\Tilde{T}}{2})}{8}\{C,\tau_{K}G_{N}\tau_{K}-G_{N}\}\Big)-\frac{\Tilde{T^{2}}}{64}\text{Tr}\Bigg(\Big(\{C,[\tau_{K},G_{N}]\}\Big)^{2}-2\{C,G_{N}\}\{C,\tau_{K}G_{N}\tau_{K}-G_{N}\}\Bigg)+O(\Tilde{T}^{3})\;.
\end{align}
For low transparency junctions this expression works well in large voltage windows. However, for energies close to a SABS, or the \textit{s} - wave gap edge, the elements of $C$ are very large and the expansion does not work. In  this case we have
\begin{align}
    \mathcal{I}\label{eq:Ireson}&\approx-\text{Tr}\Big(\tau_{K}\{G_{N},C\}^{-1}[G_{N},C]\Big)\;,\\
    \mathcal{P}\label{eq:Preson}&\approx-\text{Tr}\Bigg(\frac{1}{2}\{G_{N},C\}^{-1}\{C,\tau_{K}G_{N}\tau_{K}-G_{N}\}+\frac{1}{4}\Big(\{G_{N},C\}^{-1}\{C,[\tau_{K},G_{N}]\}\Big)^{2}\Bigg)\;,
\end{align}
which is independent of the transparency of the junction. Thus, near a SABS the voltage and differential noise do not depend on the transparency of the barrier.
\subsection{Normal metals}
First, to verify that our action reproduces well-known results in the absence of unconventional superconductivity, we consider a one-dimensional tunnel junction between two normal metals. In this case the Green's functions on both side are that of a normal metal, that is, 
$C^{R} = -C^{A} = G_{N}^{R} = -G_{N}^{A} = \tau_{3}$, $C^{K} = f_{S}\tau_{3}$ and $G^{K} = 2f_{L}\tau_{3}+2f_{T}$. We have
\begin{align}
    [G_{N},C]&=4\begin{bmatrix}
        0&0&f_{S}-(f_{L}+f_{T})&0\\
        0&0&0&f_{S}-(f_{L}-f_{T})\\
        0&0&0&0\\
        0&0&0&0
    \end{bmatrix}\; ,\\
    \text{Tr}\Big(\tau_{K}[G_{N},C]\Big)&=-8f_{T}\;,\\
    \{C,\tau_{K}G_{N}\tau_{K}-G_{N}\}&=-4\rho_{0}\otimes((1-f_{S}f_{L})\tau_{0}-f_{S}f_{T}\tau_{3})\;,\\
    \text{Tr}\Big(\{C,\tau_{K}G_{N}\tau_{K}-G_{N}\}\Big)&=-16(1-f_{S}f_{L})\;.
\end{align}
Thus, in the tunneling limit we obtain
\begin{align}
    \mathcal{I} &= 2\Tilde{T}f_{T}\;\label{eq:Inormal},\\
    \mathcal{P} &= 2\Tilde{T}(1-f_{S}f_{L})\;. \label{eq:Pnormal}
\end{align}

At zero temperature, the distribution function has two steps at $E = \pm eV$. Thus, the voltage derivative of the distribution functions consists of delta functions at $\pm|eV|$. This makes the integration trivial. It yields
\begin{align}
    \frac{\partial I}{\partial V} &= 2\frac{e^{2}}{8\pi\hbar}2\Tilde{T} = \frac{e^{2}}{2\pi\hbar}\Tilde{T}\;,\\
    \frac{\partial P_{N}}{\partial V} &= 2\frac{e^{3}}{2\pi\hbar}2\Tilde{T} = \frac{e^{3}}{\pi\hbar}\Tilde{T} \;,
\end{align}
where one factor $2$ comes from the fact that we need to sum over $\pm |eV|$ and the other from Eqs. (\ref{eq:Inormal}) and (\ref{eq:Pnormal}). 
Thus, the conductance is the product of half the conductance quantum,  and transparency, as expected because we consider a spinless metal, while 
\begin{align}
    F&= \frac{1}{2e}\frac{\frac{\partial P_{N}}{\partial V}}{\frac{\partial I}{\partial V}} = 1\;.
\end{align}
For a normal metal with electrons that do carry spin, the matrix space is twice as large and consequently the current and noise are both doubled, while the differential Fano factor remains 1.
These results are in line with Eqs. (\ref{eq:ConductanceNormal}) and (\ref{eq:NoisePowerNormal}) in the main text.

\subsection{\textit{s} - wave superconductors}
We next consider the case of a one dimensional \textit{s} - wave superconductor. For \textit{s} - wave superconductors we have for $E<\Delta_{0}$ that \begin{align}C^{R} = C^{A} &= \frac{-i}{\sqrt{\Delta_{0}^{2}-E^{2}}}\begin{bmatrix}
    E&\Delta_{0}\\-\Delta_{0}&-E
\end{bmatrix}\; ,\\C^{K} &= 0\;.\end{align} On the other hand, for $E>\Delta_{0}$ we have \begin{align}C^{R} = -C^{A} &=\frac{1}{\sqrt{E^{2}-\Delta_{0}^{2}}}\begin{bmatrix}
    E&\Delta_{0}\\-\Delta_{0}&-E
\end{bmatrix}\; ,\\C^{K}&=2\tanh{\frac{E}{k_{B}T}}C^{R}\;. \end{align}

Thus, it is convenient to separate the contributions for $|E|<\Delta_{0}$ and $|E|>\Delta_{0}$. We may expand $P$ and $I$ up to lowest order in transparency
\begin{align}
    \frac{4\pi\hbar}{e^{2}}P &= \int_{|E|<\Delta_{0}} dE \mathcal{P}_{1} + \int_{|E|>\Delta_{0}} dE \mathcal{P}_{2}\;,\label{eq:dimensionlessP}\\
    \frac{8\pi\hbar}{e}I & = \int_{|E|<\Delta_{0}}dE\mathcal{I}_{1}+\int_{|E|>\Delta_{0}} dE\mathcal{I}_{2}\label{eq:dimensionlessI}\;,
\end{align}
where the notation $\mathcal{I}_{1,2}$ and $\mathcal{P}_{1,2}$ are $\mathcal{I}$ and $\mathcal{P}$ evaluated for $|E|<\Delta_{0}$ (subscript 1) and $|E|>\Delta_{0}$ 
(subscript 2). Substitution of $C$ and $G_{N}$ into Eq. (\ref{eq:MathcalP})  gives the following equalities relevant for the calculation of the contribution to the current for $|E|<\Delta_{0}$:
\begin{align}
    [G_{N},C] &= \frac{i}{\sqrt{\Delta_{0}^{2}-E^{2}}}\begin{bmatrix}
        0&-2\Delta_{0}&0&-4f_{L}\Delta_{0}\\
        -2\Delta_{0}&0&-4f_{L}\Delta_{0}&0\\
        0&0&0&2\Delta_{0}\\
        0&0&2\Delta_{0}&0
    \end{bmatrix}\; ,\\
    \text{Tr}\Big(\tau_{K}[G_{N},C]\Big) &= 0\;,\\
    \{G_{N},C\} &= \frac{-i}{\sqrt{\Delta_{0}^{2}-E^{2}}}\begin{bmatrix}
        2E&0&4E(f_{L}+f_{T})&4f_{T}\Delta_{0}\\
        0&2E&-4f_{T}\Delta_{0}&4E(f_{L}-f_{T})\\
        0&0&-2E&0\\0&0&0&-2E
    \end{bmatrix}\; ,\\
    \{G_{N},C\}[G_{N},C]& = \frac{-1}{\Delta_{0}^{2}-E^{2}}\begin{bmatrix}
        0&4E\Delta_{0}&-8f_{T}\Delta_{0}^{2}&-8Ef_{T}\Delta_{0}\\4E\Delta_{0}&0&8Ef_{T}\Delta_{0}&8f_{T}\Delta_{0}^{2}\\0&0&0&4E\Delta_{0}\\
        0&0&4E\Delta_{0}&0
    \end{bmatrix}\; ,\\
    \text{Tr}\Big(\tau_{K}\{G_{N},C\}[G_{N},C]\Big) &=  16\frac{\Delta_{0}^{2}}{\Delta_{0}^{2}-E^{2}}f_{T}\;.
\end{align}

For the calculation of the conribution to the noise power for $|E|<\Delta_{0}$ we obtain the following equalities
\begin{align}
    \{C,\tau_{K}G_{N}\tau_{K}-G_{N}\} &=\frac{4i}{\sqrt{\Delta_{0}^{2}-E^{2}}}\begin{bmatrix}
        E&0&E(f_{L}+f_{T})&f_{T}\Delta_{0}\\0&E&-f_{T}\Delta_{0}&E(f_{L}-f_{T})\\-E(f_{L}+f_{T})&-f_{T}\Delta_{0}&-E&0\\f_{T}\Delta_{0}&-E(f_{L}-f_{T})&0&-E
    \end{bmatrix}\; , \\
    \text{Tr}\Big(\{C,\tau_{K}G_{N}\tau_{K}-G_{N}\}\Big) &= 0,\\
      \{C,[\tau_{K},G_{N}]\}^2 &=\frac{16}{\Delta_{0}^{2}-E^{2}}\rho_{0}\otimes\begin{bmatrix}
          (1-(f_{L}+f_{T})^{2})E^{2}+(f_{L}^{2}-1)\Delta_{0}^{2}&-2E\Delta_{0}f_{L}f_{T}\\2E\Delta_{0}f_{L}f_{T}&(1-(f_{L}-f_{T})^{2})E^{2}+(f_{L}^{2}-1)\Delta_{0}^{2}
      \end{bmatrix}\;, \\
    \text{Tr}\Big(\{C,[\tau_{K},G_{N}]\}^2\Big) &=\frac{1}{\Delta_{0}^{2}-E^{2}}\Big(64(1-(f_{L}^{2}+f_{T}^{2}))E^{2}-64(1-f_{L}^{2})\Delta_{0}^{2}\Big)\;,
    \end{align}
    \begin{align}
    &\{C,G_{N}\}\{C,\tau_{K}G_{N}\tau_{K}-G_{N}\}=\frac{8\Delta_{0}^{2}}{\Delta_{0}^{2}-E^{2}}\nonumber\\&\begin{bmatrix}
        E^{2}(1-2(f_{L}+f_{T})^{2})+2f_{T}^{2}\Delta_{0}^{2}&-4E\Delta_{0}f_{L}f_{T}&-E^{2}(f_{L}+f_{T})&-E\Delta_{0}f_{T}\\
        4E\Delta_{0}f_{L}f_{T}&E^{2}(1-2(f_{L}+f_{T})^{2})+2f_{T}^{2}\Delta_{0}^{2}&E\Delta_{0}f_{T}&-E^{2}(f_{L}-f_{T}),\\E^{2}(f_{L}+f_{T})&E\Delta_{0}f_{T}&E^{2}&0\\-E\Delta_{0}f_{T}&E^{2}(f_{L}-f_{T})&0&E^{2}
    \end{bmatrix}\; ,\\
    &\text{Tr}\Big(\{C,G_{N}\}\{C,\tau_{K}G_{N}\tau_{K}-G_{N}\}\Big)=\frac{1}{\Delta_{0}^{2}-E^{2}}\Big(32(1-(f_{L}^{2}+f_{T}^{2}))E^{2}+32f_{T}^{2}\Delta_{0}^{2}\Big)\;.
\end{align}
Substituting this into Eqs. (\ref{eq:IlowT}) and (\ref{eq:PlowT}) we obtain
\begin{align}
    \mathcal{I}_{1}&=\Tilde{T}^{2}\frac{\Delta_{0}^{2}}{\Delta_{0}^{2}-E^{2}}f_{T}\;,\\
    \mathcal{P}_{1}&=\Tilde{T}^{2}\frac{\Delta_{0}^{2}}{\Delta_{0}^{2}-E^{2}}(1+f_{T}^{2}-f_{L}^{2})\label{eq:P1S}\;.
\end{align}
On the other hand, for $|E|>\Delta_{0}$ we have
\begin{align}
    [G_{N},C]&=\frac{1}{\sqrt{E^{2}-\Delta_{0}^{2}}}\begin{bmatrix}0&2\Delta_{0}&-4E(f_{L}+f_{T}-f_{S})&-4\Delta_{0} f_{T}\\
    2\Delta_{0}&0&4\Delta_{0} f_{T}&-4E(f_{L}-f_{T}-f_{S})\\0&0&0&2\Delta_{0}\\0&0&2\Delta_{0}&0\end{bmatrix}\; ,\\
    \text{Tr}\Big(\tau_{K}[G_{N},C]\Big) &= -8\frac{E}{\sqrt{E^{2}-\Delta_{0}^{2}}}f_{T}\;.
\end{align}
and for the noise power
\begin{align}
    &\{C,\tau_{K}G_{N}\tau_{K}-G_{N}\}\nonumber\\&=\frac{4}{\sqrt{E^{2}-\Delta_{0}^{2}}}\begin{bmatrix}
        E(f_{S}(f_{L}+f_{T})-1)&-f_{S}(f_{L}-f_{T})\Delta_{0}&0&(f_{L}-2f_{S})\Delta_{0}\\-f_{S}(f_{L}+f_{T})\Delta_{0}&E(f_{S}(f_{L}-f_{T})-1)&(f_{L}-2f_{S})\Delta_{0}&0\\
        0&f_{L}\Delta_{0}&E(f_{S}(f_{L}+f_{T})-1)&f_{S}(f_{L}+f_{T})\Delta_{0}\\f_{L}\Delta_{0}&0&f_{S}(f_{L}-f_{T})\Delta_{0}&E(f_{S}(f_{L}-f_{T})-1)
    \end{bmatrix}\; ,\\
    &\text{Tr}\Big(\{C,\tau_{K}G_{N}\tau_{K}-G_{N}\}\Big)=-\frac{16E}{\sqrt{E^{2}-\Delta_{0}^{2}}}(1-f_{L}f_{S})\;.
\end{align}

Consequently, using Eqs. (\ref{eq:IlowT}) and (\ref{eq:PlowT}),
\begin{align}
    \mathcal{I}_{2}&=2\Tilde{T}\frac{E}{\sqrt{E^{2}-\Delta_{0}^{2}}}f_{T}\;,\\
    \mathcal{P}_{2}&=2\Tilde{T}\frac{E}{\sqrt{E^{2}-\Delta_{0}^{2}}}(1-f_{L}f_{S})\;.
\end{align}
At zero temperature, the distribution functions are step functions, with steps at $|eV| = |E|$. Therefore, the voltage derivative of these functions gives delta functions at $|eV| = |E|$, which makes the evaluation of the integral trivial. Thus, for $|eV|<\Delta_{0}$ we obtain, taking into account we have to sum the contributions of $E = eV$ and $E = -eV$,
\begin{align}
    \frac{\partial I}{\partial V}\label{eq:IBelowGap}& = \frac{e^{2}}{4\pi\hbar}\frac{\Delta_{0}^{2}}{\Delta_{0}^{2}-(eV)^{2}}\Tilde{T}^{2}\;,\\
    \frac{\partial P_{N}}{\partial V}&=\frac{e^{3}}{\pi\hbar}\frac{\Delta_{0}^{2}}{\Delta_{0}^{2}-(eV)^{2}}\Tilde{T}^{2}\;.
\end{align}

Thus, the differential Fano factor is given by
\begin{align}
    F(|eV|<\Delta_{0}) \approx 2\;.
\end{align}

On the other hand, for $|eV|>\Delta_{0}$ we have

\begin{align}
    \frac{1}{2e}\frac{\partial P_{N}}{\partial V}& = \frac{\partial I}{\partial V} =  \frac{e^{2}}{\pi\hbar}\frac{eV}{\sqrt{(eV)^{2}-\Delta_{0}^{2}}}\Tilde{T}+O(\Tilde{T}^{2})\;,
\end{align}
that is, the conductance is exactly one conductance quantum multiplied by the transmission of the junction, and
\begin{align}
    F(|eV|>\Delta_{0}) = \frac{1}{2e}\frac{\frac{\partial P_{N}}{\partial V}}{\frac{\partial I}{\partial V}} \approx 1 \label{eq:FAbovegap}\;.
\end{align}

We remark that the above solutions are only valid far enough away from the gap edge, because of the square root divergence of the elements of the Green's function near $E = \Delta_{0}$. Because the divergence is of the square root type, the approximation only fails in an energy window of order $\Tilde{T}^{2}$. Thus, we require
$||eV|-\Delta_{0}|\gg \Tilde{T}^{2}\Delta_{0}$ for these expressions to hold.

These results are in good agreement with the numerical results presented in Fig. \ref{fig:FinNoiseSdom}(a) in the main text.
\subsubsection{Near the gap edge}
For $||eV|-\Delta_{0}|\ll \Tilde{T}^{2}\Delta_{0}$ the elements of $C$ are diverging and we need to use Eqs. (\ref{eq:Ireson}) and (\ref{eq:Preson}).
For $-\Tilde{T}^{2}\Delta_{0}\ll|eV|-\Delta_{0}<0$ we have, defining $\delta E = \Delta_{0}-eV$:
\begin{align}
    C^{R}&=C^{A} \approx -i\sqrt{\frac{\Delta_{0}}{2\delta E}}\begin{bmatrix}
        1&1\\-1&-1
    \end{bmatrix}\; ,\\
    C^{K}&=0\;.
\end{align}
Consequently, we have
\begin{align}
    \{C,G_{N}\}&=-i\sqrt{\frac{2\Delta_{0}}{\delta E}}\begin{bmatrix}
        1&0&2(f_{L}+f_{T})&2f_{T}\\0&1&-2f_{T}&2(f_{L}-f_{T})\\0&0&-1&0\\0&0&0&-1
    \end{bmatrix}\; ,\\
    \{C,G_{N}\}^{-1}&=i\sqrt{\frac{\delta E}{2\Delta_{0}}}\begin{bmatrix}
        1&0&2(f_{L}+f_{T})&2f_{T}\\0&1&-2f_{T}&2(f_{L}-f_{T})\\0&0&-1&0\\0&0&0&-1
    \end{bmatrix}\;,\\
    \{C,G_{N}\}^{-1}[G_{N},C]&=\begin{bmatrix}
        0&1&-2f_{T}&-2f_{T}\\1&0&2f_{T}&2f_{T}\\0&0&0&1\\0&0&1&0
    \end{bmatrix}\; ,\\
    \text{Tr}\Big(\tau_{K}\{C,G_{N}\}^{-1}[G_{N},C]\Big)&=-4f_{T}\;.
\end{align}
For the noise power we obtain
\begin{align}
\{C,\tau_{K}G_{N}\tau_{K}-G_{N}\}&=4i\sqrt{\frac{\Delta_{0}}{2\delta E}}\begin{bmatrix}
    1&0&f_{L}+f_{T}&f_{T}\\0&1&-f_{T}&f_{L}-f_{T}\\-(f_{L}+f_{T})&-f_{T}&-1&0\\f_{T}&-(f_{L}-f_{T})&0&-1
\end{bmatrix}\;,\\
    \{C,G_{N}\}^{-1}\{C,\tau_{K}G_{N}\tau_{K}-G_{N}\}&=\begin{bmatrix}
        -2+4f_{L}^{2}+8f_{L}f_{T}&8f_{L}f_{T}&2(f_{L}+f_{T})&2f_{T}\\-8f_{L}f_{T}&-2+4f_{L}^{2}-8f_{L}f_{T}&-2f_{T}&2(f_{L}-f_{T})\\-2(f_{L}+f_{T})&-2f_{T}&-2&0\\2f_{T}&-2(f_{L}-f_{T})&0&-2
    \end{bmatrix}\; ,\\
    \text{Tr}\Big(\{C,G_{N}\}^{-1}\{C,\tau_{K}G_{N}\tau_{K}-G_{N}\}\Big)&=8(-1+f_{L}^{2})\;,\\
    \{C,G_{N}\}^{-1}\{C,[\tau_{K},G_{N}]\}&=\begin{bmatrix}2(f_{L}-f_{T})&2f_{L}&-2+4(f_{L}^{2}+f_{T}^{2}+f_{L}f_{T})&-2+4(f_{L}^{2}+f_{T}^{2})\\-2f_{L}&-2(f_{L}+f_{T})&2-4(f_{L}^{2}+f_{T}^{2})&2-4(f_{L}^{2}+f_{T}^{2}-f_{L}f_{T})\\-2&-2&-2(f_{L}+f_{T})&-2f_{L}\\2&2&2f_{L}&2(f_{L}-f_{T})\end{bmatrix}\; ,\\
    (\{C,G_{N}\}^{-1}\{C,[\tau_{K},G_{N}]\})^{2}&=4\Bigg(\frac{1}{2}(\rho_{0}+\rho_{3})\otimes\begin{bmatrix}
        f_{T} (-4 f_{L} + f_{T})& -4 f_{L} f_{T}\\4 f_{L} f_{T}&  f_{T} (4 f_{L} + f_{T})
    \end{bmatrix}\nonumber\\&+\frac{1}{2}(\rho_{0}-\rho_{3})\otimes\begin{bmatrix}
       f_{T}^2& 0 \\0&f_{T}^{2}
    \end{bmatrix}\nonumber\\&+\frac{1}{2}(\rho_{1}+i\rho_{2})\otimes\begin{bmatrix}
        -2 f_{T} (-1 + 
    2 (f_{L}^2 + f_{L} f_{T} + f_{T}^2))& -2 f_{T} (-1 + 2 f_{L}^2 + 2 f_{T}^2)\\2 f_{T} (-1 + 2 f_{L}^{2} + 2 f_{T}^{2})& 
 2 f_{T} (-1 + 2 (f_{L}^{2} - f_{L} f_{T} + f_{T}^{2}))
    \end{bmatrix}\nonumber\\&+\frac{1}{2}(\rho_{1}-i\rho_{2})\otimes\begin{bmatrix}
        2f_{T}&2f_{T}\\-2f_{T}&-2f_{T}
    \end{bmatrix}\Bigg)\;,\\
    \text{Tr}\Big((\{C,G_{N}\}^{-1}\{C,[\tau_{K},G_{N}]\})^{2}\Big)&=16f_{T}^{2}\;.
\end{align}

Thus, we have, using Eqs. (\ref{eq:Ireson}) and (\ref{eq:Preson}),
\begin{align}
    \mathcal{I}_{1}(E\approx\Delta_{0})&=4f_{T}\;,\\
    \mathcal{P}_{1}(E\approx\Delta_{0})&\approx 4(1-f_{L}^{2}-f_{T}^{2})\;.
\end{align}
That is,
\begin{align}
    \frac{\partial I}{\partial V}(|eV|\xrightarrow{}\Delta_{0}^{-})&=\frac{e^{2}}{\pi\hbar}\;,\label{eq:Iresonance}\\
    \frac{\partial P_{N}}{\partial V}(|eV|\xrightarrow{}\Delta_{0}^{-})&=0\;,\\
    F(|eV|\xrightarrow{}\Delta_{0}^{-}) = 0\;.\label{eq:Fresonsance}
\end{align}
Thus, the conductance is the conductance quantum, that is, twice the spinless conductance quantum, a perfect resonance, while the noise power vanishes.

For $0< |eV|-\Delta_{0}\ll \Tilde{T}^{2}\Delta_{0}$ we have:
\begin{align}
    C^{R}&=-C^{A}\approx \sqrt{\frac{\Delta_{0}}{2|\delta E|}}\begin{bmatrix}
        1&1\\-1&-1
    \end{bmatrix}\; \label{eq:WeakDivergence},\\
    C^{K} &= 2f_{S}C^{R}\;.
\end{align}

Thus, we have
\begin{align}
    \{C,G_{N}\}&=\sqrt{\frac{2\Delta_{0}}{|\delta E|}}\begin{bmatrix}
        1&0&0&-2(f_{L}-f_{S})\\0&1&-2(f_{L}-f_{S})&0\\0&0&1&0\\0&0&0&1
    \end{bmatrix}\; ,\\
    \{C,G_{N}\}^{-1}&=\sqrt{\frac{|\delta E|}{2\Delta_{0}}}\begin{bmatrix}
        1&0&0&2(f_{L}-f_{S})\\0&1&2(f_{L}-f_{S})&0\\0&0&1&0\\0&0&0&1
    \end{bmatrix}\; ,\\
    \{C,G_{N}\}^{-1}[G_{N},C]&=\begin{bmatrix}
        0&1&-2f_{T}&-2f_{T}\\1&0&2f_{T}&2f_{T}\\0&0&0&1\\0&0&1&0
    \end{bmatrix}\; ,\\
    \text{Tr}\Big(\tau_{K}\{C,G_{N}\}^{-1}[G_{N},C]\Big)&=-4f_{T}\;.
\end{align}
For the noise power we obtain
\begin{align}
&\{C,\tau_{K}G_{N}\tau_{K}-G_{N}\} =\nonumber\\& \sqrt{\frac{\Delta_{0}}{2|\delta E|}}\begin{bmatrix}
    -4 + 4 f_{S} (f_{L} + f_{T})& 4 f_{S} (-f_{L} + f_{T})& 0& 
 4 (f_{L} - 2 f_{S})\\ -4 f_{S} (f_{L} + f_{T})& -4 + 4 f_{L} f_{S} - 4 f_{S} f_{T}& 
 4 (f_{L} - 2 f_{S})& 0\\ 0& 4 f_{L}& -4 + 4 f_{S} (f_{L} + f_{T})& 
 4 f_{S} (f_{L} + f_{T})\\4 f_{L}& 0& 4 f_{S} (f_{L} - f_{T})& -4 + 4 f_{L} f_{S} - 4 f_{S} f_{T}
\end{bmatrix}\\
    &\{C,G_{N}\}^{-1}\{C,\tau_{K}G_{N}\tau_{K}-G_{N}\}=
    \frac{1}{2}(\rho_{0}+\rho_{3})\otimes\begin{bmatrix}
        -2 + 4 f_{L}^2 - 2 f_{L} f_{S} + 2 f_{S} f_{T}&-2f_{S}(f_{L}-f_{T})\\-2f_{S}(f_{L}+f_{T})&-2 + 4 f_{L}^2 - 2 f_{L} f_{S} - 2 f_{S} f_{T}
    \end{bmatrix}\nonumber\\&+\frac{1}{2}(\rho_{0}-\rho_{3})\otimes\begin{bmatrix}
        2(-1+f_{S}(f_{L}+f_{T}))&2 f_{S}(f_{L} + f_{T}) \\
        2  f_{S} (f_{L} - f_{T})&2 (-1 + f_{S} (f_{L} - f_{T}))
    \end{bmatrix}\nonumber\\&+\frac{1}{2}(\rho_{1}+i\rho_{2})\otimes\begin{bmatrix}
        4 (f_{L} - f_{S}) f_{S} (f_{L} - f_{T})& (4 f_{L}^2 f_{S} + 4 f_{S}^2 f_{T} - 2 f_{L} (1 + 2 f_{S} (f_{S} + f_{T})))\\ (4 f_{L}^2 f_{S} - 4 f_{S}^2 f_{T} - 2 f_{L} (1 + 2 f_{S} (f_{S} - f_{T})))&4 (f_{L} - f_{S}) f_{S} (f_{L} + f_{T})
    \end{bmatrix}\nonumber\\&+\frac{1}{2}(\rho_{1}-i\rho_{2})\otimes\begin{bmatrix}
        0&2f_{L}\\2f_{L}&0
    \end{bmatrix}\; ,\\
    &\text{Tr}\Big(\{C,G_{N}\}^{-1}\{C,\tau_{K}G_{N}\tau_{K}-G_{N}\}\Big)=8(-1+f_{L}^{2})\;,\\
    &\{C,[\tau_{K},G_{N}]\}=\sqrt{\frac{\Delta_{0}}{2|\delta E|}}\begin{bmatrix}
        4 f_{S} - 4 (f_{L} + f_{T})&
 4 (-f_{L} + f_{S})& 0& -8 f_{S} f_{T}\\4 (f_{L} - f_{S})& -4 (-f_{L} + f_{S} + 
    f_{T})& -8 f_{S} f_{T}& 0\\0& 0& 4 f_{S} - 4 (f_{L} + f_{T})& 
 4 (-f_{L} + f_{S})\\0& 0& 4 (f_{L} - f_{S})& -4 (-f_{L} + f_{S} + f_{T})
    \end{bmatrix}\\
    &\{C,G_{N}\}^{-1}\{C,[\tau_{K},G_{N}]\}=-2\rho_{0}\otimes\begin{bmatrix}
        (f_{L}+f_{T}-f_{S})&f_{L}-f_{S}\\-(f_{L}-f_{S})&- (f_{L}-f_{T}-f_{S})
    \end{bmatrix}\nonumber\\&+\frac{1}{2}(\rho_{1}+i\rho_{2})4\otimes\begin{bmatrix}
        (f_{L}-f_{S})^{2}&(f_{L}-f_{S})^{2}-f_{L}f_{T}\\-(f_{L}-f_{S})^{2}-f_{L}f_{T}&-(f_{L}-f_{S})^{2}
    \end{bmatrix}\; ,\\
    &(\{C,G_{N}\}^{-1}\{C,[\tau_{K},G_{N}]\})^{2}=4\rho_{0}\otimes\begin{bmatrix}
        f_{T}^{2}+2f_{T}(f_{L}-f_{S})&2f_{T}(f_{L}-f_{S})\\-2f_{T}(f_{L}-f_{S})&f_{T}^{2}-2f_{T}(f_{L}-f_{S})
    \end{bmatrix}\nonumber\\&+\frac{1}{2}(\rho_{1}+i\rho_{2})\otimes16\begin{bmatrix}
        -f_{T}(f_{L}-f_{S})^{2}&f_{T}^{2}f_{L}-f_{T}(f_{L}-f_{S})^{2}\\f_{T}^{2}f_{L}+f_{T}(f_{L}-f_{S})^{2}&f_{T}(f_{L}-f_{S})^{2}
    \end{bmatrix}\\
    &\text{Tr}\Big((\{C,G_{N}\}^{-1}\{C,\tau_{K},G_{N}]\})^{2}\Big)=16f_{T}^{2}\;.
\end{align}

All of the evaluated traces are equal to those obtained for the limit $-\Tilde{T}^{2}\Delta_{0}\ll |eV| - \Delta_{0}<0$.
Therefore, according to Eqs. (\ref{eq:Ireson}) and (\ref{eq:Preson}), both conductance and noise power are continuous at $|eV| = \Delta_{0}$, even though the Green's functions are not. The conductance is twice the normal state conductance, and the differential noise power vanishes. 

This is in good agreement with our numerical results in Fig. \ref{fig:FinNoiseSdom}(a), in which a local minimum in the differential Fano factor appears at $|eV| = \Delta_{0}$. We note that because the voltage window in which these values are attained is of order $\Tilde{T}^{2}\Delta_{0}$, it is hard to measure. Indeed, for low temperatures, one requires a voltage resolution much smaller than $\Tilde{T}^{2}\Delta_{0}$. Moreover, we remark that this feature is not robust against the inclusion of a \textit{p} - wave component. Indeed, in the presence of a \textit{p} - wave component, the gap edge is different for the incoming and outgoing modes, they are at $E = \Delta_{\pm}$. Thus, there is no energy for which both incoming and outgoing Green's functions diverge. Instead, if the elements of one of them diverge, e.g. at $E = \Delta_{+}$ those of the other are of order $(E^2-\Delta_{-}^{2})^{-\frac{1}{2}} = (\Delta_{+}^{2}-\Delta_{-}^{2})^{-\frac{1}{2}}$, that is, of order $r^{-\frac{1}{2}}$, where $r$ is the ratio of the \textit{s} - wave and \textit{p} - wave components.
Therefore in the surface Green's function $C$ there is no divergence, but a peak in the maximum absolute value of $C$, whose height is of order $r^{-\frac{1}{2}}$. Thus, for $ \Tilde{T}^{2}\ll r < 1$ there is no energy window in which the anticommutator dominates, and consequently no perfect resonance. Thus, this effect is not robust against perturbations.

\subsection{\textit{p} - wave superconductors}
In this part we consider an unconventional superconductor, for which \cite{tanaka2022theory}:
\begin{align}
    C &= H_{+}^{-1}(\mathbf{1}-H_{-})\;,\\
    H_{\pm} &= \frac{1}{2}(G_{S1}\pm G_{S2})\;,\\ 
    G_{S1} &= \frac{-i}{\sqrt{\Delta_{0}^{2}-E^{2}}}\begin{bmatrix}
        E&\Delta_{0}\\-\Delta_{0}&-E
    \end{bmatrix}\; ,\\
    G_{S2} &= \frac{-i}{\sqrt{\Delta_{0}^{2}-E^{2}}}\begin{bmatrix}
        E&-\Delta_{0}\\\Delta_{0}&-E
    \end{bmatrix}\; .
\end{align}
Specifically, for the 1D \textit{p} - wave superconductor, below the gap, i.e. for $|E|<\Delta_{0}$ this gives 
\begin{align}
    C^{R} = C^{A} &= \frac{1}{E}\begin{bmatrix}
        i\sqrt{\Delta_{0}^{2}-E^{2}}&-\Delta_{0}\\-\Delta_{0}&-i\sqrt{\Delta_{0}^{2}-E^{2}}
    \end{bmatrix}\; ,\\
    C^{K} &= 0\;,
\end{align}
while above the gap, i.e. for $|E|>\Delta_{0}$ we have
\begin{align}
    C^{R} = -C^{A} &= \frac{1}{E}\begin{bmatrix}
        \sqrt{E^{2}-\Delta_{0}^{2}}&-\Delta_{0}\\-\Delta_{0}&-\sqrt{E^{2}-\Delta_{0}^{2}}\end{bmatrix}\; ,\\
       C^{K}&=2\tanh{\frac{E}{2k_{B}T}} \;.
\end{align}

In discussing the required low transparency limit we need to be careful. Indeed, since $C$ has a pole at $E = 0$ the $\Tilde{T}\ll 1$ simplification of ignoring the denominator fails in this important limit. We first focus on the range $|eV|\gg \Tilde{T}\Delta_{0}$ and then consider what happens close to this resonance 
For $\Tilde{T}\Delta_{0}\ll|eV|\ll \Delta_{0}$ we are below the gap, but not close to a resonance, and may therefore use the expansion, that is, we use Eqs. (\ref{eq:IlowT}) and (\ref{eq:PlowT}). 
\begin{align}
    [G_{N},C]&=\frac{-2\Delta_{0}}{E}\begin{bmatrix}
        0&1&0&2f_{L}\\-1&0&-2f_{L}&0\\0&0&0&-1\\0&0&1&0
    \end{bmatrix}\; ,\\
    \text{Tr}\Big(\tau_{K}[G_{N},C]\Big) &= 0\;,\\
    \{G_{N},C\} &= 2i\frac{\sqrt{\Delta_{0}^{2}-E^{2}}}{E}\begin{bmatrix}
        1&0&2(f_{L}+f_{T})&2i\frac{\Delta_{0}}{\sqrt{\Delta_{0}^{2}-E^{2}}}f_{T}\\0&1&2i\frac{\Delta_{0}}{\sqrt{\Delta_{0}^{2}-E^{2}}}f_{T}&2(f_{L}-f_{T})\\0&0&-1&0\\0&0&0&-1
    \end{bmatrix}\; ,\\
    \{G_{N},C\}[G_{N},C] &= \frac{-4i\Delta_{0}\sqrt{\Delta_{0}^{2}-E^{2}}}{E^{2}}\begin{bmatrix}
        0&1&2i\frac{\Delta_{0}}{\sqrt{\Delta_{0}^{2}-E^{2}}}f_{T}&-2f_{T}\\-1&0&-2f_{T}&-2i\frac{\Delta_{0}}{\sqrt{\Delta_{0}^{2}-E^{2}}}\\0&0&0&1\\0&0&-1&0
    \end{bmatrix}\; ,\\
    \text{Tr}\Big(\tau_{K}\{G_{N},C\}[G_{N},C]\Big) &=\frac{16\Delta_{0}^{2}}{E^{2}}f_{T}\;.
\end{align}

For the calculation of noise power we require the following quantities. For the first one
\begin{align}
    &\{C,\tau_{K}G_{N}\tau_{K}-G_{N}\}=-4i\frac{\sqrt{\Delta_{0}^{2}-E^{2}}}{E}\begin{bmatrix}
        1&0&f_{L}+f_{T}&i\frac{\Delta_{0}}{\sqrt{\Delta_{0}^{2}-E^{2}}}f_{T}\\0&1&i\frac{\Delta_{0}}{\sqrt{\Delta_{0}^{2}-E^{2}}}f_{T}&f_{L}-f_{T}\\-(f_{L}+f_{T})&-i\frac{\Delta_{0}}{\sqrt{\Delta_{0}^{2}-E^{2}}}f_{T}&-1&0\\-i\frac{\Delta_{0}}{\sqrt{\Delta_{0}^{2}-E^{2}}}f_{T}&-(f_{L}-f_{T})&0&-1
    \end{bmatrix}\; ,\\
    &\text{Tr}\Big(\{C,\tau_{K}G_{N}\tau_{K}-G_{N}\}\Big)=0\;,\\
    &\{C,[\tau_{K},G_{N}]\}=\frac{-4i\sqrt{\Delta_{0}^{2}-E^{2}}}{E}\Bigg(\rho_{3}\otimes\begin{bmatrix}
        f_{L}+f_{T}&f_{L}\frac{i\Delta_{0}}{\sqrt{\Delta_{0}^{2}-E^{2}}}\\f_{L}\frac{i\Delta_{0}}{\sqrt{\Delta_{0}^{2}-E^{2}}}&-(f_{L}-f_{T})
    \end{bmatrix}+i\rho_{2}\otimes\begin{bmatrix}
        1&i\frac{\Delta_{0}}{\sqrt{\Delta_{0}^{2}-E^{2}}}\\i\frac{\Delta_{0}}{\sqrt{\Delta_{0}^{2}-E^{2}}}&-1
    \end{bmatrix}\Bigg)\;,\\
    &\{C,[\tau_{K},G_{N}]\}^{2}=\frac{16}{E^{2}}\rho_{0}\otimes\begin{bmatrix}
        -f_{T}(2f_{L}+f_{T})\Delta_{0}^{2}+E^{2}((f_{L}+f_{T})^{2}-1)&-2if_{L}f_{T}\Delta_{0}\sqrt{\Delta_{0}^{2}-E^{2}}\\-2if_{L}f_{T}\Delta_{0}\sqrt{\Delta_{0}^{2}-E^{2}}&f_{T}(2f_{L}-f_{T})\Delta_{0}^{2}+E^{2}((f_{L}-f_{T})^{2}-1)
    \end{bmatrix}\; ,\\
    &\text{Tr}\Big(\{C,[\tau_{K},G_{N}]\}^{2}\Big)=  64(f_{L}^{2}+f_{T}^{2}-1)-64\frac{\Delta_{0}^{2}}{E^{2}}f_{T}^{2}\;.    
\end{align}
and for the second one
\begin{align}
    &\{C,G_{N}\}\{C,\tau_{K}G_{N}\tau_{K}-G_{N}\}\nonumber\\&=\frac{4}{E^{2}}(\rho_{0}+\rho_{3})\otimes\begin{bmatrix}(E^{2}(2(f_{L}+f_{T})^{2}-1)+\Delta_{0}^{2}(1-2f_{L}(f_{L}+2f_{T})))&-4if_{L}f_{T}\Delta_{0}\sqrt{\Delta_{0}^{2}-E^{2}}\\-4if_{L}f_{T}\Delta_{0}\sqrt{\Delta_{0}^{2}-E^{2}}&(E^{2}(2(f_{L}-f_{T})^{2}-1)+\Delta_{0}^{2}(1-2f_{L}(f_{L}-2f_{T})))\end{bmatrix}\nonumber\\&+\frac{8}{E^{2}}i\rho_{2}\otimes\begin{bmatrix}(f_{L}+f_{T})(E^{2}-\Delta_{0}^{2})&-if_{T}\Delta_{0}\sqrt{\Delta_{0}^{2}-E^{2}}\\-if_{T}\Delta_{0}\sqrt{\Delta_{0}^{2}-E^{2}}&(f_{L}-f_{T})(E^{2}-\Delta_{0}^{2})\end{bmatrix}\nonumber\\&+\frac{4(\Delta_{0}^{2}-E^{2})}{E^{2}}(\rho_{0}-\rho_{3})\otimes\tau_{0}\;,\\
    &\text{Tr}\Big(\{C,G_{N}\}\{C,\tau_{K}G_{N}\tau_{K}-G_{N}\}\Big)=32(f_{L}^{2}+f_{T}^{2}-1)+32\frac{\Delta_{0}^{2}}{E^{2}}(1-f_{L}^{2})\;.
\end{align}
Thus, we have, using Eqs. (\ref{eq:IlowT}) and (\ref{eq:PlowT}),
\begin{align}
    \mathcal{I}_{1} &= \Tilde{T}^{2}\frac{\Delta_{0}^{2}}{E^{2}}f_{T}\;,\\
    \mathcal{P}_{1} &= \Tilde{T}^{2}\frac{\Delta_{0}^{2}}{E^{2}}(1+f_{T}^{2}-f_{L}^{2})\;.
\end{align}

We note that these expressions diverge as $E\xrightarrow{} 0$, confirming that they are not valid in this regime.

On the other hand, above the gap we have
\begin{align}
    [G_{N},C]&=\begin{bmatrix}
        0&-\frac{2\Delta_{0}}{E}&-4(f_{L}+f_{T}-f_{S})\frac{\sqrt{E^{2}-\Delta_{0}^{2}}}{E}&4\frac{\Delta_{0}}{E}f_{T}\\2\frac{\Delta_{0}}{E}&0&4\frac{\Delta_{0}}{E}f_{T}&-4(f_{L}-f_{T}-f_{S})\frac{\sqrt{E^{2}-\Delta_{0}^{2}}}{E}\\0&0&0&-\frac{2\Delta_{0}}{E}\\0&0&\frac{2\Delta_{0}}{E}&0
    \end{bmatrix}\; ,\\
    \text{Tr}\Big(\tau_{K}[G_{N},C]\Big)&=-8f_{T}\frac{\sqrt{E^{2}-\Delta_{0}^{2}}}{E}\;.
\end{align}

For the noise power we find the following quantities
\begin{align}
    &\{C,\tau_{K}G_{N}\tau_{K}-G_{N}\}=\frac{4}{E}\nonumber\\&\begin{bmatrix}
        -(1-f_{S}(f_{L}+f_{T}))\Omega&f_{S}(f_{L}-f_{T})\Delta_{0}&0&-(f_{L}-2f_{S})\Delta_{0}\\
        -f_{S}(f_{L}+f_{T})\Delta_{0}&-(1-f_{S}(f_{L}-f_{T}))\Omega&(f_{L}-2f_{S})\Delta_{0}&0\\0&-f_{L}\Delta_{0}&-(1-f_{S}(f_{L}+f_{T}))\Omega&-f_{S}(f_{L}+f_{T})\Delta_{0}\\f_{L}\Delta_{0}&0&f_{S}(f_{L}-f_{T})\Delta_{0}&-(1-f_{S}(f_{L}-f_{T}))\Omega
    \end{bmatrix}\; ,\\
    &\Omega = \sqrt{E^{2}-\Delta_{0}^{2}}\;,\\
    &\text{Tr}\Big(\{C,\tau_{K}G_{N}\tau_{K}-G_{N}\}\Big)=16\frac{\sqrt{E^{2}-\Delta_{0}^{2}}}{E}(f_{L}f_{S}-1)\;.\\
\end{align}
Thus, we have, using Eqs. (\ref{eq:IlowT}) and (\ref{eq:PlowT}),
\begin{align}
    \mathcal{I}_{2} &=2\frac{\sqrt{E^{2}-\Delta_{0}^{2}}}{E}f_{T}\;,\\
    \mathcal{P}_{2}&=2\frac{\sqrt{E^{2}-\Delta_{0}^{2}}}{E}(1-f_{L}f_{S})\;.
\end{align}

We may now calculate the conductance and the differential noise power. At zero temperature, the distribution functions are step functions with their steps at $|E| = |eV|$. Therefore, $\frac{\partial f_{T}}{\partial V}$ and $\frac{\partial f_{L}}{\partial V}$ are the combinations of two delta-like functions centered at $|E| = |eV|$. This makes the integration over energy trivial. For $\Tilde{T}\ll|eV|<\Delta_{0}$ we obtain
\begin{align}
    \frac{\partial I}{\partial V}\label{eq:IbelowgapP}&= \frac{e^{2}}{4\pi\hbar}\frac{\Delta_{0}^{2}}{(eV)^{2}}\Tilde{T}^{2}\;,\\
    \frac{\partial P_{N}}{\partial V}& = \frac{e^{3}}{\pi\hbar}\frac{\Delta_{0}^{2}}{(eV)^{2}}\Tilde{T}^{2}\;,
\end{align}
from which we may calculate the differential Fano factor as
\begin{align}
    F(\Tilde{T}\Delta_{0}\ll |eV|<\Delta_{0})= \frac{1}{2e}\frac{\frac{\partial P_{N}}{\partial V}}{\frac{\partial I}{\partial V}} = 2\;.
\end{align}
On the other hand, for $|eV|>\Delta_{0}$
\begin{align}
    \frac{\partial I}{\partial V}&=\frac{e^{2}}{2\pi\hbar}\frac{\sqrt{(eV)^{2}-\Delta_{0}^{2}}}{eV}\;,\label{eq:Iabovegap}\\
    \frac{\partial P_{N}}{\partial V}& = \frac{e^{3}}{\pi\hbar}\frac{\sqrt{(eV)^{2}-\Delta_{0}^{2}}}{eV}\;,\\
    F(|eV|\gg\Delta_{0}) & = \frac{1}{2e}\frac{\frac{\partial P_{N}}{\partial V}}{\frac{\partial I}{\partial V}}= 1\;.\label{eq:FabovegapP}
\end{align}
Again, this expression is valid only for $|eV|\gg\Delta_{0}$ as there is a divergence at $|eV| = \Delta_{0}$ because the differential conductance vanishes in the absence of a proximity effect at this voltage. 

\subsubsection{Zero voltage}
Near zero energy the elements of $C$ diverge. Because this is a first order divergence, the above formulation fails if $|eV|$ is not much larger than $\Tilde{T}\Delta_{0}$. The expression for $C$ that arises from this expression is very similar to the one that appears as the $E\xrightarrow{}\Delta_{0}^{-}$ limit of the \textit{s} - wave superconductor. 

Indeed, we have
\begin{align}
    C^{R}&=C^{A} \approx \frac{i\Delta_{0}}{E}\begin{bmatrix}
        1&i\\i&-1
    \end{bmatrix}\;,\label{eq:StrongDivergence}\\
    C^{K}&=0\;.
\end{align}
Consequently, for the current we obtain
\begin{align}
    \{C,G_{N}\} &= \frac{2i\Delta_{0}}{E}\begin{bmatrix}
        1&0&2(f_{L}+f_{T})&2if_{T}\\0&1&2if_{T}&2(f_{L}-f_{T})\\
        0&0&-1&0\\0&0&0&-1
    \end{bmatrix}\;,\\
    \{C,G_{N}\}^{-1} &= \frac{E}{2i\Delta_{0}}\begin{bmatrix}
        1&0&2(f_{L}+f_{T})&2if_{T}\\0&1&2if_{T}&2(f_{L}-f_{T})\\
        0&0&-1&0\\0&0&0&-1
    \end{bmatrix}\;,\\
    \{C,G_{N}\}^{-1}[G_{N},C]&=\begin{bmatrix}
        0&i&-2f_{T}&-2if_{T}\\-i&0&-2if_{T}&2f_{T}\\0&0&0&i\\0&0&-i&0
    \end{bmatrix}\;,\\
    \text{Tr}\Bigg(\tau_{K}\{C,G_{N}\}^{-1}[G_{N},C]\Bigg)&=-4f_{T}\;.
\end{align}
Meanwhile, for the noise power we have
\begin{align}
&\{C,\tau_{K}G_{N}\tau_{K}-G_{N}\}=-\frac{4i\Delta_{0}}{E}\begin{bmatrix}
    1&0&f_{L}+f_{T}&if_{T}\\0&1&if_{T}&f_{L}-f_{T}\\-(f_{L}+f_{T})&-if_{T}&-1&0\\-if_{T}&-(f_{L}-f_{T})&0&-1
\end{bmatrix}\;,\\
    &\{C,G_{N}\}^{-1}\{C,\tau_{K}G_{N}\tau_{K}-G_{N}\}=\begin{bmatrix}
        -2+4f_{L}^{2}+8f_{L}f_{T}&i8f_{L}f_{T}&2(f_{L}+f_{T})&2if_{T}\\8if_{L}f_{T}&-2+4f_{L}^{2}-8f_{L}f_{T}&2if_{T}&2(f_{L}-f_{T})\\-2(f_{L}+f_{T})&-2if_{T}&-2&0\\-2if_{T}&-2(f_{L}-f_{T})&0&-2
    \end{bmatrix}\; ,\\
    &\text{Tr}\Big(\{C,G_{N}\}^{-1}\{C,\tau_{K}G_{N}\tau_{K}-G_{N}\}\Big)=8(-1+f_{L}^{2})\;,\\
    &\{C,G_{N}\}^{-1}\{C,[\tau_{K},G_{N}]\}=\begin{bmatrix}2(f_{L}-f_{T})&2if_{L}&-2+4(f_{L}^{2}+f_{T}^{2}+f_{L}f_{T})&-2i+4i(f_{L}^{2}+f_{T}^{2})\\2if_{L}&-2(f_{L}+f_{T})&-2i+4i(f_{L}^{2}+f_{T}^{2})&2-4(f_{L}^{2}+f_{T}^{2}-f_{L}f_{T})\\-2&-2i&-2(f_{L}+f_{T})&-2if_{L}\\-2i&2&-2if_{L}&2(f_{L}-f_{T})\end{bmatrix}\; ,\\
    &\Big(\{C,G_{N}\}^{-1}\{C,[\tau_{K},G_{N}]\}\Big)^{2}=
    \frac{1}{2}(\rho_{0}+\rho_{3})\otimes\begin{bmatrix}
        4 f_{T} (-4 f_{L} + f_{T})& -16 i f_{L} f_{T}\\-16 i f_{L} f_{T}& 
 4 f_{T} (4 f_{L} + f_{T})
    \end{bmatrix}+\frac{1}{2}(\rho_{0}-\rho_{3})\otimes\begin{bmatrix}
        4 f_{T}^2& 0\\0& 4 f_{T}^2
    \end{bmatrix}\nonumber\\&+\frac{1}{2}(\rho_{1}+i\rho_{2})\otimes\begin{bmatrix}-8 f_{T} (-1 + 
    2 (f_{L}^2 + f_{L} f_{T} + f_{T}^2))& -8i f_{T} (-1 + 2 f_{L}^2 + 
    2 f_{T}^2)\\
         -8 i f_{T} (-1 + 2 f_{L}^2 + 2 f_{T}^2)& 
 8 f_{T} (-1 + 2 (f_{L}^2 - f_{L} f_{T} + f_{T}^2))
    \end{bmatrix}
    \nonumber\\&+\frac{1}{2}(\rho_{1}-i\rho_{2})\begin{bmatrix}
        8 f_{T}& 8 i f_{T}\\8 i f_{T}& -8 f_{T}
    \end{bmatrix}\;,
    \\
    &\text{Tr}\Big((\{C,G_{N}\}^{-1}\{C,[\tau_{K},G_{N}]\})^{2}\Big)=16f_{T}^{2}\;.
\end{align}
With this we have
\begin{align}
    \mathcal{I}_{1}(E\approx\Delta_{0})&=4f_{T}\;,\\
    \mathcal{P}_{1}(E\approx\Delta_{0})&\approx 4(1-f_{L}^{2}-f_{T}^{2})\label{eq:P1resP}\;.
\end{align}

Thus, we find that that 
\begin{align}
    \frac{\partial I}{\partial V}(|eV|\xrightarrow{}0)&=\frac{e^{2}}{\pi\hbar}\label{eq:IresonanceP}\;,\\
    \frac{\partial P_{N}}{\partial V}(|eV|\xrightarrow{}0) &= 0\;,\\
    F(|eV|\xrightarrow{}0) &= 0\;.\label{eq:FresonanceP}
\end{align}
From this we conclude there is a perfect resonance in the conductance and the differential noise vanishes at $eV = 0$.

Summarizing, in the low transparency limit, for $|eV|>\Delta_{0}$ the differential Fano factor approximately equals 1, for $\Tilde{T}\Delta_{0}\ll |eV|<\Delta_{0}$ it is approximately 2, while for $|eV|\ll \Tilde{T}\Delta_{0}$ the differential Fano factor goes to $0$. The lower $\Tilde{T}$, the sharper this drop to 0 is. Thus, for smaller $\Tilde{T}$ lower voltages and temperatures need to be used in doing the measurement.
These results are in good agreement with the numerical results presented in Fig. \ref{fig:FinNoiseSdom}(b) in the main text.

We note that  the diverge at $eV = 0$ in the \textit{p} - wave case is much stronger than the divergence at $|eV| = \Delta_{0}$ for the \textit{s} - wave case, compare Eqs. (\ref{eq:WeakDivergence}), which has $C\sim E^{-\frac{1}{2}}$ and (\ref{eq:StrongDivergence}), which has $C\sim E^{-1}$.

This has an important consequence. The resonance appears when the elements of $\Tilde{T}\{C,G\}$ are much larger than unity, that is, when the elements of $C$ are much larger than $\Tilde{T}^{-1}$. For the \textit{p} - wave superconductor this means $\frac{\Delta_{0}}{|eV|}\gg \Tilde{T}^{-1}$, i.e. $|eV|\ll \Tilde{T}\Delta_{0}$, that is, the resonance appears in a voltage window with width of order $\Tilde{T}\Delta_{0}$, while for the s-wave superconductor, where the condition is $\sqrt{\frac{2\Delta_{0}}{|\delta E|}}$ we found in the previous section the resonance appears only for $|\delta E|\ll 2\Tilde{T}^{2}\Delta_{0}$, that is, in a window a width of order $\Tilde{T}^{2}\Delta_{0}$. Consequently, for the \textit{p} - wave superconductor, the resonance is much easier to observe. 

Moreover, the divergence at $E = 0$ appears for any $r>1$, so the effect is robust against inclusion of an \textit{s} - wave component.

\clearpage
\section{1D superconductors at zero temperature}\label{sec:1Dnumerical}
In this section we discuss the dependence of the differential Fano factor on the Dynes parameter and the BTK parameter in 1D superconductor / normal metal junctions. The goal of this section is twofold. First of all, with this we motivate our choice to focus on the tunnel limit in the discussion of the tunnel junction in the main text. Next to this, many of the finite temperature features in Figs. \ref{fig:FinNoiseSdom}, \ref{fig:FinNoiseSdomISP} in the main text can be understood from the zero temperature results discussed below.

\begin{figure}[!htb]
    \centering
    \includegraphics[width = 8.6cm]{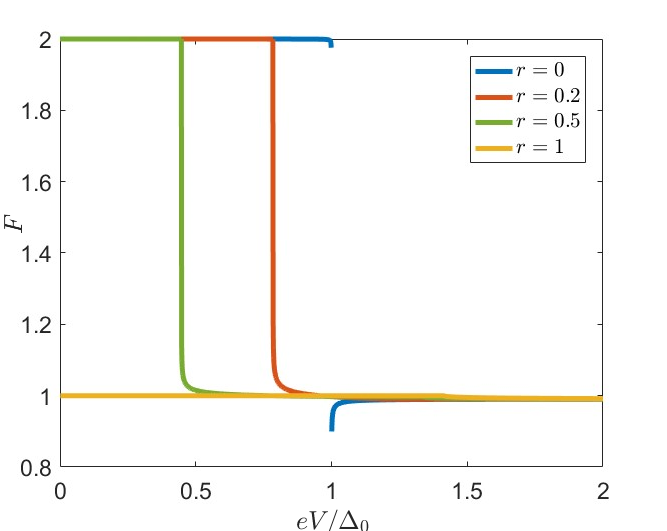}
    \includegraphics[width = 8.6cm]{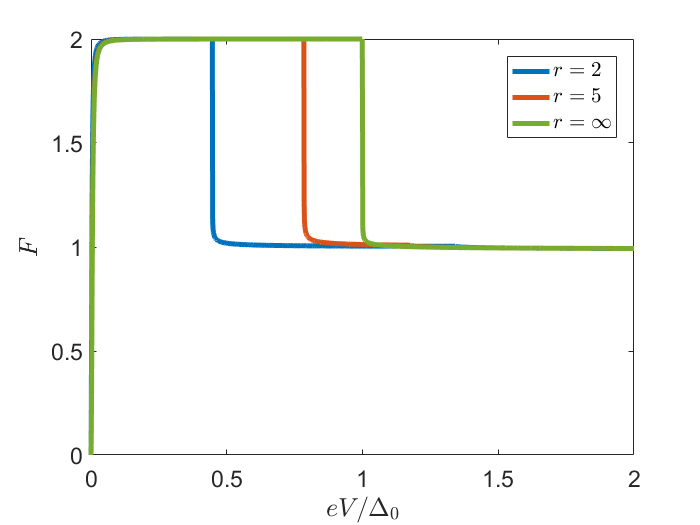}
    \includegraphics[width = 8.6cm]{figures/A.png}
    \includegraphics[width = 8.6cm]{figures/B.png}
    \caption{The dependence of the differential Fano factor on the ratio between the \textit{s} - wave and \textit{p} - wave components of the pair potential for $z = 10$ in a 1D junction. If the \textit{p} - wave component is dominant (b), the differential Fano factor converges to 0 at $eV = 0$, if it is not dominant (a) the differential Fano factor remains finite. If $r\neq 1$ there is a sharp transition between $F = 1$ and $F = 2$ at the minimum gap $|eV| = |\Delta_{-}| = \frac{|1-r|}{\sqrt{1+r^{2}}}\Delta_{0}$.  For $r = 0$ at $|eV| = \Delta_{0}$ the differential Fano factor vanishes, because of the gap edge. For illustrative purposes this is not shown in the plot.
    For $r = 0$ and $r = \infty$ this quantity equals $\Delta_{0}$, for $r = 0.2$ and $ r = 5$ it equals $0.78\Delta_{0}$, for $r = 0.5$ and $r = 2$ it equals $0.45\Delta_{0}$. For $r = 1$ the minimum gap is $0$, and therefore $F \approx 1$ for all voltages.}
    \label{fig:R1D}
\end{figure}

As discussed in the main text, our code confirms that in the low-transparency limit the differential Fano factor is $1$ for $|eV|>\Delta_{0}$, for which current is carried by quasiparticles, while for \textit{s} - wave superconductors it is $2$ for $|eV|<\Delta_{0}$, for which current is carried by Cooper pairs. Moreover, due to the ZESABS in \textit{p} - wave superconductors the differential Fano factor vanishes at zero voltage in normal metal / \textit{p} - wave superconductor junctions. For \textit{s} + \textit{p} - wave superconductors the results, shown in Fig. \ref{fig:R1D}, can be explained using these considerations as well. Indeed, if $r>1$ there is a SABS at zero energy and hence the differential Fano factor vanishes at $eV = 0$, while for $r\leq 1$ there is no ZESABS and the differential Fano factor does not vanish at $eV = 0$. Instead, because transport is carried by Cooper pairs, $F(eV = 0) = 2$ in those superconductors. 

If $|eV|>\Delta_{+} = \frac{1+r}{\sqrt{1+r^{2}}}\Delta_{0}$ the voltage is above both gaps and therefore necessarily the differential Fano factor approaches 1 if $|eV|-\Delta_{+}\gg \Tilde{T}\Delta_{0}$, while for $|eV|<|\Delta_{-}|$ the voltage is below both gaps and hence $F = 2$ for $|\Delta_{-}|-|eV|\gg \Tilde{T}\Delta_{0}$. The most interesting regime is $|\Delta_{-}|<|eV|<\Delta_{+}$. In this case the current is partially carried by Cooper pairs and partly by unpaired electrons, the quasiparticles. However, the current carried by Cooper pairs is second order in the transparency, while the current carried by normal electrons is first order in transparency. Thus, in the low transparency limit the latter dominates, and hence $F\approx 1$ within this regime.

\begin{figure}[!htb]
    \centering
    \includegraphics[width = 8.6cm]{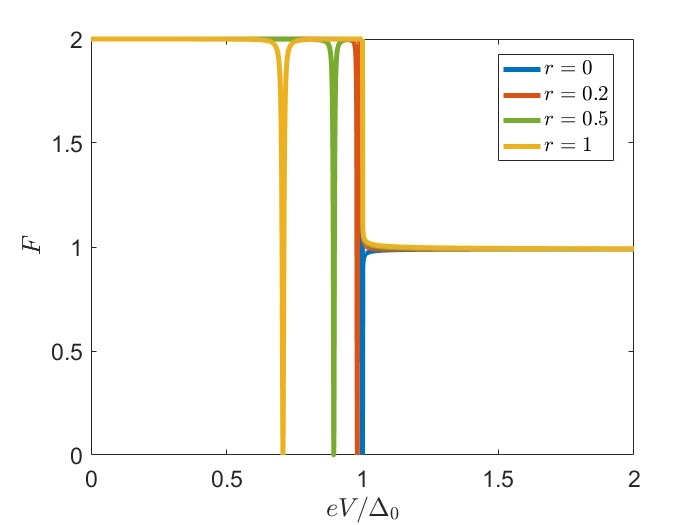}
    \includegraphics[width = 8.6cm]{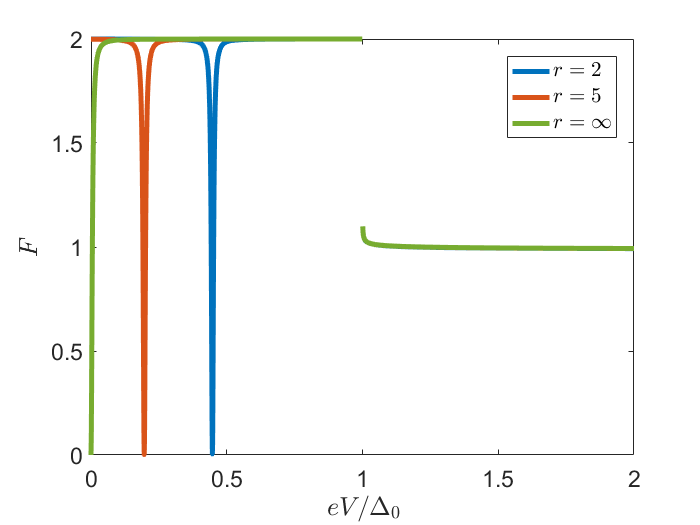}
    \includegraphics[width = 8.6cm]{figures/A.png}
    \includegraphics[width = 8.6cm]{figures/B.png}
    \caption{The differential Fano factor for 1D S / N  junctions with $z = 10$ for different ratios between the \textit{s} - wave and \textit{p} - wave component of the time-reversal broken i\textit{s} + \textit{p} - wave superconductor. Both if the \textit{s} - wave component is dominant and if the \textit{p} - wave component is dominant, the transition between $F = 1$ and $F = 2$ appears at $|eV| = \Delta_{0}$, regardless of the actual ratio of the two components. This is due to the magnitude of the gap being independent of this ratio. In the presence of a finite \textit{p} - wave component, the differential Fano factor is zero at $|eV| = \Delta_{s}$, reflecting that the ZESABS is not robust in the absence of time-reversal symmetry.}
    \label{fig:1DISP}
\end{figure}

The results for junctions with i\textit{s} + \textit{p} - wave superconductors  are quite different, as shown in Fig. \ref{fig:1DISP}. Indeed, both if the \textit{s} - wave component is dominant and if the \textit{p} - wave component is dominant, the transition between $F = 1$ and $F = 2$ appears at $|eV| = \Delta_{0}$, regardless of the actual ratio of the two components, because the magnitude of the gap is always $\Delta_{0}$, independent of the ratio between the \textit{s} - wave and \textit{p} - wave components. This ratio only changes the phase difference for the $\langle\uparrow\downarrow\rangle$ and $\langle\downarrow\uparrow\rangle$ components of the pair amplitudes in an i\textit{s} + \textit{p} - wave superconductor. In the presence of a finite \textit{p} - wave component, the differential Fano factor is zero at $|eV| = \Delta_{s} = \frac{\Delta_{0}}{\sqrt{1+r^{2}}}$, due to the noiseless mode, a SABS, which appears at this energy. Thus, this SABS is at finite energy if there is a nonzero \textit{s} - wave component. This reflects that the SABS of the superconductor is not robust against perturbations in the absence of time-reversal symmetry.

\begin{figure}[!htb]
    \centering
    \includegraphics[width = 8.6cm]{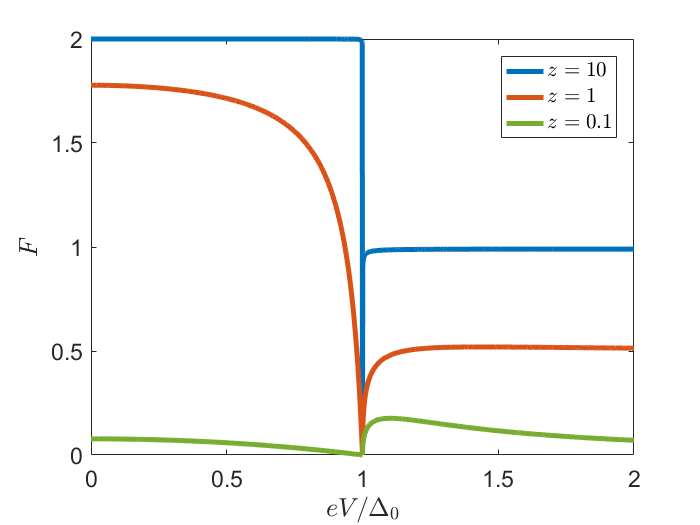}
    \includegraphics[width = 8.6cm]{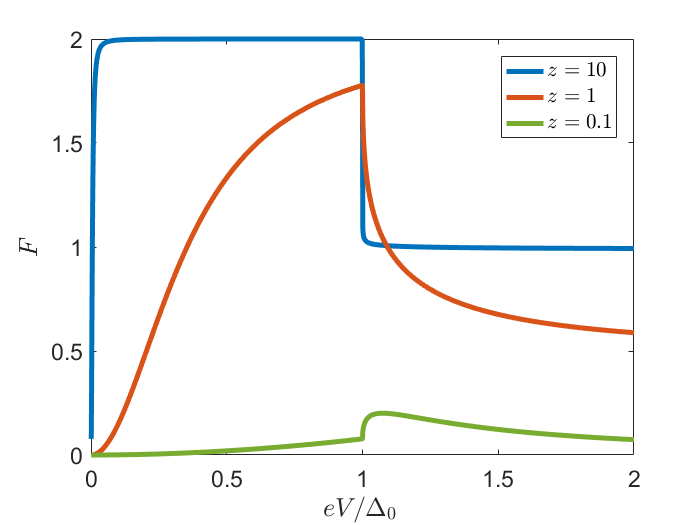}
    \includegraphics[width = 8.6cm]{figures/A.png}
    \includegraphics[width = 8.6cm]{figures/B.png}
    \caption{Dependence of the differential Fano factor in a 1D \textit{s} - wave superconductor / normal metal junction on the transparency, for \textit{s} - wave (a) and \textit{p} - wave (b) superconductors. By increasing the transparency (decreasing $z$), the differential Fano factor can be decreased, and all features are smoothened. Moreover, the results for the \textit{s} - wave and \textit{p} - wave superconductor become similar, indicating that it the high transparency limit is not suitable for distinguishing different types of superconductors.}
    \label{fig:Z1D}
\end{figure}

\begin{figure}[!htb]
    \centering
    \includegraphics[width = 8.6cm]{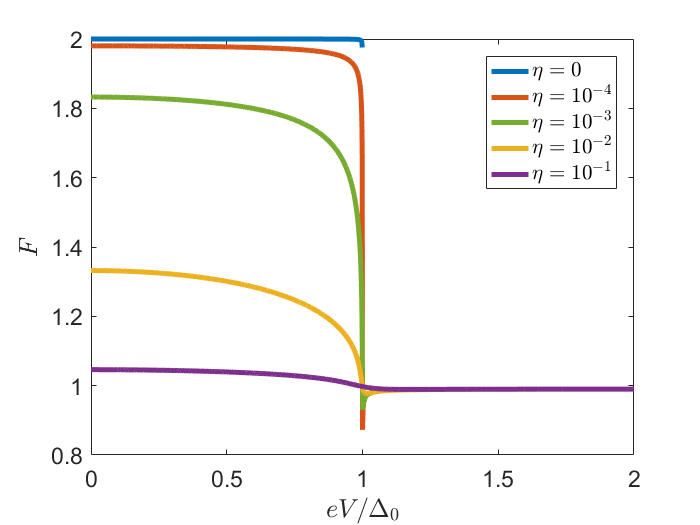}
    \includegraphics[width = 8.6cm]{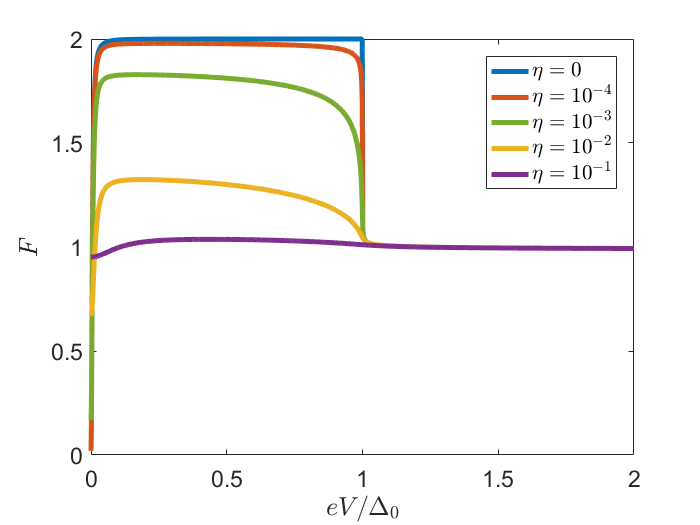}
    \includegraphics[width = 8.6cm]{figures/A.png}
    \includegraphics[width = 8.6cm]{figures/B.png}
    \caption{The voltage dependence of the differential Fano factor in a 1D superconductor / normal metal junction for different values of the Dynes parameter $\eta$ for $z = 10$. A finite Dynes parameter highly suppresses the differential Fano factor both in the \textit{s} - wave (a) and \textit{p} - wave (b) dominant case. (a): At finite Dynes parameter $F$ is closer to 1. (b): While for $\eta = 0$ the differential Fano factor satisfies $F(|eV| = 0) = 0$, at finite Dynes parameters the differential Fano factor attains a nonzero value at zero voltage.}
    \label{fig:Etas}
\end{figure}

Next we consider the influence of the transparency of the junction.
If the transparency increases, the differential Fano factor decreases, see Fig. \ref{fig:Z1D}. Moreover, its features become smoothened, there is no sharp transition at $|eV| = \Delta_{0}$. In fact the results for \textit{s} - wave and \textit{p} - wave superconductors become more similar. Therefore, for the purpose of distinguishing different types of superconductivity the low-transparency limit is the most useful regime.
We highlight the importance of the Dynes parameter $\eta$ being small. This parameter is usually associated with inelastic scattering and causes the density of states below the gap to be nonzero. Therefore, in the S / N junction this parameter is of great importance for the differential Fano factor. Indeed, as found before, for $\eta = 0$ the current is of second order in transparency because density of states vanishes below the gap. However, since for $\eta\neq 0$ the density of states does not vanish, there is a quasi-particle carried current that is linear in both the transparency and the Dynes parameter. For low enough transparencies ($\Tilde{T}\ll \eta$) this contribution to the current is the dominant one, and hence the differential Fano factor equals 1 in this limit. Indeed, as shown in Fig. \ref{fig:Etas}, for large enough $\eta$ the differential Fano factor converges to 1.
\clearpage
\section{Two dimensional superconductors}\label{sec:2Dresults}
In this section we provide additional numerical results for  two-dimensional superconductors. Our results show the dependence of the differential Fano factor on the ratio between the magnitudes of the different components of the pair potential and prove that the qualitative features do not change by varying the ratio between the \textit{s} - wave and \textit{p} - wave pair potentials made in the main text.

For two-dimensional superconductors, the Tanaka-Nazarov formalism is similar compared to 1D, but summation or integration over channels is necessary. There are a few remarks that can be made before doing any calculation. First of all, for \textit{s} - wave superconductors, the results should be similar in two dimensions and in one dimension, since for any incident angle we have Cooper pair transport for $|eV|<\Delta_{0}$ and quasiparticle transport via the continuum for $|eV|>\Delta_{0}$. For superconductors with a nonconstant gap but no SABSs (e.g. \textit{s} - wave dominant \textit{s} + \textit{p} or \textit{s} + \textit{d} -  wave superconductors), and a transparency $\Tilde{T}\ll 1$ we note that for voltages that are below the minimum gap all current is carried by Cooper pairs and $F\approx 2$. However, when the voltage exceeds the minimum gap, there are channels for which the current is carried by Cooper pairs and channels for which it is carried by quasiparticles and there is a transition between $F = 1$ and $F = 2$. If the pair potential is a smooth function of the incoming angle this transition is smooth, as shown below.

We first discuss the low-temperature limit, in which the distribution functions may be approximated by step functions.

We use the dimensionless units $\mathcal{I}$ and $\mathcal{P}$, which are defined in Eqs. (\ref{eq:dimensionlessP}) and (\ref{eq:dimensionlessI}). In these units, the differential Fano factor is the ratio of the conductance and the differential noise, without any prefactor.

In the case of $\text{\textit{p}}_{\text{x}}$ - wave superconductors, for which the lobe is directed along the normal interface, we may calculate the differential Fano factor at zero voltage analytically. Indeed, since $\Delta(\pi-\phi) = -\Delta(\phi)$ for all $\phi$ we have
\begin{align}
    \mathcal{I}_{\text{\textit{p}}_{\text{x}}} = \int_{-\frac{\pi}{2}}^{\frac{\pi}{2}} d\phi T(\phi)I_{p1D}(eV,\Delta_{0}\cos\phi)\cos\phi = \int_{-\frac{\pi}{2}}^{\frac{\pi}{2}} d\phi T(\phi)I_{p1D}(\frac{eV}{\cos\phi},\Delta_{0})\cos\phi\; , \\
    \mathcal{P}_{\text{\textit{p}}_{\text{x}}} = \int_{-\frac{\pi}{2}}^{\frac{\pi}{2}} d\phi T(\phi)P_{p1D}(eV,\Delta_{0}\cos\phi)\cos\phi = \int_{-\frac{\pi}{2}}^{\frac{\pi}{2}} d\phi T(\phi)P_{p1D}(\frac{eV}{\cos\phi},\Delta_{0})\cos\phi\;,
\end{align}
where $I_{p1D}$ and $P_{p1D}$ are the dimensionless noise power and current for a 1D \textit{p} - wave superconductor / normal metal junction respectively, calculated in Sec. \ref{sec:1Danalytical}.

From this expression we may derive a few important conclusions in the low-transparency limit. First of all, for voltages above the maximum gap $\Delta_{0}$, in the dimensionless units all modes contribute equally to noise power and current at zero temperature, i.e. $\mathcal{I}_{p1D}(\frac{eV}{\cos\phi},\Delta_{0}) = \mathcal{P}_{p1D}(\frac{eV}{\cos\phi},\Delta_{0}) $ for all $\phi$, following Eq. (\ref{eq:FabovegapP}), and hence $\mathcal{I}_{p_{x}} = \mathcal{P}_{p_{x}}$, that is, $F = 1$ as expected. Below the gap there is no sharp transition to $F = 2$, but rather a smooth one. Indeed, for all voltages there are still contributing channels for which the voltage is above the gap and thus, there is a mixture of quasiparticle transport and Cooper pair transport for voltages below the gap. Since quasiparticle transport is of lower order in transparency this contribution is larger, and $F\approx 1$ just below the gap for small transparencies. 

\begin{figure}[!htb]
    \centering
    \includegraphics[width = 8.6cm]{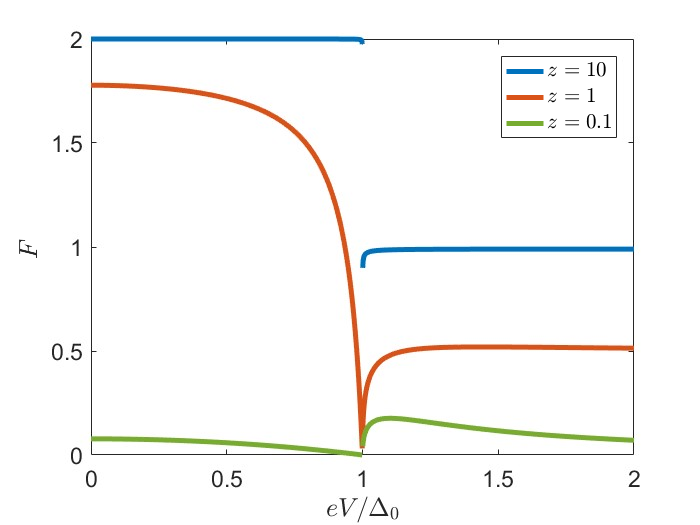}
    \includegraphics[width = 8.6cm]{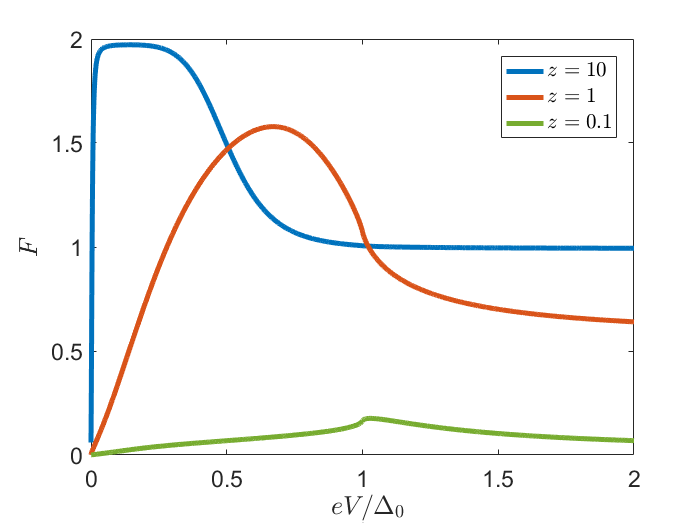}
    \includegraphics[width = 8.6cm]{figures/A.png}
    \includegraphics[width = 8.6cm]{figures/B.png}
    \caption{(a): The differential Fano factor for two-dimensional \textit{s} - wave superconductors in a junction with $z = 10$. The results are similar to the 1D case in Fig. \ref{fig:Z1D}. (b): The differential Fano factor as a function of voltage for $\text{\textit{p}}_{\text{x}}$ - wave superconductors. In the low-transparency limit the differential Fano factor converges to 0 at $eV = 0$. In contrast to the one-dimensional \textit{p} - wave superconductor in Fig. \ref{fig:Z1D} there is a gradual transition from $F = 1$ to $F = 2$.}
    \label{fig:X}
\end{figure}

However, for small voltages there are many more channels for which the pair potential is larger than the voltage than channels for which the pair potential is smaller than the voltage. Therefore, for any finite $z$, for small enough voltages there are so few channels with quasiparticle transport that Cooper pair transport dominates and $F\approx 2$ for small enough voltages, as shown in Fig. \ref{fig:X}(b). 

To understand this behavior in more detail the dependence of width of the voltage window with $F\approx 2$ on $z$ may be estimated. The contribution of channels with Cooper pair transport is suppressed compared to quasiparticle transport by a factor $z^{-2}$ because it is higher order in tunneling. However, quasiparticle transport is carried by oblique channels, and the transparency of channels with oblique incidence is suppressed by a factor $\cos^{2}\phi$, while the density of channels is suppressed by a factor $\cos\phi$. Thus, the contribution of oblique angles is suppressed by a factor $\cos^{3}\phi\approx (\frac{\pi}{2}-\phi)^{3}$. Now, if we define $\phi_{c}$ to be the angle with $|eV| = \Delta_{0}\cos\phi_{c}$, by integration over $\phi$ from $\frac{\pi}{2}$ to $\phi_{c}$ we conclude that the transport of quasiparticles is suppressed by a factor of order $(\frac{\pi}{2}-\phi_{c})^{4}$. 

Thus, both quasiparticle and Cooper pair transport have suppression factors.
The transition from quasiparticle dominated transport with $F\approx 1$ to Cooper pair dominated transprot with $F\approx 2$ happens when these two suppression factors are equal, that is, when $(\frac{\pi}{2}-\phi_{c})^{4}$ is of order $z^{-2}$. This happens for $|eV| = \Delta_{0}\cos{\Big(\frac{\pi}{2}-z^{-\frac{1}{2}}\Big)}\approx \frac{\Delta_{0}}{\sqrt{z}} = \Delta_{0}(\Tilde{T}(\phi = 0))^{\frac{1}{4}}$. Thus, the voltage window in which $F\approx 2$ is proportional to $\Tilde{T}^{\frac{1}{4}}$. Therefore, it can be significant even for large $z$.
This should be contrasted with the $O(\Delta_{0} \Tilde{T}(\phi = 0))$ dependence of the width of local minimum at $eV = 0$.
\begin{figure}[!htb]
    \centering
    \includegraphics[width = 8.6cm]{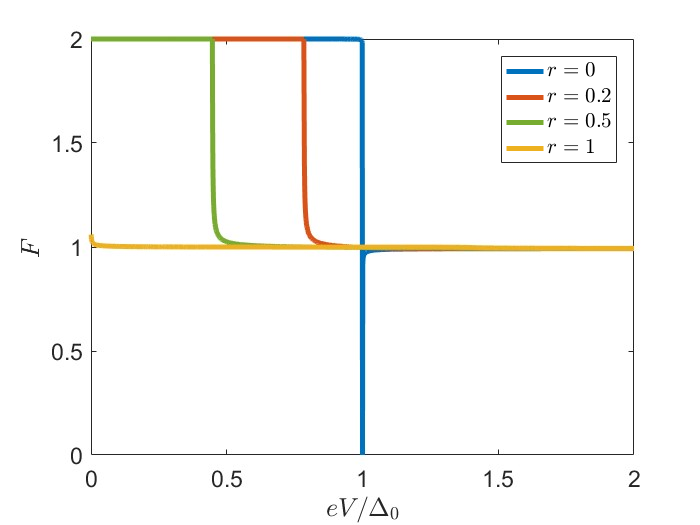}
    \includegraphics[width = 8.6cm]{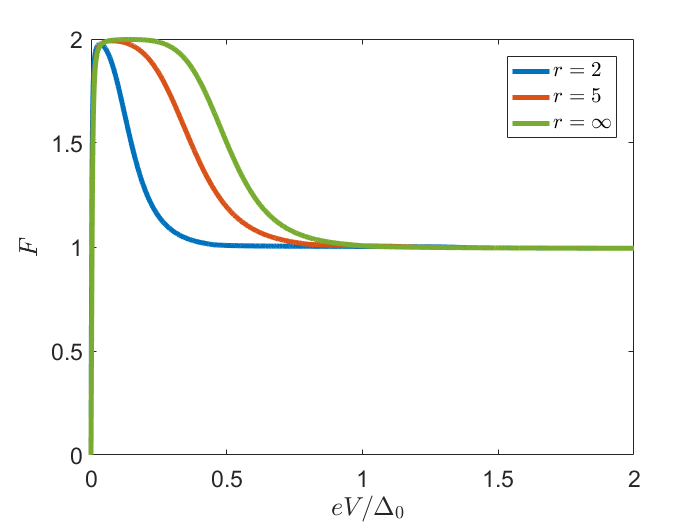}
    \includegraphics[width = 8.6cm]{figures/A.png}
    \includegraphics[width = 8.6cm]{figures/B.png}
    \caption{The differential Fano factor in junctions with $z = 10$ and with \textit{s} + $\text{\textit{p}}_{\text{x}}$ - wave superconductors. If the \textit{s} - wave component is dominant, the differential Fano factor has a transition between $F = 2$ and $F = 1$ at $|eV| = \Delta_{-}$, like for the 1D superconductor in Fig. \ref{fig:R1D}. Unlike the 1D case though, the transition is smoothened at the $F = 1$ side, due to the dependence of the gap magnitude on the angle of the channel. If the \textit{p} - wave component is dominant (b), there are dispersionless ZESABSs and hence $F(eV = 0) = 0$, even if a nonzero \textit{s} - wave component is included. The finite voltage window at which $F\approx 2$ is reduced by the inclusion of an \textit{s} - wave component.}
    \label{fig:RdepSpx}
\end{figure}
\begin{figure}[!htb]
    \centering
    \includegraphics[width = 8.6cm]{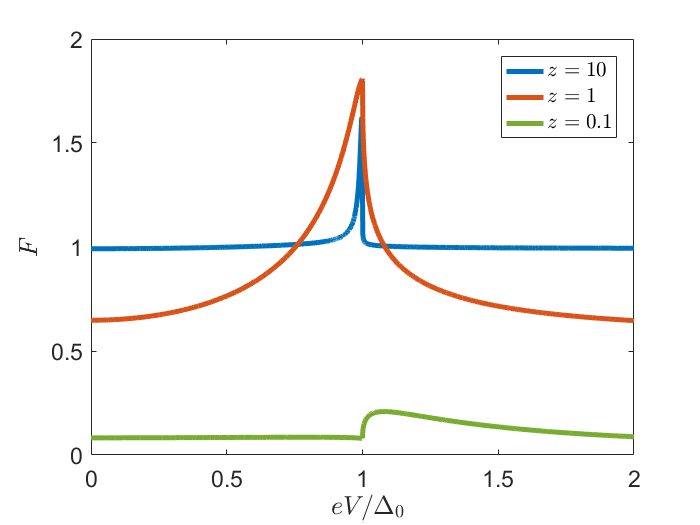}
    \caption{The differential Fano factor as a function of the BTK parameter $z$ for chiral and helical \textit{p} - wave superconductors, which give identical results. Even though the superconductors are \textit{p} - wave, the differential Fano factor does not decrease to 0 at $eV = 0$ in the low-transparency limit, because the SABSs are dispersive.}
    \label{fig:ZdepCH}
\end{figure}
\begin{figure}[!htb]
    \centering
    \includegraphics[width = 8.6cm]{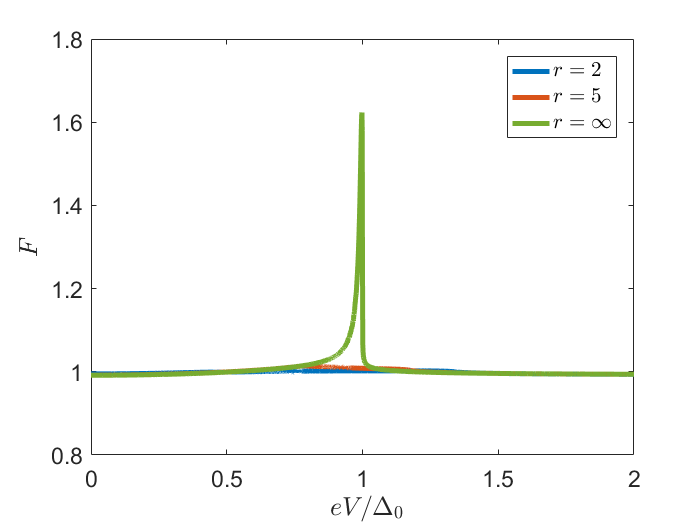}
    \includegraphics[width = 8.6cm]{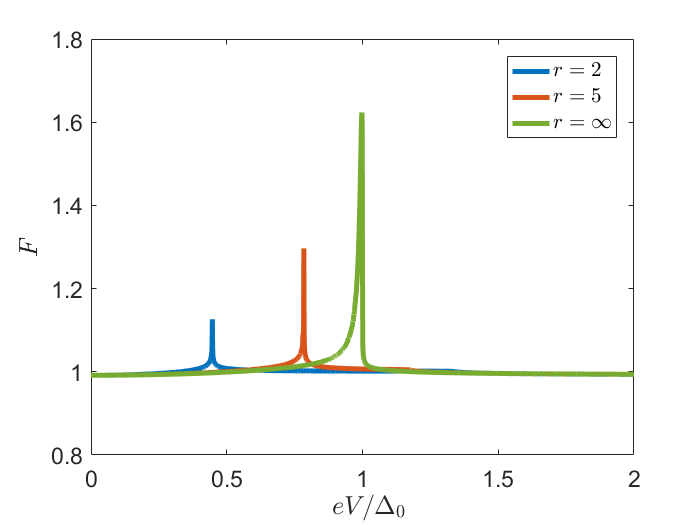}
    \includegraphics[width = 8.6cm]{figures/A.png}
    \includegraphics[width = 8.6cm]{figures/B.png}
    \caption{The differential Fano factor for two dimensional \textit{s} + \textit{p} - wave superconductors where the \textit{p} - wave is of the chiral (a) or helical (b) type if the \textit{p} - wave component is dominant.  For all $r$ the differential Fano factor converges to 1 on both sides of the peak. For chiral \textit{p} - wave superconductors the peak disappears almost completely upon the inclusion of an \textit{s} - wave component of the pair potential, on the other hand, for the helical type superconductor the peak is suppressed and shifts to $|eV| = |\Delta_{-}|$. The difference appears because for \textit{s} + helical \textit{p} - wave superconductor the gap has the same magnitude for all channels while for \textit{s} + chiral \textit{p} - wave superconductors the gap is channel dependent. We used $z  = 10$ in the calculation.}
    \label{fig:CHPdom}
\end{figure}
\begin{figure}[!htb]
    \centering
    \includegraphics[width = 8.6cm]{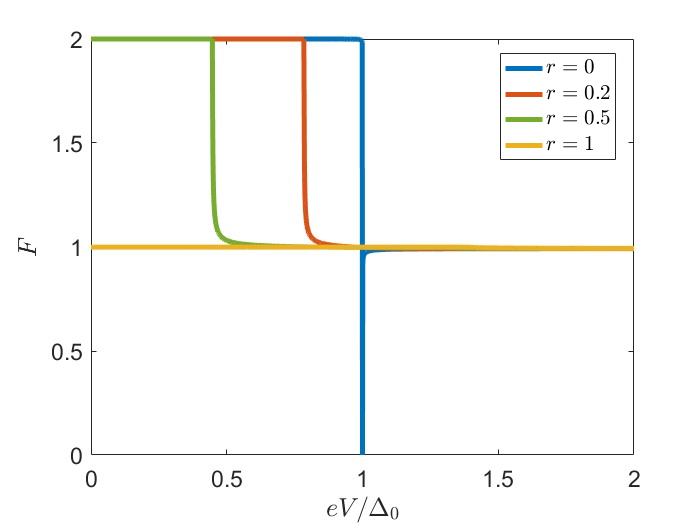}
    \includegraphics[width = 8.6cm]{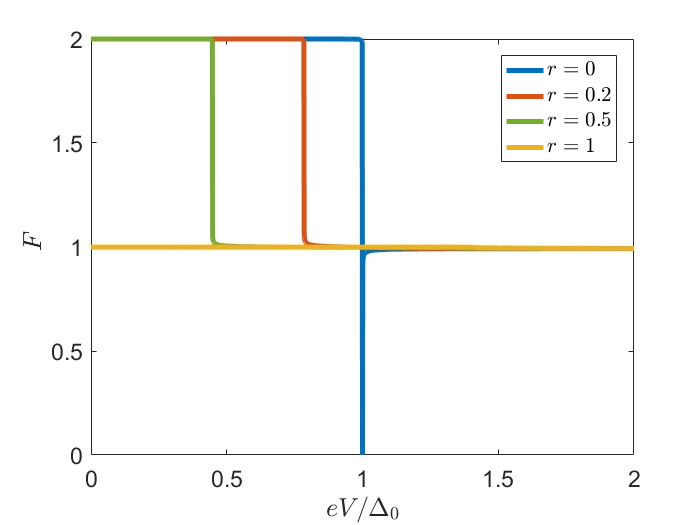}
    \includegraphics[width = 8.6cm]{figures/A.png}
    \includegraphics[width = 8.6cm]{figures/B.png}
    \caption{The differential Fano factor for junctions with $z = 10$ and two dimensional \textit{s} + \textit{p} - wave superconductors where the \textit{p} - wave is of the chiral (a) or helical (b) type if the \textit{s} - wave component is dominant. The inclusion of a \textit{p} - wave component shifts the transition between $F = 1$ and $F = 2$ down to $|eV| = \frac{1-r}{\sqrt{1+r^{2}}}\Delta_{0}$, and at $r = 1$ it is constant, $F \approx 1$ except close to $eV = 0$. For chiral \textit{p} - wave superconductors the transition between $F = 1$ and $F = 2$ is smoothened due to the nonconstant gap magnitude when \textit{s} - wave and chiral \textit{p} - wave potentials are mixed. On the other hand, for \textit{s} + helical \textit{p} - wave superconductors the gap magnitude is constant and the transition between the two regimes remains sharp. We used $z = 10$ in the calculation.}
    \label{fig:CHSdom}
\end{figure}

We may also consider a mixture of \textit{s} - wave and $\text{\textit{p}}_{\text{x}}$ - wave pair potentials in a  \textit{s} + $\text{\textit{p}}_{\text{x}}$ - wave superconductor. As shown in Fig. \ref{fig:RdepSpx}(a), adding a small $\text{\textit{p}}_{\text{x}}$ - wave component to an \textit{s} - wave superconductor suppresses the voltage at which $F$ changes from 1 to 2, due to the suppression of the gap magnitude for each angle. This voltage converges to 0 as $r$ approaches 1. Thus, for $r\approx 1$ the voltage window with $F \approx 2$ is very small. Moreover, due to the mechanism described above for $\text{\textit{p}}_{\text{x}}$ - wave superconductors the transition between $F = 1$ and $F = 2$ is not sudden but smooth for any $r\neq 0$. Similarly, if a small \textit{s} - wave component is added to a $\text{\textit{p}}_{\text{x}}$ - wave superconductor, the voltage window in which $F\approx 2$ decreases, see Fig. \ref{fig:RdepSpx}(b). The ZESABS however is robust against inclusion of this component \cite{tanaka2022theory} as long as the \textit{p} - wave component is larger than the \textit{s} - wave component, and hence the differential Fano factor vanishes at zero voltage even in the presence of a small \textit{s} - wave component.

For chiral and helical \textit{p} - wave superconductors the differential Fano factor is qualitatively different, see Fig. \ref{fig:ZdepCH}. For such superconductors the incident and reflected angle do not have opposite pair potential for all channels, but only for the channel at normal incidence, and hence the SABSs have a dispersion \cite{furusaki2001spontaneous,matsumoto1999quasiparticle,read2000paired,volovik2003universe,iniotakis2007andreev,tanaka2009theory}. The lack of a dispersionless ZESABSs implies that $F(eV = 0)\neq 0$. Moreover, for any voltage below $\Delta_{0}$, there are two types of channels.  First, there are channels for which the voltage is below the gap and not near a SABS, for those the current is carried by Cooper pairs and hence the contribution to the noise power is twice the contribution to the current. Second, there are the channels that are close to the SABS and hence contribute significantly to the current, but almost nothing to the noise power. On the other hand, if $|eV|>\Delta_{0}$ there is only one type of transport, quasiparticle transport via the continuum.

For both  helical and  chiral \textit{p} - wave superconductors, the differential Fano factor  has a sharp peak at $|eV| = \Delta_{0}$. This peak quickly decays to $F = 1$ both for voltages below and above the gap. The results for the two types of superconductors are exactly the same, because a helical \textit{p} - wave superconductor may be considered as the sum of two spinless chiral superconductors with opposite chirality. However, the change in the differential Fano factor upon inclusion of an \textit{s} - wave component is significantly different for the two types of superconductors.
For the chiral \textit{p} - wave superconductor, upon inclusion of an \textit{s} - wave component the magnitude of the gap becomes channel dependent. Therefore the peak is smeared out and almost completely disappears, as shown in Fig \ref{fig:CHPdom}(a). 
On the other hand, for \textit{s} + helical \textit{p} - wave superconductors the gap magnitude remains independent of angle even upon inclusion of an \textit{s} - wave component.  Upon inclusion of an \textit{s} - wave component the peak remains and is shifted to $|eV| = \frac{r-1}{\sqrt{1+r^{2}}}\Delta_{0}$. This comes along with a suppression of the width and height of the peak in $F$, as shown in Fig. \ref{fig:CHPdom}(b). If the \textit{s} - wave is dominant on the other hand, the results are very similar regardless of the type of the \textit{p} - wave used, and the results are also very similar to those in the \textit{s} + $\text{\textit{p}}_{\text{x}}$ - wave or 1D \textit{s} + \textit{p} - wave superconductors, as shown in Fig. \ref{fig:CHSdom}.

\subsection{i\textit{s} + \textit{p} - wave superconductors}
\begin{figure}[!htb]
    \centering
    \includegraphics[width = 8.6cm]{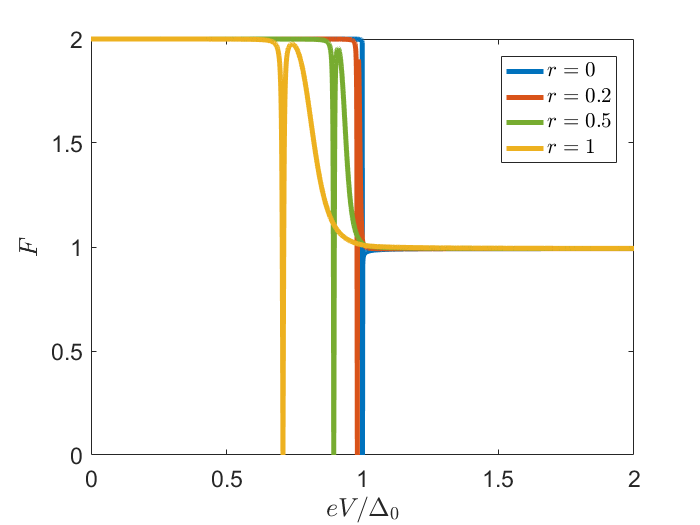}
    \includegraphics[width = 8.6cm]{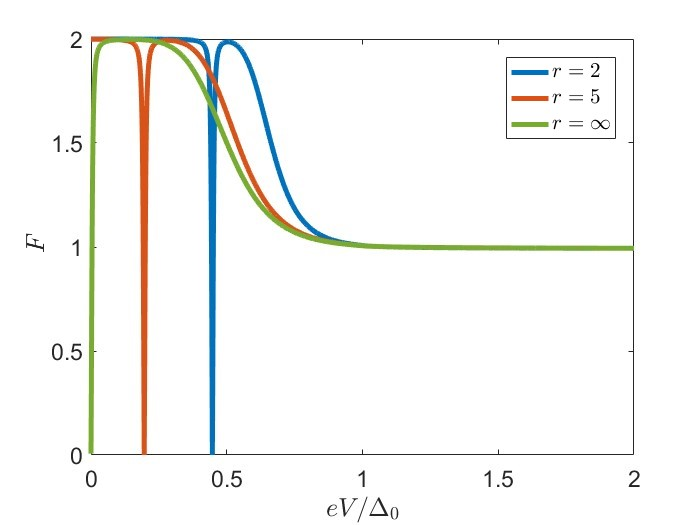}
    \includegraphics[width = 8.6cm]{figures/A.png}
    \includegraphics[width = 8.6cm]{figures/B.png}
    \caption{The differential Fano factor in junctions with i\textit{s} + $\text{\textit{p}}_{\text{x}}$  - wave superconductors and $z = 10$. Both if the \textit{s} - wave component is dominant (a) and if the \textit{p} - wave component is dominant (b) the differential Fano factor has a sharp local minimum for $|eV|\approx \Delta_{s}$, correlated with the existence of a ZESABS. For voltages smaller than the energy of the SABS the differential Fano factor equals 2. For larger voltages the differential Fano factor almost reaches two again but then gradually decreases to one. Notably, the ZESABS of the $\text{\textit{p}}_{\text{x}}$ - wave superconductor immediately shifts to nonzero energies upon inclusion of an \textit{s} - wave component. These results are in sharp contrast with the results for time-reversal symmetric \textit{s} + $\text{\textit{p}}_{\text{x}}$ superconductors in Fig. \ref{fig:RdepSpx}.}
    \label{fig:RdepISPx}
\end{figure}

We may also consider the differential Fano factor for time reversal and inversion symmetry broken superconductors in two dimensions. The results are shown in Fig.  \ref{fig:RdepISPx} for i\textit{s} + $\text{\textit{p}}_{\text{x}}$ - wave superconductors. The results are easily understood from comparison to the 1D superconductors, see Fig. \ref{fig:1DISP}. Indeed, for a nonzero \textit{p} - wave component there is a sharp local minimum in the differential Fano factor at $|eV| = \Delta_{s}$ due to the appearance of an SABS. Similar to the 1D i\textit{s} + \textit{p} - wave superconductor we see that there is only a ZESABS if the \textit{s} - wave components vanishes, the ZESABS is not robust in the absence of time-reversal symmetry. If $r>0$, for voltages below the energy of the SABS the differential Fano factor equals two. For higher voltages it reaches almost two but then decays because the transition between $F = 1$ and $F = 2$ is gradual in superconductors in which the gap magnitude is channel dependent as explained in the discussion of \textit{s} + $\text{\textit{p}}_{\text{x}}$ - wave superconductors.

\begin{figure}[!htb]
    \centering
    \includegraphics[width = 8.6cm]{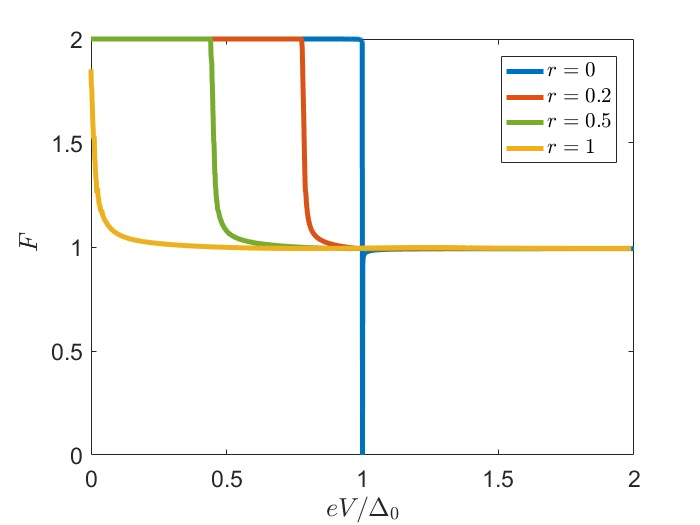}
    \includegraphics[width = 8.6cm]{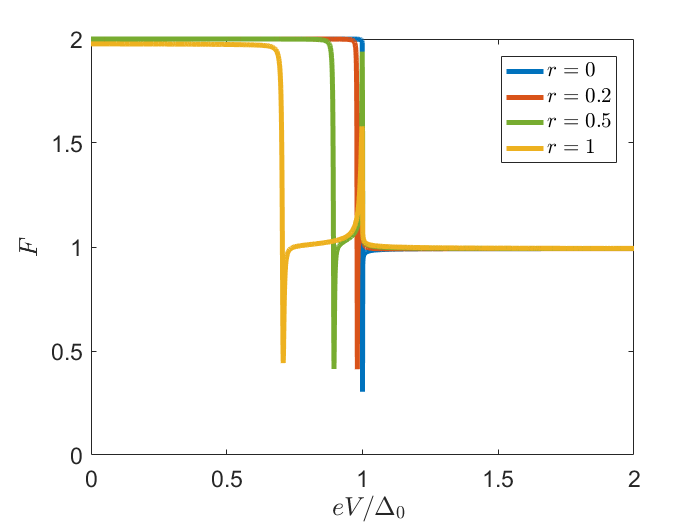}
    \includegraphics[width = 8.6cm]{figures/A.png}
    \includegraphics[width = 8.6cm]{figures/B.png}
    \caption{The differential Fano factor in \textit{s} - wave dominant i\textit{s} + \textit{p} - wave superconductors with chiral (a) or helical (b)  \textit{p} - wave components. In both cases the inclusion of a \textit{p} - wave component reduces the voltage at which $F$ transitions between 2 and 1. For i\textit{s} + chiral \textit{p} - wave superconductors the differential Fano factor changes smoothly between 1 and 2, but for i\textit{s} + helical \textit{p} - wave this transition remains sharp. The BTK parameter was set to $z = 10$.}
    \label{fig:ispCHS}
\end{figure}

Next we may consider i\textit{s} + chiral and i\textit{s} + helical \textit{p} - wave superconductors. If the \textit{s} - wave component is dominant the results are similar in both cases, see Fig. \ref{fig:ispCHS}, and also similar to \textit{s} + chiral and \textit{s} + helical \textit{p} - wave superconductors, as seen from the comparison of Fig. \ref{fig:ispCHS} to Fig. \ref{fig:CHSdom}. The only difference is that the voltage window in which $F = 2$ is larger for 2D i\textit{s} +  helical \textit{p} - wave superconductors than for 2D \textit{s} +  helical \textit{p} - wave superconductors, because the minimum gap is suppressed less in the former case. For i\textit{s} + helical \textit{p} - wave superconductors the transition between $F = 1$ and $F = 2$ is accompanied by a sharp local minimum in the differential Fano factor, because the gap magnitude is independent of channel. This local minimum is absent for i\textit{s} + chiral \textit{p} - wave superconductors because the gap magnitude is channel dependent in i\textit{s} + chiral \textit{p} - wave superconductors.
 
\begin{figure}[!htb]
    \centering
    \includegraphics[width = 8.6cm]{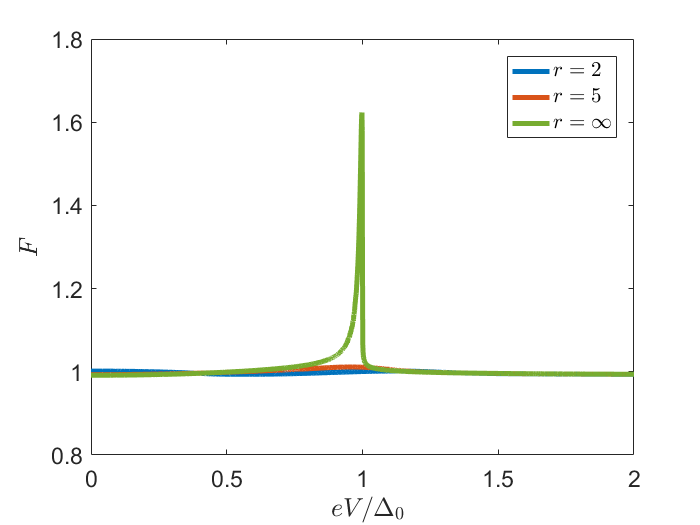}
    \includegraphics[width = 8.6cm]{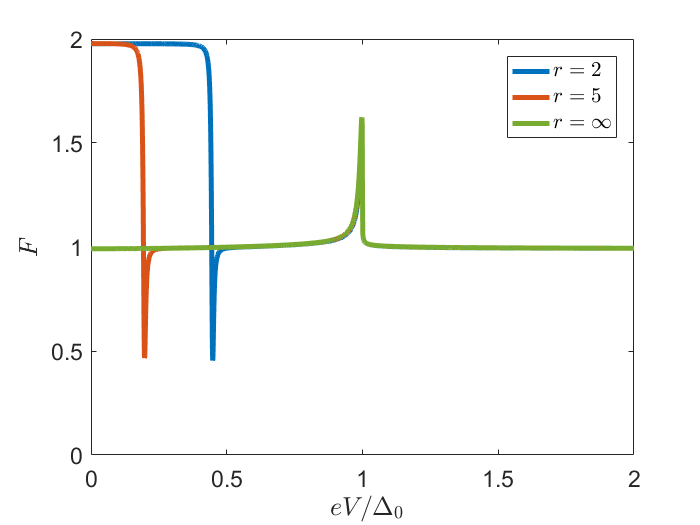}
    \includegraphics[width = 8.6cm]{figures/A.png}
    \includegraphics[width = 8.6cm]{figures/B.png}
    \caption{The differential Fano factor in junctions with \textit{p} - wave dominant i\textit{s} + \textit{p} - wave superconductors. For chiral \textit{p} - wave superconductors (a) the results are very similar to those obtained with \textit{s} +  chiral \textit{p} - wave superconductors in Fig. \ref{fig:CHPdom}, since time-reversal symmetry is broken in both of them. On the other hand, for the i\textit{s} + helical \textit{p} - wave superconductors (b) there is a significant change compared to \textit{s} + helical \textit{p} - wave superconductors in Fig. \ref{fig:CHPdom}. Indeed, there is a voltage window in which $F = 2$, a feature that is usually observed for \textit{s} - wave dominant superconductors. This is due to the opening of a gap in the spectrum of SABSs. The BTK parameter was set to $z = 10$.}
    \label{fig:ispCHP}
\end{figure}

 For \textit{p} - wave dominant i\textit{s} + \textit{p} - wave superconductors there is a large difference between the chiral and helical variants, as shown in Fig. \ref{fig:ispCHP}. For i\textit{s} + chiral superconductors the results are similar compared to the \textit{s} + chiral \textit{p} - wave superconductors in Fig. \ref{fig:CHPdom}, because both are time-reversal broken. Indeed, the phase between the two contributions to the pair potential is different for each modes, the only difference between \textit{s} + chiral and i\textit{s} + chiral \textit{p} - wave superconductors is whether the zero phase difference appears for normal or oblique incidence. On the other hand, for \textit{p} - wave dominant time-reversal broken i\textit{s} + helical \textit{p} - wave superconductors the results, shown in Fig. \ref{fig:ispCHP} are significantly different compared to the \textit{p} - wave dominant time-reversal symmetric \textit{s} + helical \textit{p} - wave superconductors.  Indeed, for i\textit{s} + helical \textit{p} - wave superconductors the peak at $|eV| = \Delta_{0}$ does not shift, but only decays upon the inclusion of an \textit{s} - wave component. Moreover there is a voltage window with $F = 2$ in this case, which does not appear in normal metal / \textit{s} + helical \textit{p} - wave superconductor junctions, see Fig. \ref{fig:CHPdom}. This voltage window appears because the time-reversal symmetry breaking destroys the topological protection of the helical \textit{p} - wave superconductor and introduces a gap in the spectrum of SABSs, the energy of the SABS is given by $E_{BS} = \Delta_{0}\sqrt{1-\frac{r}{1+r^{2}}\cos^{2}\phi}$. For i\textit{s} + chiral \textit{p} - wave superconductors on the other hand, such gap in the SABS spectrum does not open and therefore there is no voltage window with $F = 2$ in Fig. \ref{fig:ispCHP}. Thus, in \textit{p} - wave dominant time-reversal broken non-centrosymmetric superconductors, chiral and helical phases are easily distinguished through the differential Fano factor.

\clearpage
\section{Finite temperature}\label{sec:FiniteTemperature}
In this section we discuss the implementation of the calculation of the differential Fano factor at finite temperatures, and thus provide details on the method used to produce Figs. \ref{fig:FinNoiseSdom} - \ref{fig:BWMain} in the main text. We show that the temperature dependence of the differential Fano factor can be determined via the calculation of 5 quantities, $N_{1,2,3,4,5}$, defined in Eqs. (\ref{eq:N1def}) - (\ref{eq:N5def}) below, that are evaluated at zero temperature, and a single integration over energy, greatly reducing the cost of computation compared to a brute force approach. Next to this we show additional results of the temperature dependence for \textit{s} + \textit{p} - wave superconductors with different ratios of the magnitudes of the \textit{s} - wave and \textit{p} - wave contributions to the pair potential, thereby showing that the results depend only quantitatively on the exact value of $r$. Qualitatively they depend only on which one is dominant. Additionally, we provide results for the chiral \textit{d} - wave superconductor to support our claim that also for dispersive SABSs the suppression of the differential Fano factor should be attributed only to the presence of SABSs, not to the odd-parity of the \textit{p} - wave superconductor. The results are summarized in Table \ref{tab:mixtures} for mixed (i)\textit{s} + \textit{p} - wave. For other types of superconductors, such as \textit{d} - wave superconductors or $\text{\textit{p}}_{\text{x}}$ - wave superconductors, the results are equivalent to one of the results in Table \ref{tab:mixtures}.

\begin{table}[ht]
    \centering
    \begin{tabular}{|c|c|c|c|c|c|}
    \hline
         Type&ZBCP&$F_{00}$&$F(T>0,|eV|\approx k_{B}T)<0?$ &Minima for $T>0$&Fig.\\
         \hline
         1D \textit{s} + \textit{p}, $r<1$&\xmark&2&\xmark&-&\ref{fig:FinNoiseSdomAppendix}\\
         1D \textit{s} + \textit{p}, $r>1$&\cmark&0&\cmark&$|eV|\approx k_{B}T$&\ref{fig:FinNoisePdom}\\
         1D i\textit{s} + \textit{p}, $r<1$&\xmark&2&\cmark&$|eV|\approx\Delta_{s}$&\ref{fig:FinNoiseSdomISPAppendix}\\
         1D i\textit{s} + \textit{p}, $r>1$&\xmark&2&\cmark&$|eV|\approx\Delta_{s}$&\ref{fig:FinNoisePdomISP}\\
          2D \textit{s} + helical \textit{p}, $r<1$ & \xmark&2&\xmark&-&\ref{fig:FanoSH}\\
          2D \textit{s} + helical \textit{p}, $r>1$ & \cmark&1&\xmark&$|eV| = |\Delta_{-}|$&\ref{fig:FanoHS}\\
          2D i\textit{s} + helical \textit{p}, $r<1$ & \xmark&2&\xmark&$|eV| = \Delta_{s},|eV| = \Delta_{0}$&\ref{fig:FanoiSH}\\
          2D i\textit{s} + helical \textit{p}, $r>1$ & \xmark&2&\xmark&$|eV| = \Delta_{s},|eV| = \Delta_{0}$&\ref{fig:FanoHiS}\\
          2D \textit{s} + chiral \textit{p}, $r<1$ & \xmark&2&\xmark&-&\ref{fig:FanoSC}\\
          2D \textit{s} + chiral \textit{p}, $r>1$ & \cmark&1&\xmark&$|eV| = |\Delta_{-}|$&\ref{fig:FanoCS}\\
          2D i\textit{s} + chiral \textit{p}, $r<1$ & \xmark&2&\xmark&$|eV| = |\Delta_{-}|$&\ref{fig:FanoiSC}\\
          2D i\textit{s} + chiral \textit{p}, $r>1$ & \xmark&1&\xmark&$|eV| = |\Delta_{-}|$&\ref{fig:FanoCiS}\\
          3D \textit{s} + B-W, $r<1$&\xmark&1&\xmark&-&\ref{fig:SBW1}\\
          3D \textit{s} + B-W, $r>1$&\cmark&2&\xmark&$|eV| = |\Delta_{-}|$&\ref{fig:SBW2}\\
          3D i\textit{s} + B-W, $r<1$&\xmark&2&\xmark&$|eV| = \Delta_{0}$&\ref{fig:SBWI}\\
          3D i\textit{s} + B-W, $r>1$&\xmark&2&\xmark&$|eV| = \Delta_{0}$&\ref{fig:BWI}\\
          \hline
    \end{tabular}
    \caption{The conductance and differential Fano factor features for mixed (i)\textit{s} + \textit{p} - wave superconductors in low-transparency tunnel junctions for several types of non-centrosymmetric superconductors.}
    \label{tab:mixtures}
\end{table}

The reason that calculations at finite temperatures are significantly more difficult than for zero temperature is as follows. At zero temperature, the distribution functions reduce to step functions, at each energy either all states are occupied, or all of them are empty. Therefore, the voltage enters the expression for noise and current only in the integration bounds. In the integrand for $|E|<|eV|$ we have $f_{L} = 0$ and $f_{T} = \text{sign}(E)$, while in the integrand for $|E|>|eV|$ we have $f_{L} = \text{sign}(E)$ and $f_{T} = 0$. Thus, the voltage derivative can be extracted from the integrand evaluated only at $|E| =  |eV|$ and for two different values of  $(f_{L},f_{T})$ in the normal metal, namely $(0,1)$ and $(1,0)$. 

On the other hand, at finite temperatures this sharp contrast disappears, the distribution functions are not step functions anymore, but  instead they are continuous functions. Therefore, the voltage enters in a nontrivial way in the integrand and energy integration is required at each voltage and temperature, and the expressions for current and in a brute force approach noise need to be evaluated at intermediate values of $(f_{L},f_{T})$. This leads to a large increase of computational cost. 

However, the computational cost can be reduced by exploiting the known dependence of current and noise power on the distribution functions. For the current, we know that it is necessarily linear in the transverse distribution function, due to the causality structure and the absence of any supercurrent.
Therefore, using that $\frac{\partial f_{T}}{\partial V} = \frac{1}{2k_{B}T}\Big(\cosh^{-2}(\frac{E+eV}{k_{B}T})+\cosh^{-2}(\frac{E-eV}{k_{B}T})\Big)$, we find that
\begin{align}
    \sigma(eV,T) = \frac{1}{2}\int_{-\infty}^{\infty}\sigma(E,T = 0)\Bigg(\cosh^{-2}(\frac{E+eV}{k_{B}T})+\cosh^{-2}(\frac{E-eV}{k_{B}T})\Bigg) dE\;.
\end{align}
Thus, the conductance at any nonzero temperature can be calculated from the zero temperature conductance via a single numerical energy integration, and it is not necessary to do any extra matrix calculations. This significantly reduces the computational cost. Moreover, using this approach no extra numerical errors are made due to discretization of $\frac{\partial f_{T}}{\partial V}$.

A similar, but slightly more elaborate, procedure can be used for the noise power. 
As seen from Eq. (\ref{eq:PNdef}) in the main text, all the terms in the noise expression that either have no $\tau_{K}$'s or have have the form $X_{1}\tau_{K}X_{2}\tau_{K}X_{3}$, where $X_{1,2,3}$ are matrices depending on $C,G_{N}$, but do not contain $\tau_{K}$. The trace of terms without $\tau_{K}$'s does not involve the Keldysh component and hence does not depend on any distribution function. Consequently, the voltage dependent terms all have the form $\begin{bmatrix}
    X_{l}^{R}&X_{l}^{K}\\0&X_{l}^{A}
\end{bmatrix}$, where only $X_{l}^{K}$ depends on the distribution functions, and it is linear in those distribution functions. Now $X_{1}\tau_{K}X_{2}\tau_{K}X_{3} = \begin{bmatrix}
    X_{1}^{R}\tau_{3}X_{2}^{A}\tau_{3}X_{3}^{R}+X_{1}^{K}\tau_{3}X_{2}^{K}\tau_{3}X_{3}^{R}&X_{1}^{R}\tau_{3}X_{2}^{A}\tau_{3}X_{3}^{K}+X_{1}^{R}\tau_{3}X_{2}^{R}\tau_{3}X_{3}^{A}+X_{1}^{K}\tau_{3}X_{2}^{K}\tau_{3}X_{3}^{K}\\X_{1}^{A}\tau_{3}X_{2}^{K}\tau_{3}X_{3}^{R}&X_{1}^{A}\tau_{3}X_{2}^{R}\tau_{3}X_{3}^{A}+X_{1}^{A}\tau_{3}X_{2}^{K}\tau_{3}X_{3}^{K}\end{bmatrix}$. Thus, $\text{Tr}X_{1}\tau_{K}X_{2}\tau_{K}X_{3} = \text{Tr}(X_{1}^{R}\tau_{3}X_{2}^{A}\tau_{3}X_{3}^{R}+X_{1}^{A}\tau_{3}X_{2}^{R}\tau_{3}X_{3}^{A})+\text{Tr}(X_{1}^{K}\tau_{3}X_{2}^{K}\tau_{3}X_{3}^{R}+X_{1}^{A}\tau_{3}X_{2}^{K}\tau_{3}X_{3}^{K})$. The first of these traces is  independent of voltage and thus can be ignored for our purposes. The second trace has terms with two Keldysh components, and is thus quadratic in the distribution functions. 

Thus, all voltage dependent terms in $P_{N}$ are of second order in the distribution functions. Since $f_{S}^{2}$ is independent of voltage, we may express the voltage dependent part of $P_{N}$ as the sum of 5 terms:
\begin{align}\label{eq:PviaN}
    \frac{\partial P_{N}}{\partial V} = \frac{\partial}{\partial V}\int_{-\infty}^{\infty} dE N_{1}f_{S}f_{L}+N_{2}f_{S}f_{T}+N_{3}f_{L}^{2}+N_{4}f_{T}^{2}+N_{5}f_{L}f_{T}\;,
\end{align}
where $N_{1,2,3,4,5}$ are coefficients that should be determined using Eq. (\ref{eq:PNdef}) in the main text. The coefficients $N_{1,2,3,4,5}$ may depend on energy, but not on voltage of temperature, since they do not depend on the distribution functions.

Eq. (\ref{eq:PviaN}) is valid for any choice of the distribution functions $f_{L},f_{T},f_{S}$, thus what remains is to determine the coefficients $N_{1,2,3,4,5}$. This can be done by evaluating $P_{N}$ for specific values of $f_{S},f_{L},f_{T}$.
Indeed, since this expression is a second order polynomial in $f_{S},f_{L}, f_{T}$ its coefficients $N_{i}$ can be uniquely determined by solving the Keldysh equation at the following five combinations of distribution functions: $(f_{S},f_{L},f_{T}) \in\{(1,0,0),(1,1,0),(1,\frac{1}{2},0), (1,0,1),(1,0,\frac{1}{2}),(1,1,1)\}$.

This can be understood as follows. The expression for noise power, Eq. (\ref{eq:PNdef}) always involves an integral over energy. Defining the integrand as $\Tilde{\mathcal{P}}_{N}$, we may write
\begin{align}\label{eq:PNdefSupp1}
    \frac{\partial P_{N}}{\partial V} = \frac{\partial}{\partial V}\int_{-\infty}^{\infty} dE \Tilde{\mathcal{P}}_{N}(E,f_{L},f_{T})\;.
\end{align}
For the tunnel junction we have, \textit{c.f.} Eq. (\ref{eq:PNdef}) in the main text,
\begin{align}\label{eq:PNdefSupp2}
    \Tilde{\mathcal{P}}_{N}&=-\int_{-\frac{\pi}{2}}^{\frac{\pi}{2}} \cos(\phi)d\phi\frac{e^{2}N_{\text{ch}}}{8\pi\hbar}\text{Tr}\Bigg(\frac{\Tilde{T}(\phi)}{2}(4-2\Tilde{T}(\phi)+\Tilde{T}(\phi)\{C,G_{N}\})^{-1}(\{C,\tau_{K}G_{N}\tau_{K}\}-\{C,G_{N}\})\nonumber\\&+\frac{\Tilde{T}(\phi)^{2}}{4}\Big((4-2\Tilde{T}(\phi)+\Tilde{T}(\phi)\{C,G_{N}\})^{-1}\{C,[\tau_{K},G_{N}]\}\Big)^{2}\Bigg)\; ,
\end{align}
where $C$ depends on $E$ and $G_{N}$ on $f_{L},f_{T}$. 

Because both Eqs. (\ref{eq:PviaN}) and (\ref{eq:PNdefSupp2}) are expressions for $P_{N}$ that are valid for any choice of the distribution functions, we have 
\begin{align}
    &-\int_{-\infty}^{\infty} dE\int_{-\frac{\pi}{2}}^{\frac{\pi}{2}} \cos(\phi)d\phi\frac{e^{2}N_{\text{ch}}}{8\pi\hbar}\text{Tr}\Bigg(\frac{\Tilde{T}(\phi)}{2}(4-2\Tilde{T}(\phi)+\Tilde{T}(\phi)\{C,G_{N}\})^{-1}(\{C,\tau_{K}G_{N}\tau_{K}\}-\{C,G_{N}\})\nonumber\\&+\frac{\Tilde{T}(\phi)^{2}}{4}\Big((4-2\Tilde{T}(\phi)+\Tilde{T}(\phi)\{C,G_{N}\})^{-1}\{C,[\tau_{K},G_{N}]\}\Big)^{2}\Bigg)\nonumber\\ &= \int_{-\infty}^{\infty} dE N_{1}(E)f_{S}f_{L}+N_{2}(E)f_{S}f_{T}+N_{3}(E)f_{L}^{2}+N_{4}(E)f_{T}^{2}+N_{5}(E)f_{L}f_{T}\;,
\end{align}
for any choice of $f_{S},f_{L},f_{T}$. Thus,
\begin{align}
    &-\int_{-\frac{\pi}{2}}^{\frac{\pi}{2}} \cos(\phi)d\phi\frac{e^{2}N_{\text{ch}}}{8\pi\hbar}\text{Tr}\Bigg(\frac{\Tilde{T}(\phi)}{2}(4-2\Tilde{T}(\phi)+\Tilde{T}(\phi)\{C,G_{N}\})^{-1}(\{C,\tau_{K}G_{N}\tau_{K}\}-\{C,G_{N}\})\nonumber\\&+\frac{\Tilde{T}(\phi)^{2}}{4}\Big((4-2\Tilde{T}(\phi)+\Tilde{T}(\phi)\{C,G_{N}\})^{-1}\{C,[\tau_{K},G_{N}]\}\Big)^{2}\Bigg) \nonumber\\&=  N_{1}(E)f_{S}f_{L}+N_{2}(E)f_{S}f_{T}+N_{3}(E)f_{L}^{2}+N_{4}(E)f_{T}^{2}+N_{5}(E)f_{L}f_{T}\;,
\end{align}
for any choice of $f_{S},f_{L},f_{T}$.

The 5 coefficients $N_{1,2,3,4,5}$ may be determined uniquely by evaluation of this expression at five combinations of $(f_{L},f_{S},f_{T})$. Which five combinations are chosen does not matter, as long as they give linearly independent equations. 
We choose the following five sets of distribution functions: $(f_{S},f_{L},f_{T}) \in\{(1,0,0),(1,1,0),(1,\frac{1}{2},0), (1,0,1),(1,0,\frac{1}{2}),(1,1,1)\}$. The five equations that are obtained are
\begin{align}
    &\mathcal{\Tilde{P}}_{N}(E,f_{S} = 1, f_{L} = 1, f_{T} = 0) = N_{1}+N_{3}\;,\\
    &\mathcal{\Tilde{P}}_{N}(E,f_{S} = 1, f_{L} = \frac{1}{2}, f_{T} = 0) = \frac{1}{2}N_{1}+\frac{1}{4}N_{3}\;,\\
    &\mathcal{\Tilde{P}}_{N}(E,f_{S} = 1, f_{L} = 0, f_{T} = 1) = N_{2}+N_{4}\;,\\
    &\mathcal{\Tilde{P}}_{N}(E,f_{S} = 1, f_{L} = 0, f_{T} = \frac{1}{2}) = \frac{1}{2}N_{2}+\frac{1}{4}N_{4}\;,\\
    &\mathcal{\Tilde{P}}_{N}(E,f_{S} = 1,f_{L} = 1, f_{T} = 1) = N_{1}+N_{2}+N_{3}+N_{4}+N_{5}\;.
\end{align}
From this set of equations we may uniquely determine $N_{1,2,3,4,5}$:
\begin{align}
    N_{1}(E) &= 4\Tilde{\mathcal{P}}_{N}(E,f_{S} = 1,f_{L} = \frac{1}{2},f_{T} = 0)-\Tilde{\mathcal{P}}_{N}(E,f_{S} = 1,f_{L} = 1,f_{T} = 0)\; ,\label{eq:N1def}\\
    N_{2}(E) &= 4\Tilde{\mathcal{P}}_{N}(E,f_{S} = 1,f_{L} = 0,f_{T} = \frac{1}{2})-\Tilde{\mathcal{P}}_{N}(E,f_{S} = 1,f_{L} = 0,f_{T} = 1)\; ,\\
    N_{3}(E) &= 2\Tilde{\mathcal{P}}_{N}(E,f_{S} = 1,f_{L} = 1,f_{T} = 0)-4\Tilde{\mathcal{P}}_{N}(E,f_{S} = 1,f_{L} = \frac{1}{2},f_{T} = 0)\; ,\\
    N_{4}(E) &= 2\Tilde{\mathcal{P}}_{N}(E,f_{S} = 1,f_{L} = 0,f_{T} = 1)-4\Tilde{\mathcal{P}}_{N}(E,f_{S} = 1,f_{L} = 0,f_{T} = \frac{1}{2})\; ,\\
    N_{5}(E) &= \Tilde{\mathcal{P}}_{N}(E,f_{S} = 1,f_{L} = 1,f_{T} = 1)-(N_{1}+N_{2}+N_{3}+N_{4})\; .\label{eq:N5def}
\end{align}
The expressions for $\Tilde{P}_{N}$ can be evaluated numerically. 
With this, we have determined the coefficients $N_{1,2,3,4,5}$, which depend on energy, but, importantly, are independent of temperature and voltage. Thus, we may compute the noise for any choice of the distribution functions, without altering these coefficients. 

We may now consider a physical system in which we set $f_{S} = \tanh{\frac{E}{2k_{B}T}}$, $f_{L} = \frac{1}{2}\Big(\tanh{\frac{E+eV}{2k_{B}T}}+\tanh{\frac{E-eV}{2k_{B}T}}\Big)$ and $f_{T} = \frac{1}{2}\Big(\tanh{\frac{E+eV}{2k_{B}T}}-\tanh{\frac{E-eV}{2k_{B}T}}\Big)$. For such a system, the differential noise power may be written as,
\begin{align}\label{eq:dPNdVCalcs}
    \frac{\partial P_{N}}{\partial V} = \int_{-\infty}^{\infty} dE N_{1}\frac{\partial f_{L}}{\partial V}\tanh{\frac{E}{2k_{B}T}}+N_{2}\frac{\partial f_{T}}{\partial V}\tanh{\frac{E}{2k_{B}T}}+2N_{3}f_{L}\frac{\partial f_{L}}{\partial V}+2N_{4}f_{T}\frac{\partial f_{T}}{\partial V}+N_{5}(f_{L}\frac{\partial f_{T}}{\partial V}+f_{T}\frac{\partial f_{L}}{\partial V})\;, 
\end{align}
where $\frac{\partial f_{L}}{\partial V} = \frac{e}{2k_{B}T}\Big(\cosh^{-2}(\frac{E+eV}{2k_{B}T})-\cosh^{-2}(\frac{E-eV}{2k_{B}T})\Big)$, $\frac{\partial f_{T}}{\partial V} = \frac{e}{2k_{B}T}\Big(\cosh^{-2}(\frac{E+eV}{2k_{B}T})+\cosh^{-2}(\frac{E-eV}{2k_{B}T})\Big)$. 
The resulting calculations are more heavy than those at zero temperature, because we need to evaluate the noise for each of the five different combinations ($\{(1,1,0),(1,\frac{1}{2},0),(1,0,1),(1,0,\frac{1}{2}),(1,1,1)\}$) and perform a numerical integral over energy. However, since we do not need to do numerical differential or evaluation of $\mathcal{P}_{N}$ for any other values, it constitutes a large reduction of computational cost and numerical errors compared to the naive approach of brute force calculation of $\frac{\partial P_{N}}{\partial V}$.
In the following subsections we use Eqs. (\ref{eq:N1def}) - (\ref{eq:dPNdVCalcs}) to calculate the noise power for several distinct USC/normal metal junctions.
\subsection{1D}
The differential Fano factor for low voltages is very sensitive to temperature. Indeed, the main contribution at zero temperature is due to the change of $f_{L}^{2}$ and $f_{T}^{2}$, which jump from 0 to 1 at $|eV| = |E|$. However, for any finite temperature, we find that $\frac{\partial(f_{L,T}^{2})}{\partial V}\xrightarrow[]{}0,$ as $V \xrightarrow{} 0$, and in this limit as well $\frac{\partial f_{L}}{\partial V}\xrightarrow{}0$, while $\frac{\partial f_{T}}{\partial V}$ remains finite as $V\xrightarrow{}0$. 
Therefore, at finite temperature, the differential noise power is severely suppressed for $|eV|\ll k_{B}T$, while the conductance is not, as has been predicted before for normal metals \cite{blanter2000shot}. Notably, this does not mean that the noise power is smaller at finite temperatures. Indeed, if $T\neq 0$ there is noise even in the absence of any current, thermal noise \cite{blanter2000shot}. Thus, the ratio of noise power to current diverges at $eV = 0$ at finite temperatures. The differential Fano factor on the other hand goes to zero for any system \cite{blanter2000shot}, because the distribution functions are analytic for all $k_{B}T>0$.
\begin{figure}[!htb]
    \centering
    \includegraphics[width = 8.6cm]{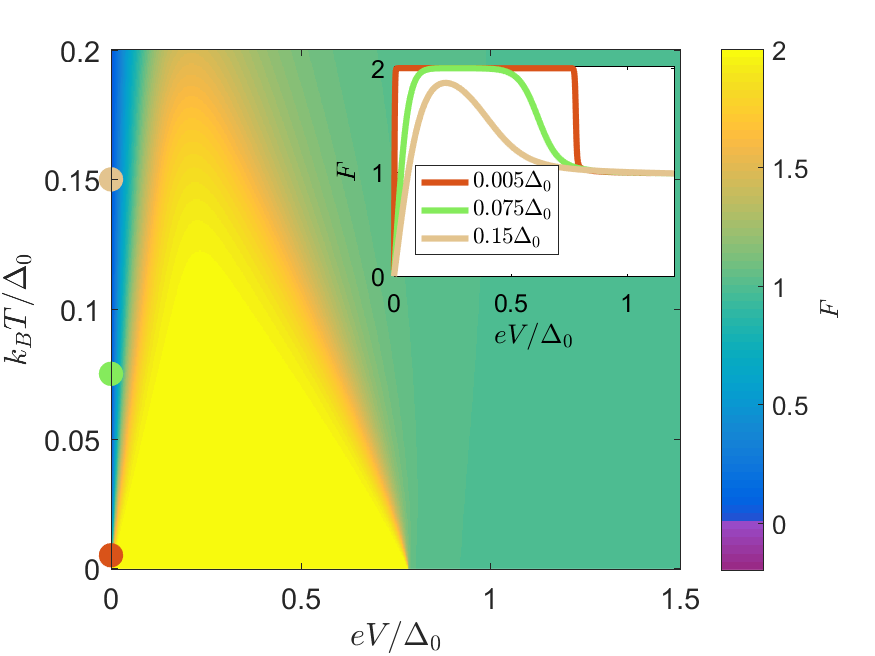}
    \includegraphics[width = 8.6cm]{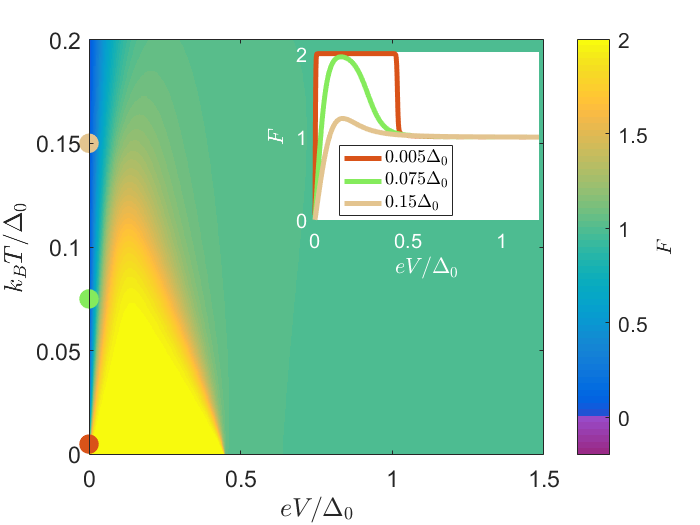}
    \includegraphics[width = 8.6cm]{figures/A.png}
    \includegraphics[width = 8.6cm]{figures/B.png}
    \caption{The differential Fano factor as a function of voltage and temperature in 1D \textit{s} - wave dominant \textit{s} + \textit{p} - wave superconductors, $r = 0.2$ (a) and $r = 0.5$ (b). At finite temperatures, a local minimum at $eV = 0$ with a width of $k_{B}T$ develops, with $F = 0$ at $eV = 0$. For larger $r$ the differential Fano factor goes to 1 for smaller voltages. We use $z = 10$.} \label{fig:FinNoiseSdomAppendix}
\end{figure}
As shown in Fig. \ref{fig:FinNoiseSdomAppendix}, if the \textit{s} - wave component of the pair potential is dominant, the differential Fano factor vanishes at zero voltage and  approaches 2 for $\Delta_{0}\gg|eV|\gg k_{B}T$. Therefore, while at zero temperature the local minimum in the differential Fano factor indicates a \textit{p} - wave dominant pair potential, as was found in Fig. \ref{fig:R1D} in Appendix \ref{sec:1Dnumerical}, at finite temperature the zero voltage differential Fano factor is zero for any material, including \textit{s} - wave superconductors.

\begin{figure}[!htb]
    \centering
    \includegraphics[width = 8.6cm]{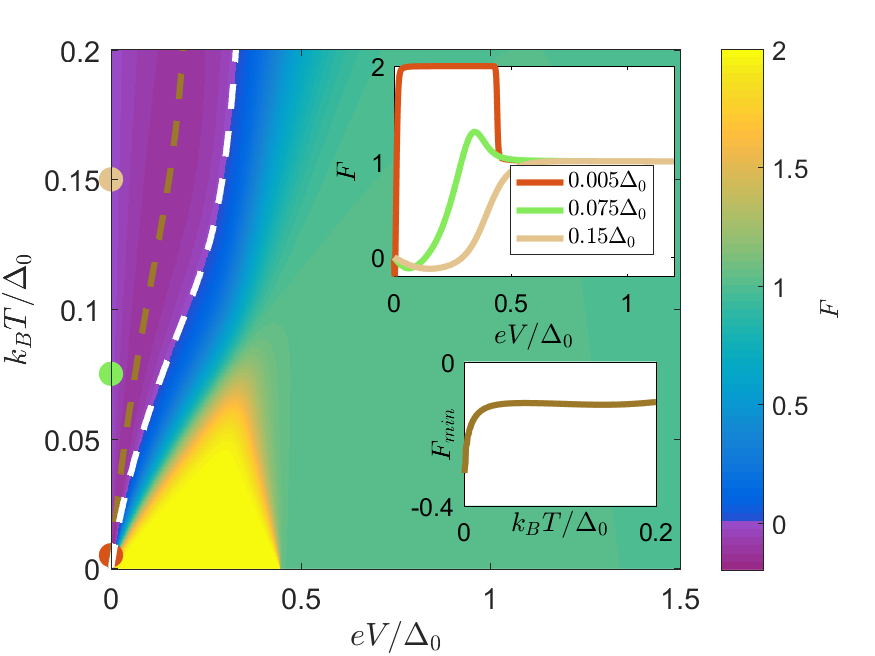}
    \includegraphics[width = 8.6cm]{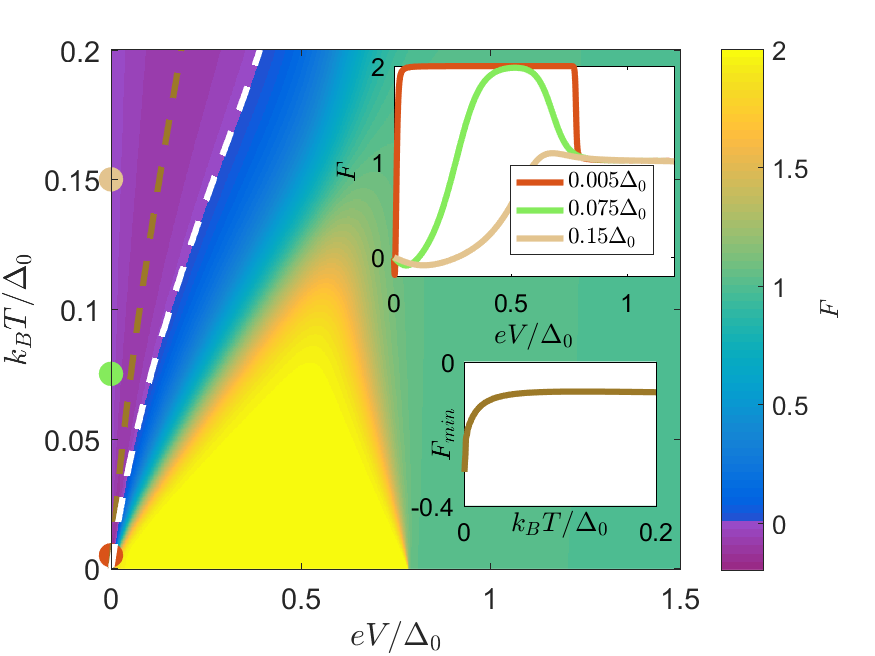}
    \includegraphics[width = 8.6cm]{figures/A.png}
    \includegraphics[width = 8.6cm]{figures/B.png}
    \caption{The differential Fano factor in \textit{p} - wave dominant 1D \textit{s} + \textit{p} - wave superconductors, $r = 2$ (a) and $r = 5$ (b). At finite temperatures, a local minimum at $eV = 0$ with a width of $k_{B}T$ develops, which reaches $F = 0$ at $eV = 0$, and is negative for $0<|eV|\lesssim k_{B}T$. A dashed line is used to indicate at which voltage the differential Fano factor attains a minimum. The corresponding minimum differential Fano factor, which is negative, is shown in the lower inset. The upper inset shows line cuts at specific temperatures, indicated with dots on the vertical axis. We used $z = 10$.}
    \label{fig:FinNoisePdom}
\end{figure}

If the \textit{p} - wave component is dominant, see Fig. \ref{fig:FinNoisePdom}, the dependence of $F$ on voltage and temperature is more complicated. In fact, there is a voltage and temperature window for which $F<0$, that is, the noise power decreases with increasing voltage. As elaborated in the main text, the appearance of a negative differential noise power can be explained with the help of \cite{lesovik1993negative}. 
Here, we link this appearance to our zero temperature temperature expressions in Appendix \ref{sec:1Danalytical}.

In Appendix \ref{sec:1Danalytical}, Eqs. (\ref{eq:P1S}) and (\ref{eq:P1resP}), we have seen that at zero energy, only the $f_{L}^{2}$ and $f_{T}^{2}$ terms appear in the noise expressions of both \textit{s} - wave and \textit{p} - wave superconductors. Thus, for small energies, these two contributions are dominating and need to be considered.

To understand the finite temperature behavior we therefore need to consider $\frac{\partial (f_{L})^{2}}{\partial V}$ and $\frac{\partial(f_{T})^{2}}{\partial V}$ near $E = 0$ at finite temperatures and small voltages. We know that $f_{L}$ is an even function of voltage, while $f_{T}$ is an odd function of voltage. Thus, $f_{L}^{2}$ and $f_{T}^{2}$ are even functions of voltage and consequently their energy derivative vanishes at $E = 0$.

However, the leading order for small $E/k_{B}T$ is different for both. Indeed, even though $f_{L}$ is even, for any finite 
 temperature it still vanishes at $E = 0$ and is therefore of second order in $E/k_{B}T$. Therefore, $f_{L}^{2}$ has a zero of fourth order. On the other hand, $f_{T}$ is first order in $E$, and therefore, $f_{T}^{2}$ is second order. Thus, at finite temperatures and low energies, the $f_{T}^{2}$ term dominates the voltage dependence.

Keeping in mind that $\frac{\partial f_{L}^{2}}{\partial V}$ and $\frac{\partial f_{T}^{2}}{\partial V}$ have opposite sign, for \textit{s} - wave superconductors both $f_{L,T}$ contributions have the same sign, both contribute positively to $\frac{\partial P_{N}}{\partial V}$ and there differential Fano factor remains nonnegative.  
However, for \textit{p} - wave superconductors, the differential Fano factor vanishes at zero temperature, because the contribution of the $f_{T}^{2}$ term comes with a negative sign. Thus, at finite temperature, for which this term dominates, the differential Fano factor, is negative. 

\begin{figure}[!htb]
    \centering
    \includegraphics[width = 8.6cm]{figures/FinNoise1DIR0p2WhitePurple3.png}
    \includegraphics[width = 8.6cm]{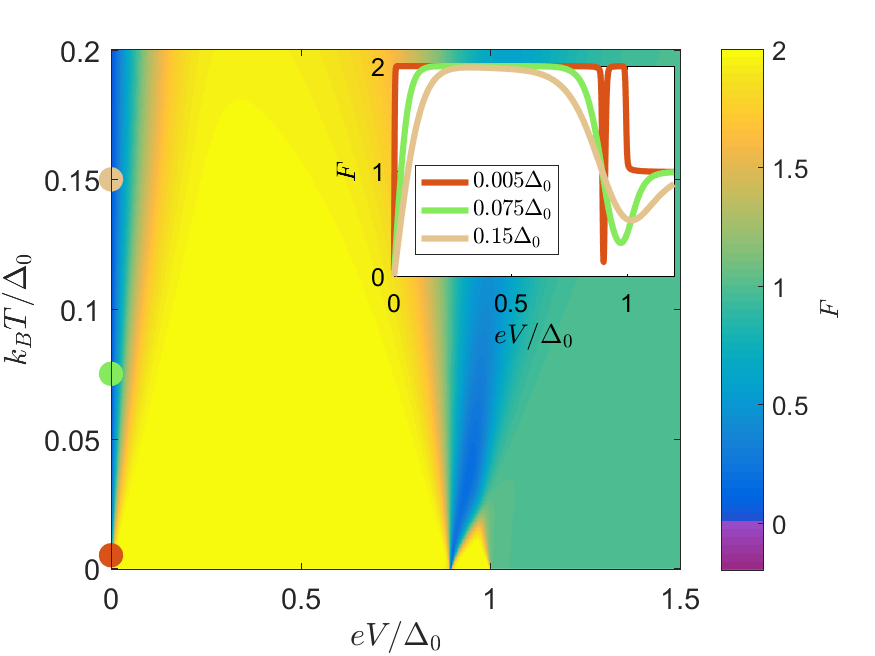}
    \includegraphics[width = 8.6cm]{figures/A.png}
    \includegraphics[width = 8.6cm]{figures/B.png}
    \caption{The differential Fano factor in \textit{s} - wave dominant one-dimensional  i\textit{s} + \textit{p} - wave superconductors, with $r = 0.2$ (a) and $r = 0.5$ (b). The differential noise power is highly suppressed for small voltages and voltages for which there is a SABS at $|eV| = \Delta_{s}$. These local minima become broader with increasing temperature. The BTK parameter was set to $z = 10$.}
    \label{fig:FinNoiseSdomISPAppendix}
\end{figure}

\begin{figure}[!htb]
    \centering
    \includegraphics[width = 8.6cm]{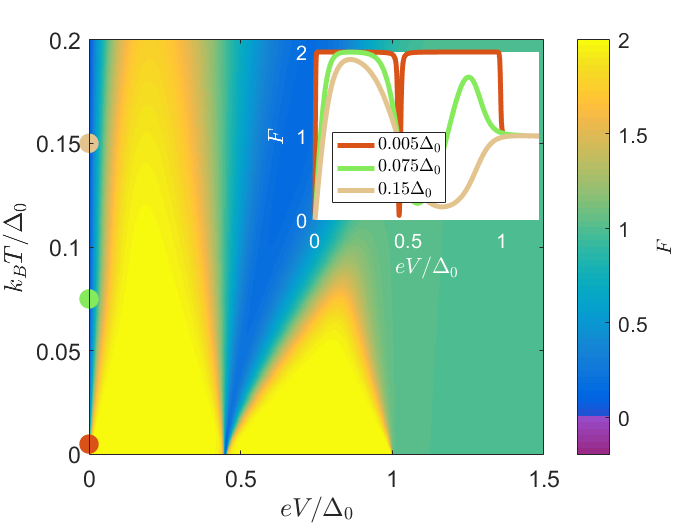}
    \includegraphics[width = 8.6cm]{figures/FinNoise1DIR5WhitePurple3.png}
    \includegraphics[width = 8.6cm]{figures/A.png}
    \includegraphics[width = 8.6cm]{figures/B.png}
    \caption{The differential Fano factor in \textit{p} - wave dominant one-dimensional i\textit{s} + \textit{p} - wave superconductors, $r = 2$ (a) and $r = 5$ (b).  For very small but finite temperatures, there is a voltage window for which the differential Fano factor is negative, close to the SABS at $|eV| = \Delta_{s}$. However, this is hard to observe in experiment. For $eV \xrightarrow{} 0$ the differential Fano factor converges to 0. The BTK parameter was set to $z = 10$.}
    \label{fig:FinNoisePdomISP}
\end{figure}

 From the numerical results in Fig. \ref{fig:FinNoisePdom} we find that there is a window of order $k_{B}T$ in which the differential Fano factor is negative.  Similar calculations were performed for i\textit{s} + \textit{p} - wave superconductors. The results are shown in Figs. \ref{fig:FinNoiseSdomISPAppendix} and \ref{fig:FinNoisePdomISP}. The differential Fano factor is non-negative close to $eV = 0$. However, for very low temperature it is negative at finite voltages close to the SABS energy. This confirms the relation to channels with a  high effective transparency in a narrow window. An important difference with \textit{p} - wave superconductors however is that this only appears for very small temperatures, as the temperature is increased further the differential noise power becomes positive, and therefore the negative differential Fano factor is hard to measure in these materials. This difference can be attributed to the fact that $f_{L}^{2}$ and $f_{T}^{2}$ have significantly different behavior at nonzero voltages compared to zero voltage. Indeed, for $eV\neq 0$, $f_{L,T}$ are not symmetric or antisymmetric around $E = eV$, and hence the above analysis does not follow through and the differential noise for finite temperatures depends on both $f_{L}^{2}$ and $f_{T}^{2}$.
\clearpage
\subsection{2D}
Also for the two-dimensional superconductors the temperature dependence of the differential Fano factor was investigated. In this section we focus on the noncentrosymmetric $\textit{s} + p$ - wave superconductors, the anapole $i\textit{s} + p$ - wave superconductors and \textit{d} - wave superconductors. For the \textit{p} - wave component we consider $p_{x}$, chiral $p$ and helical $p$ - wave variants, for the $\textit{d}$ - wave we consider the $\text{\textit{d}}_{\text{x}^{2}-\text{y}^{2}}$, $\text{\textit{d}}_{\text{xy}}$ and chiral $d$ - wave cases.

\subsubsection{Noncentrosymmetric superconductors}
\begin{figure}[!htb]
    \centering
    \includegraphics[width = 8.6cm]{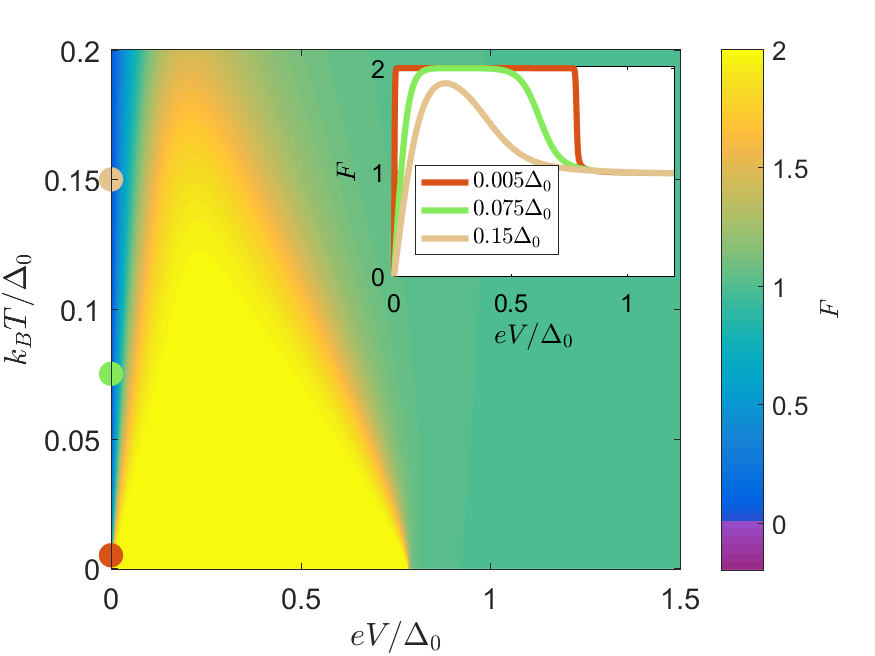}
    \includegraphics[width = 8.6cm]{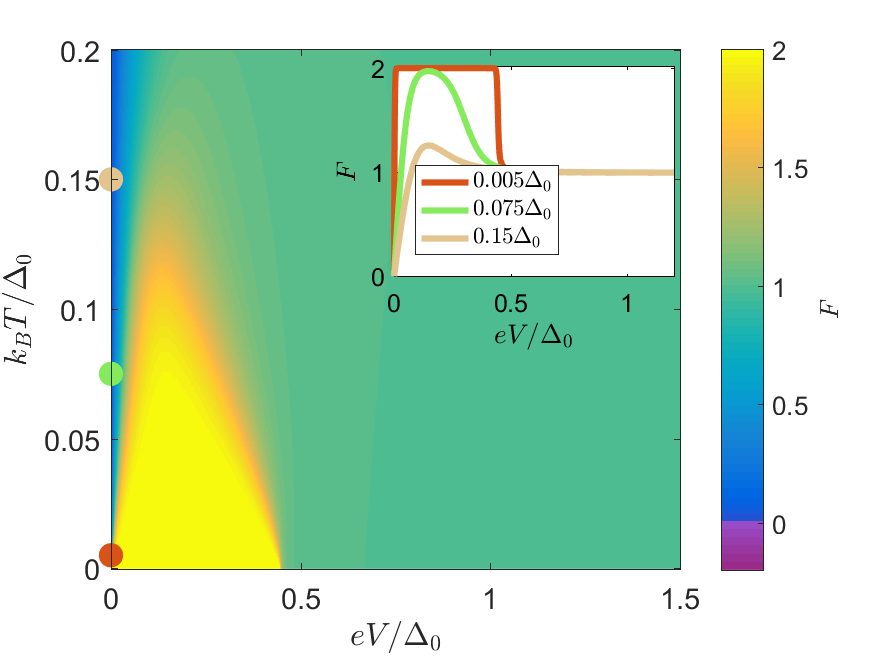}
    \includegraphics[width = 8.6cm]{figures/A.png}
    \includegraphics[width = 8.6cm]{figures/B.png}
    \caption{The differential Fano factor as a function of voltage and temperature using \textit{s} - wave dominant \textit{s} + $\text{\textit{p}}_{\text{x}}$ - wave superconductors with (a): $r =0.2$ and (b): $r = 0.5$. The results are very similar to the 1D junctions with \textit{s} + \textit{p} - wave superconductors in Fig. \ref{fig:FinNoiseSdomAppendix}, though the gap size depends on the channel, which means that $F$ is already smooth at zero temperature. The BTK parameter was set to $z = 10$.}
    \label{fig:FanoSPX}
\end{figure}

\begin{figure}[!htb]
    \centering
    \includegraphics[width = 8.6cm]{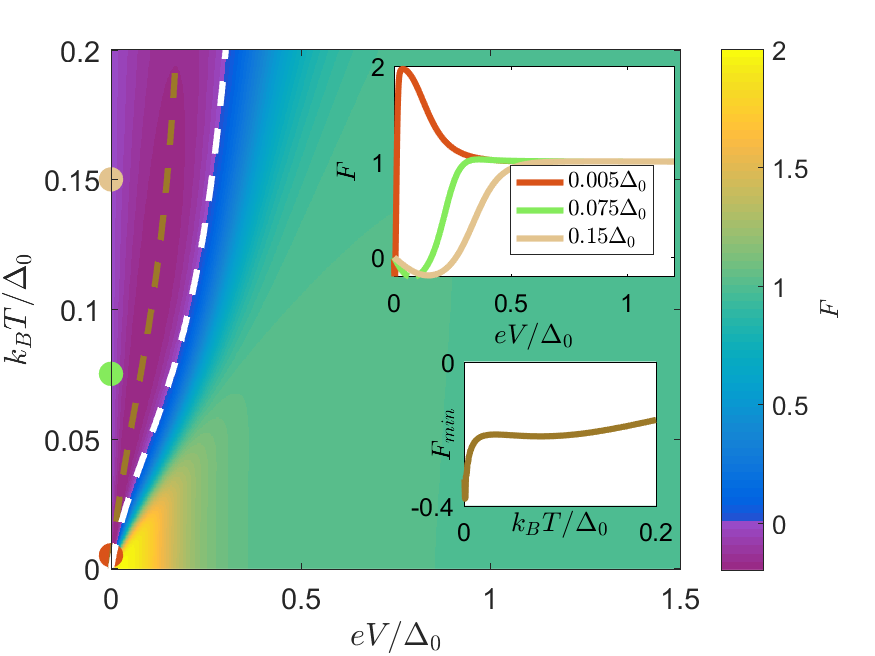}
    \includegraphics[width = 8.6cm]{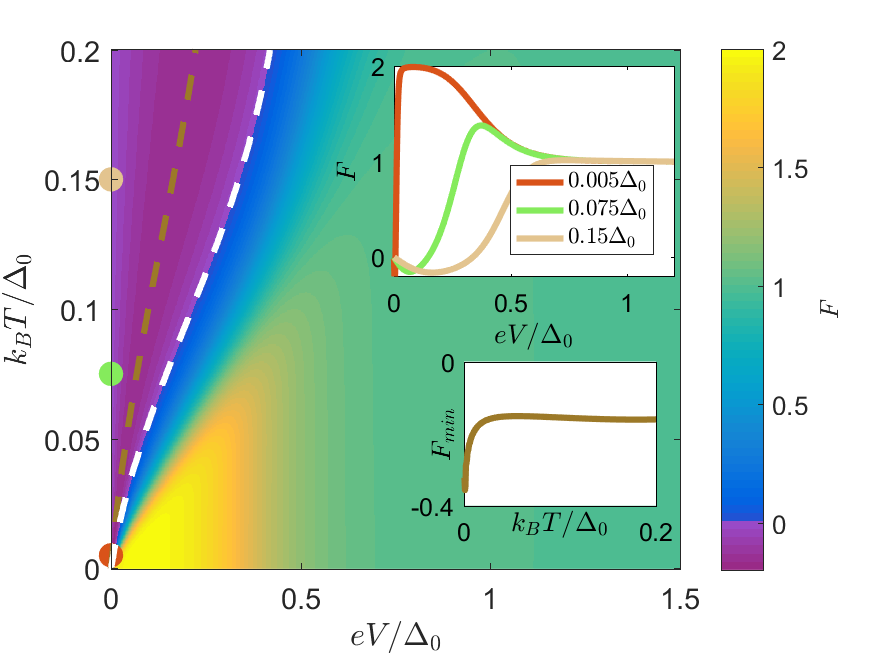}
    \includegraphics[width = 8.6cm]{figures/A.png}
    \includegraphics[width = 8.6cm]{figures/B.png}
    \caption{The differential Fano factor as a function of voltage and temperature for junctions with \textit{p} - wave dominant \textit{s} + $\text{\textit{p}}_{\text{x}}$ - wave superconductors with (a): $r =2$ and (b): $r = 5$. The results are similar to the 1D junctions with \textit{s} + \textit{p} - wave superconductors in Fig. \ref{fig:FinNoisePdom}, though the gap size depends on the channel, which means that $F$ is already smooth at zero temperature. Due to this, the voltage window with $F\approx 2$ is small and disappears for larger temperatures, for which the differential Fano factor just approaches 1. The BTK parameter was set to $z = 10$.}
    \label{fig:FanoPXS}
\end{figure}

 For the $\text{\textit{s}} + \text{\textit{p}}_{\text{x}}$ - wave superconductor, shown in Figs. \ref{fig:FanoSPX} and \ref{fig:FanoPXS}, the results are very similar to the 1D \textit{s} + \textit{p} - wave case presented before in Figs. \ref{fig:FinNoiseSdomAppendix} and \ref{fig:FinNoisePdom}. Indeed, the only difference is that the transition between $F = 2$ and $F = 1$ is smoother in this case. This  decreases the width of the $F\approx 2$ plateau for all temperatures. This is expected from the zero temperature results in Appendix \ref{sec:2Dresults}, Fig. \ref{fig:RdepSpx}.

\begin{figure}[!htb]
    \centering
    \includegraphics[width = 8.6cm]{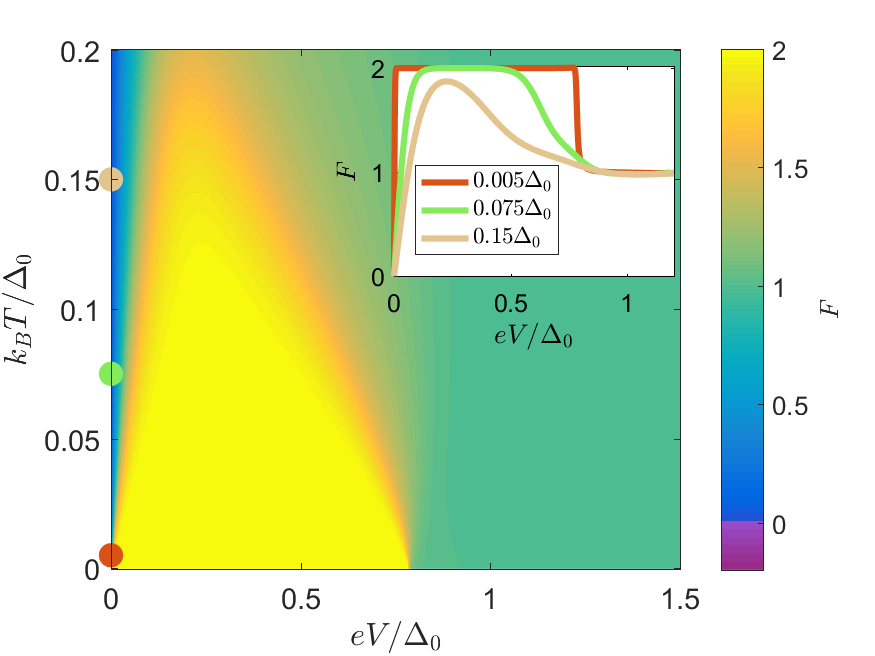}
    \includegraphics[width = 8.6cm]{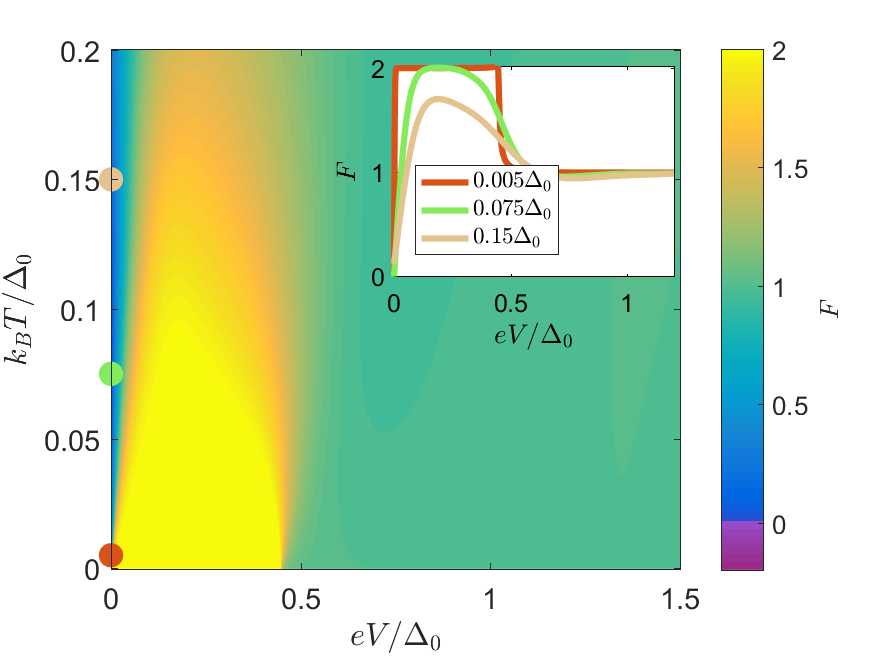}
    \includegraphics[width = 8.6cm]{figures/A.png}
    \includegraphics[width = 8.6cm]{figures/B.png}
    \caption{The differential Fano factor as a function of temperature and voltage using \textit{s} - wave dominant \textit{s} + chiral \textit{p} - wave superconductors, with (a): $r =0.2$ and (b): $r =0.5$. The results are qualitatively similar to those of an \textit{s} - wave dominant 1D superconductor in Fig. \ref{fig:FinNoiseSdomAppendix}, due to the absence of SABSs. The BTK parameter was set to $z = 10$.}
    \label{fig:FanoSC}
\end{figure}

\begin{figure}[!htb]
    \centering
    \includegraphics[width = 8.6cm]{figures/FinNoiseHelicalR0p2WhitePurple3.png}
    \includegraphics[width = 8.6cm]{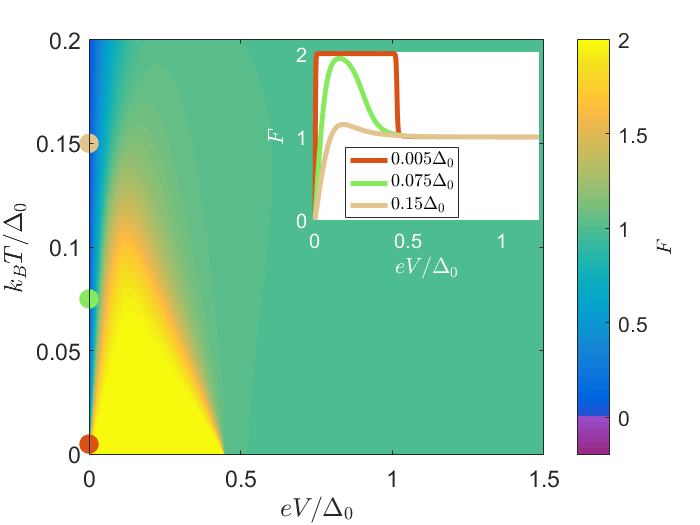}
    \includegraphics[width = 8.6cm]{figures/A.png}
    \includegraphics[width = 8.6cm]{figures/B.png}
    \caption{The differential Fano factor as a function of temperature and voltage using \textit{s} - wave dominant \textit{s} + helical \textit{p} - wave superconductors, with (a): $r =0.2$ and (b): $r =0.5$. The results are qualitatively similar to those of an  \textit{s} - wave dominant 1D superconductor in Fig. \ref{fig:FinNoiseSdomAppendix}, due to the absence of SABSs. The BTK parameter was set to $z = 10$.}
    \label{fig:FanoSH}
\end{figure}

\begin{figure}[!htb]
    \centering
    \includegraphics[width = 8.6cm]{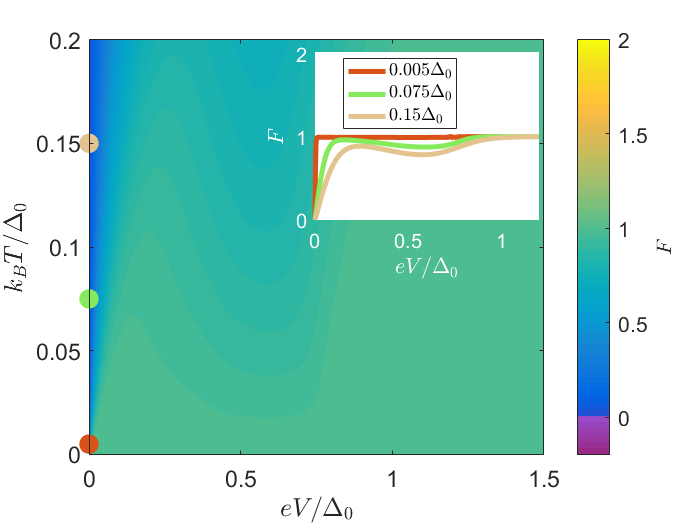}
    \includegraphics[width = 8.6cm]{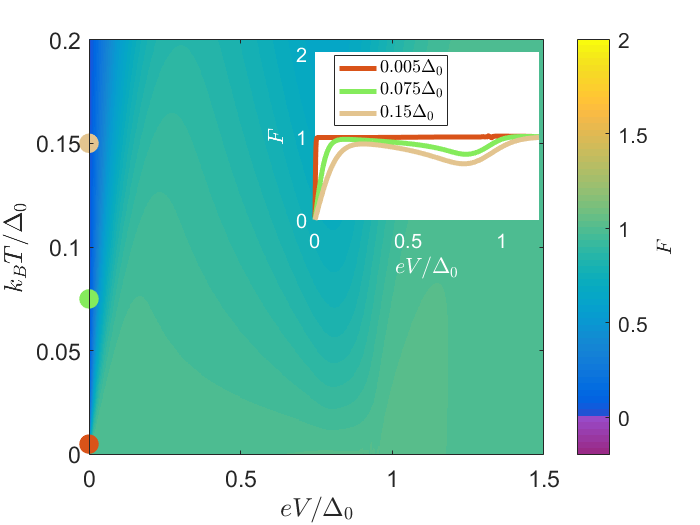}
    
    \includegraphics[width = 8.6cm]{figures/A.png}
    \includegraphics[width = 8.6cm]{figures/B.png}
    \caption{The differential Fano factor as a function of temperature and voltage using a \textit{p} - wave dominant \textit{s} + chiral \textit{p} - wave superconductor, with (a): $r =2$ and (b): $r =5$. The differential Fano factor vanishes at $eV = 0$ and at finite temperatures an additional local minimum appears for $|eV|\approx|\Delta_{-}|$, which equals $\approx 0.45\Delta_{0}$ for $r = 2$ and $\approx 0.78\Delta_{0}$ for $r = 5$. This minimum appears because below this energy there exist SABSs, while above this energy there is only transport from the continuum. This local minimum is not sharp, because the gap is not constant as a function of angle. Therefore, there is a finite voltage window with contributions from both the continuum and the SABSs.}
    \label{fig:FanoCS}
\end{figure}

\begin{figure}[!htb]
    \centering
    \includegraphics[width = 8.6cm]{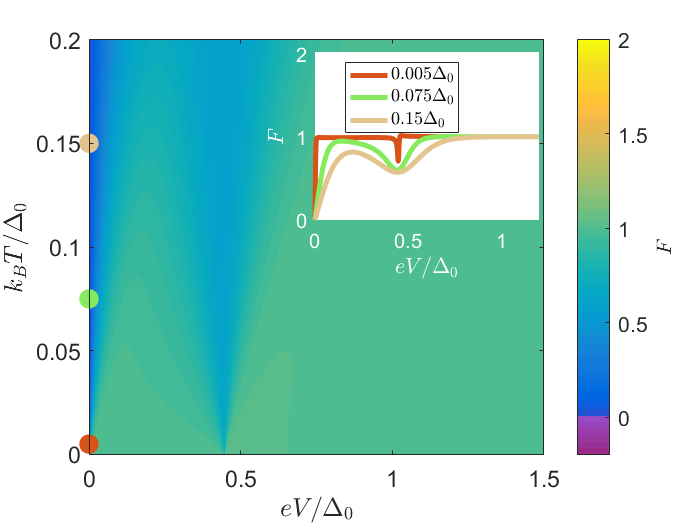}
    \includegraphics[width = 8.6cm]{figures/FinNoiseHelicalR5WhitePurple3.png}
    \includegraphics[width = 8.6cm]{figures/A.png}
    \includegraphics[width = 8.6cm]{figures/B.png}
    \caption{The differential Fano factor as a function of temperature and voltage using a \textit{p} - wave dominant \textit{s} + helical \textit{p} - wave superconductor, with (a): $r =2$ and (b): $r =5$. The differential Fano factor vanishes at $eV = 0$ and at finite temperatures an additional local minimum appears for $|eV|\approx|\Delta_{-}|$, $\approx 0.45\Delta_{0}$ for $r = 2$ and $\approx 0.78\Delta_{0}$ for $r = 5$. This minimum appears because below this energy there exist SABSs, while above this energy  transport from the continuum dominates. The local minimum is sharper than for the \textit{s} + chiral \textit{p} - wave superconductor displayed in Fig. \ref{fig:FanoCS}, because the gap magnitude is constant and equal to $|\Delta_{-}|$. Thus, noiseless surface Andreev SABS transport appears only for $|eV|<|\Delta_{-}|$ and continuum transport only appears for $|eV|>|\Delta_{-}|$, there is no voltage window in which both of these processes are important. The BTK parameter was set to $z = 10$.}
    \label{fig:FanoHS}
\end{figure}

If the \textit{s} - wave component is dominant, the results for the \textit{s} + helical and \textit{s} + chiral \textit{p} - wave superconductors are also similar to that of an \textit{s} - wave superconductor, as shown in Figs. \ref{fig:FanoSC} and \ref{fig:FanoSH}. Indeed, there is a plateau with $F = 2$ below the lowest gap, and for nonzero temperatures the differential Fano factor vanishes at zero voltage.
If the \textit{p} - wave component is dominant however, the results are different for the \textit{s} +  chiral and \textit{s} +  helical \textit{p} - wave superconductors than for the 1D \textit{p} - wave superconductor, and a negative differential Fano factor is absent, as shown in Figs. \ref{fig:FanoCS} and \ref{fig:FanoHS}. This is due to the dispersion of the SABSs. Indeed, we saw already at zero temperature that due to this, the differential Fano factor at zero voltage and zero temperature is nonzero, it is in fact 1. For this reason, at finite temperatures, $F$ does not become negative, but instead vanishes at zero voltage, like for normal metals and \textit{s} - wave superconductors. For voltages much larger than $k_{B}T$ the differential Fano factor goes back to $F = 1$. However, for nonzero temperatures a local minimum exists close to $|eV| = |\Delta_{-}|$. The temperature at which this local minimum becomes prominent is lower if the \textit{p} - wave component is dominant, see Fig \ref{fig:FanoHS}. The differential Fano factor remains nonnegative, but the minimum differential Fano factor is significantly smaller than 1, and therefore this feature can be verified in experiment.

The main difference between the \textit{s} + chiral and  \textit{s} + helical \textit{p} - wave superconductors is that in the helical case the local minimum around $|eV| = |\Delta_{-}|$ appears sharper and more prominent for any nonzero temperature, as shown in Fig. \ref{fig:FanoHS}. Thus, while for zero temperature the noise power spectra of the \textit{s} + chiral and \textit{s} +  helical \textit{p} - wave superconductors are almost indistinguishable, see Fig. \ref{fig:CHSdom}, at finite temperatures a difference arises. For even larger temperatures, for which this local minimum is broad and shallow, they are similar again, due to the similarity in the SABS spectrum of the two types of superconductors.
\clearpage
\subsubsection{Anapole superconductors}
\begin{figure}[!htb]
    \centering
    \includegraphics[width =8.6cm]{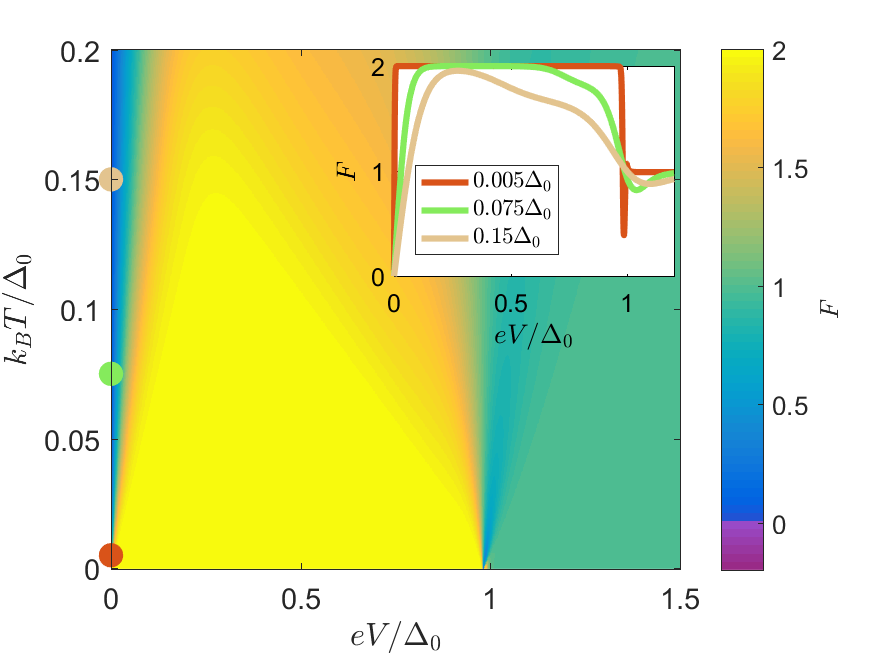}
    \includegraphics[width = 8.6cm]{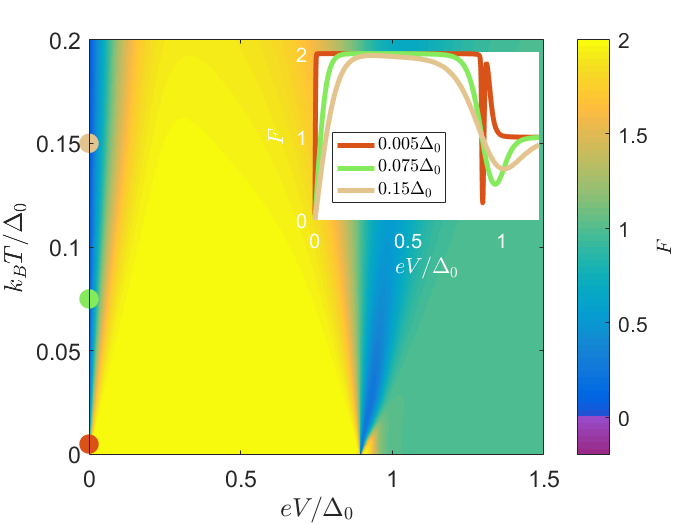}
    \includegraphics[width = 8.6cm]{figures/A.png}
    \includegraphics[width = 8.6cm]{figures/B.png}
    \caption{The differential Fano factor for \textit{s} - wave dominant i\textit{s} + $\text{\textit{p}}_{\text{x}}$ - wave superconductors, with (a): $r =0.2$ and (b): $r = 0.5$. Results are very similar to those of the 1D i\textit{s} + \textit{p} - wave superconductor in Fig. \ref{fig:FinNoiseSdomISPAppendix}. The BTK parameter was set to $z = 10$.}
    \label{fig:FanoiSPX}
\end{figure}

\begin{figure}[!htb]
    \centering
    \includegraphics[width = 8.6cm]{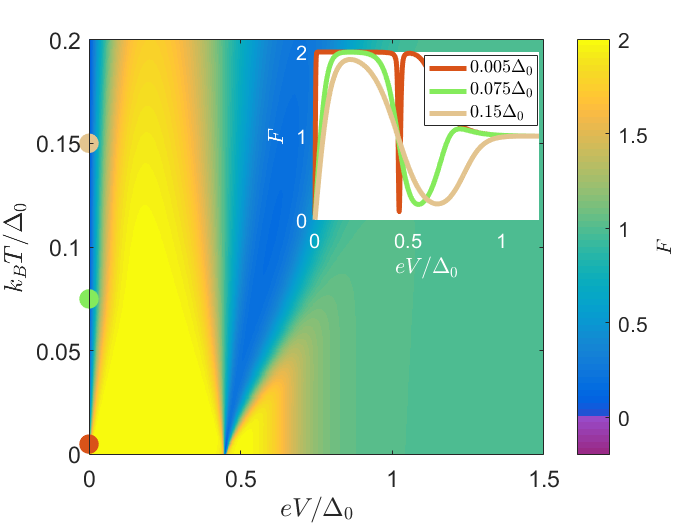}
    \includegraphics[width = 8.6cm]{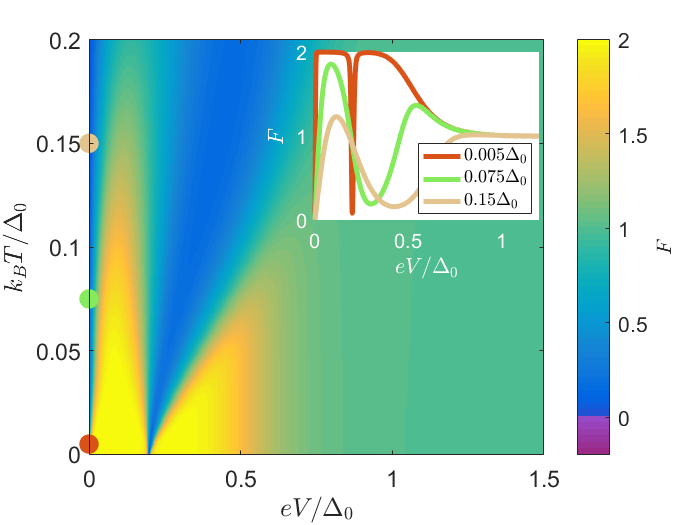}
        \includegraphics[width = 8.6cm]{figures/A.png}
    \includegraphics[width = 8.6cm]{figures/B.png}
    \caption{The differential Fano factor for \textit{p} - wave dominant i\textit{s} + $\text{\textit{p}}_{\text{x}}$ - wave superconductors, with (a): $r = 2$ and (b): $r = 5$. The results are similar to those of the 1D i\textit{s} + \textit{p} - wave superconductor in Fig. \ref{fig:FinNoisePdomISP}, however the transition between $F = 2$ and $F =1$ is much smoother for the i\textit{s} + $\text{\textit{p}}_{\text{x}}$ - wave superconductors. The BTK parameter was set to $z = 10$.}
    \label{fig:FanoPxiS}
\end{figure}

\begin{figure}[!htb]
    \centering
    \includegraphics[width = 8.6cm]{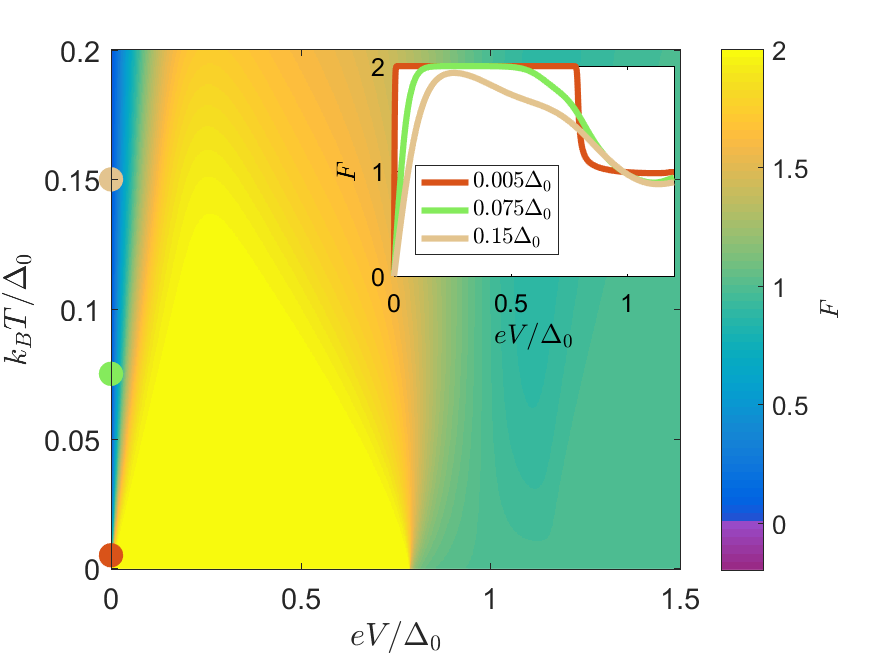}
    \includegraphics[width = 8.6cm]{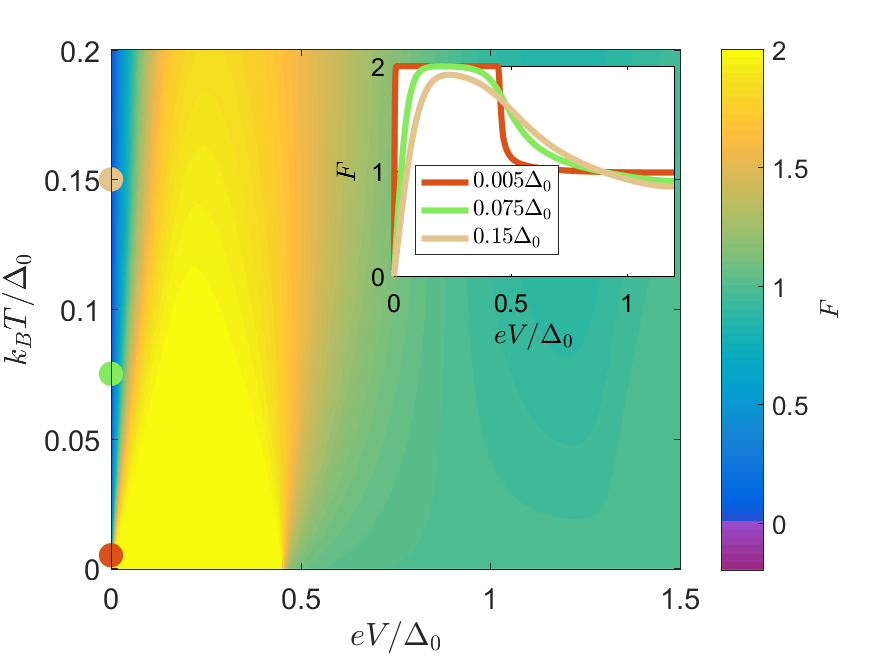}
    \includegraphics[width = 8.6cm]{figures/A.png}
    \includegraphics[width = 8.6cm]{figures/B.png}
    \caption{The differential Fano factor for \textit{s} - wave dominant i\textit{s} + chiral \textit{p} - wave superconductors, with (a): $r =0.2$ and (b): $r = 0.5$. The results are very similar to those for the \textit{s} - wave dominant \textit{s} + chiral \textit{p} - wave superconductors in Fig. \ref{fig:FanoSC}, because the phase is channel dependent. The BTK parameter was set to $z = 10$.}
    \label{fig:FanoiSC}
\end{figure}

\begin{figure}[!htb]
    \centering
    \includegraphics[width = 8.6cm]{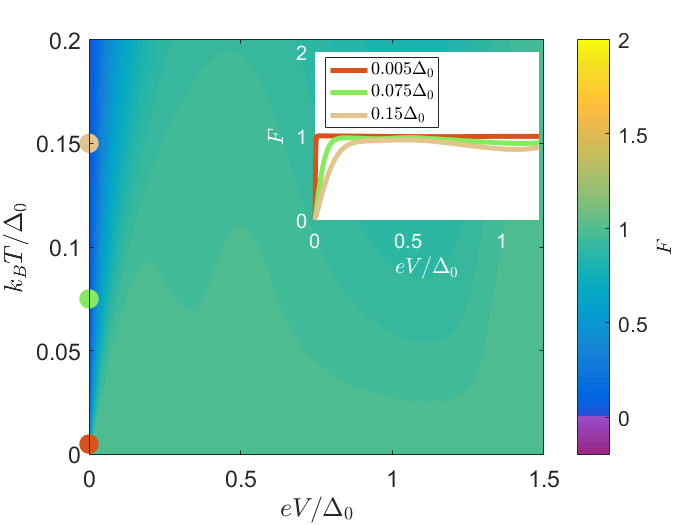}
    \includegraphics[width = 8.6cm]{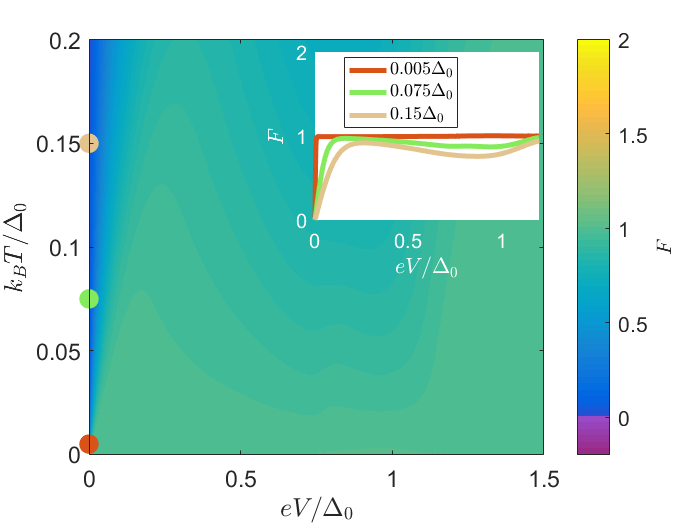}
    \includegraphics[width = 8.6cm]{figures/A.png}
    \includegraphics[width = 8.6cm]{figures/B.png}
    \caption{The differential Fano factor for \textit{p} - wave dominant i\textit{s} + chiral \textit{p} - wave superconductors, with (a): $r =2$ and (b): $r = 5$. Results are very similar to those of \textit{p} - wave dominant \textit{s} + chiral \textit{p} - wave superconductors in Fig. \ref{fig:FanoCS}, i.e. the relative phase between \textit{s} - wave and \textit{p} - wave components has no large importance, because this phase difference is channel dependent. The BTK parameter was set to $z = 10$.}
    \label{fig:FanoCiS}
\end{figure}

\begin{figure}[!htb]
    \centering
    \includegraphics[width = 8.6cm]{figures/FinNoiseHelicalIR0p2WhitePurple3.png}
    \includegraphics[width = 8.6cm]{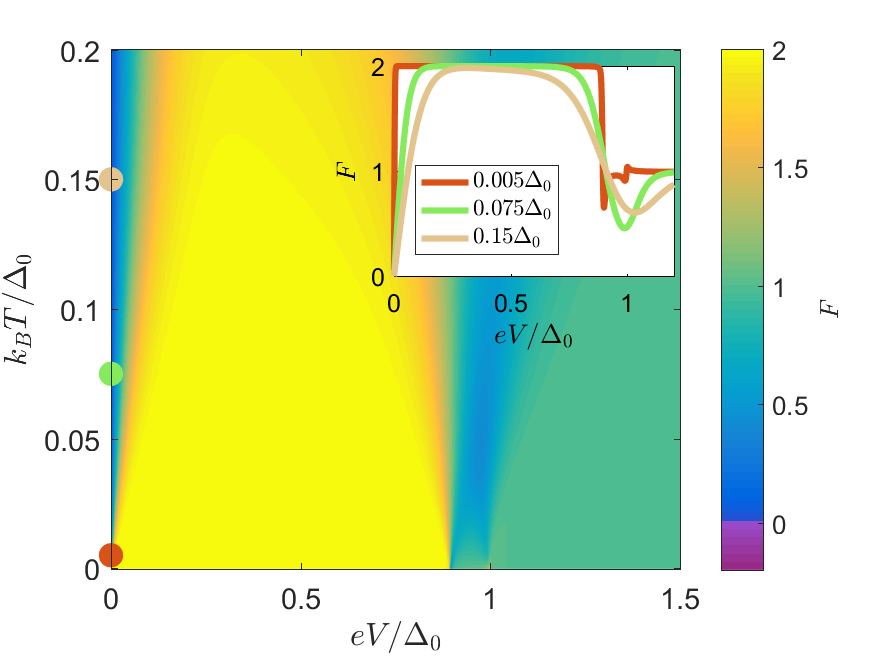}
    \includegraphics[width = 8.6cm]{figures/A.png}
    \includegraphics[width = 8.6cm]{figures/B.png}
    \caption{The differential Fano factor for \textit{s} - wave dominant i\textit{s} + helical \textit{p} - wave superconductors, with (a): $r =0.2$ and (b): $r = 0.5$.There are additional local minima for $|eV| = \Delta_{s}= \frac{\Delta_{0}}{\sqrt{1+r^{2}}}$ and $\Delta_{0}$, because there are SABSs for $\Delta_{s}<|eV|< \Delta_{0}$. For $r = 0.2$ these two local minima are not distinguishable within the resolution. For $r = 0.5$ one may distinguish two local minima at very low temperatures, these correspond to the lowest and highest energies for which there exists a SABS. The BTK parameter was set to $z = 10$.}
    \label{fig:FanoiSH}
\end{figure}

\begin{figure}[!htb]
    \centering
    \includegraphics[width  = 8.6cm]{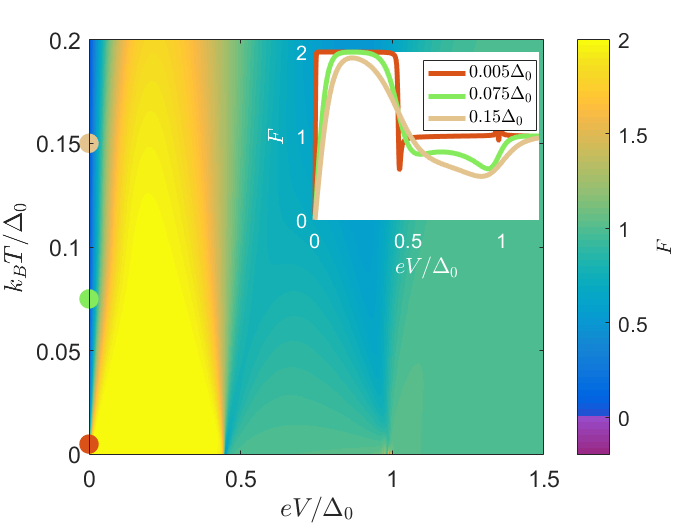}
    \includegraphics[width  = 8.6cm]{figures/FinNoiseHelicalIR5WhitePurple3.png}
    \includegraphics[width = 8.6cm]{figures/A.png}
    \includegraphics[width = 8.6cm]{figures/B.png}
    \caption{The differential Fano factor for \textit{p} - wave dominant i\textit{s} + helical \textit{p} - wave superconductors, with (a): $r =2$ and (b): $r = 5$. The SABSs spectrum now is gapped due to the broken time reversal symmetry. Consequently there are no SABSs for $|E|<\Delta_{s} = \frac{\Delta_{0}}{\sqrt{1+r^{2}}}$. For $r = 2$ this equals approximately $0.45\Delta_{0}$, for $r = 5$ it is $0.20\Delta_{0}$. Moreover, all SABSs have lower energy than the bulk gap. Thus, SABSs appear only for $\Delta_{s}<|E|<\Delta_{0}$, and thus the effective transparency is enhanced within this window. For this reason, next to the usual local minimum at zero voltage, additional local minima appear at $|eV| = \Delta_{s}$ and $|eV| = \Delta_{0}$.  Between $eV = 0$ and $|eV| = \Delta_{s}$, the differential Fano factor goes approximately to 2 because transport is only carried by Cooper pairs due to the gap in the spectrum of SABSs. For $\Delta_{s}\ll |eV|\ll\Delta_{0}$ the balance between noiseless transport via SABSs and Cooper pair transport gives $F\approx 1$, while for $|eV|>\Delta_{0}$ we also have $F\approx 1$, but now because there is only quasiparticle transport from the continuum. The BTK parameter was set to $z = 10$.}
    \label{fig:FanoHiS}
\end{figure}

For the time-reversal broken mixture of even and odd-parity superconductivity the results using a two-dimensional $i\text{\textit{s}} + \text{\textit{p}}_{\text{x}}$ - wave or a one-dimensional $i$ \textit{s} + \textit{p} - wave superconductor are very similar, as shown in Figs. \ref{fig:FanoiSPX} and \ref{fig:FanoPxiS} compared to Fig. \ref{fig:FinNoiseSdomISPAppendix}. There is a minimum when the voltage approximately equals the SABS energy $|eV| = \Delta_{s}$, which broadens with increasing temperature.
For i\textit{s} + chiral \textit{p} - wave superconductors, if the \textit{s} - wave component is dominant, the results for nonzero temperatures are still very similar to those for an \textit{s} - wave superconductor below the gap, as shown in Fig. \ref{fig:FanoiSC}. Above the gap however, there is a suppression of the differential Fano factor that is broader and shallower as the strength of the \textit{p} - wave superconductor becomes larger. This again emphasizes that the finite temperature  effects of the differential Fano factor can be used to distinguish types of superconductors that can not be distinguished based on the zero temperature differential Fano factor.
If the \textit{p} - wave component is dominant, see Fig. \ref{fig:FanoCiS}, the results are very similar to that of the \textit{s} + chiral \textit{p} - wave superconductor in Fig. \ref{fig:FanoCS}, the relative phase between the \textit{s} - wave and \textit{p} - wave components has a small influence, though the local minimum of $F$ for $|eV|\approx\Delta_{0}$ is slightly less pronounced for i\textit{s} + chiral \textit{p} - wave superconductors. In all cases $F$ is nonnegative due to the absence of dispersionless edge modes.
Just as for zero temperature, the results at finite temperature for the i\textit{s} + helical \textit{p} - wave case are significantly different compared to the i\textit{s} + chiral \textit{p} - wave case and the time-reversal symmetric \textit{s} + helical \textit{p} - wave case, as shown in Figs. \ref{fig:FanoiSH} and \ref{fig:FanoHiS}. Indeed, the peak with $F = 2$ for $0<|eV|<|\Delta_{-}|$ is accompanied with three local minima where $F<1$, namely around $eV  = 0$, $|eV| = \Delta_{s}$ and $|eV| = \Delta_{0}$. The first one is the usual finite temperature suppression to $F = 0$ that exists in all cases. The latter two are local minima that are due to SABSs, which become broader with increasing temperature, and merge for large temperatures. For $k_{B}T = 0.2\Delta_{0}$ the differential Fano factor is substantially smaller than 1, $F\approx 0.5$ at the local minimum. The local minimum around $|eV| = \Delta_{0}$ is slightly stronger than the other one for low temperatures. The differential Fano factor is nonnegative in the whole domain.
\clearpage
\subsubsection{\textit{d} - wave superconductors}
\begin{figure}[!htb]
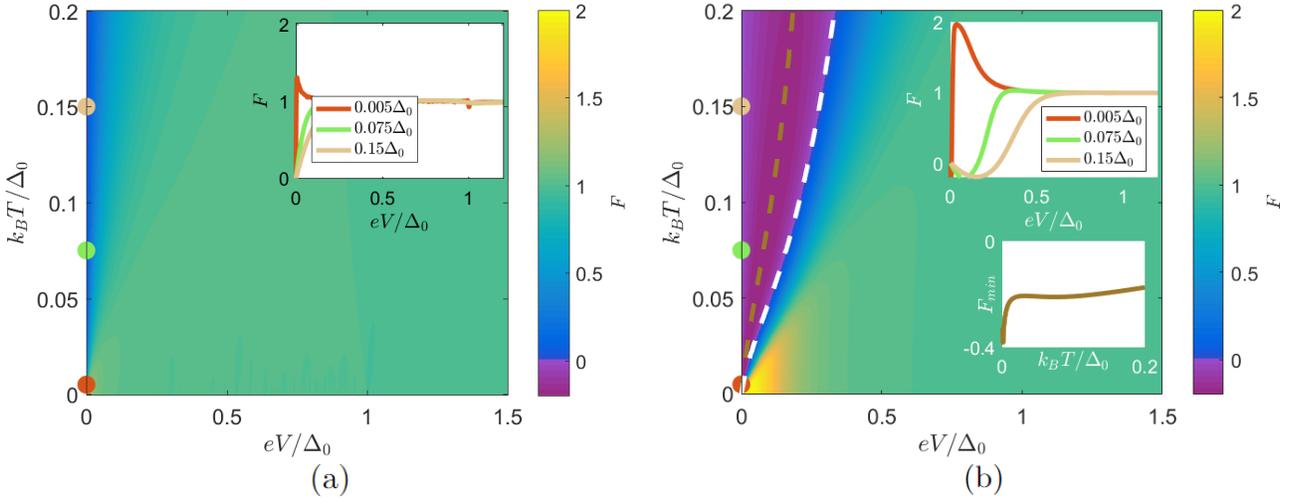

    \centering
    \includegraphics[width = 8.6cm]{figures/FinNoiseDxsqWhitePurple3.png}
    \includegraphics[width = 8.6cm]{figures/FinNoiseDxyWhitePurple3.png}
    \includegraphics[width = 8.6cm]{figures/A.png}
    \includegraphics[width = 8.6cm]{figures/B.png}
    \caption{(a): The differential Fano factor in an S / N junction in which the superconductor is a \textit{d} - wave superconductor with its lobe oriented along the normal of the interface, i.e. it is a $\text{\textit{d}}_{\text{x}^{2}-\text{y}^{2}}$ - wave superconductors. There are no SABSs, however, there exist quasiparticles for any energies, it is a nodal gap superconductor. Hence the differential Fano factor is determined by the relative amount of channels for which there is Cooper pair transport and it approaches two only for low voltages. 
    (b): The differential Fano factor in an S / N junction in which the normal to the interface corresponds to a node direction of the pair potential, i.e. a $\text{\textit{d}}_{\text{xy}}$ - wave superconductor. Because of the dispersionless ZESABSs the differential Fano factor is negative at finite temperatures, similar to the 1D \textit{p} - wave in Fig. \ref{fig:FinNoisePdom} and 2D $\text{\textit{p}}_{\text{x}}$ - wave in Fig. \ref{fig:FanoPXS}. There is a smooth transition between $F\approx 2$ and $F\approx 1$. The BTK parameter was set to $z = 10$.}
    \label{fig:FanoD1SM}
\end{figure}

The differential Fano factor was calculated as well for junctions with spin singlet \textit{d} - wave superconductors. First we consider the time-reversal symmetric case, which has four nodes on the Fermi surface. For this type of superconductor the results highly depend on the orientation of the lobes of the \textit{d} - wave. If the normal of the interface corresponds to a lobe direction, which is termed a $\text{\textit{d}}_{\text{x}^{2}-\text{y}^{2}}$ - wave superconductor, Fig. \ref{fig:FanoD1SM}(a), the incoming and reflected angle always experience the same gap, which depends on the angle. Therefore, there are no SABS and at zero temperature we have $F(eV = 0) \approx 2$ and then $F$ gradually decreases to $F \approx 1$, which is reached at $|eV| = \Delta_{0}$. At finite temperatures, the differential Fano factor is suppressed to $0$ at $eV = 0$, creating a local minimum that has a width of order $k_{B}T$, just as for the \textit{s} - wave case. 

On the other hand, if the normal to the interface corresponds to a nodal direction of the \textit{d} - wave, that is, it is a $\text{\textit{d}}_{\text{xy}}$ - wave superconductor, the incoming and reflected angles have opposite pair potential for all angles, and therefore there is a flat band of ZESABSs \cite{tanaka1995theory,tanaka2000interface,covington1997observation,alff1997spatially,wei1998directional,biswas2002evidence,chesca2006observation,hu1994midgap,kashiwaya2000tunneling,matsumoto1995boundary}. As shown in Fig. \ref{fig:FanoD1SM}(b), at zero temperature this leads to $F(eV = 0) = 0$, while at finite temperature $F(0<|eV|\lesssim k_{B}T)<0$, just as for the one-dimensional \textit{p} - wave superconductor. 
Since \textit{d} - wave superconductors are spin singlet superconductors, and the interface is not spin-active, there are only spin singlet pairs, also at the SABS. Thus, our results for $\text{\textit{d}}_{\text{xy}}$ - wave superconductors in Fig. \ref{fig:FanoD1SM}(b) show that negative differential Fano factors in tunnel junctions can not be used to prove the existence of spin triplet pair potentials, only to prove the existence of dispersionless ZESABSs. 

\begin{figure}[!htb]
    \centering
    \includegraphics[width=8.6cm]{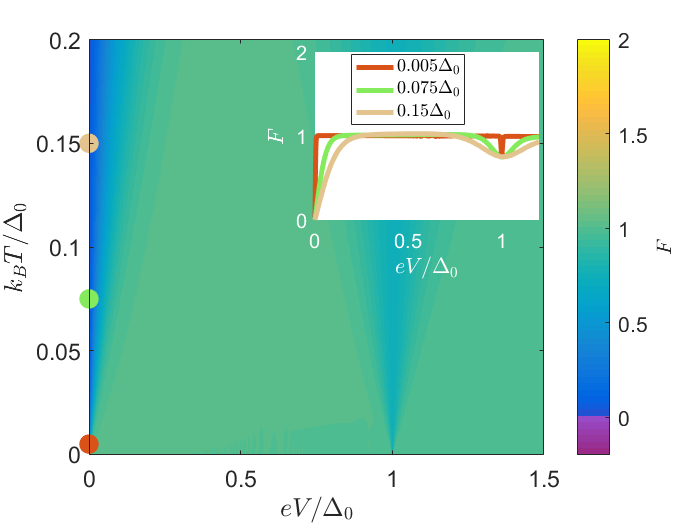}
    \includegraphics[width=8.6cm]{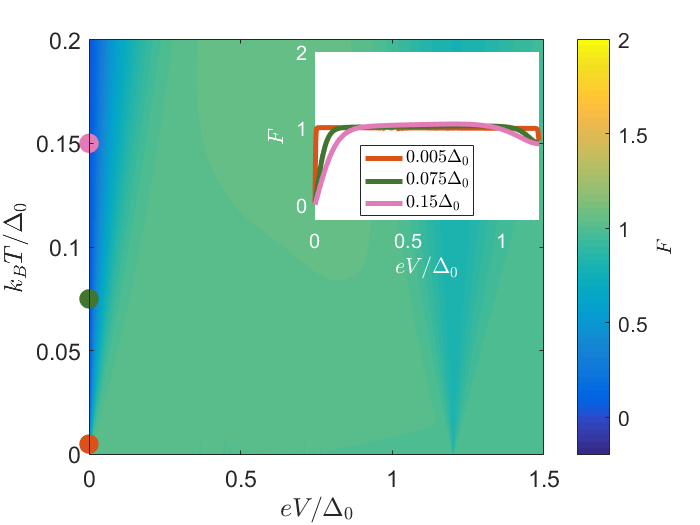}
    \includegraphics[width = 8.6cm]{figures/A.png}
    \includegraphics[width = 8.6cm]{figures/B.png}
    \caption{(a): The differential Fano factor in an S / N junction in which the superconductor is a chiral \textit{d} - wave superconductor. There are dispersive SABSs, and therefore $F \approx 1$ even below the gap, and the differential noise  power remains positive. Results are very similar to those of chiral \textit{p} - wave superconductors, showing that for the noise power it does not matter whether the pair amplitudes are spin singlet or spin triplet. There is a clear local minimum for $eV  = \Delta_{0}$. (b): Upon inclusion of a small \textit{s} - wave component ($\Delta(\phi) = \Delta_{0}(\frac{1}{5}+e^{2i\phi})$), the local minimum shifts due to the increase of the maximum gap, but the results do not qualitatively change. The BTK parameter was set to $z = 10$.}
    \label{fig:ChiralD}
\end{figure}

Similarly, we calculated the differential Fano factor in S / N junctions with chiral \textit{d} - wave superconductors, in which $\Delta(\phi) = \Delta_{0}e^{2i\phi}$. Such superconductors have dispersive SABSs \cite{kashiwaya2014tunneling}. The results are shown in Fig. \ref{fig:ChiralD}(a). Just as for chiral \textit{p} - wave superconductors, the differential Fano factor in chiral \textit{d} - wave superconductors does not reach 2, but remains close to 1 due to the interplay between noiseless transport via SABSs and Cooper pair transport, which take place at the same energies. At finite temperatures a local minimum appears at $|eV| = \Delta_{0}$ because there are SABSs for all energies below $\Delta_{0}$, but there exists no SABSs for $|E|>\Delta_{0}$. This again confirms that the differential Fano factor only depends on the presence of SABSs, not on the spin structure of the underlying pair potential. Upon inclusion of a small \textit{s} - wave component the local minimum shifts to higher voltages because of the increase of the maximum gap, see Fig. \ref{fig:ChiralD}(b). Unlike for the \textit{s} + chiral \textit{p} - wave, for the \textit{s} + chiral \textit{d} - wave superconductor the magnitude of the gap is the same for opposite spins, because both \textit{s} and \textit{d} - wave components are spin singlet. Thus, the local minimum remains equally sharp as without the \textit{s} - wave component for \textit{s} + chiral \textit{d} - wave superconductors.

\clearpage
\subsection{3D}
\begin{figure}[htb]
    \centering
    \includegraphics[width =  8.6cm]{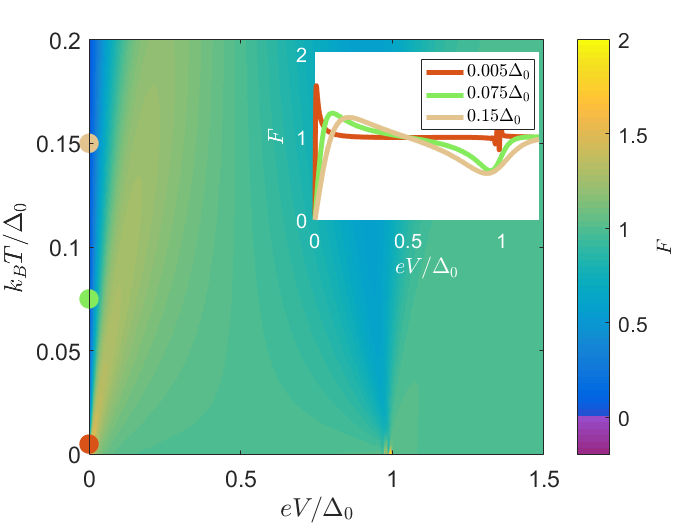}
    \caption{The differential Fano factor in the junction between a normal metal and a B-W superconductor as a function of voltage and temperature. Near zero energy Cooper pair transport dominates over SABS transport and therefore the differential Fano factor reaches almost two. For larger in-gap voltages there is an almost perfect balance between quasiparticle transport and Cooper pair transport and hence $F = 1$. At finite temperatures a minimum with a width of order $k_{B}T$ develops. Just as for the 2D superconductors this feature can be attributed to the difference in effective transparency below and above the gap. The BTK parameter was set to $z = 10$.}
    \label{fig:BW}
\end{figure}

In three-dimensions, we may, next to the generalizations of the 1D and 2D superconductors,  consider the B-W superconductor, for which the pair potential reads 
\begin{align}
    \Delta(\phi,\theta) &= \Delta_{0}\vec{d}(\phi,\theta)\;,\\
    \vec{d}(\phi,\theta) &= (\cos\phi,\sin\phi\cos\theta,\sin\phi,\sin\theta) \;.   
\end{align}
For such superconductor the ZESABSs form a point in the two-dimensional in-plane momentum space. As shown in Sec. \ref{sec:ZeroFano}, due to this at zero temperature and zero voltage Cooper pair transport completely dominates over SABS mediated transport and therefore the differential Fano factor is two. The voltage and temperature dependence is shown in Fig. \ref{fig:BW}. The region with $F\approx 2$ can only be observed for low temperatures and $|eV|$ of the order of $k_{B}T$. The reason for this is that for any nonzero energy, the SABSs form a circle instead of a point in momentum space. Thus, at finite energies the perfect balance between SABS transport and Cooper pair transport, which also appeared for 2D helical \textit{p} - wave superconductors, is restored and hence $F\approx 1$ for voltages below the gap that are not close to $eV = 0$. For $|eV|\approx\Delta_{0}$ a minimum with width of order $k_{B}T$ develops. This minimum may be understood in a fashion similar to the one appearing for helical \textit{p} - wave superconductors. For each energy below $\Delta_{0}$ there are incident angles with a resonance, and hence the effective transparency is increased, while above $\Delta_{0}$ such resonances are absent.

\begin{figure}[htb]
    \centering
    \includegraphics[width=8.6cm]{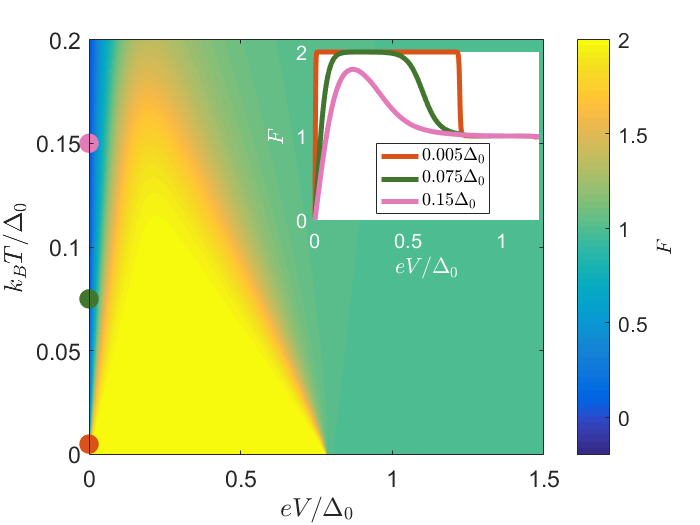}
    \includegraphics[width=8.6cm]{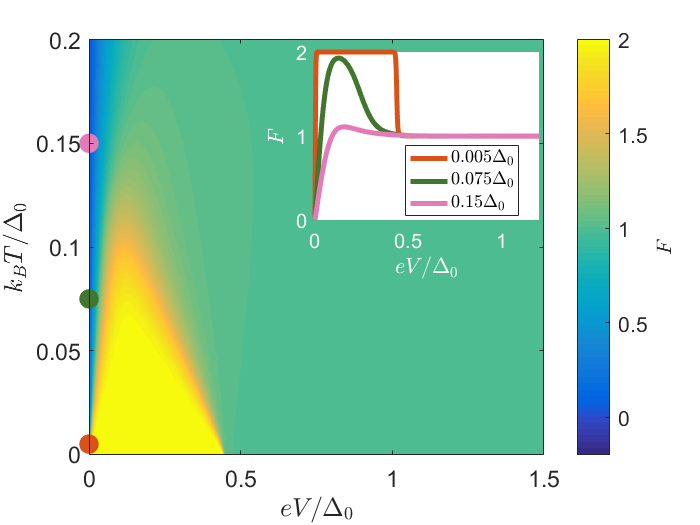}
    \includegraphics[width = 8.6cm]{figures/A.png}
    \includegraphics[width = 8.6cm]{figures/B.png}
    
    \caption{The differential Fano factor in S / N tunnel junctions with \textit{s} - wave dominant \textit{s} + B-W wave superconductors  for (a) $r = 0.2$ and (b) $r = 0.5$. the differential Fano factor is 2 below the minimum gap and transitions to 1 above this value. The BTK parameter was set to $z = 10$.}
    \label{fig:SBW1}
\end{figure}

We may also consider mixed parity \textit{s} + B-W superconductors. The results for the \textit{s} - wave dominant case are shown in Fig. \ref{fig:SBW1}. The results are qualitatively similar to those of the 1D and 2D \textit{s} - wave superconductor, with for low temperatures $F\approx 2$ below the minimum gap magnitude and $F\approx 1$ above the minimum gap magnitude. By increasing temperature the window with $F\approx 2$ decreases like for the other superconductors, due to the dominance of quasiparticle transport of order $\Tilde{T}$ over Cooper pair transport of order $\Tilde{T}^{2}$.

\begin{figure}[htb]
    \centering
    \includegraphics[width=8.6cm]{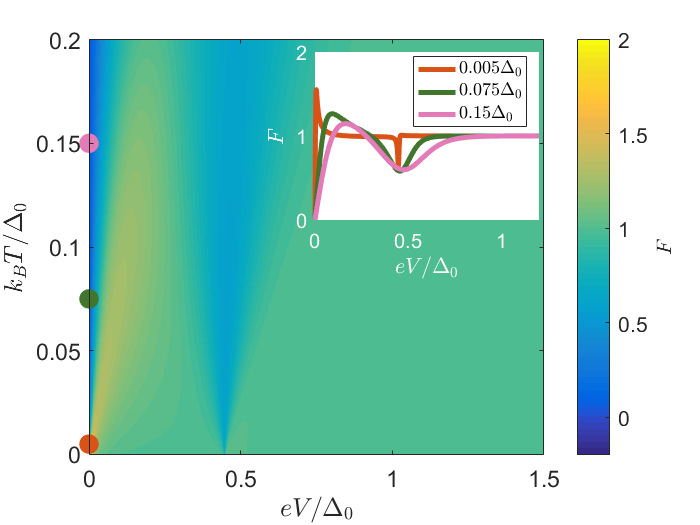}
    \includegraphics[width=8.6cm]{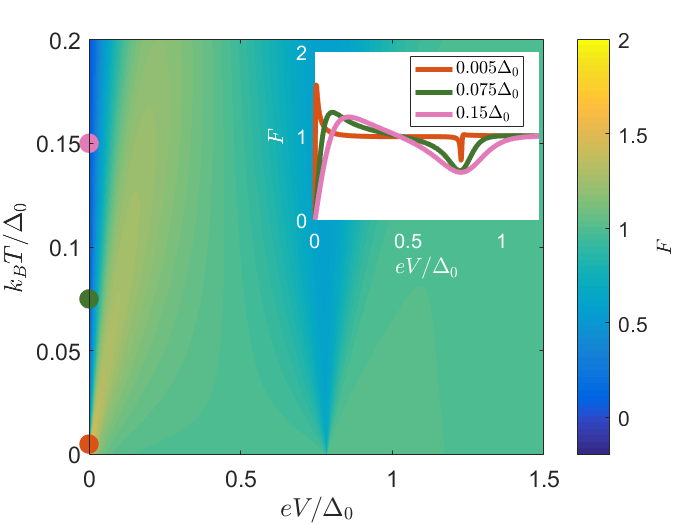}
    \includegraphics[width = 8.6cm]{figures/A.png}
    \includegraphics[width = 8.6cm]{figures/B.png}
    
    \caption{The differential Fano factor in S / N tunnel junctions with \textit{p} - wave dominant \textit{s} + B-W wave superconductors for (a) $r = 2$ and (b) $r = 5$. The results are qualitatively similar to those of a B-W superconductor, but the local minimum shifts to $|E|\approx |\Delta_{-}|$, which in (a) equals $0.45\Delta_{0}$ and in (b) $0.78\Delta_{0}$. Moreover, for smaller gaps the window with $F>1$ is smaller. The BTK parameter was set to $z = 10$.}
    \label{fig:SBW2}
\end{figure}

For the \textit{p} - wave dominant \textit{s} + B-W superconductors the results are presented in Fig. \ref{fig:SBW2}. The differential Fano factor is qualitatively similar to that of B-W superconductors, but the local minima that appears for nonzero temperatures shifts from $|eV| = \Delta_{0}$ to $|eV| = |\Delta_{-}|$. Just as for the 2D helical superconductors the similarity between the \textit{p} - wave and \textit{p} - wave dominant \textit{s} + \textit{p} - wave superconductors is protected by topology \cite{schnyder2015topological}.

\begin{figure}[htb]
    \centering
    \includegraphics[width=8.6cm]{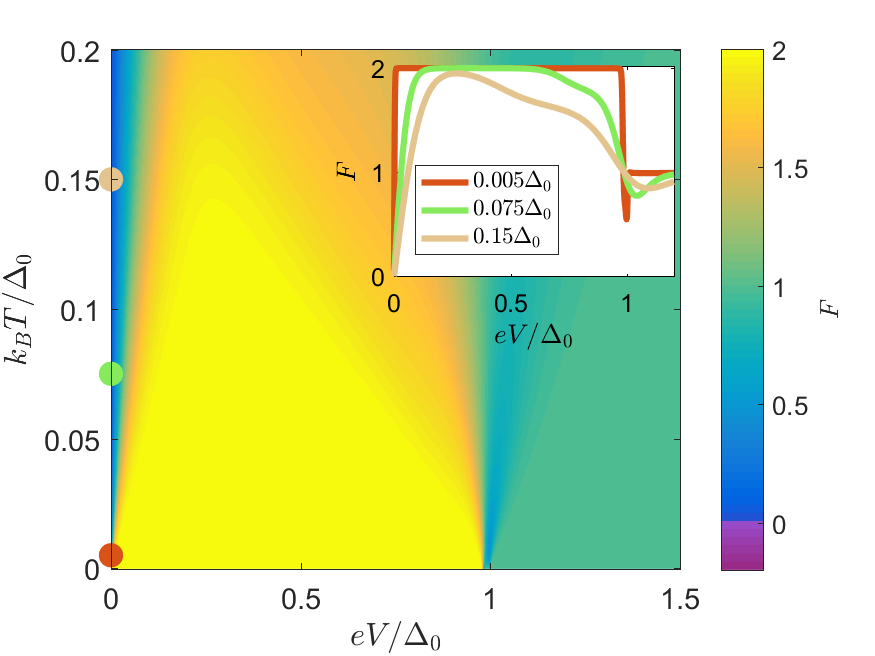}
    \includegraphics[width=8.6cm]{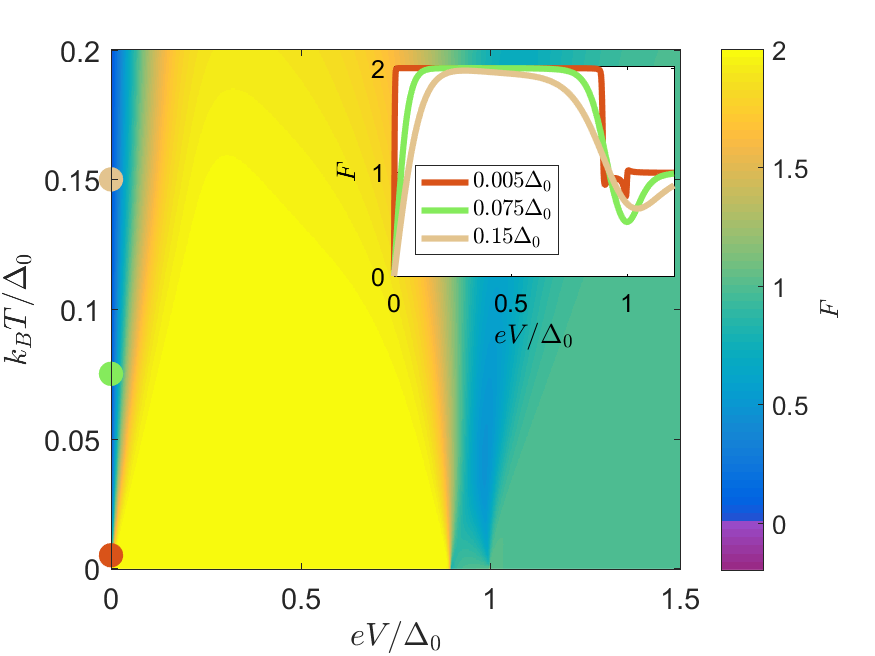}
    
    \includegraphics[width = 8.6cm]{figures/A.png}
    \includegraphics[width = 8.6cm]{figures/B.png}
    
    \caption{The differential Fano factor in S / N tunnel junctions with i\textit{s} + B-W superconductor with $r = 0.2$ (a) and $r = 0.5$(b). Due to the breaking of topological protection, a gap in the SABS spectrum opens and $F = 2$ for voltages below $\Delta_{s}$, which is close to $\Delta_{0}$. Results are very similar to those of 2D helical \textit{p} - wave superconductors in Fig. \ref{fig:FanoiSH}. The BTK parameter was set to $z = 10$.}
    \label{fig:SBWI}
\end{figure}

\begin{figure}[htb]
    \centering
    \includegraphics[width=8.6cm]{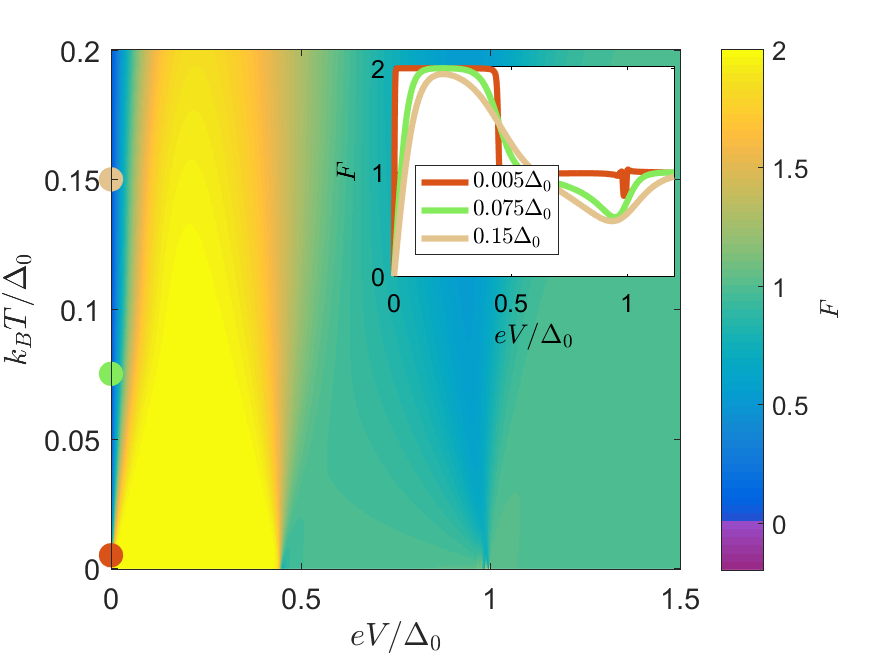}
    \includegraphics[width=8.6cm]{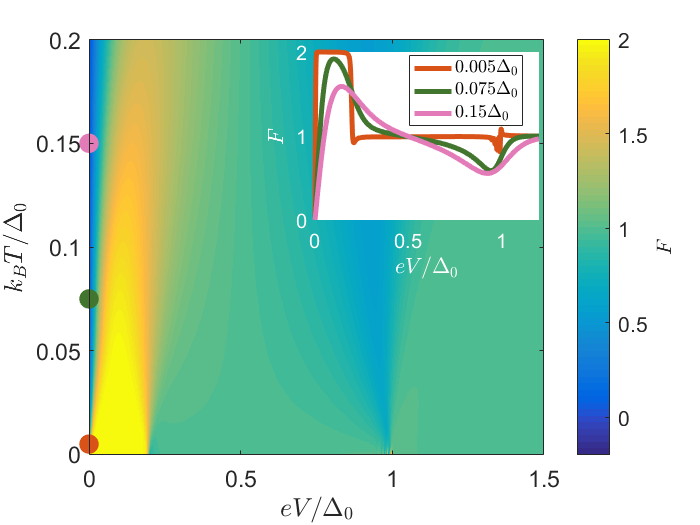}
   
    \includegraphics[width = 8.6cm]{figures/A.png}
    \includegraphics[width = 8.6cm]{figures/B.png}
    
    \caption{The differential Fano factor in S / N tunnel junctions with i\textit{s} + B-W superconductor with $r = 2$(a) and  $r = 5$(b). Due to the breaking of topological protection, a gap in the SABS spectrum opens and $F = 2$ for small voltages. Results are very similar to those of 2D helical \textit{p} - wave superconductors in Fig. \ref{fig:FanoHiS}, though now the local minimum at $|eV| = \Delta_{s}$, which is $ 0.45\Delta_{0}$ for (a) and $ 0.20\Delta_{0}$ for (b), is much weaker because of the smaller weight of the lowest energy SABS. The BTK parameter was set to $z = 10$.}
    \label{fig:BWI}
\end{figure}

By breaking of this protection via time-reversal symmetry breaking in i\textit{s} + B-W superconductors, a gap opens in the SABS spectrum and consequently $F \approx 2$ for $|eV|<\Delta_{s}$, as shown in Figs. \ref{fig:SBWI} and \ref{fig:BWI}. Thus, the results are very similar to those for i\textit{s} + helical \textit{p} - wave superconductors in Figs. \ref{fig:FanoiSH} and Figs. \ref{fig:FanoHiS}. A difference however, is that for the i\textit{s} + B-W superconductor, the ZESABS at $E = \Delta_{s}$ is a single SABS in the 2D space of in-plane momenta, that is, its contribution is small compared to the Cooper pair contribution. Therefore, the change in effective transparency at $|eV| = \Delta_{s}$ is rapid, but smooth in the B-W superconductor. Consequently, as shown in Fig. \ref{fig:BWI}, there is no sharp dip in the differential Fano factor at $|eV| = \Delta_{s}$, only a transition from $F = 2$ to $F = 1$, unlike for the i\textit{s} + helical \textit{p} - wave superconductor (see Fig. \ref{fig:FanoHiS}).
\clearpage
\section{Proof zero voltage differential Fano factor}\label{sec:ZeroFano}
In this Appendix we proof that at zero temperature, the zero voltage differential Fano factor becomes $F_{00} = 1$ in the low transparency limit $\Tilde{T}\ll 1$ for helical and chiral \textit{p} - wave superconductors. Next to this, we prove that for superconductors with SABSs that have a nonlinear dispersion around normal incidence, $F_{00} = \frac{1}{n}$, where $n$ is the order of the dispersion. With this we provide the details of the calculation of the values presented in Table \ref{tab:Dispersions} in the main text. In the rest of this section for simplicity we write any conductance in units of $\frac{e^{2} N_{\text{ch}} }{8\pi\hbar}$, any noise power in terms of $\frac{e^{3} N_{\text{ch}}}{4\pi\hbar}$ and any action in terms of $\frac{t_{0} N_{\text{ch}} }{8\pi\hbar}$. In these units the action reads
\begin{align}
    S = -\frac{1}{2}\int_{-\infty}^{\infty} dE\int_{-\frac{\pi}{2}}^{\frac{\pi}{2}}\cos\phi d\phi \text{Tr ln}(4-2\Tilde{T}+\Tilde{T}\{C,G_{\chi}\}) \;.
\end{align}
Moreover, in these units, we may write the differential Fano factor as
\begin{align}    \frac{\frac{\partial P_{N}}{\partial V}}{\frac{\partial I}{\partial V}}\;,
\end{align} without any prefactor.
 \subsection{Helical and chiral \textit{p}}
For the helical and chiral \textit{p} - wave superconductors, we may decompose the problem into  two different spin sectors. Indeed, for helical \textit{p} - wave superconductors with $\Delta_{1}(\phi) = -\Delta_{2}(\phi) = \Delta_{0}$ and $\vec{d} = (\cos\phi,\sin\phi,0)$ there are only $\uparrow\uparrow$ and $\downarrow\downarrow$ correlations and hence, we may focus on only one spin sector, and the results are the same for the other spin-sector. Similarly, for chiral \textit{p} - wave superconductors with $\Delta_{1}(\phi) = -\Delta_{2}(\phi) = e^{i\phi}$ and $\vec{d} = (0,0,1)$ one may decompose the $\uparrow\downarrow$ and $\downarrow\uparrow$ sectors in exactly the same way. Moreover, the $\downarrow\uparrow$ block of the chiral \textit{p} - wave superconductor is equivalent to the $\uparrow\uparrow$ block of the helical \textit{p} - wave superconductor, and the $\uparrow\downarrow$ block of the chiral \textit{p} - wave superconductor is equivalent to the $\downarrow\downarrow$ block of the helical \textit{p} - wave superconductor.

This implies that the results are necessarily the same for pure helical and pure chiral \textit{p} - wave superconductors, only upon inclusion of an \textit{s} - wave they may start to differ. Therefore, we may consider the two types of superconductors using the same formulas. We have below the gap (i.e. for $|E|<\Delta_{0}$)
\begin{align}
    C^{R} = C^{A} = 
        \frac{iE\sqrt{\Delta_{0}^{2}-E^{2}}+i\Delta_{0}^{2}\cos\phi\sin\phi}{(E^{2}-\Delta_{0}^{2}\sin^{2}\phi)}\tau_{3}-\frac{\Delta(E\cos\phi+\sqrt{\Delta_{0}^{2}-E^{2}}\sin\phi)}{(E^{2}-\Delta_{0}^{2}\sin^{2}\phi)}\tau_{1}\;,
\end{align}
and since the superconductor is in equilibrium, below the gap we find that $C^{K} = (C^{R}-C^{A})\tanh{\frac{E}{2k_{B}T}} = \mathbf{0}$, where $\mathbf{0}$ denotes the zero matrix in Nambu-spin space. 

In the normal metal, in the absence of a proximity effect, we have
\begin{align}
    G_{N}^{R} = -G_{N}^{A} = \tau_{3}\;.
\end{align}
In the absence of any voltage, there is no current or noise power at zero temperature. If a voltage is applied, then for $E<|eV|$ we have
$f_{L} = 0, f_{T} = 1$ and hence $G_{N}^{K} = 2\mathbf{1}$, while for $E>|eV|$ we have $f_{L} = 1, f_{T} = 0$ and hence $G_{N}^{K} = 2\tau_{3}$. At zero temperature energies larger than $|eV|$ do not contribute to any observable, and consequently we may focus on $|E|<|eV|$

Using the counting field we may write for $|E|<|eV|$ the following counting field dependent Green's function:
\begin{align}
    G_{\chi} = e^{-i\frac{\chi}{2}\tau_{K}}G_{N} e^{i\frac{\chi}{2}\tau_{K}}&= \begin{bmatrix}
        e^{i\chi}\tau_{3}&1+e^{i\chi}\\1-e^{i\chi}&-e^{i\chi}\tau_{3}
    \end{bmatrix}\; ,
\end{align}
and hence the anticommutator that we require in the calculation of the action reads
\begin{align}
    \{C,G_{\chi}\} = \begin{bmatrix}
        e^{i\chi}\{C^{R},\tau_{3}\}&2(1+e^{i\chi})C^{R}\\2(1-e^{i\chi})C^{R}&-e^{i\chi}\{C^{R},\tau_{3}\}
    \end{bmatrix}\;.
\end{align}
In the low temperature limit the distribution functions are step functions, and hence the voltage derivative only appears in the integral limits, and not in the integrand. Therefore, the voltage derivative follows immediately from the sum of the integrands at $|E| = |eV|$ and equals, keeping in mind that the contributions of $E = -eV$ and $E = eV$ are equal,
\begin{align}
    \frac{\partial S(\chi)}{\partial V} &=- \int_{-\frac{\pi}{2}}^{\frac{\pi}{2}} \cos\phi d\phi\text{Tr ln}(4-2\Tilde{T}+\Tilde{T}\{C,G_{\chi}\})\approx-\int_{-\frac{\pi}{2}}^{\frac{\pi}{2}} \cos\phi d\phi\text{Tr ln}(1+\frac{\Tilde{T}}{4}\{C,G_{\chi}\})\nonumber\\&=-\int_{-\frac{\pi}{2}}^{\frac{\pi}{2}} \cos\phi d\phi \sum_{m = 1}^{\infty}\frac{1}{m}(-1)^{m+1}\text{Tr}\Big((\frac{\Tilde{T}}{4}\{C(E = eV),G_{\chi}\})^{n}\Big)\;,
\end{align}
where the term $-2\Tilde{T}$ was ignored because it contains $\Tilde{T}$ without being compensated by a large term such as $\text{Tr}\Big(\{C^{R},\tau_{3}\}\Big)$, which means it provides a correction that is negligible in the tunneling limit.
The term for $m = 1$ is traceless, because the retarded and advanced components cancel out exactly. The $m = 2$ term equals $\frac{\Tilde{T}^{2}}{16}\Big(B^{2}e^{2i\chi}+4(1-e^{2i\chi})\Big)\mathbf{1}\approx \frac{\Tilde{T}^{2}}{16}B^{2}e^{2i\chi}\mathbf{1}$, where the second term was because it is of second order in transparency but is not compensated by the poles of the Green's function, while $B = \frac{1}{2}\text{Tr}\Big(\{C^{R},\tau_{3}\}\Big)= \frac{2i(E\sqrt{\Delta_{0}^{2}-E^{2}}+\Delta_{0}^{2}\cos\phi\sin\phi)}{(E^{2}-\Delta_{0}^{2}\sin^{2}\phi)}$ contains poles at $|E| = \Delta_{0}|\sin\phi|$. Thus, the second order contribution is proportional to the identity matrix in Keldysh-Nambu spin space. Therefore, all odd order terms have zero trace, while all even order terms are proportional to the identity matrix in Keldysh-Nambu-spin space. From this we conclude that
\begin{align}\label{eq:dSdVforAnalytical}
    \frac{\partial S(\chi)}{\partial V} = -4\int_{-\frac{\pi}{2}}^{\frac{\pi}{2}} \cos\phi d\phi \sum_{m = 1}^{\infty}\frac{1}{2m}(\frac{1}{16}\Tilde{T}^{2}B^{2}e^{2i\chi})^{m} = -2\int_{-\frac{\pi}{2}}^{\frac{\pi}{2}}\cos\phi d\phi\text{ln}(1-\frac{1}{16}\Tilde{T}^{2}B^{2}e^{2i\chi})\;,
\end{align}
where the factor 4 in front arises because the trace is taken in Nambu-Keldysh space.
Thus, we find that
\begin{align}\label{eq:Conductancedistributios}
    \frac{\partial I}{\partial V} = i\frac{\partial }{\partial V}\frac{\partial S}{\partial \chi}|_{\chi = 0} = -\frac{1}{4}\int_{-\frac{\pi}{2}}^{\frac{\pi}{2}} 
   \cos\phi d\phi\frac{\Tilde{T}^{2}B^{2}}{1-\frac{1}{16}\Tilde{T}^{2}B^{2}}\;,
\end{align}
Before solving the integral, we first consider the impact of the magnitude of $B$ on the conductance. For small $B$, the current is second order in tunneling, while for large $B$, specifically, $B\gg \Tilde{T}^{-1}$, the current is zeroth order in tunneling. These are the resonant modes. 
For helical or chiral superconductors $B$ has a single pole (for $E = 0$ it is at $\phi = 0$, for $E\neq 0$ at nonzero angles). Near the pole we have $B\propto (\phi-\phi_{p})^{-1}$, where $\phi_{p}$ is the angle at which the pole appears. Thus, the condition $B\gg \Tilde{T}$ holds as long as $|\phi-\phi_{p}|\ll \Tilde{T}$, which implies the obtained conductance should be of first order in $\Tilde{T}$, just like that of a normal metal.

We may solve the conductance integral analytically for $E = 0$, which determines the conductance at $eV = 0$. Indeed, in this case $B = -2i\cot\phi$, and hence the expression becomes
\begin{align}\label{eq:CurrToRefTo}
    \frac{\partial I}{\partial V} = \int_{-\frac{\pi}{2}}^{\frac{\pi}{2}}\frac{\Tilde{T}^{2}\cot^{2}\phi}{1+\frac{1}{4}\Tilde{T}^{2}\cot^{2}\phi}\cos\phi d\phi = \int_{-\frac{\pi}{2}}^{\frac{\pi}{2}}\frac{\Tilde{T}^{2}}{\tan^{2}\phi+\frac{1}{4}\Tilde{T}^{2}}\cos\phi d\phi =2\int_{0}^{\frac{\pi}{2}}\frac{\Tilde{T}^{2}}{\tan^{2}\phi+\frac{1}{4}\Tilde{T}^{2}}\cos\phi d\phi \;,
\end{align}
where in the last line we used that the integrand is even in $\phi$.
Using transformation of variables to $u = \sin\phi$, we obtain:
\begin{align}\label{eq:Iintermsofu}
   \frac{\partial I}{\partial V} &= 2\int_{0}^{1}\frac{\Tilde{T}^{2}}{\frac{u^{2}}{1-u^{2}}+\frac{1}{4}\Tilde{T}^{2}}du = 
    2\int_{0}^{1}\frac{\Tilde{T}^{2}(1-u^{2})}{u^{2}+\frac{1}{4}\Tilde{T}^{2}(1-u^{2})}du = 
    2\int_{0}^{1}\frac{a^{2}(1-u^{2})^3}{u^{2}+\frac{1}{4}a^{2}(1-u^{2})^{3}}du\;,
\end{align}
where we defined $a = z^{-2}$, so that, since we are in the low transparency limit $a\ll 1$, $\Tilde{T} \approx a\cos^2{\phi}$.
Moreover, in the limit $a\ll 1$ the above integral has only significant contributions when $u$ is of the same order or smaller than $a$. Indeed, for $u\lesssim a$, a window with width of order $a$, the integrand is of order 1, so this window has a contribution of $O(a)$. On the other hand, if $u\gg a$, a window with width of order 1, the conductance is of order $a^{2}$, which implies that the contribution of this window is $O(a^{2})$. Thus, to lowest order in $a$ we may ignore any terms that are products of $u^{2}$ and $a^{2}$, and hence
\begin{align}\label{eq:Nis1caseI}    
 \frac{\partial I}{\partial V}\approx  2\int_{0}^{1}du\frac{a^{2}}{\frac{1}{4}a^{2}+u^{2}} +O(a^{2})= 4a\tan^{-1}\frac{2}{a}+O(a^2) = 2\pi a+O(a^{2})\;,
\end{align}
that is, in the low transparency limit ($a\ll 1$) the conductance is approximately given by
\begin{align}\label{eq:ConductanceHelical}
   \frac{\partial I}{\partial V}\approx\frac{2\pi}{z^{2}}\;.
\end{align}
Note it is of first order in transparency, as $\Tilde{T}\propto z^{-2}$, just like the normal state conductance. 
Meanwhile, the noise power is given by
\begin{align}\label{eq:PNccotnributios}
    \frac{\partial P_{N}}{\partial V} = \frac{\partial^{2}\frac{\partial S}{\partial V}}{\partial\chi^{2}}& = \int_{-\frac{\pi}{2}}^{\frac{\pi}{2}}\cos\phi d\phi\frac{-\frac{1}{2}\Tilde{T}^{2}B^{2}}{(1-\frac{1}{16}\Tilde{T}^{2}B^2)^{2}}\;.
\end{align}
We may again first consider the channels separately. By comparing Eqs. (\ref{eq:Conductancedistributios}) and (\ref{eq:PNccotnributios}), we note that channels with $B$ of order 1 or smaller, have a differential noise contribution that is twice as large as the current contribution, that is, the noise power is of second order in tunneling and twice as large as the current. On the other hand, channels with $B\gg \Tilde{T}^{-1}$ have $P_{N}(\phi)\approx 0$, that is, they do not contribute to the noise power (resonance). Thus, after summation over channels we should obtain a nonzero noise power, but a Fano factor that is less than 2. 

At zero energy, where we found $B = -2i\cot\phi$ we may express the averaged noise power as
\begin{align}\label{eq:NoiseToRefTo}
   \frac{\partial P_{N}}{\partial V} &= 2\int_{-\frac{\pi}{2}}^{\frac{\pi}{2}}\frac{\Tilde{T}^{2}\cot^{2}\phi}{(1+\frac{1}{4}\Tilde{T}^{2}\cot^{2}\phi)^2}\cos\phi d\phi = 4\int_{0}^{\frac{\pi}{2}}\frac{\Tilde{T}^{2}\cot^{2}\phi}{(1+\frac{1}{4}\Tilde{T}^{2}\cot^{2}\phi)^2}\cos\phi d\phi\;,
\end{align}
where in the last equality we used that the integrand is even in $\phi$.
Like for the evaluation of the integrals, we substitute $u = \sin\phi$ and $\Tilde{T} = a\cos^{2}\phi = a(1-u^{2})$ we obtain
\begin{align}\label{eq:Pintermsofu}
    \frac{\partial P_{N}}{\partial V}& = 4\int_{0}^{1}\frac{\Tilde{T}^{2}\frac{1-u^{2}}{u^{2}}}{(1+\frac{1}{4}\Tilde{T}^{2}\frac{1-u^{2}}{u^{2}})^{2}}du= 4\int_{0}^{1}\frac{\Tilde{T}^{2}u^{2}(1-u^{2})}{(u^{2}+\frac{1}{4}\Tilde{T}^{2}(1-u^{2}))^{2}}du= 4\int_{0}^{1}\frac{a^{2}u^{2}(1-u^{2})^{3}}{(u^{2}+\frac{1}{4}a^{2}(1-u^{2})^{3})^{2}}du\;.
\end{align}
In the limit $a\ll 1$ we may again ignore all terms of higher order in $a,u$ (again because for $a\ll u$ the integrand is of order $1$ while for $u\gg a$ it is of order $a^{2}$) and obtain
\begin{align}\label{eq:NoiseHelical}
    \frac{\partial P_{N}}{\partial V}&= 4\int_{0}^{1}\frac{a^{2}u^{2}}{(u^{2}+\frac{1}{4}a^{2})^{2}}du = -\frac{2a^{2}}{1+\frac{1}{4}a^{2}}+4a\tan^{-1}\frac{2}{a}\approx 2\pi a = \frac{2\pi}{z^{2}}\;.
\end{align}
That is,
\begin{align}\label{eq:FHelical}
    F &= \frac{\frac{\partial P_{N}}{\partial V}}{\frac{\partial I}{\partial V}} = \frac{\frac{2\pi}{z^{2}}}{\frac{2\pi}{z^{2}}} = 1+O(z^{-4})\;.
\end{align}

This derivation holds for two dimensional superconductors in which the SABSs have a dispersion or three dimensional superconductors in which the SABSs have a dispersion along one direction. 
\subsection{B-W superconductors}
We may also consider three dimensional superconductors in which there is a ZESABS only for one mode. 
To this end we consider the B-W superconductor, in which $\Delta_{1}(\phi) = -\Delta_{2}(\phi) = \Delta_{0}$ and $\vec{d} = (\cos\phi,\sin\phi\cos\theta,\sin\phi,\sin\theta)$. In this case, the angular integration is accompanied by $\frac{1}{\pi}\int_{0}^{\frac{\pi}{2}}d\phi \int_{0}^{2\pi}d\theta\cos\phi\sin\phi$ instead of the 2D value $\frac{1}{2}\int_{-\frac{\pi}{2}}^{\frac{\pi}{2}}\cos\phi d\phi = \int_{0}^{\frac{\pi}{2}}\cos\phi d\phi$, that is, the action reads
\begin{align}
    S = -\frac{1}{\pi}\int_{-\infty}^{\infty} dE \int_{0}^{2\pi}d\theta\int_{0}^{\frac{\pi}{2}}d\phi \cos\phi\sin\phi\text{Tr ln}(4-2\Tilde{T}+\Tilde{T}\{C(\theta,\phi),G_{\chi}\})
\end{align}
We again consider an interface with normal in the x-direction. The resulting expressions are very similar to those for the helical \textit{p} - wave case. Indeed, by rotational symmetry around the x-axis, the contribution to the action can not depend on $\theta$, and consequently the integral over $\theta$ yields a numerical factor $2\pi$, while we may choose a specific $\theta$ to do our calculations. We choose $\theta = 0$, and consequently the action can be written as
\begin{align}
    S = -2\int_{-\infty}^{\infty} dE \int_{0}^{\frac{\pi}{2}}d\phi\cos\phi\sin\phi\text{Tr ln}(4-2\Tilde{T}+\Tilde{T}\{C(\theta = 0,\phi),G_{\chi}\})
\end{align}
We may exploit that $\vec{d}(\theta = 0,\phi) = (\cos\phi,\sin\phi,0)$, which is exactly the d-vector we encountered for the helical \textit{p} - wave superconductor. 
Thus, if we consider a B-W superconductor, the only change compared to the helical \textit{p} - wave superconductor is the addition of an extra factor $2\sin\phi$ in the integral which translates to an extra factor $2u$ after substitution of $u 
= \sin\phi$. Via Eq. (\ref{eq:Iintermsofu}) this reads 
\begin{align}
    \frac{\partial I}{\partial V}&=4\int_{0}^{1}\frac{a^{2}(1-u^{2})^{3}}{u^{2}+\frac{1}{4}a^{2}(1-u^{2})^{3}} udu\;.
\end{align}
Thus, the portion for $u\ll a$ contributes of order $a a = a^{2}$, which means we may not immediately deduce this is the leading term. However, we may follow another approach in this case. 
Indeed, we may substitute $v = u^2$ and $dv = 2udu$  to obtain
\begin{align}
    \frac{\partial I}{\partial V} =  2\int_{0}^{1}\frac{a^{2}(1-v)^{3}}{v+\frac{1}{4}a^{2}(1-v)^{3}} dv\;.
\end{align}
Now we may select $b$ such that $a^2\ll b\ll 1$. Notice that this is always possible if $a\ll 1$, e.g. $b = a$. Here we keep $b$ general to show that the end result does not depend on a specific choice of $b$. Then for $v<b$ we may set $1-v\approx 1$, while for $v>b$ we may ignore any terms of order $a^2$ compared to $v$. Thus, we obtain
\begin{align}\label{eq:ConductanceBW}
    \frac{\partial I}{\partial V}\approx 2\int_{0}^{b}\frac{a^{2}}{v+\frac{1}{4}a^{2}}dv +2\int_{b}^{1}\frac{a^{2}(1-v)^{3}}{v}dv\;.
\end{align}
Similarly, for the noise power, by adding a $u$ in Eq. (\ref{eq:Pintermsofu}) 
and separating the integrals for $v<b$ and $v>b$ with the corresponding approximations, we obtain
\begin{align}
    \frac{\partial P_{N}}{\partial V}\approx 8\int_{0}^{b}\frac{a^{2}u^{2}}{(u^{2}+\frac{1}{4}a^{2})^{2}}udu+8\int_{b}^{1}\frac{a^{2}(1-u^{2})^{3}}{u^{2}}udu\;.
\end{align}
Again, inserting $v = u^{2}$ and $dv = 2u du$ we find
\begin{align}\label{eq:NoiseBW}
    \frac{\partial P_{N}}{\partial V}\approx 4\int_{0}^{b}\frac{a^{2}v}{(v+\frac{1}{4}a^{2})^{2}}dv+4\int_{b}^{1}\frac{a^{2}(1-v)^{3}}{v}dv\;.
\end{align}
The integrals from $b$ to $1$ are the same in the expressions for $\frac{\partial I}{\partial V}$ and $\frac{\partial P_{N}}{\partial V}$, but in the expression for the differential noise power there is a factor 2 in front. The first integral in the expression for $\frac{\partial I}{\partial V}$ evaluates to
\begin{align}
    2a^{2}\text{ln}(\frac{4b+a^{2}}{a^{2}})\approx 2a^{2}\text{ln}\frac{4b}{a^{2}}\;.
\end{align}
Meanwhile the first integral in the expression for $\frac{\partial P_{N}}{\partial V}$ reads
\begin{align}
    4a^{2}\left(-\frac{4b}{a^{2}+4b}+\text{ln}\frac{4b+a^{2}}{a^{2}}\right)\approx 4a^{2}\text{ln}\frac{4b}{a^{2}}\;,
\end{align}
where we used that $|\text{ln}\frac{4b}{a^{2}}|\gg 1$. Hence also the first integrals have the same result. Since both integrals have an additional factor 2 in front in Eq. (\ref{eq:NoiseBW}) for the differential conductance, we find
\begin{align}\label{eq:FBW}
    F = \frac{\frac{\partial P_{N}}{\partial V}}{\frac{\partial I}{\partial V}}\approx 2+O(|\text{ln} a|^{-1})\;.
\end{align}
To evaluate the zero bias conductance, we note that the first integral equals $2a^{2}\text{ln}\frac{4b}{a^{2}}$. This is much smaller than $a$. Moreover, in the second integral, the leading order term is of order $-2a^{2}\text{ln}b$. Since $1\gg b\gg a$ we conclude that also this integral is much smaller than $a$. Thus, the zero bias conductance decreases faster than $a$ when $a\xrightarrow{}0$. On the other hand, it is well-known that $\sigma_{N}$ scales linearly with transparency \cite{blanter2000shot}, that is, it is of order $a$. Consequently, $\sigma(eV = 0)/\sigma_{N}\ll 1$ in tunnel junctions with B-W superconductors.

\subsection{Two-dimensional chiral \textit{d}}
To confirm that $F = 1$ for linear dispersions around $E_{BS} = 0$, independent of the coefficient in front of the linear term in $E_{BS}$, we repeated the calculation for the chiral \textit{d} - wave superconductor, for which the dispersion $\frac{d E_{BS}}{d\phi}$ is a factor 2 larger than for the chiral or helical \textit{p} - wave superconductors.
We may use a similar approach for the chiral \textit{d} - wave case. 
At $E = 0$, the bulk Green's function of the chiral $\textit{d}$ - wave superconductor reads, focusing on a single spin sector since all matrices are proportional to the identity in spin space,
\begin{align}
    G_{S}(\phi) = \frac{1}{i\Delta_{0}}\begin{bmatrix}
        0&\Delta_{0}e^{2i\phi}\\-\Delta_{0}e^{-2i\phi}&0
    \end{bmatrix} = \frac{1}{i}\begin{bmatrix}
        0&e^{2i\phi}\\-e^{-2i\phi}&0
    \end{bmatrix}\;.
\end{align}
From this we find
\begin{align}
    H_{+} = \frac{1}{2}(G_{S}(\phi)+G_{S}(\pi-\phi)) = \frac{1}{2i}\begin{bmatrix}
        0&e^{2i\phi}+e^{-2i\phi}\\-(e^{2i\phi}+e^{-2i\phi})&0
    \end{bmatrix} = \cos2\phi\tau_{2}\;,
\end{align}
while
\begin{align}
    H_{-} = \frac{1}{2}(G_{S}(\phi)-G_{S}(\pi-\phi)) = \frac{1}{2i}\begin{bmatrix}
        0&e^{2i\phi}-e^{-2i\phi}\\-(e^{2i\phi}-e^{-2i\phi})&0
    \end{bmatrix} = \sin2\phi\tau_{1}
\end{align}
Thus, for the chiral \textit{d} - wave superconductor the boundary condition contains
\begin{align}
    C(E = 0) &  = H_{+}^{-1}(\mathbf{1}-H_{-})= i\tan{2\phi}\tau_{3}+(\cos{2\phi})^{-1}\tau_{2}\;,
\end{align}
 that is, $B = 2i\tan{2\phi}$, which has poles at $\phi = \pm\frac{\pi}{4}$, that is, $\sin\phi = \frac{1}{\sqrt{2}}$. Because of symmetry between $(-\frac{\pi}{2},0)$ and $(0,\frac{\pi}{2})$ we may write, analogous to Eq. (\ref{eq:CurrToRefTo}) but replacing $\cot\phi$ by $\tan2\phi$ and using that $\tan2\phi = \frac{\sin\phi\cos\phi}{\frac{1}{2}-\sin^{2}\phi}$
 \begin{align}
     \frac{\partial I}{\partial V} &= 2\int_{0}^{\frac{\pi}{2}} \frac{\Tilde{T}^{2}\tan^{2}2\phi}{1+\frac{1}{4}\Tilde{T}^{2}\tan^{2}2\phi}\cos\phi d\phi = 2\int_{0}^{\frac{\pi}{2}}\frac{\Tilde{T}^{2}\sin^{2}\phi\cos^{2}\phi}{(\sin^{2}\phi-\frac{1}{2})^{2}+\frac{1}{4}\Tilde{T}^{2}\sin^{2}\phi\cos^{2}\phi}\cos\phi d\phi\nonumber\\&= 2\int_{0}^{1}\frac{\Tilde{T}^{2}u^{2}(1-u^{2})}{(u^{2}-\frac{1}{2})^{2}+\frac{1}{4}\Tilde{T}^{2}u^{2}(1-u^{2})}du \approx 2\int_{0}^{1}\frac{a^{2}u^{2}(1-u^{2})^{3}}{(u^{2}-\frac{1}{2})^{2}+\frac{1}{4}a^{2}u^{2}(1-u^{2})^{3}}du\;,
 \end{align}
 where in the last line we used $\Tilde{T} \approx a\cos^{2}\phi = a(1-u^{2})$.
The most important contribution to this integral is not for $u\approx 0$, but for $u^{2}\approx\frac{1}{2}$, that is, for $u\approx \frac{1}{\sqrt{2}}$. Thus, in the integrand we may replace $u^{2}\approx 1-u^{2}\approx \frac{1}{2}$ and $u^{2}-\frac{1}{2} = (u-\frac{1}{\sqrt{2}})(u+\frac{1}{\sqrt{2}}) \approx \sqrt{2}(u-\frac{1}{\sqrt{2}})$. With this we find
\begin{align}
    \frac{\partial I}{\partial V}&\approx2\int_{0}^{1}\frac{a^{2}\frac{1}{2}\frac{1}{8}}{(u-\frac{1}{\sqrt{2}})^{2}2+\frac{1}{4}a^{2}\frac{1}{2}\frac{1}{8}} du= \int_{0}^{1}\frac{a^{2}}{16(u-\frac{1}{\sqrt{2}})^{2}+\frac{1}{8}a^{2}}du\nonumber\\&= \frac{\frac{a^{2}}{16}}{\frac{a}{8\sqrt{2}}}\Big(\tan^{-1}\frac{8}{a}+\tan^{-1}\frac{8(\sqrt{2}-1)}{a}\Big) = \frac{\pi}{\sqrt{2}} a\;.
\end{align}
 
The noise power is obtained by replacing $\cot^{2}\phi$ by $\tan^{2} 2\phi$ in Eq. (\ref{eq:NoiseToRefTo}) in a similar fashion:
\begin{align}
    \frac{\partial P_{N}}{\partial V} &= 4 \int_{0}^{\frac{\pi}{2}}\frac{\Tilde{T}^{2}\tan^{2}2\phi}{(1+\frac{1}{4}\Tilde{T}^{2}\tan^{2}2\phi)^{2}}\cos\phi d\phi = 4\int_{0}^{\frac{\pi}{2}}\frac{\Tilde{T}^{2}\frac{\sin^{2}\phi\cos^{2}\phi}{(\frac{1}{2}-\sin^{2}\phi)^{2}}}{(1+\frac{1}{4}\frac{\sin^{2}\phi\cos^{2}\phi}{(\frac{1}{2}-\sin^{2}\phi)^{2}})^{2}}\cos\phi d\phi\nonumber\\& = 4\int_{0}^{\frac{\pi}{2}}\frac{\Tilde{T}^{2}\sin^{2}\phi\cos^{2}\phi(\frac{1}{2}-\sin^{2}\phi)^{2}}{((\frac{1}{2}-\sin^{2}\phi)^{2}+\frac{1}{4}\sin^{2}\phi\cos^{2}\phi)^{2}}\cos\phi d\phi = 4\int_{0}^{1} \frac{\Tilde{T}^{2}u^{2}(1-u^{2})(u^{2}-\frac{1}{2})^{2}}{((\frac{1}{2}-u^{2})^{2}+\frac{1}{4}\Tilde{T}^{2}u^{2}(1-u^{2}))^{2}} du\nonumber\\&=
    4\int_{0}^{1} \frac{a^{2}u^{2}(1-u^{2})^{3}(\frac{1}{2}-u^{2})^{2}}{((\frac{1}{2}-u^{2})^{2}+\frac{1}{4}a^{2}u^{2}(1-u^{2})^{3})^{2}} du \approx 4\int_{0}^{1} \frac{a^{2}\frac{1}{2}\frac{1}{8}2(u-\frac{1}{\sqrt{2}})^{2}}{(2(u-\frac{1}{\sqrt{2}})^{2}+\frac{1}{4}a^{2}\frac{1}{2}\frac{1}{8})^{2}}du\nonumber\\&=\int_{0}^{1}\frac{\frac{1}{2}a^{2}(u-\frac{1}{\sqrt{2}})^{2}}{(2(u-\frac{1}{\sqrt{2}})^{2}+\frac{1}{64}a^{2})^{2}}du =\int_{0}^{1}\frac{\frac{1}{8}a^{2}(u-\frac{1}{\sqrt{2}})^{2}}{((u-\frac{1}{\sqrt{2}})^{2}+\frac{1}{128}a^{2})^{2}}du\nonumber\\& =\frac{a^{2}}{2}\Bigg(\frac{ -16  (-64 + 64\sqrt{2} + a^2)}{ (64 + a^2) (64 (3 - 2 \sqrt{2}) + a^2)}+
    \frac{\sqrt{2}}{a}\Big(\tan^{-1}\frac{8}{a}+\tan^{-1}\frac{8(\sqrt{2}-1)}{a}\Big)\Bigg)
    \approx \frac{a^{2}}{2}\frac{\sqrt{2}}{a}\pi = \frac{\pi}{\sqrt{2}}a \;,
\end{align}
where in the last line we kept the result of the integral up to first order in $a$.
Thus, to lowest order in transparency,
\begin{align}
    F\approx 1\;.
\end{align}
Thus, chiral \textit{d} - wave superconductors and chiral \textit{p} - wave superconductors can not be distinguished by the zero voltage differential Fano factor. 

\subsection{Analytical proof of fractional Fano factors}\label{sec:AnaFrac}
While linear dispersions always give $F\approx 1$, in the presence of dispersive SABSs for which the dispersion around normal incidence is nonlinear, the differential Fano factor is smaller, because there are more angles for which the effective transparency is high.
For such superconductors, the relation between the dispersion and  $F_{00}$ can be shown analytically in certain limits. We assume that the temperature is zero, that $a$ is very small and assume that the dispersion is proportional to $\phi^{n}$ for angles close to normal incidence. Moreover, we assume that the SABSs only cross zero once. In that case, only angles close to normal incidence give a considerable contribution. Moreover, since SABSs always appear as first order poles, the expressions are very similar to the case $n = 1$, but now with $B \approx  -2id\phi^{-n} =  -2idu^{-n}$ for small $u$, where $d$ is the proportionality constant which depends on the specific pair potential under consideration. That this proportionality constant does not influence the differential Fano factor can be deduced from the fact that in Eq. (\ref{eq:dSdVforAnalytical}) $B^{2}$ is multiplied with the square of the transparency $\Tilde{T}^{2}$. Consequently, the proportionality constant can be absorbed in $\Tilde{T}$ and hence to lowest order in $\Tilde{T}$ it does not influence the differential Fano factor, unless $a^{-1}\lesssim d$. Since $a^{-1}$ is a very large number, we may assume $d\ll a^{-1}$.  
Since the contribution of small $u$ dominates the integral, we may to lowest order in transparency obtain the expression for the conductance by replacing $u$ by $u^{n}$ and $a$ by $a d$ in Eq. (\ref{eq:Nis1caseI}),
\begin{align}
   \frac{\partial I}{\partial V}\approx 2\int_{0}^{1}du\frac{a^{2}d^{2}}{\frac{1}{4}a^{2}d^{2}+u^{2n}} \approx 8\left(\frac{ad}{2}\right)^{\frac{1}{n}}\int_{0}^{\infty} dw \frac{1}{1+w^{2n}}\;.\label{eq:INn}
\end{align}
where $w = u\left(\frac{2}{ad}\right)^{\frac{1}{n}}$ and the upper limit of integration $(\frac{2}{ad})^{\frac{1}{n}}$ is approximated by $\infty$ because $ad\ll 1$.
Similarly, for the differential noise power we obtain, in analogy to Eq. (\ref{eq:NoiseHelical})
\begin{align}
    \frac{\partial P_{N}}{\partial V} = 4\int_{0}^{\infty}\frac{a^{2}d^{2}u^{2n}}{(u^{2n}+\frac{1}{4}a^{2}d^{2})^{2}}du = 16\left(\frac{a d}{2}\right)^{\frac{1}{n}}\int_{0}^{\infty}\frac{w^{2n}}{(1+w^{2n})^{2}}dw = 16\left(\frac{a d}{2}\right)^{\frac{1}{n}}\int_{0}^{\infty}\frac{2nw^{2n-1}}{(1+w^{2n})^{2}}\frac{w}{2n}dw\;.
\end{align}
By performing integration by parts on $\frac{2nw^{2n-1}}{(1+w^{2n})^{2}}$ and $\frac{w}{2n}$ we obtain
\begin{align}\label{eq:PNn}
    \frac{\partial P_{N}}{\partial V} \approx  \left(\frac{a d}{2}\right)^{\frac{1}{n}}\frac{8}{n}\int_{0}^{\infty}\frac{1}{(1+w^{2n})}dw\;.
\end{align}
From this we conclude that
\begin{align}
    F_{00} = \frac{1}{n}\;,
\end{align}
in agreement with our numerical calculations.

The expansion of this theory to three-dimensional superconductors is particularly simple in this case. Indeed, an extra factor $\sin\phi$ appears in the initial integral, which gives an extra factor $u$ in both of the above integrals. They therefore read

\begin{align}
   \frac{\partial I}{\partial V}\approx 2\int_{0}^{1}du\frac{a^{2}d^{2}}{\frac{1}{4}a^{2}d^{2}+u^{2n}}u \approx 8\left(\frac{ad}{2}\right)^{\frac{2}{n}}\int_{0}^{\infty} dw \frac{1}{1+w^{2n}}w = 4\left(\frac{ad}{2}\right)^{\frac{2}{n}}\int_{0}^{\infty} d\Tilde{w} \frac{1}{1+\Tilde{w}^{n}} \;.\label{eq:INn3D}
\end{align}
where we used the substitution $\Tilde{w} = w^{2}$.
The noise reads
\begin{align}
    \frac{\partial P_{N}}{\partial V} = 4\int_{0}^{\infty}\frac{a^{2}d^{2}u^{2n}}{(u^{2n}+\frac{1}{4}a^{2}d^{2})^{2}} u du = 16\left(\frac{a d}{2}\right)^{\frac{2}{n}}\int_{0}^{\infty}\frac{w^{2n}}{(1+w^{2n})^{2}}w dw = 16\left(\frac{a d}{2}\right)^{\frac{2}{n}}\int_{0}^{\infty}\frac{2nw^{2n-1}}{(1+w^{2n})^{2}}\frac{w}{2n} w dw\;.
\end{align}
By performing integration by parts on $\frac{2nw^{2n-1}}{(1+w^{2n})^{2}}$ and $\frac{w}{2n}$ we obtain
\begin{align}\label{eq:PNn3D}
    \frac{\partial P_{N}}{\partial V} \approx  \left(\frac{a d}{2}\right)^{\frac{2}{n}}\frac{16}{n}\int_{0}^{\infty}\frac{1}{(1+w^{2n})} w dw  = \left(\frac{a d}{2}\right)^{\frac{2}{n}}\frac{8}{n}\int_{0}^{\infty}\frac{1}{(1+\Tilde{w}^{n})} d\Tilde{w}\;.
\end{align}

This gives almost identical expressions as above, but then with a factor $\frac{1}{2}$ in front of both current and noise power, which thus cancels out, and moreover the substitution $n\xrightarrow{}\frac{n}{2}$. We conclude that
\begin{align}
    F_{00} = \frac{1}{\frac{n}{2}} = \frac{2}{n}\;.
\end{align}
\subsection{Numerical calculation of fractional differential Fano factors}
In this subsection we show how to construct superconductors with nonlinear dispersion and display the numerical values of the differential Fano factor in those cases.

This is achieved by ensuring that the dispersion of the SABS vanishes at $\phi = 0$. For example, via combinations of chiral \textit{p} - wave and chiral \textit{f} -  wave superconductors, the pair potential $\Delta_{0}
(e^{i\phi}-\frac{1}{3}e^{3i\phi})$ can be constructed. For almost normal incidence, $\phi\approx 0$, we have 
\begin{align}
\Delta(\phi)&\approx \Delta_{0}\Bigg(\Big(1+i\phi-\frac{1}{2}\phi^{2}-\frac{i}{6}\phi^{3}+O(\phi^{4})\Big)-\Big(\frac{1}{3}+i\phi-\frac{3}{2}\phi^{2}-\frac{3}{2}i\phi^{3}+O(\phi^{4})\Big)\Bigg)\nonumber\\& = \Delta_{0}(\frac{2}{3}+\phi^{2}+\frac{4}{3}i\phi^{3})+O(\phi^{4})\;.
\end{align}
Now, since this is still a chiral odd-parity state, we have $\Delta(\pi-\phi) = \Delta_{0}(-e^{-i\phi}+\frac{1}{3}e^{-3i\phi}) = -\Delta_{0}^{*}(\phi)$.
Thus we have
\begin{align}
    H_{+} = \frac{1}{2}\frac{1}{\sqrt{E^{2}-|\Delta(\phi)|^{2})}}\Big(\begin{bmatrix}
        E&\Delta(\phi)\\-\Delta^{*}(\phi)&-E
    \end{bmatrix}+\begin{bmatrix}
        E&-\Delta^{*}(\phi)\\\Delta(\phi)&-E
    \end{bmatrix}\Big) = \frac{1}{\sqrt{E^{2}-|\Delta(\phi)|^{2})}}\begin{bmatrix}
        E&i\text{Im}\Delta(\phi)\\i\text{Im}\Delta(\phi)&-E
    \end{bmatrix}\;.
\end{align}
SABSs can appear when one of the eigenvalues of $H_{+}$ equals 0, which is for $E = \pm \text{Im}\Delta(\phi)$. For the former case the corresponding eigenvector is $\begin{bmatrix}
    1\\i
\end{bmatrix}$, for the latter case it is $\begin{bmatrix}
    1\\-i
\end{bmatrix}.$
To understand whether this SABS exists we need to consider $\mathbf{1}-H_{-}$ at $E = \pm i\text{Im}\Delta(\phi)$. It equals
\begin{align}
    \begin{bmatrix}1&i\text{sign Re}\Delta(\phi)\\-i\text{sign Re}\Delta(\phi)&1\end{bmatrix}\;.
\end{align}
Also this matrix has a zero eigenvalue with eigenvector $\begin{bmatrix}
    1\\i\text{sign Re}\Delta(\phi)
\end{bmatrix}$.
For small enough $\phi$ we have $\text{sign Re}\Delta(\phi) = 1$. 
Thus, for $E = \text{Im}\Delta(\phi)$ the two matrices have the same eigenvector with eigenvalue $0$ and hence $C = H_{+}^{-1}(\mathbf{1}-H_{-})$ does not have a pole and there is no SABS in this channel at this energy. On the other hand, for $E = -\text{Im}(\Delta(\phi))$ the eigenvector of $H_{+}$ with eigenvalue has a nonzero eigenvalue for $\mathbf{1}-H_{-}$ and consequently, $C$ has a pole at this energy.

Therefore the SABS energy $E_{BS}(\phi)$ is for small $\phi$ given by
\begin{align}
    E_{BS}(\phi) = -\text{Im}(\Delta(\phi)) \approx -\frac{7}{2}\Delta_{0}\phi^{3}\;,
\end{align}
that is, the dispersion vanishes at $\phi = 0$. We found numerically that in this case
\begin{align}
    F_{3}(k_{B}T\ll |eV|\ll \frac{\Delta_{0}}{z^{2}}) \approx \frac{1}{3}\;,
\end{align}
where the approximate sign $\approx$ has been used to indicate that for any finite temperature there exist small deviations from the zero temperature value.
For nonzero voltages the differential Fano factor approaches $1$ within a small voltage window of order $\Tilde{T}\Delta_{0}$, because for nonzero $\phi$ there is a finite dispersion.

A similar procedure can be used to construct a pair potential for which also the $\phi^{3}$-component of the dispersion vanishes, using $\Delta(\phi) = \Delta_{0}(e^{i\phi}-\frac{1}{2}e^{3i\phi}+\frac{1}{10}e^{5i\phi})$. In this case for small angles the dispersion is of fifth order, $\Delta(\phi)\propto\phi^{5}$ and we found numerically that
\begin{align}
    F_{5}(k_{B}T\ll |eV|\ll \frac{\Delta_{0}}{z^{2}}) \approx \frac{1}{5}\;.
\end{align}
Again for nonzero voltages the differential Fano factor returns to $1$, due to $\frac{dE_{BS}}{d\phi}$ being nonzero for nonzero angles.

Similarly, we constructed the pair potentials $\Delta_{7}(\phi) = \Delta_{0}(e^{i\phi}-\frac{3}{5}e^{3i\phi}+\frac{1}{5}e^{5i\phi}-\frac{1}{35}e^{7i\phi})$ and $\Delta_{9}(\phi) = \Delta_{0}(e^{i\phi}-\frac{2}{3}e^{3i\phi}+\frac{2}{7}e^{5i\phi}-\frac{1}{14}e^{7i\phi}+\frac{1}{126}e^{9i\phi})$ so that for small $\phi$ we have $\Delta_{7}(\phi)\propto\phi^{7}$ and $\Delta_{9}(\phi)\propto\phi^{9}$, for which we found that the corresponding differential Fano factors $F_{7,9}$ equal
\begin{align}
    F_{7}(k_{B}T\ll|eV|\ll\Delta_{0}) &= \frac{1}{7}\;,\\
    F_{9}(k_{B}T\ll|eV|\ll\Delta_{0}) &= \frac{1}{9}\;.
\end{align}

Similarly, we constructed analogies of the helical \textit{p} - wave superconductor to higher order dispersions. For example, to obtain third order dispersion one defines
\begin{align}
    \Tilde{\vec{d}}(\phi) = (\cos\phi-\frac{1}{3}\cos3\phi,\sin\phi-\frac{1}{3}\sin\phi,0)\;.
\end{align}
We note that this d-vector is not normalized. For numerical implementation we first implement normalization. The magnitude of $\Tilde{\vec{d}}(\phi)$ then needs to be absorbed in the magnitude of the pair potential. The generalization to higher order dispersions is entirely similar and we find the same results for the generalizations of the helical as for the generalizations of the chiral \textit{p} - wave superconductor, showing the robustness of our results.

Thus, if $E_{BS}\propto\phi^{n}$ for small angles, then
\begin{align}
    F_{n}(k_{B}T\ll |eV|\ll \frac{\Delta_{0}}{z^{2}}) \approx \frac{1}{n}\;.
\end{align}
This is consistent with the observation that $F = 0$ for dispersionless ZESABSs. One should always keep in mind that for any finite temperature $F_{n}(T>0,eV = 0) = 0$. Therefore, these features can only be found for $k_{B}T\ll |eV|\ll \max_{\phi}\Tilde{T}(\phi)\Delta_{0}$. 

Moreover, one should take into account that for linear dispersion in which the coefficient in front of the linear term of order $\Tilde{T}$ or smaller, the differential Fano factor also deviates from 1. Indeed, in Eqs. (\ref{eq:INn}) and (\ref{eq:PNn}) we only kept the leading order term in $a$, which is proportional to $(ad)^{\frac{1}{n}}$. If $d$ is of the same order as $a$, we should also take into account terms of order $a^{\frac{2}{n}}$, and the results are significantly altered. It has been numerically verified there is a continuous transition between $F = 1$ and $F = \frac{1}{3}$ and between $F = \frac{1}{3}$ and $F = \frac{1}{5}$ if one lowers the coefficient of the linear and third order term to 0. 
\subsubsection{3D}
Also in three dimensions one may alter the dispersion of the SABSs. If there is dispersion only in one direction, the results are equivalent to the 2D analog, as discussed before. Now assume that there is only a single ZESABS, but the dispersion is proportional to $\phi^{n}$, where $\phi$ denotes the angle made with the normal of the interface. Since we do not need to break the in-plane rotational symmetries, we may generalize the B-W superconductor in entirely the same way as the helical \textit{p} - wave superconductor. We analyzed the same cases as for the 2D superconductors and found
\begin{align}
    F \approx \frac{2}{n}\;.
\end{align}
Thus, for any generalizations of the B-W pair potential the differential Fano factor is twice as high as for the corresponding generalization of the helical \textit{p} - wave superconductor, due to the smaller relative contribution of the ZESABS.
\clearpage
\section{Extension of the Tanaka-Nazarov boundary condition}\label{sec:Extension}
In this section we motivate our extension of the Tanaka-Nazarov boundary condition to describe junctions whose transmission eigenvalues are rapidly fluctuation functions of the angle of incidence and therefore are well described by eigenvalue distribution functions. We separate the two cases of (i) junctions that conserve in-plane momentum, such as the double barrier junction and (ii) junctions that do not in-plane momentum, such as the diffusive barrier and the chaotic cavity, and show that in the latter case an additional integral needs to be introduced.  In  the first two subsections of this Appendix we derive the full counting action for those cases and we do not restrict the pair potential of the superconductor.  With this we derive Eqs. (\ref{eq:WithTwoIntegrals}) and (\ref{eq:WithOneIntegral}) in the main text. In the third subsection we consider the differential Fano factor that can be calculated from those actions for the special case of a helical \textit{p} - wave superconductor.
\subsection{Conserved in-plane momentum}\label{sec:DBAction}
In the double barrier junction
there are two ballistic interfaces whose normal vector points in the same direction. Therefore the in-plane momentum of the electrons is conserved in double barrier junction. To derive the full counting action, we first consider the discrete form of the full counting statistics action for the Tanaka-Nazarov boundary condition
\begin{align}
    S = -\frac{t_{0}}{8\pi\hbar}\sum_{n = 1}^{N_{\text{ch}}}\text{Tr ln}(4-2\Tilde{T}_{n}+\Tilde{T}_{n}\{G_{\chi},C_{n}\})\;,
\end{align}
where $N_{\text{ch}}$ is the number of channels and $C_{n}$ is the surface Green's function for channel $n$. Each channel is characterized by its angle of incidence $\phi_{n}$.
Now assume that there are many channels, $N_{\text{ch}}\gg 1$. We may divide the set $M$ of all channels in $N'$ pieces $M_{m}$, $m = 1...N'$ that all contain $\frac{N_{\text{ch}}}{N'}$ channels. 
\begin{figure}[b]
    \centering
    \includegraphics[width=4.3cm]{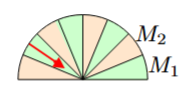}
    \caption{Partitioning of the channels into subsets with similar incoming momentum. For illustrative purposes $N^{\prime} = 8$ is chosen.}
    \label{fig:partitionincoming}
\end{figure}
There are a few requirements we pose to this subdivision.
First of all, we require that there are many channels within every subset, and that there are many subsets, $1\ll N'\ll N_{\text{ch}}$. We require that all channels are contained in one of the subsets, that is, $M = \cup_{m}M_{m}$, but that no channel is in two different subsets at the same time, that is, $M_{m}\cap M_{m'} = \emptyset$ for all $m\neq m'$. We choose the sets $\{M_{m}\}$ such that $|\phi_{n^{m}_{1}}-\phi_{n^{m}_{2}}|\ll \frac{\pi}{2}$ for any choice of channels $n^{m}_{1,2}\in M_{m}$, which is possible because $N'\ll N_{\text{ch}}$. 

We index the channels within a subset $M_{m}$ by $n^{m}_{i}$, where the upper index indicates the subset it belongs to and $i$ runs from 1 to $N'$ through the subset.
Because we chose the subsets such that the incident momentum is almost the same for all channels within a subset, the surface Green's functions $C_{n^{m}_{i}}$ of the modes $n^{m}_{i}\in M_{m}$ all satisfy $C_{n^{m}_{i}}\approx C_{n^{m}_{1}}=:C_{m}$  and hence we may write
\begin{align}
    S &\approx \sum_{m} S_{m}\;,\\
    S_{m}&= -\frac{t_{0}}{8\pi\hbar}\sum_{n\in M_{m}}\text{Tr ln} \Big(4-2\Tilde{T}_{n}+\Tilde{T}_{n}\{G_{\chi},C_{m}\}\Big)\;.
\end{align}
Within our approximation, in the summation over the many channels in $M_{m}$ only the transparency depends on the channel, the Green's functions on both sides are independent of $n$. Therefore, for each segment we may use the same approach as used for incorporation of the Dorokhov distribution in Nazarov's theory \cite{belzig2003full}. The transmission eigenvalues can be estimated to a good approximation by a continuous distribution function $\rho_{m}(\Tilde{T})$ for the eigenvalues, $0\leq\Tilde{T}\leq 1$ in this segment,
and hence we may write
\begin{align}
    S_{m}&\approx -\frac{t_{0}N_{\text{ch}}}{8\pi\hbar N^{\prime}}\int_{0}^{1} d\Tilde{T} \rho_{m}(\Tilde{T})\text{Tr ln}\Big(4-2\Tilde{T}+\Tilde{T}\{G_{\chi},C_{m}\}\Big)\;.
\end{align}
Now, if there are many resonances in the double barrier junction, the transmission eigenvalues fluctuate rapidly as a function of incident angle, and each segment has many modes for which $\Tilde{T}\approx 1$ and many modes for which $\Tilde{T}\approx 0$. To a good approximation, the proportion of modes with high transparency is approximately the same for each segment.   

Therefore, we may assume $\rho_{m}(\Tilde{T})$ is the same for each segment, and hence is equal to the distribution function $\rho(\Tilde{T})$ of transmission eigenvalues of the junction. Hence, we may take this distribution function out of the sum over channels and obtain
\begin{align}
    S&\approx-\frac{t_{0}N_{\text{ch}}}{8\pi\hbar N^{\prime}} \int_{0}^{1} d\Tilde{T}\rho(\Tilde{T}) \sum_{m} \text{Tr ln}\Big(4-2\Tilde{T}+\Tilde{T}\{G_{\chi},C_{m}\}\Big)\;.
\end{align}
Now, since $M_{m}$ contains those channels with incoming angle approximately equal to $\phi_{m}$, in the continuum limit we may replace $\sum_{m}$ by $\frac{N_{\text{ch}}}{2}\int_{-\frac{\pi}{2}}^{\frac{\pi}{2}} d\phi \cos\phi$ as is usually done for the Tanaka-Nazarov boundary condition.  
With this we obtain the expression 
\begin{align}
    S&= -\frac{t_{0}N_{\text{ch}}}{16\pi\hbar}\int_{0}^{1}d\Tilde{T}\rho(\Tilde{T})\int_{-\frac{\pi}{2}}^{\frac{\pi}{2}}d\phi\cos\phi\text{Tr ln}\Big(4-2\Tilde{T}+\Tilde{T}\{G_{\chi},C(\phi)\}\Big)\;.
\end{align}
This reproduces Eq. (\ref{eq:WithOneIntegral}) in the main text.
\subsection{Nonconserved in-plane momentum}\label{sec:DiffusiveAction}
Next to junctions in which in-plane momentum is conserved, we may consider junctions in which in-plane momentum is not conserved, such as in a diffusive barrier junction, where the momentum is completely randomized by scattering on impurities. Another example is the chaotic cavity, where the in-plane momentum is not conserved upon scattering on those parts of the walls of the cavity for which the normal to the interface is not in the transport direction. Again we consider the many channels approximation. Because the channels depend on both the incoming and outgoing momentum, we require the object
\begin{align}
    \mathcal{C}(\phi,\Tilde{\phi}) = H_{+}(\phi,\Tilde{\phi})^{-1}(\mathbf{1}-H_{-}(\phi,\Tilde{\phi}))\;,\\
    H_{\pm}(\phi,\Tilde{\phi}) = G_{S}(\phi)\pm G_{S}(\pi-\Tilde{\phi})\;.
\end{align}

The first part of the derivation remains similar. We define $S_{m}$ as in the previous case for the $N'\ll N_{\text{ch}}$ different segments sorted on incoming angle, and each $S_{m}$ has $\frac{N_{\text{ch}}}{N^{\prime}}$ modes. Because the incoming and outgoing momentum are not related to each other in diffusive barrier junctions and chaotic cavities, the outgoing momentum is different for each of the channels within $M_{m}$ and hence $\mathcal{C}$ still varies significantly within each segment. 

\begin{figure}[b]
    \centering
    \includegraphics[width=8.6cm]{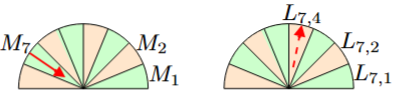}
    \caption{Partitioning of the channels into subsets with similar incoming and outgoing momentum. For illustrative purposes $N^{\prime} = \Tilde{N} = 8$ is chosen.}
    \label{fig:partitionincomingandoutgoing}
\end{figure}
Therefore, we subdivide the sets $M_{m}$ further into $\Tilde{N} $subsets $L_{l}^{m}$ that contain $\frac{N_{\text{ch}}}{\Tilde{N}N^{\prime}}$ channels, so that within each $L_{l}^{m}$ not only the incident angles $\phi_{n^{l}_{i}}$ are similar, but also the angles of the outgoing momenta $\Tilde{\phi}_{n^{l}_{i}}$, that is, not only $|\phi_{n^{l}_{i}}-\phi_{n^{l}_{j}}|\ll\frac{\pi}{2}$, but also  $|\Tilde{\phi}_{n^{l}_{j}}-\Tilde{\phi}_{n^{l}_{j}}|\ll\frac{\pi}{2}$ for all pairs $n^{l}_{i,j}\in L_{l}^{m}$. We need each subset $L_{l}^{m}$ to contain many channels, $\frac{N_{\text{ch}}}{\Tilde{N}N^{\prime}}\gg 1$, but the restriction on the outgoing momenta requires that this is only a small portion of the channels in $M_{m}$, that is, $\Tilde{N}\ll N'$. Thus, for this formalism to work we need to use $1\ll \Tilde{N}\ll N'\ll N_{\text{ch}}$.

We label the modes within each $L_{l}^{m}$ by $n_{i}^{m,l}$, where the upper indices are indicative of the subset $L_{l}^{m}$ and the lower index $i$ runs from 1 to $\Tilde{N}$ through this subset. All incoming and outgoing momenta are similar within $L_{l}^{m}$. 
Consequently the surface Green's functions are similar for all modes $n^{m,l}_{i}$ in $L_{l}^{m}$. That is, they satisfy $\mathcal{C}_{n^{m,l}_{i}}\approx \mathcal{C}_{n^{m,l}_{1}}=:\mathcal{C}_{m,l}$. Hence, we may write
\begin{align}
    S &=\sum_{m} S_{m}\;,\\
    S_{m} &= \sum_{l}S_{m,l}\;,\\
    S_{m,l} &= -\frac{t_{0}}{8\pi\hbar}\sum_{n\in L_{l}^{m}}\text{Tr ln} (4-2\Tilde{T}_{n}+\Tilde{T}_{n}\{G_{\chi},\mathcal{C}_{m,l}\})\;.
\end{align}
Within the summation over channels in $L_{l}^{m}$ the only channel dependent quantity is the transmission.
Since $L_{l}^{m}$ contains many channels, we may follow the same procedure as in the previous subsection.  The sum over transmission eigenvalues is well described by the integral over the transmission eigenvalues $\rho_{m,l}(\Tilde{T})$, i.e.
\begin{align}
    S_{m,l}&\approx-\frac{t_{0}N_{\text{ch}}}{8\pi\hbar N^{\prime}\Tilde{N}}\int_{0}^{1}d\Tilde{T}\rho_{m,l}(\Tilde{T})\text{Tr ln}(4-2\Tilde{T}+\Tilde{T}\{G_{\chi},\mathcal{C}_{m,l}\})\;.
\end{align}
Now, for both the diffusive interface and the chaotic cavity, the distribution  does not depend on the direction of either incoming or outgoing momentum, and hence $\rho_{m,l}(\Tilde{T})  = \rho(\Tilde{T})$ for all $l,m$. We may thus extract this outside of the sum over $l,m$ and obtain
\begin{align}
   S&\approx -\frac{t_{0}N_{\text{ch}}}{8\pi\hbar N^{\prime}\Tilde{N}}\int_{0}^{1} d\Tilde{T}\rho(\Tilde{T}) \sum_{m}\sum_{l} \text{Tr ln}(4-2\Tilde{T}+\Tilde{T}\{G_{\chi},\mathcal{C}_{m,l}\})\;.
\end{align}
Now, since there are many subsets $L_{l}^{m}$ in each $M_{m}$ we may replace the summation over $l$ in the usual way by an integration over the angle of outgoing momentum, $\sum_{l}\xrightarrow{}\frac{\Tilde{N}}{2}\int_{-\frac{\pi}{2}}^{\frac{\pi}{2}}d\Tilde{\phi}\cos\Tilde{\phi}$. We obtain
\begin{align}
    S\approx -\frac{t_{0}N_{\text{ch}}}{16\pi\hbar N^{\prime}}\int_{0}^{1} d\Tilde{T}\rho(\Tilde{T}) \sum_{m}\int_{-\frac{\pi}{2}}^{\frac{\pi}{2}} d\Tilde{\phi}\cos\Tilde{\phi}\text{Tr ln}(4-2\Tilde{T}+\Tilde{T}\{G_{\chi},\mathcal{C}_{m}(\Tilde{\phi})\})\;.
\end{align}
Because there are also many subsets $M_{m}$, we may replace the sum over $m$ also in the usual way, $\sum_{m}\xrightarrow{}\frac{N^{\prime}}{2}\int_{-\frac{\pi}{2}}^{\frac{\pi}{2}}d\phi\cos\phi$. With this we obtain the desired result in the continuum limit,
\begin{align}
    S\approx -\frac{t_{0}N_{\text{ch}}}{32\pi\hbar}\int_{0}^{1}d\Tilde{T}\rho(\Tilde{T})\int_{-\frac{\pi}{2}}^{\frac{\pi}{2}}d\Tilde{\phi}\cos\Tilde{\phi}\int_{-\frac{\pi}{2}}^{\frac{\pi}{2}}d\phi\cos\phi \text{Tr ln}(4-2\Tilde{T}+\Tilde{T}\{G_{\chi},\mathcal{C}(\phi,\Tilde{\phi})\})\;.
\end{align}
This is Eq. (\ref{eq:WithTwoIntegrals}) in the main text.
\subsection{Analytical differential Fano factors}\label{sec:AnalyticalDiffFano}
In this section we show that for the double barrier junction with a helical \textit{p} - wave superconductor, the zero voltage differential Fano factor can be obtained analytically. Moreover, we also calculate the cases $p = 1,2$ using the single integral formalism, Eq. (\ref{eq:WithOneIntegral}). This does not correspond to the actual diffusive and chaotic cavity junctions, but we find nevertheless that the values obtained are close the values in Table \ref{tab:FanoDiffusive} that were calculated numerically using Eq. (\ref{eq:WithTwoIntegrals}).

In the spinless normal state the current is given by $\frac{\partial I(\Tilde{T})}{\partial V} = \frac{e^{2}}{2\pi\hbar}\Tilde{T}$ and $\frac{\partial P(\Tilde{T})}{\partial V} = \frac{e^{3}}{\pi\hbar}\Tilde{T}(1-\Tilde{T})$.
Meanwhile in the spinless \textit{s} - wave superconductor $\frac{\partial I(\Tilde{T})}{\partial V} = \frac{e^{2}}{8\pi\hbar}\frac{8\Tilde{T}^{2}}{(\Tilde{T}-2)^{2}}$ and $\frac{\partial P(\Tilde{T})}{\partial V} = \frac{e^{3}}{4\pi\hbar}\frac{64\Tilde{T}^{2}(1-\Tilde{T})}{(\Tilde{T}-2)^{4}}$. Since we are interested in the differential Fano factor, which is the ratio of differential noise and conductance, the normalization constant of $\rho(\Tilde{T})$ is irrelevant, it drops out, and hence we may use the non-normalized $\rho(\Tilde{T}) =\Tilde{T}^{-\frac{p}{2}}(1-\Tilde{T})^{-\frac{1}{2}}$. We find below that the integrals $\int_{0}^{1}d\Tilde{T}\rho(\Tilde{T}) \frac{\partial I(\Tilde{T})}{\partial V}(\Tilde{T})$ and $\int_{0}^{1}d\Tilde{T}\rho(\Tilde{T})\frac{\partial P(\Tilde{T})}{\partial V}$ converge and can be calculated analytically.

For the helical \textit{p} - wave superconductor we have, in Nambu-spin space
\begin{align}
    C^{R}(E = 0,\phi) = \begin{bmatrix}
         i \cot\phi& 0& 0& -\csc\phi\\ 0& -i \cot\phi& \csc\phi&
   0\\0& \csc\phi& i\cot\phi& 0\\-\csc\phi& 0& 0& 
  -i \cot\phi
\end{bmatrix}\; ,
\end{align}
and $C^{A} = C^{R}$, $C^{K} = \mathbf{0}$.

We may focus on each of the different spin sectors separately. Since they give the same results, therefore, here we discuss only one of the sectors for clarity of presentation. Within this spin sector we have
\begin{align}
    C^{R}(E= 0,\phi) = \begin{bmatrix}
        -i\cot\phi&\csc\phi\\\csc\phi&i\cot\phi
    \end{bmatrix}.
\end{align}
For each angle, the contributions to the current and noise power can be found using the following quantities, denoting short $C$ for $C(E = 0,\phi)$:
\begin{align}
    &\{C,[\tau_{K},G_{N}]\} = \begin{bmatrix}4 i (f_{L} + f_{T}) \cot\phi& -4 f_{L} \csc\phi& 
  4 i \cot\phi& -4 \csc\phi\\-4 f_{L} \csc\phi& 
  4 i (-f_{L} + 
     f_{T}) \cot\phi& -4 \csc\phi& -4 i \cot\phi\\-4 i \cot\phi& 4 \csc\phi& -4 i (f_{L} + f_{T}) \cot\phi& 
  4 f_{L} \csc\phi\\4 \csc\phi& 4 i \cot\phi& 
  4 f_{L} \csc\phi& 4 i (f_{L} - f_{T}) \cot\phi\end{bmatrix}\;,\\
    &\{C,G_{N}\} = -2i\cot\phi \rho_{3}+\frac{1}{2}(\rho_{1}+i\rho_{2})\otimes\begin{bmatrix}
        -4i(f_{L}+f_{T})\cot\phi&4 f_{T} \csc\phi\\4f_{T}\csc\phi&-4i(f_{L}-f_{T})\cot\phi
    \end{bmatrix}\;,\\
    &\left((4-2\Tilde{T})\mathbf{1}+\Tilde{T}\{C,G_{N}\}\right)^{-1} = \frac{1}{(2-\Tilde{T})^{2}+\Tilde{T}^{2}\cot^{2}\phi}\Big((1-\frac{1}{2}\Tilde{T}\mathbf{1}+\frac{i}{2}\Tilde{T}\cot\phi \rho_{3})\nonumber\\&+\frac{\Tilde{T}}{2}(\rho_{1}+i\rho_{2})\otimes\begin{bmatrix}
        i(f_{L}+f_{T})\cot\phi&-f_{T}\csc\phi\\-f_{T}\csc\phi&i(f_{L}-f_{T})\cot\phi\end{bmatrix}\Big)\\
        &\tau_{K}G_{N}\tau_{K}C+C\tau_{K}G_{N}\tau_{K} = 2i\cot\phi \rho_{3}+\frac{1}{2}(\rho_{1}-i\rho_{2})\otimes\begin{bmatrix}
        -4i(f_{L}+f_{T})\cot\phi&4 f_{T} \csc\phi\\4f_{T}\csc\phi&-4i(f_{L}-f_{T})\cot\phi
    \end{bmatrix}\;,\\
        &\tau_{K}G_{N}\tau_{K}C+C\tau_{K}G_{N}\tau_{K}-CG_{N}-G_{N}C = 4i\cot\phi\rho_{3}+i\rho_{2}\otimes\begin{bmatrix}
            4i(f_{L}+f_{T})\cot\phi&-4f_{T}\csc\phi\\-4f_{T}\csc\phi&4i(f_{L}-f_{T})\cot\phi
    \end{bmatrix}\;.
\end{align}
We find that
\begin{align}
    2J_{K}(\phi) &= 
    ((4-2\Tilde{T})\mathbf{1}+\Tilde{T}\{C(E = 0,\phi),G_{N}\})^{-1}\Tilde{T}(C\tau_{K}G_{N}-CG_{N}\tau_{K}+\tau_{K}G_{N}C-G_{N}\tau_{K}C) \;,\nonumber\\&=\frac{1}{2}(\rho_{0}+\rho_{3})\otimes\begin{bmatrix}
        \frac{(2 \Tilde{T} ((f_{L} + f_{T}) \cot\phi (-i (-2 + \Tilde{T}) + T \cot\phi) - 
   2 f_{T} \Tilde{T} \csc^2\phi))}{(-2 + \Tilde{T})^2 + 
 \Tilde{T}^2 \cot^{2}\phi}& \frac{2 i f_{L} \Tilde{T}}{
\Tilde{T} \cos\phi + i (-2 + \Tilde{T}) \sin\phi}\\\frac{2 i f_{L} \Tilde{T}}{\Tilde{T} \cos\phi + 
 i (-2 + \Tilde{T}) \sin\phi}& \frac{-2 (f_{L} - f_{T}) \Tilde{T} \cot\phi (-i (-2 + \Tilde{T}) + \Tilde{T} \cot\phi) - 4 f_{T} \Tilde{T}^2 \csc^2\phi}{(-2 + \Tilde{T})^2 + 
 \Tilde{T}^2 \cot^2\phi}
    \end{bmatrix}\nonumber\\&+\frac{\Tilde{T}}{2-\Tilde{T}+i\Tilde{T}\cot\phi}\frac{1}{2}(\rho_{0}-\rho_{3})\otimes\begin{bmatrix}
        -2i(f_{L}+f_{T})\cot\phi&2f_{L}\csc\phi\\2f_{L}\csc\phi&2i(f_{L}-f_{T})\cot\phi
    \end{bmatrix}\nonumber\\&+\frac{\Tilde{T}}{2-\Tilde{T}+i\Tilde{T}\cot\phi}\frac{1}{2}(\rho_{1}-i\rho_{2})\otimes\begin{bmatrix}
        -2i\cot\phi&2\csc\phi\\2\csc\phi&2i\cot\phi
    \end{bmatrix}\nonumber\\&+\frac{1}{(2-\Tilde{T})^{2}+\Tilde{T}^{2}\cot^{2}\phi}\frac{1}{2}(\rho_{1}+i\rho_{2})\otimes\Bigg(2 \Tilde{T}\csc\phi \Big(-2 + \Tilde{T} + i (-1 + 2 f_{L}^2 + 2 f_{T}^2) \Tilde{T} \cot\phi \Big)\tau_{1}\nonumber\\&+\Big(4f_{L}f_{T}\Tilde{T}^{2}(\cot^{2}\phi-1)\Big)\tau_{0}+2 \Tilde{T} \cot\phi \Big(-i (-2 + \Tilde{T}) + (-1 + 2 f_{L}^2 + 2 f_{T}^2) \Tilde{T} \cot\phi\Big)\tau_{3}\Bigg)
    \\
    -2\mathcal{P}(\phi) &= ((4-2\Tilde{T})\mathbf{1}+\Tilde{T}\{C(E = 0,\phi),G_{N}\})^{-1}\Tilde{T}(\tau_{K}G_{N}\tau_{K}C+C\tau_{K}G_{N}\tau_{K}-CG_{N}-G_{N}C)\;,\nonumber\\&=\frac{1}{(2-\Tilde{T})^{2}+\Tilde{T}^{2}\cot^{2}\phi}\frac{1}{2}(\rho_{0}+\rho_{3})\otimes\Big((2 \Tilde{T} (-i (-2 + \Tilde{T}) \cot\phi + (-1 + 2 f_{L}^{2} + f_{T}^2) \Tilde{T} \cot^2\phi - 
   2 f_{T}^2 \Tilde{T} \csc\phi)\tau_{0}\nonumber\\&+8f_{L}f_{T}\Tilde{T}^{2}\cot^{2}\phi\tau_{3}+8if_{L}f_{T}\Tilde{T}^{2}\cot\phi\csc\phi\tau_{1}\Big)+\frac{\Tilde{T}}{2-\Tilde{T}+i\Tilde{T}\cot\phi}\Bigg(\frac{1}{2}(\rho_{0}-\rho_{3})\otimes(-2i\cot\phi\tau_{0})\nonumber\\&+\frac{1}{2}(\rho_{1}-i\rho_{2})\otimes\begin{bmatrix}
        -2i(f_{L}+f_{T})\cot\phi&-2f_{T}\csc\phi\\-2f_{T}\csc\phi&2i(f_{L}-f_{T})\cot\phi
    \end{bmatrix}\nonumber\\&+\frac{1}{2}(\rho_{1}+i\rho_{2})\otimes\begin{bmatrix}
        2i(f_{L}+f_{T})\cot\phi&-2f_{T}\csc\phi\\-2f_{T}\csc\phi&-2i(f_{L}-f_{T})\cot\phi
    \end{bmatrix}\Bigg)
\end{align}
At zero temperature the current and the noise are given by
\begin{align}
    I &= -\frac{eN_{\text{ch}}}{16\pi\hbar}\int_{0}^{1}d\Tilde{T}\rho(\Tilde{T})\int_{-\frac{\pi}{2}}^{\frac{\pi}{2}}\cos\phi d\phi\Bigg(\int_{E<|eV|}dE\text{Tr}(J_{K}(\phi,f_{T} = 1,f_{L} = 0)\Bigg)\nonumber\\&-\frac{eN_{\text{ch}}}{16\pi\hbar}\int_{-\frac{\pi}{2}}^{\frac{\pi}{2}}\Bigg(\int_{E>|eV|}dE\text{Tr}(J_{K}(\phi,f_{T} = 0,f_{L} = 1)\Bigg)\\
    P_{N}& = \int_{0}^{1}d\Tilde{T}\rho(\Tilde{T})\frac{eN_{\text{ch}}}{16\pi\hbar}\int_{-\frac{\pi}{2}}^{\frac{\pi}{2}}\cos\phi d\phi\Bigg(\int_{E<|eV|}dE\text{Tr}\Big((\mathcal{P}(\phi,f_{T} = 1,f_{L} = 0)-J_{K}^{2}(\phi,f_{T} = 1,f_{L} = 0)\Big)\nonumber\\&-\int_{E>|eV|}dE\text{Tr}\Big(\mathcal{P}(\phi,f_{T} = 0,f_{L} = 1)-J_{K}^{2}(\phi,f_{T} = 0,f_{L} = 1)\Big)\Bigg)
\end{align}
Upon taking the voltage derivative we only need to evaluate these quantities at $E = eV$. 
Moreover, all of the above quantities evaluate to 0 for $f_{T} = 0, f_{L} = 1$, which leaves us with only the integral for $|E|<|eV|$. Thus, keeping in mind that $E = eV$ and $E = -eV$ contribute equally to the current and noise, we only need
\begin{align}
    j_{K}(\phi)  &= -\text{Tr}\Big(J_{K}(\phi,f_{T} = 1, f_{L} = 0)\Big)= \frac{4 \Tilde{T}^2}{2 (\Tilde{T}-1) \cos (2 \phi )+(\Tilde{T}-2) \Tilde{T}+2}\;,\\
    p_{N}(\phi) &= \text{Tr}\Big(\mathcal{P}(\phi,f_{T} = 1, f_{L} = 0)\Big)-\text{Tr}\Big(J_{K}^{2}(\phi,f_{T} = 1, f_{L} = 0)\Big) = \frac{4 \Tilde{T}^2}{2 (\Tilde{T}-1) \cos (2 \phi )+(\Tilde{T}-2) \Tilde{T}+2}\nonumber\\&-\frac{4 \Tilde{T}^2 (-2 + \Tilde{T} (2 + \Tilde{T}) - 2 (-1 + \Tilde{T}) \cos2\phi)}{(2 + (-2 + \Tilde{T}) \Tilde{T} + 2 (-1 + \Tilde{T}) \cos2\phi)^2}\;.
\end{align}
Using Mathematica we find
\begin{align}
    \frac{\partial}{\partial V}\frac{I}{\frac{eN_{\text{ch}}}{16\pi\hbar}} &=2\int_{0}^{1}\int_{-\frac{\pi}{2}}^{\frac{\pi}{2}} d\phi \cos{\phi} d\Tilde{T}\rho(\Tilde{T})j_{K}(\phi) =\int_{0}^{1}\int_{-\frac{\pi}{2}}^{\frac{\pi}{2}} d\phi \cos{\phi} d\Tilde{T}\rho(\Tilde{T})\frac{8 \Tilde{T}^2}{2 (\Tilde{T}-1) \cos (2 \phi )+(\Tilde{T}-2) \Tilde{T}+2} \nonumber\\
    &  =\int_{0}^{1} d\Tilde{T}8\frac{\text{acos}(\frac{\Tilde{T}}{2-\Tilde{T}})}{1-\Tilde{T}}\Tilde{T}^{1-\frac{p}{2}}\;,\\
    \frac{\partial}{\partial V}\frac{P_{N}}{\frac{e^{2}N_{\text{ch}}}{8\pi\hbar}} &=2\int_{0}^{1}\int_{-\frac{\pi}{2}}^{\frac{\pi}{2}} d\Tilde{T}d\phi \cos{\phi} d\Tilde{T}\rho(\Tilde{T})p_{N}(\phi) = \int_{0}^{1} \int_{-\frac{\pi}{2}}^{\frac{\pi}{2}} d\Tilde{T} d\phi \cos\phi d\Tilde{T}\rho(\Tilde{T})\Bigg(\frac{8 \Tilde{T}^2}{2 (\Tilde{T}-1) \cos (2 \phi )+(\Tilde{T}-2) \Tilde{T}+2}\nonumber\\&-\frac{8 \Tilde{T}^2 (-2 + \Tilde{T} (2 + \Tilde{T}) - 2 (-1 + \Tilde{T}) \cos2\phi)}{(2 + (-2 + \Tilde{T}) \Tilde{T} + 2 (-1 + \Tilde{T}) \cos2\phi)^2}\Bigg)\nonumber\\&=\int_{0}^{1} d\Tilde{T}\frac{16\Tilde{T}^{2-\frac{p}{2}}}{(2-\Tilde{T})^{2}\sqrt{1-\Tilde{T}}}+8\frac{\text{acos}(\frac{T}{2-\Tilde{T}})}{1-\Tilde{T}}\Tilde{T}^{1-\frac{p}{2}}\;.
\end{align}

For $p = 2$ these integrals can be evaluated using Mathematica to $32G$ and $32G-16$, where $G\approx 0.91$ is the Catalan constant, giving \begin{align}F = \frac{\frac{\partial}{\partial V}\frac{I}{\frac{eN_{\text{ch}}}{16\pi\hbar}} }{\frac{\partial}{\partial V}\frac{P_{N}}{\frac{e^{2}N_{\text{ch}}}{8\pi\hbar}}} = 1-\frac{1}{2G}\;.\end{align}
In general we may say that $F = 1-\frac{1}{2A}$ where
\begin{align}
    A = 4\frac{\int_{0}^{1} d\Tilde{ T}\frac{\text{acos}(\frac{\Tilde{T}}{\Tilde{T}-\Tilde{T}})}{1-\Tilde{T}}\Tilde{T}^{1-\frac{p}{2}}}{\int_{0}^{1} d\Tilde{T}\frac{-16\Tilde{T}^{2-\frac{p}{2}}}{(2-\Tilde{T})^{2}\sqrt{1-\Tilde{T}}}}\;.
\end{align}
By using this expressions we find the differential Fano factor for double barrier listed in Table \ref{tab:FanoDiffusive}. For the diffusive barrier and chaotic cavities the above procedure, as explained, is not exact and therefore we used numerical calculations to obtain the differential Fano factor in those types of junctions.
\clearpage
\section{Full counting statistics}\label{sec:FullCountingStatistics}
In this section we show that for one-dimensional systems, the voltage derivative of the action for FCS at $eV = 0$ can be obtained analytically for all transparencies. We show that for the 1D \textit{p} - wave superconductor the voltage derivative of the action at zero temperature and zero voltage is given by Eq. (\ref{eq:dSdVP}) in the main text, that is the action is independent of transparency and all even order current correlation functions vanish. For the \textit{s} - wave superconductor the transparency does appear and generally all current correlation functions are nonzero. 

The voltage derivative of the FCS for voltages below the gap can be calculated as follows. 
If we write the energy integral outside the trace, the action is given by
\begin{align}
    S(\chi) = -\frac{t_{0}}{8\pi\hbar}\int_{-\infty}^{\infty} dE \text{Tr ln}((4-2\Tilde{T})\mathbf{1}+\Tilde{T}\{G_{\chi},C\})\label{eq:Sdef}\;,
\end{align}
where $G_{\chi} = e^{-\frac{i\chi}{2}\tau_{K}}G_{N}e^{\frac{i\chi}{2}\tau_{K}}$, $G_{N}$ is the quasiclassical Green's function of a normal metal and $C$ is the quasiclassical Green's function of the superconductor. Because we are interested in the voltage derivative of the current correlation functions near zero voltage, we evaluate this expression only for $|E|<\Delta_{0}$. In that case the retarded surface Green's function $C^{R}$ satisfies $C^{R} = -\tau_{3}\left(C^{R}\right)^{\dagger}\tau_{3}$, that is, the retarded and advanced components are related by $C^{R} = C^{A}$ and hence $C^{K}  = \tanh{\frac{E}{2k_{B}T}}(C^{R}-C^{A}) = 0$. For the 1D \textit{s} - wave superconductor this quantity is given by
\begin{align}
    C^{R} = \frac{1}{i\sqrt{\Delta_{0}^{2}-E^{2}}}\begin{bmatrix}
        E&\Delta_{0}\\-\Delta_{0}&-E
    \end{bmatrix}\; ,
    \label{eq:CdefappGS}
\end{align}
while for the 1D \textit{p} - wave superconductor it is given by
\begin{align}
    C^{R} = \frac{1}{E}\begin{bmatrix}
        i\sqrt{\Delta_{0}^{2}-E^{2}}&-\Delta_{0}\\-\Delta_{0}&-i\sqrt{\Delta_{0}^{2}-E^{2}}\label{eq:CdefappGP}
    \end{bmatrix}\;.
\end{align}
The appearance of $E^{-1}$ in front indicates the presence of a SABS at $E = 0$ \cite{tanaka2022theory}. Moreover, $C^{R}(E) = -C^{R}(-E) = -C^{A}(-E)$, which means the pairing is odd-frequency.
We denote the action for the normal metal / \textit{s} - wave superconductor junction by $S_{s}$ and the action for the normal metal / \textit{p} - wave superconductor junction by $S_{p}$.

The Green's function as a function of the counting field is given by
\begin{align}
G_{\chi} &= \rho_{3}\otimes\begin{bmatrix}
        \cos\chi+i(f_{L}+f_{T})\sin\chi&0\\0&-\cos\chi+i(f_{L}-f_{T})\sin\chi
    \end{bmatrix}\nonumber\\&+\rho_{1}\otimes\begin{bmatrix}
        f_{L}+f_{T}&0\\0&f_{T}-f_{L}
    \end{bmatrix}\nonumber\\&+i\rho_{2}\otimes\begin{bmatrix}
        (f_{L}+f_{T})\cos\chi+i\sin\chi&0\\0&(f_{T}-f_{L})\cos\chi+i\sin\chi
    \end{bmatrix}\;.
\end{align}

For the \textit{s} - wave superconductor we have
\begin{align}
    \{G_{\chi},C\}&=\frac{-2}{\sqrt{\Delta_{0}^{2}-E^{2}}}\Bigg(\rho_{3}\otimes\begin{bmatrix}
        iE\cos\chi-E(f_{L}+f_{T})\sin\chi&-f_{L}\Delta_{0}\sin\chi\\f_{L}\Delta_{0}\sin\chi&iE\cos\chi+E(f_{L}-f_{T})\sin\chi
    \end{bmatrix}\nonumber\\&+\rho_{1}\otimes\begin{bmatrix}
        iE(f_{L}+f_{T})&i\Delta_{0} f_{T}\\-i\Delta_{0} f_{T}&iE(f_{L}-f_{T})
    \end{bmatrix}\nonumber\\&+i\rho_{2}\otimes\begin{bmatrix}
        -E\sin\chi+iE(f_{L}+f_{T})\cos\chi&-\Delta_{0}\sin{\chi}+i\Delta_{0} f_{T}\cos\chi\\\Delta_{0}\sin\chi-i\Delta_{0} f_{T}\cos\chi&E\sin\chi+iE(f_{L}-f_{T})\cos\chi
    \end{bmatrix}\Bigg)\;.
\end{align}

For \textit{p} - wave superconductors we have
\begin{align}
    \{G_{\chi},C\}&=\frac{2}{E}\Bigg(\rho_{3}\otimes\begin{bmatrix}
        i\sqrt{\Delta_{0}^{2}-E^{2}}\cos\chi-\sqrt{\Delta_{0}^{2}-E^{2}}(f_{L}+f_{T})\sin\chi&-if_{L}\Delta_{0}\sin\chi\\-if_{L}\Delta_{0}\sin\chi&i\sqrt{\Delta_{0}^{2}-E^{2}}\cos\chi+\sqrt{\Delta_{0}^{2}-E^{2}}(f_{L}-f_{T})\sin\chi
    \end{bmatrix}\nonumber\\&+\rho_{1}\otimes\begin{bmatrix}
        i\sqrt{\Delta_{0}^{2}-E^{2}}(f_{L}+f_{T})&-\Delta_{0} f_{T}\\-\Delta_{0}f_{T}&i\sqrt{\Delta_{0}^{2}-E^{2}}(f_{L}-f_{T})
    \end{bmatrix}\nonumber\\&+i\rho_{2}\otimes\begin{bmatrix}
        -\sqrt{\Delta_{0}^{2}-E^{2}}\sin\chi+i\sqrt{\Delta_{0}^{2}-E^{2}}(f_{L}+f_{T})\cos\chi&-i\Delta_{0}\sin{\chi}-\Delta_{0} f_{T}\cos\chi\\-i\Delta_{0}\sin\chi-\Delta_{0} f_{T}\cos\chi&\sqrt{\Delta_{0}^{2}-E^{2}}\sin\chi+i\sqrt{\Delta_{0}^{2}-E^{2}}(f_{L}-f_{T})\cos\chi
    \end{bmatrix}\Bigg)\;.
\end{align}
Recognizing that the trace of the logarithm of any matrix is equal to the logarithm of the determinant of this matrix, we calculate the action using Mathematica after substitution of Eqs. (\ref{eq:CdefappGS}) and (\ref{eq:CdefappGP}) into Eq. (\ref{eq:Sdef}). 

Thus, we may write,  using that $\text{Tr ln}(4-2\Tilde{T}\mathbf{1}+\Tilde{T}\{G,C\})= \text{ln det}(4-2\Tilde{T}\mathbf{1}+\Tilde{T}\{G,C\})$ and using Mathematica for the evaluation of the resulting determinant
\begin{align}
    S_{s}(\chi)&=-\frac{t_{0}}{8\pi\hbar}\int_{|E|>\Delta_{0}} dE \text{Tr ln}((4-2\Tilde{T})\mathbf{1}+\Tilde{T}\{G_{\chi},C\})\nonumber\\& -\frac{t_{0}}{8\pi\hbar} \int_{|E|<\Delta_{0}}dE 2\text{ln}\Bigg(\frac{2 \left(8 E^2 (-1 + \Tilde{T}) + (8 + 
       \Tilde{T} (-8 + (1 + f_{L}^2 - f_{T}^2) \Tilde{T})) \Delta_{0}^{2}\right)}{E^2 - \Delta_{0}^{2}}\nonumber\\& +\frac{ 
    2\Tilde{T}^2 \Delta_{0}^{2}\left( (1 - f_{L}^2 + f_{T}^2) \cos\big(2 \chi\big) + 
       2 i f_{T} \sin\Big(2 \chi\big)\right)}{E^2 - \Delta_{0}^{2}}\Bigg)\;,\\
       S_{p}(\chi) &= -\frac{t_{0}}{8\pi\hbar}\int_{|E|>\Delta_{0}} dE \text{Tr ln}((4-2\Tilde{T})\mathbf{1}+\Tilde{T}\{G_{\chi},C\})\nonumber\\& -\frac{t_{0}}{8\pi\hbar} \int_{|E|<\Delta_{0}}dE 2\text{ln}\Bigg(\frac{2 \left(-8 E^2 (-1 + \Tilde{T}) + (1 + f_{L}^2 - f_{T}^2) \Tilde{T}^2 \Delta_{0}^{2}\right) +2 
    \Tilde{T}^2 \Delta_{0}^{2} \left((1 - f_{L}^2 + f_{T}^2) \cos\big(2 \chi\big) + 
       2 i f_{T} \sin\big(2 \chi\big)\right)}{E^2}\Bigg)\;.
\end{align}
In both cases, the first integral, which involves only energies larger than the gap, does not depend on voltage for $k_{B}T\ll |eV|\ll \Delta_{0}$, because $f_{L} \approx 1-O(e^{-\frac{k_{B}T}{\Delta_{0}}})$ and $f_{T} \approx 0$ within this integral. The second integral, which involves the below gap energies however, does depend on voltage. At zero temperature we have $f_{L} = 1$ and $f_{T} = 0$ for $|E|>|eV|$, while $f_{L} = 0$ and $f_{T} = 1$ for $|E|<|eV|$. Thus, 

\begin{align}
    S_{s}(\chi) &= -\frac{t_{0}}{8\pi\hbar}\int_{|E|>\Delta_{0}} dE \text{Tr ln}((4-2\Tilde{T})\mathbf{1}+\Tilde{T}\{G_{\chi},C\})\nonumber\\& -\frac{t_{0}}{8\pi\hbar} \int_{|E|<|eV|}dE 2\text{ln}\Bigg(\frac{2 \left(8 E^2 (-1 + \Tilde{T}) + (8 -8\Tilde{T}) \Delta_{0}^{2}\right)+4\Tilde{T}^2 \Delta_{0}^{2} e^{2i\chi}}{E^2 - \Delta_{0}^{2}}\Bigg)\nonumber\\
   &-\frac{t_{0}}{8\pi\hbar} \int_{|eV|<|E|<\Delta_{0}}dE2\text{ln}\Bigg(\frac{2 \left(8 E^2 (-1 + \Tilde{T}) + (8 + 
       \Tilde{T} (-8 + 2 \Tilde{T})) \Delta_{0}^{2}\right)}{E^2 - \Delta_{0}^{2}}\Bigg)\;,\\
    S_{p}(\chi) &= -\frac{t_{0}}{8\pi\hbar}\int_{|E|>\Delta_{0}} dE\text{Tr ln}((4-2\Tilde{T})\mathbf{1}+\Tilde{T}\{G_{\chi},C\})\nonumber\\& -\frac{t_{0}}{8\pi\hbar} \int_{|E|<|eV|}dE 2\text{ln}\Bigg(\frac{2 \left(-8 E^2 (-1 + \Tilde{T}) \right) + 4
    \Tilde{T}^2 \Delta_{0}^{2} e^{2i\chi}}{E^2}\Bigg) \nonumber\\& -\frac{t_{0}}{8\pi\hbar} \int_{|eV|<|E|<\Delta_{0}}dE2\text{ln}\Bigg(\frac{2 \left(-8 E^2 (-1 + \Tilde{T})\right) + 4 \Tilde{T}^2 \Delta_{0}^{2}}{E^2}\Bigg)\;.
\end{align}

In this expression, the voltage appears only in the integral limits, and consequently, taking the derivative we obtain the following expression:

\begin{align}
    \frac{\partial S_{s}}{\partial V} &= -\frac{et_{0}}{2\pi\hbar}\text{ln}\frac{-8(eV)^{2}(1-\Tilde{T})+\Delta_{0}^{2}(8(1-\Tilde{T})+2\Tilde{T}^{2}e^{2i\chi})}{-8(eV)^{2}(1-\Tilde{T})+\Delta_{0}^{2}(8(1-\Tilde{T})+2\Tilde{T}^{2})}\;,\\
    \frac{\partial S_{p}}{\partial V}&=-\frac{et_{0}}{2\pi\hbar}\text{ln}\frac{8(eV)^{2}(1-\Tilde{T})+2\Tilde{T}^{2}\Delta_{0}^{2}e^{2i\chi}}{8(eV)^{2}(1-\Tilde{T})+2\Tilde{T}^{2}\Delta_{0}^{2}}\;.
\end{align}
If we next take the limit $V\xrightarrow{}0$ we obtain

\begin{align}
    \frac{\partial S_{s}}{\partial V}&=-\frac{et_{0}}{2\pi\hbar}\text{ln}\frac{(8(1-\Tilde{T})+2\Tilde{T}^{2}e^{2i\chi})}{(8(1-\Tilde{T})+2\Tilde{T}^{2})}\label{eq:dsdVSsupp}\;,\\
    \frac{\partial S_{p}}{\partial V}&=-\frac{et_{0}}{2\pi\hbar}\text{ln}e^{2i\chi} = -\frac{iet_{0}}{\pi\hbar}\chi\label{eq:dSdVPsupp}\;.
\end{align}
which coincides with Eq. (\ref{eq:dSdVP}) in the main text.

From these equations physical observables can be calculated by taking derivatives with respect to $\chi$ and then putting $\chi = 0$. The zeroth order vanishes as required by the Keldysh technique \cite{belzig2001fullSN}, the lowest two orders are related to conductance and differential noise power via
\begin{align}
    \frac{\partial I_{s,p}}{\partial V} &= i\frac{e}{t_{0}}\frac{\partial}{\partial\chi}\frac{\partial S_{s,p}}{\partial V}(\chi = 0)\;,\\
    \frac{\partial P_{N(s,p)}}{\partial V} &=\frac{2e^{2}}{t_{0}} \frac{\partial^{2}}{\partial\chi^{2}}\frac{\partial S_{s,p}}{\partial V}(\chi = 0)\;.
\end{align}
Using these expressions we find, in agreement with the expressions in Appendix \ref{sec:1Danalytical} that for the \textit{s} - wave superconductor in the low-transparency limit we have $\frac{\partial S_{s}}{\partial V}\approx-\frac{et_{0}\Tilde{T}^{2}e^{2i\chi}}{8\pi\hbar}$ and consequently $\frac{\partial I}{\partial V}(\chi = 0) = \frac{\Tilde{T}^{2}}{4}\frac{e^{2}}{\pi\hbar}$ and $\frac{\partial P_{N}}{\partial V} = \Tilde{T}^{2}\frac{e^{3}}{\pi\hbar}$, while for the \textit{p} - wave superconductors we have, irrespective of the transparency, $\frac{\partial I}{\partial V} = \frac{e^{2}}{\pi\hbar}$ and $\frac{\partial P_{N}}{\partial V} = 0$.

Next to this, from Eqs. (\ref{eq:dsdVSsupp}) and (\ref{eq:dSdVPsupp}) we may immediately draw a few conclusions. 
First of all, for \textit{p} - wave superconductors, the expression is linear in $\chi$, which means that not only the noise power, but also all current correlation functions of order higher than 1 vanish, the current is constant in time, it always attains the maximum value. Secondly, because we consider the limit $eV\xrightarrow{}0$ and the surface Green's function $C$ of any \textit{p} - wave dominant \textit{s} +  \textit{p} - wave superconductor is equivalent to that of a \textit{p} - wave superconductor at $eV = 0$ \cite{tanaka2009theory}, these conclusions pertain to any \textit{p} - wave dominant \textit{s} + \textit{p} - wave superconductor. Lastly, if we take the limit $\Tilde{T}\xrightarrow{}1$ of the expression for the \textit{s} - wave superconductor, we obtain the same expression as obtained for the \textit{p} - wave superconductor, the action is linear in $\chi$. By substitution it was verified that this is the case for the full transparency limit of any superconductor. Thus, at zero temperature, higher order current correlation functions can not be used to distinguish SABS transport from fully transparent barriers.
\end{document}